\documentclass[onecolumn,prx,longbibliography,nofootinbib, superscriptaddress]{revtex4-2}

\usepackage[dvips]{graphicx} 
\usepackage{amsfonts}
\usepackage{amssymb}
\usepackage{amscd}
\usepackage{amsmath}    
\usepackage{amsthm}
\usepackage{appendix}
\usepackage{bbm}
\usepackage{dsfont}
\usepackage{enumerate}
\usepackage{enumitem}
\usepackage{epsfig}
\usepackage{float}
\usepackage[colorlinks, linkcolor=red, anchorcolor=blue, citecolor=green]{hyperref}
\usepackage{makecell}
\usepackage{subfigure}
\usepackage{subfloat}
\usepackage{xcolor}		
\usepackage{physics}
\usepackage[most]{tcolorbox}
\usepackage{tabularx}
\usepackage{booktabs}
\usepackage{mathrsfs}

\usepackage{MnSymbol}
\usepackage{tikz}
\usetikzlibrary{arrows.meta,positioning}

\tikzset{
  vertex/.style={circle,draw,line width=0.9pt,minimum size=7.5mm,inner sep=0pt},
  edge/.style={-{Stealth[length=2.5mm,width=1.8mm]},line width=0.9pt},
  weight/.style={font=\scriptsize,fill=white,inner sep=1.3pt,align=center},
  every loop/.style={min distance=10mm,looseness=7}
  faint/.style={opacity=0.2}
}
\usepackage{pgfplots}
\usetikzlibrary{arrows}
\usetikzlibrary{shapes,fadings,snakes}
\usetikzlibrary{decorations.pathmorphing,patterns}
\usetikzlibrary{calc}
\usetikzlibrary{positioning}
\definecolor{amber}{rgb}{1.0, 0.75, 0.0}

\graphicspath{{./figure/}}

\newtheorem{theorem}{Theorem}

\newtheorem{lemma}{Lemma}
\newtheorem{corollary}{Corollary}

\newtheorem{definition}{Definition}
\newtheorem{proposition}{Proposition}

\newtheorem{example}{Example}

\newenvironment{informaltheorem}[1]
{\par\medskip\noindent\textbf{Theorem~\ref{#1} (informal).}\ \itshape}
{\par\medskip}

\newenvironment{informalcorollary}[1]
{\par\medskip\noindent\textbf{Corollary~\ref{#1} (informal).}\ \itshape}
{\par\medskip}

\newenvironment{informallemma}[1]
{\par\medskip\noindent\textbf{Lemma~\ref{#1} (informal).}\ \itshape}
{\par\medskip}

\newcommand{\dket}[1]{\vert #1 \rangle\!\rangle}
\newcommand{\dbra}[1]{\langle\!\langle #1 \vert} 
\newcommand{\dbraket}[2]{\langle\!\langle #1 \vert #2 \rangle\!\rangle}
\newcommand{\dketbra}[2]{\vert #1 \rangle\!\rangle\langle\!\langle #2 \vert}

\newcommand{\pt}{\mathrm{pt}}

\newtcolorbox[auto counter]{mybox}[2][]{
	enhanced,
	breakable,
	colback=blue!5!white,
	colframe=blue!75!black,
	fonttitle=\bfseries,
	title=Box \thetcbcounter: #2,#1
}

\begin{document}
\title{SPAM-robust learnability of quantum dynamics}

\author{Zitai Xu}
\email{zt3927@umd.edu}
\affiliation{Joint Center for Quantum Information and Computer Science, University of Maryland, College Park, Maryland 20742, USA}

\author{Guoding Liu}
\email{lgd22@mails.tsinghua.edu.cn}
\affiliation{Center for Quantum Information, Institute for Interdisciplinary Information Sciences, Tsinghua University, Beijing, 100084 China}

\author{Senrui Chen}
\email{csenrui@gmail.com}
\affiliation{Institute for Quantum Information and Matter, Caltech, Pasadena, CA 91125, USA}

\begin{abstract}
    Accurate characterization of quantum dynamics is essential for the calibration and control of quantum devices. Unknown state preparation and measurement (SPAM) errors, however, introduce gauge ambiguities that limit what can be learned. A central question is therefore which properties are learnable despite SPAM errors, and whether physical structure in the dynamics can restrict these ambiguities and expand the set of learnable properties. In this work, we address both questions within a unified framework for quantum channels and continuous-time evolution. First, for general quantum channels and Lindbladian dynamics, we use the cycle structure of a generalized pattern-transfer graph to fully characterize all learnable parameter combinations and show achievability by constructing explicit SPAM-robust learning protocols, establishing a baseline where continuous gauge freedom prevents many individual parameters from being identified. Second, we show that coherent structure drastically restricts this gauge ambiguity, reducing continuous gauge freedom to discrete or even trivial choices. This restriction manifests across both discrete- and continuous-time settings, including purely Hamiltonian evolution, noisy quantum channels close to a target unitary, and Lindbladian dynamics close to a target Hamiltonian. In particular, when the dynamics preserves or remains close to a coherent Hamiltonian form, all coherent coefficients and broad classes of dissipative error parameters become individually learnable. These results establish a systematic route to exploiting physical structure for SPAM-robust characterization, providing a foundation for the reliable calibration and control of quantum devices.
\end{abstract}

\maketitle

\tableofcontents

\section{Introduction}
Quantum technologies promise advances in computation~\cite{Shor1997Polynomial,Grover1996,Harrow2009}, secure communication~\cite{BennettBrassard2014}, sensing~\cite{Giovannetti_2006,Gottesman_2012}, and simulation~\cite{Bernien2017QuantumSimulator,Georgescu2014QuantumSimulation}. Realizing these applications relies on the execution of highly accurate quantum operations. Because of imperfect control and environmental coupling, noise benchmarking and characterization are essential for diagnosing errors, calibrating devices, and improving overall performance~\cite{BlumeKohout2017Demonstration,Bartling2025Universal,Fan2025,Hangleiter2024Robustly,Guo_2025}. 

However, unknown state preparation and measurement (SPAM) errors pose a fundamental obstacle to rigorous characterization. Changes in the description of the dynamics can be fully compensated by inverse changes in the preparation and measurement operations, leaving all accessible statistics mathematically unchanged~\cite{Nielsen2021gatesettomography}. This gauge freedom strictly limits what can be learned about a quantum system, even given unlimited experimental data. The central questions in benchmarking are therefore: which properties remain intrinsically learnable, and what practical and physical structures can restrict this gauge freedom to make more information accessible?

These questions arise across two complementary paradigms: digital quantum processors modeled by quantum channels, i.e., completely positive trace-preserving (CPTP) maps, and analog simulators described by continuous-time Hamiltonians or Lindbladians. In the digital setting, the learnability of Pauli noise on Clifford gates has been successfully characterized using the cycle space of a pattern transfer graph~\cite{Learnability}, with recent extensions to mid-circuit measurements~\cite{Zhang2025} and structured gate-set Pauli noise~\cite{Chen2026}. However, universal computation requires non-Clifford operations whose errors generally lack Pauli form. Similarly, in the continuous-time setting, while Hamiltonian-learning protocols can tolerate SPAM errors below a constant threshold~\cite{Huang_2023} and concurrent work has classified the individually learnable coefficients of local Lindbladians~\cite{flammia2026gaugelindbladian}, a complete picture is still missing. Specifically, combinations of parameters can remain learnable even when their constituent coefficients are gauge-ambiguous. A unified theory must therefore encompass arbitrary noisy operations and characterize these learnable combinations in both digital and continuous settings.

In this work, we provide a general characterization of SPAM-robust learnability for both arbitrary quantum channels and Lindbladian evolutions, assuming access to noiseless single-qubit Clifford control. We completely identify the learnable multiplicative combinations of nonzero Pauli transfer matrix (PTM) elements. We show that these combinations correspond to cycles of a generalized pattern transfer graph, along which all gauge factors exactly cancel. Furthermore, we construct explicit, SPAM-robust protocols based on Pauli character projection~\cite{Helsen2019} to extract these quantities, establishing both what can be learned and how to learn it.

While individual parameters in general quantum dynamics often retain continuous gauge ambiguities, we show that coherent structure restricts the allowed gauge transformations and makes additional parameters learnable. This constraint is highly relevant in practice, as quantum gates and many analog simulators are explicitly designed to implement unitary evolution. For Hamiltonian dynamics, we first explain how the reversibility of unitary evolution restricts the gauge ambiguity in the Hamiltonian coefficients to discrete sign flips. We then show that every coefficient magnitude, as well as every odd-weight coefficient, is learnable. Furthermore, within any connected group of overlapping Hamiltonian terms, the even-weight coefficients retain at most a single common sign ambiguity, making all their relative signs learnable. Finally, we construct SPAM-robust protocols that estimate these coefficient magnitudes with Heisenberg scaling while recovering all accessible sign information.

Finally, we demonstrate that coherent structure also enhances learnability when characterizing noisy dynamics near a known target operation. Because gauge-equivalent descriptions must remain close to the same target, the ambiguity in the leading-order errors is strictly constrained. For quantum channels near a known unitary, we provide a graph-theoretic characterization of first-order learnability, complementing existing analyses of small gate-set errors~\cite{demendoza2026gaugeinvarianttheorysmallmarkovian} and first-order learnable error rates of Pauli channels~\cite{Learnability}. For Lindbladian errors around an ideal Hamiltonian whose pattern transfer graph connects all nonzero supports, we prove that every coherent error coefficient and broad classes of dissipative coefficients become individually learnable to first order. Consequently, prior knowledge of the target dynamics restricts the admissible gauges and unlocks additional physical error parameters for device calibration.

The remainder of this paper is organized as follows. Section~\ref{sec:summary} summarizes our main results, and Secs.~\ref{sec:prior-work} and~\ref{sec:preliminary} review related work and introduce the preliminaries. Section~\ref{sec:gauge-learnability} formulates gauge freedom and learnability, followed by the analysis of general quantum channels in Sec.~\ref{sec:CPTP} and Hamiltonian and Lindbladian evolution in Secs.~\ref{sec:Hamiltonian} and~\ref{sec:Lindbladian}. Section~\ref{sec:first-order} develops the first-order theory, and Sec.~\ref{sec:case} presents case studies. We conclude in Sec.~\ref{sec:outlook} with a discussion of open questions and future directions.

\section{Summary of results}
\label{sec:summary}

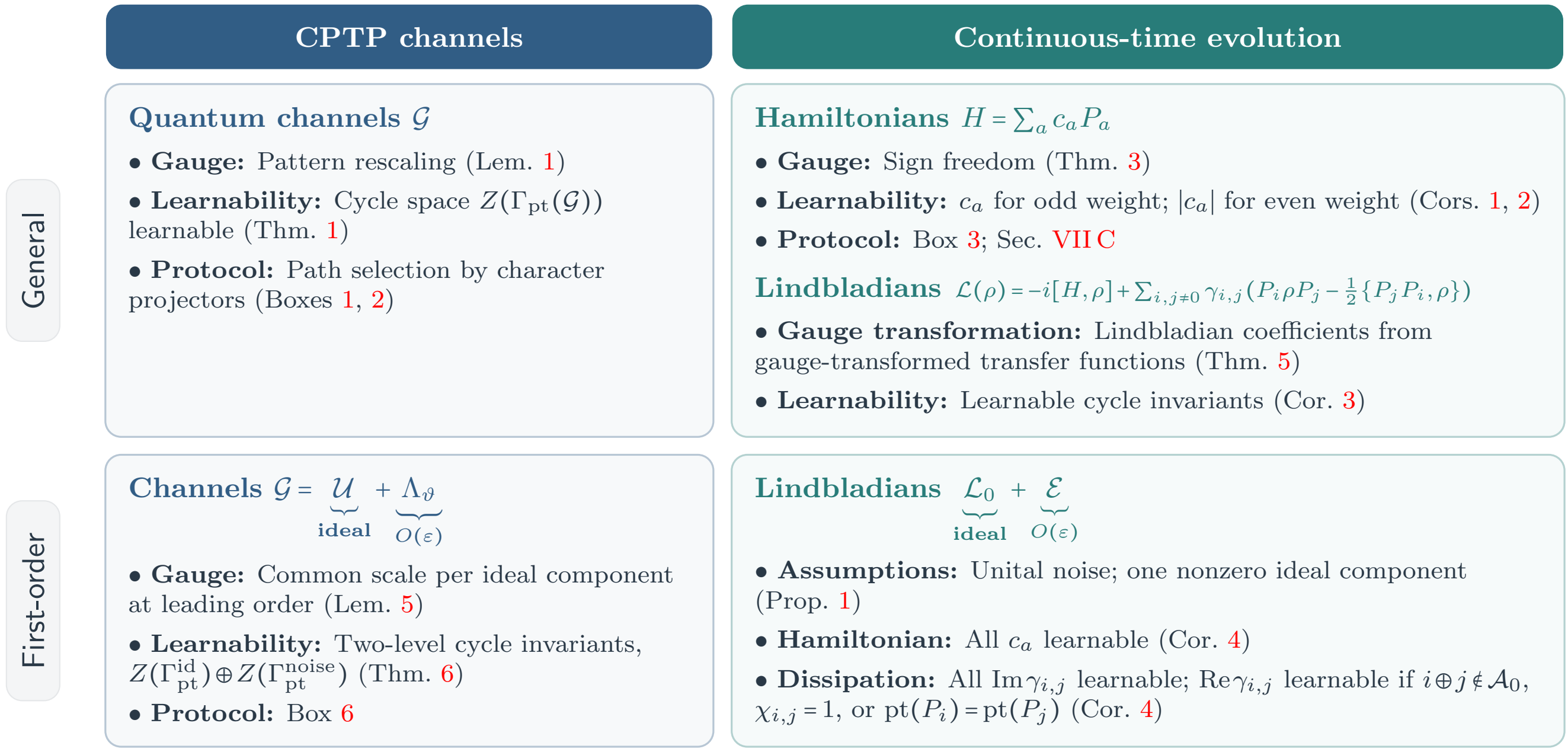
\begin{figure}[htbp!]
    \centering
\begingroup
\usetikzlibrary{backgrounds,fit}
\definecolor{overviewBlue}{HTML}{315D85}
\definecolor{overviewTeal}{HTML}{287C79}
\definecolor{overviewInk}{HTML}{293846}
\definecolor{overviewMuted}{HTML}{647480}
\thinmuskip=2mu
\medmuskip=2mu
\thickmuskip=3mu
\begin{tikzpicture}[
 x=\linewidth,y=1cm,font=\footnotesize,text=overviewInk,
 card/.style={rounded corners=5pt,line width=0.6pt,anchor=north west,
   text width=0.365\linewidth,align=flush left,inner xsep=7pt,inner ysep=7pt},
 blue card/.style={card,text width=0.35\linewidth},
 teal card/.style={card,text width=0.49\linewidth},
 frame/.style={rounded corners=5pt,line width=0.6pt,inner sep=0pt},
 banner/.style={rounded corners=5pt,font=\small\bfseries,text=white,
   text width=0.365\linewidth,align=center,inner xsep=7pt,
   minimum height=0.72cm,anchor=south west},
 row/.style={font=\normalsize\sffamily\mdseries,rotate=90,align=center,
   text=overviewInk,inner xsep=10pt,inner ysep=5pt}
]
 \node[banner,fill=overviewBlue,text width=0.35\linewidth] at (0.07,0.16) {CPTP channels};
 \node[banner,fill=overviewTeal,text width=0.49\linewidth] at (0.46,0.16) {Continuous-time evolution};

 \node[blue card] (general-cptp) at (0.07,0) {
  {\small\bfseries\color{overviewBlue}Quantum channels $\mathcal G$}\par\vspace{4pt}
  \textbullet\ \textbf{Gauge:} Pattern rescaling (Lem.~\ref{lemma:gauge_schur})\par\vspace{3pt}
  \textbullet\ \textbf{Learnability:} Cycle space $Z(\Gamma_{\pt}(\mathcal G))$ learnable (Thm.~\ref{thm:general_cptp_learnability})\par\vspace{3pt}
  \textbullet\ \textbf{Protocol:} Path selection by character projectors (Boxes~\ref{box:oriented_cycle},~\ref{box:unoriented_cycle})
 };
 \node[teal card] (general-continuous) at (0.46,0) {
  {\small\bfseries\color{overviewTeal}Hamiltonians $H=\sum_a c_aP_a$}\par\vspace{4pt}
  \textbullet\ \textbf{Gauge:} Sign freedom (Thm.~\ref{thm:Hamiltonian_gauge_space})\par\vspace{3pt}
  \textbullet\ \textbf{Learnability:} $c_a$ for odd weight; $|c_a|$ for even weight (Cors.~\ref{cor:odd-Hamiltonian},~\ref{cor:even-Hamiltonian})\par\vspace{3pt}
  \textbullet\ \textbf{Protocol:} Box~\ref{box:odd-Pauli}; Sec.~\ref{subsec:Hamiltonian-protocol}\par\vspace{6pt}
  \mbox{{\small\bfseries\color{overviewTeal}Lindbladians}\enspace{\color{overviewTeal}\scriptsize$\mathcal L(\rho)=-i[H,\rho]+\sum_{i,j\neq0}\gamma_{i,j}(P_i\rho P_j-\tfrac12\{P_jP_i,\rho\})$}}\par\vspace{4pt}
  \textbullet\ \textbf{Gauge transformation:} Lindbladian coefficients from gauge-transformed transfer functions (Thm.~\ref{thm:Lindbladian-reconstruction})\par\vspace{3pt}
  \textbullet\ \textbf{Learnability:} Learnable cycle invariants (Cor.~\ref{cor:Lindbladian-function-cycle})
 };
 \coordinate (general-bottom) at (current bounding box.south);
 \coordinate (general-left-bottom) at (general-cptp.south east |- general-bottom);
 \coordinate (general-right-bottom) at (general-continuous.south east |- general-bottom);
 \begin{scope}[on background layer]
  \node[frame,fill=overviewBlue!4,draw=overviewBlue!35,
    fit=(general-cptp.north west)(general-left-bottom)] (general-left-frame) {};
  \node[frame,fill=overviewTeal!4,draw=overviewTeal!35,
    fit=(general-continuous.north west)(general-right-bottom)] {};
 \end{scope}
 \coordinate (lower-row) at ([yshift=-0.20cm]general-bottom);
 \node[blue card] (first-order-cptp) at (0.07,0 |- lower-row) {
  {\small\bfseries\color{overviewBlue}Channels $\mathcal G=\underbrace{\mathcal U}_{\text{ideal}}+\underbrace{\Lambda_{\vartheta}}_{O(\varepsilon)}$}\par\vspace{4pt}
  \textbullet\ \textbf{Gauge:} Common scale per ideal component at leading order (Lem.~\ref{lemma:first-order-component-gauge})\par\vspace{3pt}
  \textbullet\ \textbf{Learnability:} Two-level cycle invariants, $Z(\Gamma_{\pt}^{\mathrm{id}})\oplus Z(\Gamma_{\pt}^{\mathrm{noise}})$ (Thm.~\ref{thm:first-order-learnable-decomposition})\par\vspace{3pt}
  \textbullet\ \textbf{Protocol:} Box~\ref{box:first_order_cycle}
 };
 \node[teal card] (first-order-continuous) at (0.46,0 |- lower-row) {
  {\small\bfseries\color{overviewTeal}Lindbladians $\underbrace{\mathcal L_0}_{\text{ideal}}+\underbrace{\mathcal E}_{O(\varepsilon)}$}\par\vspace{4pt}
  \textbullet\ \textbf{Assumptions:} Unital noise; one nonzero ideal component (Prop.~\ref{prop:first-order-connected-Lindbladian})\par\vspace{3pt}
  \textbullet\ \textbf{Hamiltonian:} All $c_a$ learnable (Cor.~\ref{cor:first-order-individual-Lindbladian-coefficients})\par\vspace{3pt}
  \textbullet\ \textbf{Dissipation:} All $\operatorname{Im}\gamma_{i,j}$ learnable; $\operatorname{Re}\gamma_{i,j}$ learnable if $i\oplus j\notin\mathcal A_0$, $\chi_{i,j}=1$, or $\pt(P_i)=\pt(P_j)$ (Cor.~\ref{cor:first-order-individual-Lindbladian-coefficients})
 };
 \coordinate (first-order-bottom) at (current bounding box.south);
 \coordinate (first-order-left-bottom) at (first-order-cptp.south east |- first-order-bottom);
 \coordinate (first-order-right-bottom) at (first-order-continuous.south east |- first-order-bottom);
 \begin{scope}[on background layer]
  \node[frame,fill=overviewBlue!4,draw=overviewBlue!35,
    fit=(first-order-cptp.north west)(first-order-left-bottom)] (first-order-left-frame) {};
  \node[frame,fill=overviewTeal!4,draw=overviewTeal!35,
    fit=(first-order-continuous.north west)(first-order-right-bottom)] {};
 \end{scope}
 \node[row,rounded corners=5pt,fill=overviewInk!5,draw=overviewInk!12,
   minimum height=0.48cm] at (0.025,0 |- general-left-frame.center) {General};
 \node[row,rounded corners=5pt,fill=overviewInk!5,draw=overviewInk!12,
   minimum height=0.48cm] at (0.025,0 |- first-order-left-frame.center)
   {First-order};
\end{tikzpicture}
\endgroup
    \caption{Overview of SPAM-robust learnability results for quantum channels and continuous-time dynamics. First-order results characterize learnability for weak perturbations of a known ideal unitary or Hamiltonian.}
    \label{fig:results-overview}
\end{figure}

We develop a unified framework for SPAM-robust learnability of general quantum channels and continuous-time dynamics generated by time-independent Hamiltonians and Lindbladians. Throughout, we assume noiseless single-qubit Clifford control and fixed SPAM errors dependent on the preparation and measurement settings. We characterize learnability by identifying the gauge freedom allowed by each physical model and determining the parameter functions invariant under these transformations. These gauge-invariant functions are precisely the SPAM-robust learnable quantities. Figure~\ref{fig:results-overview} summarizes the main results.

Gauge freedom arises because an experiment probes the evolution together with imperfect state preparation and measurement. An invertible map $T$, which we call a gauge, acts simultaneously on the state, measurement, and every operation in the circuit. Using the Pauli-Liouville representation, writing the circuit as a sequence of operations $\mathcal A_1,\ldots,\mathcal A_m$, this action leaves its outcome probability unchanged:
\begin{equation}\label{eq:summary-gauge-invariance}
    p=\dbra{M}\mathcal A_m\cdots\mathcal A_1\dket{\rho}
    =\bigl(\dbra{M}T^{-1}\bigr)
    \bigl(T\mathcal A_mT^{-1}\bigr)\cdots
    \bigl(T\mathcal A_1T^{-1}\bigr)
    \bigl(T\dket{\rho}\bigr).
\end{equation}
The factors $T^{-1}T$ cancel between successive operations, so a change in the description of the evolution can be compensated by corresponding changes in the unknown preparation and measurement. A gauge is valid for a given model only when its action preserves the physical and structural assumptions defining that model. Learnability therefore depends both on the invariants of a given gauge freedom and on the constraints that determine which gauges are admissible. Keeping the control and SPAM assumptions fixed allows us to isolate how changes in physical structure alter the accessible information. Section~\ref{sec:gauge-learnability} develops the notions of gauge freedom and learnability in detail.

\textbf{General quantum channel. }We begin with a general quantum channel to establish what remains learnable without imposing additional structure on the unknown dynamics. For a quantum channel $\mathcal{G}$, the principal restriction on the gauge freedom comes from the noiseless single-qubit Clifford controls. Any valid gauge must leave these trusted operations unchanged, and Lemma~\ref{lemma:gauge_schur} shows that it must therefore be diagonal in the Pauli basis,
\begin{equation}\label{eq:summary-diagonal-gauge}
    T=\sum_P \lambda_{\pt(P)}\dketbra{P}{P},
\end{equation}
where the pattern $\pt(P)\in\{0,1\}^n$ is the $n$-bit string indicating the support of $P$: its $q$-th bit is $1$ if $P$ acts nontrivially on qubit $q$ and is $0$ otherwise. Thus, all Pauli operators with the same support pattern are rescaled by the same nonzero real number. This is the common starting point for our analysis. Since noiseless single-qubit Clifford control is assumed throughout this work, the same restriction applies in every setting considered below.

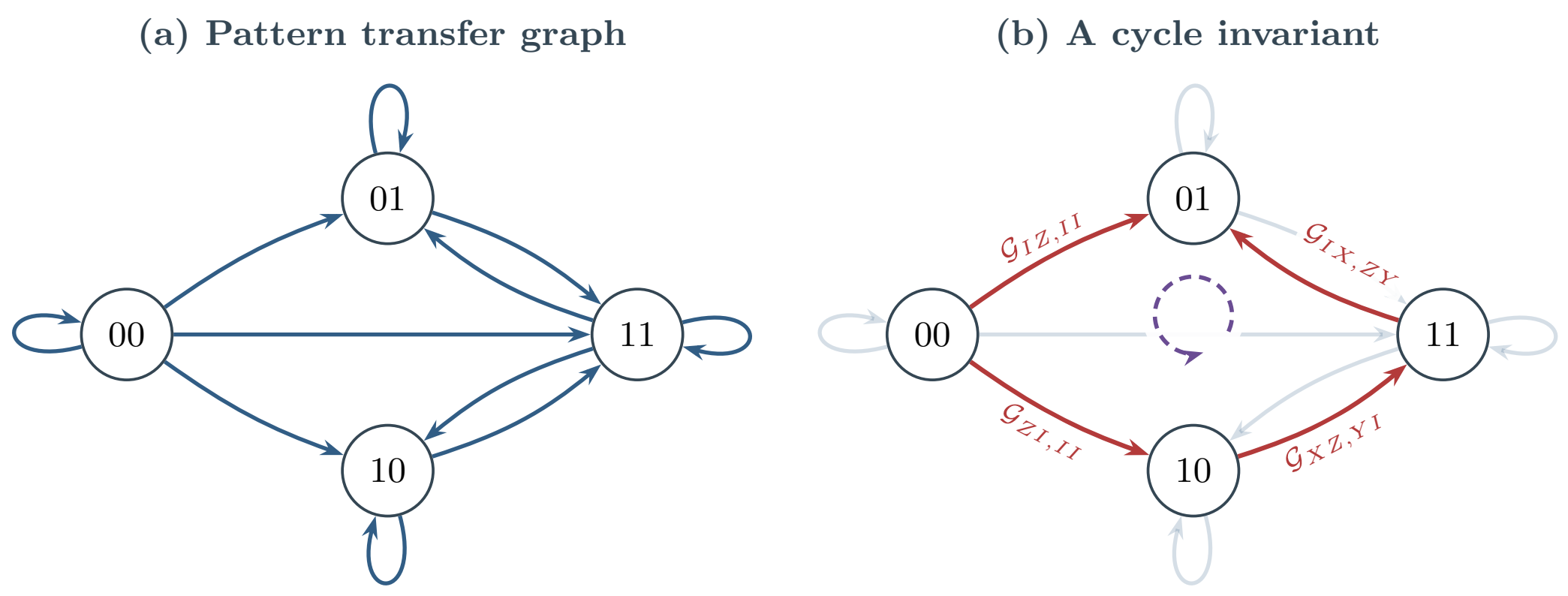
\begin{figure}[htbp]
    \centering
    \begingroup
\definecolor{transferGraph}{HTML}{315D85}
\definecolor{highlightGraph}{HTML}{B33A3A}
\definecolor{traversalGraph}{HTML}{6A4C93}
\definecolor{graphInk}{HTML}{344653}
\begin{tikzpicture}[
    x=1cm,
    y=1cm,
    pattern vertex/.style={
        circle,
        draw=graphInk,
        fill=white,
        line width=0.7pt,
        minimum size=8mm,
        inner sep=0pt,
        font=\small
    },
    transfer edge/.style={
        -{Stealth[length=2mm,width=1.4mm]},
        draw=transferGraph,
        line width=1pt
    },
    faded edge/.style={
        transfer edge,
        opacity=0.2
    },
    highlighted edge/.style={
        -{Stealth[length=2.1mm,width=1.5mm]},
        draw=highlightGraph,
        line width=1.15pt
    },
    edge label/.style={
        font=\scriptsize,
        text=highlightGraph,
        fill=white,
        fill opacity=0.92,
        text opacity=1,
        inner sep=1.5pt
    },
    panel title/.style={
        font=\small\bfseries,
        text=graphInk
    }
]
    \node[panel title] at (2.50,4.08) {(a) Pattern transfer graph};

    \node[pattern vertex] (p00) at (0.25,1.45) {$00$};
    \node[pattern vertex] (p01) at (2.55,2.65) {$01$};
    \node[pattern vertex] (p10) at (2.55,0.25) {$10$};
    \node[pattern vertex] (p11) at (4.75,1.45) {$11$};

    \draw[transfer edge] (p00) to[loop left,looseness=7,min distance=8mm] (p00);
    \draw[transfer edge] (p01) to[loop above,looseness=7,min distance=8mm] (p01);
    \draw[transfer edge] (p10) to[loop below,looseness=7,min distance=8mm] (p10);
    \draw[transfer edge] (p11) to[loop right,looseness=7,min distance=8mm] (p11);
    \draw[transfer edge] (p01) to[bend left=11] (p11);
    \draw[transfer edge] (p11) to[bend left=11] (p01);
    \draw[transfer edge] (p10) to[bend right=11] (p11);
    \draw[transfer edge] (p11) to[bend right=11] (p10);
    \draw[transfer edge] (p00) to[bend left=8] (p01);
    \draw[transfer edge] (p00) -- (p11);
    \draw[transfer edge] (p00) to[bend right=8] (p10);

    \begin{scope}[xshift=7.1cm]
        \node[panel title] at (2.50,4.08) {(b) A cycle invariant};

        \node[pattern vertex] (q00) at (0.25,1.45) {$00$};
        \node[pattern vertex] (q01) at (2.55,2.65) {$01$};
        \node[pattern vertex] (q10) at (2.55,0.25) {$10$};
        \node[pattern vertex] (q11) at (4.75,1.45) {$11$};

        \draw[faded edge] (q00) to[loop left,looseness=7,min distance=8mm] (q00);
        \draw[faded edge] (q01) to[loop above,looseness=7,min distance=8mm] (q01);
        \draw[faded edge] (q10) to[loop below,looseness=7,min distance=8mm] (q10);
        \draw[faded edge] (q11) to[loop right,looseness=7,min distance=8mm] (q11);
        \draw[faded edge] (q01) to[bend left=11] (q11);
        \draw[faded edge] (q11) to[bend left=11] (q01);
        \draw[faded edge] (q10) to[bend right=11] (q11);
        \draw[faded edge] (q11) to[bend right=11] (q10);
        \draw[faded edge] (q00) to[bend left=8] (q01);
        \draw[faded edge] (q00) -- (q11);
        \draw[faded edge] (q00) to[bend right=8] (q10);

        \draw[highlighted edge] (q00) to[bend left=8]
            node[edge label,above,sloped,pos=0.48] {$\mathcal G_{IZ,II}$} (q01);
        \draw[highlighted edge] (q11) to[bend left=11]
            node[edge label,above,sloped,pos=0.48,xshift=2mm,yshift=2mm]
            {$\mathcal G_{IX,ZY}$} (q01);
        \draw[highlighted edge] (q10) to[bend right=11]
            node[edge label,below,sloped,pos=0.48] {$\mathcal G_{XZ,YI}$} (q11);
        \draw[highlighted edge] (q00) to[bend right=8]
            node[edge label,below,sloped,pos=0.48] {$\mathcal G_{ZI,II}$} (q10);

        \begin{scope}[shift={(2.55,1.62)}]
            \fill[white,opacity=0.94] (0,0) circle[radius=4.5mm];
            \draw[-{Stealth[length=1.9mm,width=1.4mm]},draw=traversalGraph,
                densely dashed,line width=1.05pt]
                (-25:3.4mm) arc[start angle=-25,end angle=285,radius=3.4mm];
        \end{scope}
    \end{scope}
\end{tikzpicture}
\endgroup
    \caption{Pattern transfer graph of an illustrative noisy two-qubit $ZZ$ rotation. (a) displays one arrow for each pair of patterns connected by nonzero PTM elements, with parallel edges combined as one edge for  visual clarity. (b) highlights four particular elements and fixes the counterclockwise traversal shown at the center. Along this closed route, weights traversed with their arrows are multiplied and a weight traversed against its arrow is inverted, giving the invariant $\mathcal G_{ZI,II}\mathcal G_{XZ,YI}\mathcal G_{IX,ZY}/\mathcal G_{IZ,II}$.}
    \label{fig:summary-noisy-zz-pattern-graph}
\end{figure}

Writing the map in the Pauli basis, let $G_{ij}=2^{-n}\operatorname{Tr}[P_i\mathcal{G}(P_j)]$ denote its Pauli transfer matrix elements. To build intuition for the resulting invariants, we associate a pattern transfer graph with the channel, as illustrated in Fig.~\ref{fig:summary-noisy-zz-pattern-graph}. Its vertices are the Pauli patterns, and every nonzero Pauli transfer matrix element $\mathcal G_{ij}$ gives a directed edge from the input pattern $\pt(P_j)$ to the output pattern $\pt(P_i)$. Distinct matrix elements are retained as distinct edges. Under the gauge in Eq.~\eqref{eq:summary-diagonal-gauge}, the weight of this edge transforms as
\begin{equation}\label{eq:summary-edge-gauge-action}
    (T\mathcal{G}T^{-1})_{ij}=\frac{\lambda_{\pt(P_i)}}{\lambda_{\pt(P_j)}}\mathcal G_{ij}.
\end{equation}
To see how gauge-invariant quantities arise, choose a closed path through the graph and a direction in which to traverse it, as illustrated in Fig.~\ref{fig:summary-noisy-zz-pattern-graph}(b). Multiply the weight of an edge when the route follows its direction and the inverse weight when it runs against the direction. We call the product obtained in this way a \emph{cycle quantity}. Every gauge factor cancels in such a product, so every cycle quantity is learnable. It turns out that these cycle quantities characterize all learnable multiplicative functions, as summarized in the following theorem.

\begin{informaltheorem}{thm:general_cptp_learnability}
    For a general quantum channel, a multiplicative function of nonzero Pauli transfer matrix elements is learnable if and only if it can be expressed as a product of powers of cycle quantities. 
\end{informaltheorem}
Any multiplicative function outside this class is not gauge invariant and is therefore unlearnable in the presence of SPAM errors. Equivalently, after representing multiplicative combinations additively through logarithms, the formal statement of Theorem~\ref{thm:general_cptp_learnability} identifies the learnable combinations with the cycle space of the pattern transfer graph. This gives a complete characterization of learnable multiplicative combinations even when individual matrix elements remain ambiguous.

\textbf{Channel learning protocols. }To extract the information identified by the gauge analysis, we use Pauli character projection to select a prescribed path through the Pauli transfer matrix. For Pauli operators $Q$ and $R$, let $\chi(Q,R)$ be $+1$ when they commute and $-1$ when they anticommute, and let $\mathcal R$ be the unitary channel corresponding to the Pauli gate $R$. Character orthogonality gives
\begin{equation}\label{eq:summary-character-projection}
    \mathbb E_{R\sim\mathsf P_n}
    \bigl[\chi(Q,R)\mathcal R\bigr]
    =\dketbra{Q}{Q}.
\end{equation}
Thus, sampling Pauli gates and weighting the observed outcomes by their characters cancels every Pauli component except the selected one. Applying this projection successively isolates the product of the transfer-matrix elements along a chosen path $\gamma$ in the pattern transfer graph. Noiseless local Clifford gates connect consecutive Pauli operators whenever they have the same pattern. Denoting the weight of an edge by $g_e$, the resulting signal has the simple form
\begin{equation}\label{eq:summary-projected-path-signal}
    W_\gamma=A\prod_{e\in\gamma}g_e,
\end{equation}
where all the SPAM error factors are absorbed into $A$. For an oriented cycle, repeating the projected cycle to depth $m$ gives
\begin{equation}\label{eq:summary-oriented-cycle-signal}
    F(m)=A r^m,\qquad r=\prod_{e\in c}g_e.
\end{equation}
Since the SPAM errors enter only through the depth-independent amplitude $A$, fitting the rate $r$ yields the cycle product. The complete protocol is presented in Box~\ref{box:oriented_cycle}. For an unoriented cycle, we instead combine projected path signals as a ratio, using common endpoint settings so that their SPAM factors cancel. Box~\ref{box:unoriented_cycle} presents the corresponding construction. Thus, Pauli character projection provides an explicit SPAM-robust protocol for every learnable cycle quantity.

\textbf{Hamiltonian evolution. }Preserving coherent structure imposes consistency conditions absent from a general noisy channel. We now show how these conditions reduce the ambiguity in nonidentity Pauli coefficients to discrete sign choices. Consider a time-independent Hamiltonian $H=\sum_a c_aP_a$, where $a$ runs over the nonzero terms which specify the interaction structure. Let $\mathcal U_t^{(H)}$ denote the unitary channel obtained by evolving under $H$ for time $t$. We call $H$ and $H'=\sum_a c_a'P_a$ gauge equivalent if there exists a valid  gauge $T$ such that
\begin{equation}\label{eq:summary-Hamiltonian-gauge-equivalence}
    \mathcal U_t^{(H')}=T\mathcal U_t^{(H)}T^{-1},
\end{equation}
holds for all $t\geq 0$. A Hamiltonian coefficient is learnable only if it remains invariant in all gauge equivalent Hamiltonians. Recall from Eq.~\eqref{eq:summary-diagonal-gauge} that $T$ assigns one gauge factor $\lambda_u$ to each Pauli pattern $u$.
In the meantime, one single Hamiltonian coefficient induces several edges in the pattern transfer graph with same weights. Gauge transformation can rescale these weights, but they must be identical to preserve Hamiltonian form. This therefore imposes stronger gauge constraints. 
Lemma~\ref{lemma:Hamiltonian_gauge} expresses this observation explicitly. Suppose $T$ is a valid gauge preserving Hamiltonian form, then for every nonzero Hamiltonian term $c_aP_a$, there exists $\mu_a$ such that
\begin{equation}\label{eq:summary-Hamiltonian-gauge-action}
    \frac{\lambda_{\pt(P_aP_b)}}{\lambda_{\pt(P_b)}}=\mu_a,
\end{equation}
for every $P_b$ satisfying $\{P_{a},P_{b}\}=0$. Here $\mu_a$ may depend on the Hamiltonian term $P_a$, but not on the choice of $P_b$. Replacing $P_b$ by $P_aP_b$ reverses the ratio, giving $\mu_a=\mu_a^{-1}$ and hence $\mu_a\in\{+1,-1\}$. These $\mu_a$ factors completely characterize the ambiguity in the Hamiltonian coefficients induced by gauge transformation of $T$, as summarized below.

\begin{informaltheorem}{thm:Hamiltonian_gauge_space}
    Gauge-equivalent Hamiltonians $H$ and $H'$ can differ only in the signs of their Pauli coefficients:
    \begin{equation}
        c_a'=\mu_a c_a,
        \qquad \mu_a\in\{+1,-1\}.
    \end{equation}
    The allowed sign changes can be represented by assigning signs $\lambda_u\in\{+1,-1\}$ to the Pauli patterns, subject to Eq.~\eqref{eq:summary-Hamiltonian-gauge-action}. Thus, every coefficient has a learnable magnitude, while the remaining sign choices are constrained by the Hamiltonian terms.
\end{informaltheorem}
The remaining sign ambiguity depends on the parity of the Pauli weight. Every odd-weight term generates a nonzero transfer between two Pauli operators with the same pattern, forming a self-loop in the pattern transfer graph. Therefore all odd weight coefficients are guaranteed to be learnable.

\begin{informalcorollary}{cor:odd-Hamiltonian}
    Every odd-weight Hamiltonian coefficient is individually learnable: no valid gauge can change either its magnitude or its sign.
\end{informalcorollary}
For even-weight terms, the remaining sign choices are constrained by how the Hamiltonian terms overlap. Two terms are connected if they act on a common qubit or are linked by a chain of nonzero Hamiltonian terms with overlapping supports. Such a chain may include both odd- and even-weight terms. It turns out that if two even-weight terms are connected, they must share the same sign flip.

\begin{informalcorollary}{cor:even-Hamiltonian}
    Every even-weight coefficient has a learnable magnitude. Within each connected group of Hamiltonian terms, the even-weight coefficients can only change by a simultaneous sign flip,
    \begin{equation}\label{eq:summary-Hamiltonian-even-sign}
        c_a'=\sigma_{C(a)}c_a,
        \qquad \sigma_{C(a)}\in\{+1,-1\},
    \end{equation}
    where $C(a)$ denotes the connected group containing $P_a$. These sign flips may be chosen independently for different disconnected groups.
\end{informalcorollary}
Consequently, the relative signs, pairwise products, and ratios of even-weight coefficients within the same connected group are learnable. In particular, when all nonzero Hamiltonian terms are connected, the Hamiltonian is determined up to a single collective sign flip of its even-weight terms. Thus, coherent structure leaves only a discrete obstruction to identifying these coefficients, without strengthening the SPAM assumptions.

\textbf{Hamiltonian learning protocols. }To recover the information made learnable by coherent structure, we combine Pauli character projection with fast Pauli control. As in the channel protocol, character projection isolates a closed path whose repeated signal is insensitive to SPAM errors. Hamiltonian reshaping now allows us to select cycles associated with particular coefficients. For any target coefficient $c_a$, we first isolate its Pauli term by averaging $H$ over Pauli gates that commute with $P_a$,
\begin{equation}\label{eq:summary-Hamiltonian-reshaping}
    \mathbb E_{R\in\operatorname{Cent}(P_a)}
    \bigl[RHR^\dagger\bigr]=c_aP_a.
\end{equation}
This Hamiltonian reshaping can be implemented with fast Pauli control, and reduces the relevant evolution to that generated by $c_aP_a$. Choose any Pauli $P_b$ that anticommutes with $P_a$ and let it evolve under the reshaped Hamiltonian for time $\tau$, denote $Q=-iP_aP_b$, one obtains, 
\begin{equation}\label{eq:summary-Hamiltonian-Pauli-rotation}
    \mathcal{U}_{\tau}^{(c_aP_a)}\dket{P_b}
    =\cos(2c_a\tau)\dket{P_b}
    +\sin(2c_a\tau)\dket{Q}.
\end{equation}
Applying character projector in Eq.~\eqref{eq:summary-character-projection} to selects $P_b$ before and after the evolution forms a cycle with weight $\cos(2c_a\tau)$. Repeating this cycle to depth $m$ gives
\begin{equation}\label{eq:summary-Hamiltonian-cosine-signal}
    F_{\cos}(m)=A_{\cos}\cos^m(2c_a\tau).
\end{equation}
The unknown SPAM errors enter only through the depth-independent amplitude $A_{\cos}$, so fitting the signal across depths extracts $\cos(2c_a\tau)$. This construction applies to both odd- and even-weight terms and provides access to $|c_a|$ by varying the evolution time.

For odd-weight $P_a$, we can also extract the sine branch in Eq.~\eqref{eq:summary-Hamiltonian-Pauli-rotation} by choosing $P_b$ to have a different nonidentity Pauli on every qubit in the support of $P_a$ and identity elsewhere. Then $P_b$ anticommutes with $P_a$, and $Q$ has the same pattern as $P_b$. Selecting the $Q$ branch after the evolution and applying a noiseless local Clifford that maps $Q$ back to $P_b$ closes a cycle with weight $\sin(2c_a\tau)$. Repeating this cycle gives
\begin{equation}\label{eq:summary-Hamiltonian-cycle-signals}
    F_{\sin}(m)=A_{\sin}\sin^m(2c_a\tau),
\end{equation}
from which the sine is extracted in the same way. Together, the sine and cosine estimates at increasing evolution times allow robust phase estimation (Sec.~\ref{subsec:robust-phase-estimation}) to recover the odd-weight coefficient $c_a$ with Heisenberg scaling. Box~\ref{box:odd-Pauli} presents the full protocol.

To determine the relative signs of connected even-weight coefficients, the required cycles may involve several Hamiltonian terms. Their construction and the corresponding protocols are given in Sec.~\ref{subsec:Hamiltonian-protocol}, in particular Box~\ref{box:even-Pauli-cycle}. While these protocols recover the learnable coefficient information one term at a time, Sec.~\ref{subsec:Hamiltonian-structural-learning} further shows how the learning can be parallelized when the Hamiltonian has additional sparse and $k$-local structure.

\textbf{General Lindbladian evolution. }General Lindbladian dynamics provides a contrasting case: a valid gauge must preserve Lindblad form, but can mix the coherent and dissipative contributions. This allows us to distinguish the restrictions imposed by coherent structure from those imposed by continuous-time evolution alone. In the Pauli representation, a time-independent Lindbladian takes the standard form
\begin{equation}\label{eq:summary-Lindbladian-standard-form}
    \mathcal L(\rho)
    =-i\sum_{a\neq0}c_a[P_a,\rho]
    +\sum_{i,j\neq0}\gamma_{i,j}
    \left(P_i\rho P_j-\frac{1}{2}\{P_jP_i,\rho\}\right).
\end{equation}
Here the real coefficients $c_a$ describe coherent Hamiltonian evolution, while the positive semidefinite matrix $\gamma=(\gamma_{i,j})$, often called the Kossakowski matrix, describes dissipation.

The gauge acts directly on the Pauli-Liouville matrix entries of $\mathcal L$, whereas the quantities we want to learn are $c$ and $\gamma$. To connect these two descriptions, we introduce the Pauli transfer function $F_{a}(b)$ as,
\begin{equation}\label{eq:summary-Lindbladian-transfer-function}
    F_a(b)=\omega(a,b)^*\mathcal L_{a\oplus b,b},
\end{equation}
with $\omega(a,b)\in\{\pm1,\pm i\}$ defined by $P_aP_b=\omega(a,b)P_{a\oplus b}$. The matrix entry $\mathcal{L}_{i,j}$ describes the transfer from $P_j$ to $P_i$. Thus, $F_a$ groups transfers obtained by multiplying the same Pauli $P_a$. More specifically, $b$ labels the input Pauli operator $P_b$, while subscript $a$ fixes an multiplying Pauli $P_{a}$. $F_{a}(b)$ then describes the transfer from $P_b$ to $P_{a\oplus b}$, up to a known phase $\omega(a,b)^{*}$.  

This definition has a simple relation with the coefficients $c$ and $\gamma$. Define the Pauli Fourier transformation as
\begin{equation}\label{eq:summary-Lindbladian-Fourier-transform}
    \widehat F_a(k)=\frac{1}{4^n}\sum_b\chi(P_k,P_b)F_a(b).
\end{equation}
The Lindbladian coefficients are then recovered as follows.
\begin{informallemma}{lemma:Lindbladian-Fourier-reconstruction}
    Each dissipative coefficient is a Fourier coefficient up to a known phase, while each Hamiltonian coefficient is obtained from the difference of two Fourier coefficients:
    \begin{equation}\label{eq:summary-Lindbladian-Fourier-reconstruction}
        c_a=\frac{i}{2}\bigl(\widehat F_a(0)-\widehat F_a(a)\bigr),
        \qquad
        \gamma_{i,j}=\omega(i,j)^*\widehat F_{i\oplus j}(j).
    \end{equation}
\end{informallemma}

As in Eq.~\eqref{eq:summary-edge-gauge-action}, the gauge transforms an entry $\mathcal{L}_{i,j}$ to $(T\mathcal{L}T^{-1})_{i,j}=\mathcal{L}_{i,j}\lambda_{\pt(P_i)}/\lambda_{\pt(P_j)}$. Thus the $F_{a}(b)$ transforms as
\begin{equation}\label{eq:summary-Lindbladian-F-gauge}
    F'_a(b)
    =\frac{\lambda_{\pt(P_{a\oplus b})}}
    {\lambda_{\pt(P_b)}}F_a(b),
\end{equation}
where $F'_a$ is the Pauli transfer function of the gauge transformed Lindbladian $\mathcal{L}'=T\mathcal{L}T^{-1}$. 
Applying the Fourier relations to $F'$ then reconstructs the transformed Hamiltonian and dissipative coefficients explicitly.

\begin{informaltheorem}{thm:Lindbladian-reconstruction}
    The coefficients of the gauge-transformed Lindbladian $\mathcal L'=T\mathcal LT^{-1}$ are
    \begin{equation}\label{eq:summary-Lindbladian-transformed-coefficients}
    \begin{aligned}
        c'_a
        &=\frac{i}{2\cdot4^n}\sum_b
        \bigl(1-\chi(P_a,P_b)\bigr)
        \frac{\lambda_{\pt(P_{a\oplus b})}}{\lambda_{\pt(P_b)}}F_a(b),\\
        \gamma'_{i,j}
        &=\frac{\omega(i,j)^*}{4^n}\sum_b
        \chi(P_j,P_b)
        \frac{\lambda_{\pt(P_{i\oplus j\oplus b})}}{\lambda_{\pt(P_b)}}F_{i\oplus j}(b).
    \end{aligned}
    \end{equation}
    The transformed Hamiltonian coefficients $c_a'$ are real, and the transformed dissipative matrix $\gamma'=(\gamma'_{i,j})$ is Hermitian.
\end{informaltheorem}

The remaining requirement for the transformed generator to describe Lindbladian evolution is that $\gamma'$ needs to be positive semidefinite. When the original dissipative matrix $\gamma$ is strictly positive, sufficiently small changes in the gauge preserve this property. Thus, requiring Lindblad form does not, in general, restrict coefficient ambiguity to the discrete sign choices found for Hamiltonians.

As in the general channel case, products and ratios of $F_a(b)$ along cycles of the pattern transfer graph are learnable. However, recovering the Hamiltonian and dissipative coefficients $c$ and $\gamma$ requires a Fourier transform that mixes $F_a(b)$ over all inputs $b$, making a direct characterization of their learnability difficult.  Knowledge of a coherent target provides an additional constraint; we next show how it makes more of these coefficients learnable in the weak-error regime.

\textbf{First-order learnability. }Coherent structure can also restrict gauge ambiguity when it specifies a known target rather than the exact implemented dynamics. This is the setting of calibration, where the objective is to identify small coherent and dissipative errors around an intended evolution. Requiring gauge-equivalent descriptions to remain close to the same target imposes additional constraints, which can make error parameters learnable to first order.

We model the implemented channel as
\begin{equation}\label{eq:summary-first-order-model}
    \mathcal G=\mathcal U+\Lambda_{\vartheta}.
\end{equation}
Here $\mathcal U$ is the known ideal unitary channel, and $\vartheta=(\vartheta_1,\ldots,\vartheta_d)$ collects the unknown noise parameters, with each $\vartheta_k=O(\varepsilon)$. The perturbation has entries $(\Lambda_{\vartheta})_{i,j}=O(\varepsilon)$,  which is also first order small. 

Within this perturbative model, we call two parameter vectors $\vartheta$ and $\vartheta'$ gauge equivalent if there exists a valid gauge $T$ such that
\begin{equation}\label{eq:theta-gauge-equivalency}
    \mathcal U+\Lambda_{\vartheta'}=T(\mathcal U+\Lambda_{\vartheta})T^{-1}.
\end{equation}
A linear function $f_x(\vartheta)=\langle x,\vartheta\rangle$ is first-order learnable if its values on any two gauge-equivalent parameter vectors $\vartheta$ and $\vartheta'$ agree up to second-order corrections:
\begin{equation}\label{eq:summary-first-order-learnability}
    f_x(\vartheta')-f_x(\vartheta)=O(\varepsilon^2).
\end{equation}
Thus, the quantity is unchanged by the gauge to first order. For experiments with first order accuracy, gauge equivalent descriptions therefore yield a consistent value and are regarded as learnable.

The known ideal channel determines an ideal pattern transfer graph containing only the nonzero PTM entries of $\mathcal U$. The edge weights in this ideal graph should remain invariant in the constant order and can vary only in first order, forcing the gauge to also be identical in the constant order as summarized below.

\begin{informallemma}{lemma:first-order-component-gauge}
    Within every connected component $C$ of the ideal pattern transfer graph, all pattern gauge parameters agree at leading order:
    \begin{equation}\label{eq:summary-first-order-component-gauge}
        \lambda_v=\kappa_C+O(\varepsilon),
        \qquad v\in C,
    \end{equation}
    where $\kappa_C\neq0$ may differ between connected components.
\end{informallemma}

Indeed, an ideal edge has a nonzero zeroth-order weight. Preserving that weight forces the ratio of the gauge parameters at its two endpoints to be $1+O(\varepsilon)$, and the same relation propagates along every path within an ideal connected component.

\begin{figure}[htbp]
    \centering
    \begingroup
\definecolor{idealGraph}{HTML}{287C79}
\definecolor{noiseGraph}{HTML}{D97932}
\definecolor{graphInk}{HTML}{344653}
\definecolor{graphMuted}{HTML}{83919A}
\begin{tikzpicture}[
    x=1cm,
    y=1cm,
    pattern vertex/.style={
        circle,
        draw=graphInk,
        fill=white,
        line width=0.65pt,
        minimum size=7.2mm,
        inner sep=0pt,
        font=\small
    },
    component vertex/.style={
        circle,
        draw=graphInk,
        fill=white,
        line width=0.7pt,
        minimum size=13mm,
        inner sep=1.5pt,
        align=center,
        font=\small
    },
    ideal edge/.style={
        -{Stealth[length=2.1mm,width=1.5mm]},
        draw=idealGraph,
        line width=1.05pt
    },
    noise edge/.style={
        -{Stealth[length=2.1mm,width=1.5mm]},
        draw=noiseGraph,
        line width=1.05pt
    },
    component boundary/.style={
        draw=graphMuted,
        dashed,
        rounded corners=5pt,
        line width=0.55pt
    },
    panel title/.style={
        font=\small\bfseries,
        text=graphInk,
        align=center
    }
]
    \node[panel title] at (2.35,3.82) {(a) Ideal graph and noise-induced edges};

    \node[pattern vertex] (p00) at (0.15,1.35) {$00$};
    \node[pattern vertex] (p01) at (2.45,2.55) {$01$};
    \node[pattern vertex] (p10) at (2.45,0.15) {$10$};
    \node[pattern vertex] (p11) at (4.55,1.35) {$11$};

    \draw[component boundary] (-0.43,0.77) rectangle (0.73,1.93);
    \draw[component boundary]
        (1.70,-0.48) -- (3.05,-0.48) -- (5.20,0.79) --
        (5.20,1.91) -- (3.05,3.18) -- (1.70,3.18) -- cycle;

    \draw[ideal edge] (p00) to[loop left,looseness=7,min distance=8mm] (p00);
    \draw[ideal edge] (p01) to[loop above,looseness=7,min distance=8mm] (p01);
    \draw[ideal edge] (p10) to[loop below,looseness=7,min distance=8mm] (p10);
    \draw[ideal edge] (p11) to[loop right,looseness=7,min distance=8mm] (p11);
    \draw[ideal edge] (p01) to[bend left=11] (p11);
    \draw[ideal edge] (p11) to[bend left=11] (p01);
    \draw[ideal edge] (p10) to[bend right=11] (p11);
    \draw[ideal edge] (p11) to[bend right=11] (p10);

    \draw[noise edge] (p00) to[bend left=8] (p01);
    \draw[noise edge] (p00) to[bend right=8] (p10);
    \draw[noise edge] (p01) to[bend left=30] (p11);
    \draw[noise edge] (p10) to[bend right=30] (p11);

    \node[font=\scriptsize,text=graphMuted] at (6.15,1.72) {contract};
    \node[font=\Large,text=graphMuted] at (6.15,1.34) {$\Longrightarrow$};

    \node[panel title] at (9.25,3.82) {(b) Noise-induced component graph};

    \node[component vertex] (c0) at (7.65,1.35) {$\{00\}$};
    \node[component vertex,minimum size=18mm] (cnz) at (10.55,1.35)
        {$\{01,10,11\}$};

    \draw[noise edge] (c0) to[bend left=13] (cnz);
    \draw[noise edge] (c0) to[bend right=13] (cnz);
    \draw[noise edge] (cnz) to[loop right,looseness=7,min distance=11mm] (cnz);

    \draw[ideal edge] (3.55,-0.92) -- (4.20,-0.92);
    \node[anchor=west,font=\scriptsize,text=graphInk] at (4.30,-0.92) {ideal edge};
    \draw[noise edge] (6.35,-0.92) -- (7.00,-0.92);
    \node[anchor=west,font=\scriptsize,text=graphInk] at (7.10,-0.92) {noise-induced edge};
\end{tikzpicture}
\endgroup
    \caption{Ideal and noise-induced pattern transfer graph for a two-qubit $ZZ$ rotation with first-order coherent, Pauli, and amplitude-damping noise. (a) shows transfers between Pauli patterns: green arrows represent nonzero PTM entries of the ideal channel, while orange arrows represent entries that are zero in the ideal channel and become nonzero due to noise. The dashed regions mark the two connected components of the ideal graph, $\{00\}$ and $\{01,10,11\}$. (b) replaces each connected component with a single vertex and retains only the noise-induced edges. Transfers within a component become self-loops, while transfers from the identity pattern $00$ to nonzero patterns become edges between the two vertices.}
    \label{fig:summary-first-order-graphs}
\end{figure}
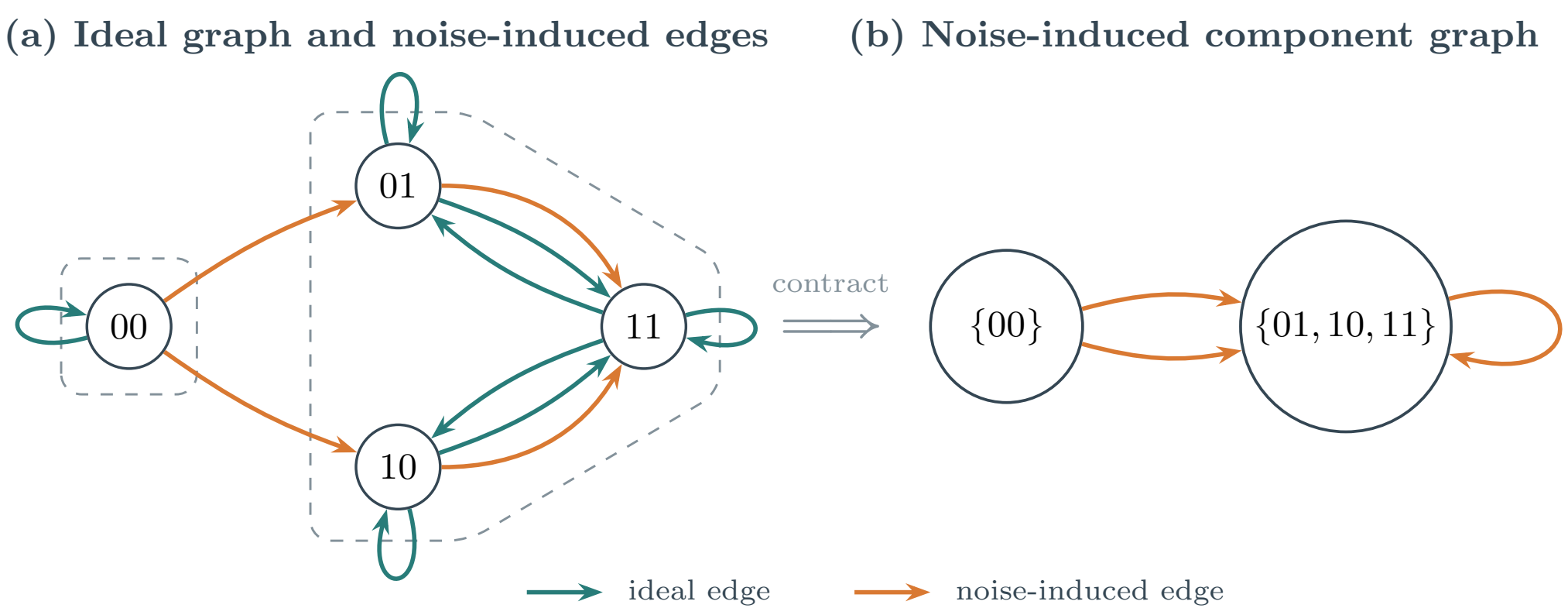

At first order, the noise can either change a PTM entry already present in the ideal channel or create a new $O(\varepsilon)$ small PTM entry where the ideal channel has zero weight. The first kind is represented on the ideal pattern transfer graph. To represent the second, we contract each ideal connected component to a single vertex and retain only the edges with $O(\varepsilon)$ weights, which are created by the noise. We call the result the noise-induced component transfer graph. Figure~\ref{fig:summary-first-order-graphs} illustrates this construction for a noisy two-qubit $ZZ$ rotation. 

We assign one parameter to each edge of the two graphs. Let $\mathcal G_e$ denote the unknown weight of an edge $e$. For an edge in the ideal graph with known weight $\mathcal U_e$, write $\mathcal G_e=\mathcal U_e(1+\theta_e^{\mathrm{id}})$. For a noise-induced edge with nonzero first-order weight, write $\mathcal G_e=\varepsilon\exp(\theta_e^{\mathrm{noise}}/\varepsilon)$. For simplicity, we consider positive noise-induced entries $\mathcal{G}_e$ in this summary. Negative entries are treated using signed logarithms in Sec.~\ref{sec:first-order}. By construction, parameters of $\theta^{\mathrm{id}},\theta^{\mathrm{noise}}$ are in $O(\varepsilon)$. A linear function of these parameters has the form
\begin{equation}
    f_x(\theta)
    =\langle x_{\mathrm{id}},\theta^{\mathrm{id}}\rangle
    +\langle x_{\mathrm{noise}},\theta^{\mathrm{noise}}\rangle,
\end{equation}
Each vector $x=(x_{\mathrm{id}},x_{\mathrm{noise}})$ corresponds to a unique linear function $f_x$. The first-order learnable functions can therefore be  characterized by the subspace of their corresponding vectors $x$. We use  $\mathfrak L^{(1)}$ to denote the learnable subspace such that for any $x\in\mathfrak L^{(1)}$, the function $f_{x}$ associated with $x$ is first order learnable.

The learnable subspace is determined by closed paths in the ideal and noise-induced graphs. Choose a closed path with a traverse direction,  allowing edges to be followed in either direction. This path defines a vector for edges: assign $+1$ to each edge followed along the traverse direction, $-1$ to each edge followed against that direction, and zero to all other edges not on the path. The span of these vectors is called the \emph{cycle space} of the graph. The next theorem characterizes the learnable subspace in terms of cycle spaces.

\begin{informaltheorem}{thm:first-order-learnable-decomposition}
    The vectors $x$ corresponding to first-order learnable linear functions form the space
    \begin{equation}\label{eq:summary-first-order-learnable-decomposition}
        \mathfrak L^{(1)}=Z_{\mathrm{id}}\oplus Z_{\mathrm{noise}}.
    \end{equation}
    Here $Z_{\mathrm{id}}$ and $Z_{\mathrm{noise}}$ are the cycle spaces of the ideal pattern transfer graph and the noise-induced component transfer graph, respectively.
\end{informaltheorem}

This cycle-space characterization shows how a known coherent target can make individual error parameters learnable even when the implemented dynamics is noisy. A noise-induced transfer between two different patterns in the same ideal connected component becomes a self-loop after contraction. Since a self-loop is itself a cycle, the corresponding parameter is learnable to first order.

\textbf{First-order Lindbladian characterization. }Knowledge of a coherent target also allows the generator-level characterization to be translated into statements about physical error coefficients. We apply the first-order framework to a known ideal Hamiltonian $H_0$. Let $\mathcal L_0(\rho)=-i[H_0,\rho]$ be its generator. We write the implemented generator as $\mathcal L=\mathcal L_0+\mathcal E$, where the error generator $\mathcal E=O(\varepsilon)$ contains both coherent and dissipative contributions:
\begin{equation}\label{eq:summary-first-order-Lindbladian-model}
    \mathcal E(\rho)
        =-i\sum_a\widetilde c_a[P_a,\rho]
        +\sum_{i,j\neq0}\gamma_{i,j}
        \left(P_i\rho P_j-\frac12\{P_jP_i,\rho\}\right).
\end{equation}
We assume that $\mathcal E$ is unital and that the ideal pattern transfer graph generated by $H_0$ is connected on all $2^n-1$ nonzero patterns. The connectedness condition is satisfiable for a broad class of experimentally relevant Hamiltonians, even for the simple two-qubit $ZZ$ interaction, while unitality excludes noise-induced transfers from the identity pattern, so all noise-induced transfers lie within the connected nonzero component and become self-loops after contraction. Under these assumptions, the general first-order framework yields a direct characterization of the individually learnable physical coefficients.

\begin{informalcorollary}{cor:first-order-individual-Lindbladian-coefficients}
    Every coherent error coefficient $\widetilde c_a$ is individually first-order learnable. A dissipative coefficient $\gamma_{i,j}$ is individually first-order learnable whenever
    \begin{equation}\label{eq:summary-first-order-individual-gamma}
        i\oplus j\notin\mathcal A_0,
        \qquad
        [P_i,P_j]=0,
        \qquad\text{or}\qquad
        \pt(P_i)=\pt(P_j),
    \end{equation}
    where $\mathcal A_0$ is the Pauli support of the ideal Hamiltonian $H_0$. If none of these conditions holds, $\operatorname{Im}\gamma_{i,j}$ remains individually learnable, while $\operatorname{Re}\gamma_{i,j}$ need not be. Thus, every possible first-order unlearnable direction is confined to this exceptional real off-diagonal sector of the dissipative matrix.
\end{informalcorollary}

A complete characterization of the learnable linear combinations and gauge variations within the exceptional real sector is given in Appendix~\ref{app:collective-Lindbladian-learnability}. Thus, even when the implemented dynamics includes dissipation, knowledge of a coherent target restricts the remaining ambiguity in its error coefficients.

\textbf{Case studies. }Section~\ref{sec:case} illustrates how ideal-gate structure and restrictions on noise mechanisms shape learnability in concrete models. We first consider a non-Clifford gate followed by Pauli noise as a simple setting for isolating the role of the ideal gate. This example shows how non-Clifford dynamics can make individual Pauli-noise parameters learnable through self-loops in the pattern transfer graph, and gives explicit non-Clifford examples for which the entire Pauli channel can be reconstructed.

We then study noisy two-qubit $ZZ$ rotations, an experimentally relevant family of entangling gates. For a channel model combining coherent overrotation, relaxation, and dephasing, we show that all error parameters can be determined despite unknown SPAM errors, up to a parametrization redundancy. A complementary example treats $ZZ$ evolution with general first-order unital Lindbladian noise. Here we characterize all learnable parameter combinations and identify the two remaining gauge freedom, demonstrating how knowledge of the ideal interaction makes a broad noise model accessible.

Finally, we study how the physical nature of the noise affects learnability. In a Lindbladian model, different noise processes are described by different jump operators. We consider three models: $Z$-type jumps describing local and correlated dephasing, weight-one jumps describing processes such as relaxation and collective decay, and their combination when both these processes exist. For each model, we discuss the learnability of Hamiltonian and dissipative parameters and their combinations in the presence of unknown SPAM errors. These examples show that restrictions on the allowed noise processes provide a further source of gauge constraints, complementing those supplied by coherent dynamics and known targets.

\section{Prior works}
\label{sec:prior-work}

Accurate characterization of quantum dynamics is essential for diagnosing and improving quantum devices. In practice, however, experimental data are affected by errors not only in the evolution of interest but also in state preparation and measurement (SPAM). Unlike statistical fluctuations, systematic SPAM errors do not vanish with repeated sampling and can substantially bias standard process tomography. This difficulty motivated early work on self-consistent quantum tomography~\cite{Merkel2013,Stark2014,BlumeKohout2013GST}. The seminal gate set tomography~\cite{Nielsen2021gatesettomography} placed gauge freedom at the center of quantum-gate characterization. A simultaneous change of representation of the states, measurements, and gates leaves every circuit outcome probability unchanged, so a gate set can be reconstructed only up to such gauge transformations. Subsequent work~\cite{Gu2021RandomizedLGST,Rudinger2022,Brieger2023} developed efficient protocols and extended the approach to other settings. These approaches typically address the gauge ambiguity through numerical gauge optimization, with less emphasis on explicitly characterizing which properties remain invariant under all admissible gauge transformations.

An explicit characterization of gauge-invariant information was obtained in Ref.~\cite{Learnability} for Clifford gates subject to Pauli noise, assuming perfect single-qubit control. This work introduced a graph-theoretic framework that identifies the learnable linear combinations of logarithmic Pauli fidelities with the cycle space of a pattern transfer graph. Subsequent works~\cite{Zhang2025,Chen2026} extended this framework to mid-circuit measurements and gate sets. However, the scope of these works is restricted to Pauli noise models and does not encompass the characterization of noisy non-Clifford gates or general quantum channels.

Characterizing and mitigating noise in non-Clifford gates is a problem that  attracted substantial interest, with methods developed for randomized benchmarking, noise tailoring and error mitigation~\cite{Carignan_Dugas_2015,Cross_2016,Helsen2019,Onarati2019,Hines2023,Dihedral,Ye2026,Layden2026}. These approaches primarily aim to estimate selected properties of the noise or reduce its effects on computation, without systematically characterizing and learning the full gauge-invariant information available in the presence of SPAM errors. Exceptions include e.g. \cite{endo2018practical} that incorporates gate set tomography with quantum error mitigation to rigorously deal with the gauge ambiguity. However, without a clear analytical understanding of the gauge parameters, such method usually requires huge number of experiments and not easily scalable. Our results can potentially make gauge-consistent mitigation of non-Clifford noise more practical. Note that, recent works have experimentally demonstrated scalable gauge-consistent error mitigation for Clifford gates~\cite{chen2026disambiguating}.

In this work, we extend the graph-theoretic framework of Ref.~\cite{Learnability} to characterize the gauge-invariant information of more general quantum channels under perfect single-qubit Clifford control. Our results recover the characterization for Clifford gates subject to Pauli noise as a special case and apply more broadly to noisy non-Clifford gates and general noise models. We also develop SPAM-robust protocols to learn these gauge-invariant quantities, using character projections inspired by character benchmarking~\cite{Helsen2019,Ye2026} to isolate products of channel matrix elements along paths in the pattern transfer graph. Together, these results establish both what information about a general noisy gate is learnable and how to extract it in the presence of SPAM errors.

Our framework also applies to continuous-time evolution, where the quantities of interest are the Hamiltonian and Lindbladian coefficients governing the dynamics. Our characterization of learnability and the accompanying protocols connect to the broader study of Hamiltonian and Lindbladian learning~\cite{Huang_2023,Yu2023,frança2022efficientrobustestimationmanyqubit,Hu2025,Ma_2026,brahmachari2026learninghamiltoniansheisenberglimit,liu2025robustlindbladianestimationquantum,Sinha2026,ivashkov2026ansatzfreelearninglindbladiandynamics,liu2026learningmanybodyhamiltoniansusing,gong2026out}. Recent works achieve SPAM-robust Hamiltonian learning using single-qubit control, assuming that the SPAM error is below a constant threshold~\cite{Huang_2023,Ma_2026,brahmachari2026learninghamiltoniansheisenberglimit,liu2026learningmanybodyhamiltoniansusing}. Our results show that, when this small-SPAM assumption is relaxed, gauge ambiguities can prevent the exact identification of some Hamiltonian coefficients. We characterize these ambiguities and construct protocols to extract the learnable Hamiltonian information even when the SPAM error exceeds the thresholds assumed in these works. Gauge freedom has also been considered in Lindbladian learning~\cite{Sinha2026}. That work focuses on learning protocols rather than a full characterization of the admissible gauges and their invariants. Also they require multi-qubit control for learning Lindbladian coefficients. More recently, a concurrent work Ref.~\cite{flammia2026gaugelindbladian} studies SPAM-robust learning of local Lindbladians, identifying gauge-invariant coefficients and developing efficient learning protocols. These results characterize learnability within specific physical models, but leave open how physical structure constrains gauge freedom across different classes of dynamics. Moreover, identifying which individual coefficients are learnable does not fully characterize the information accessible through their combinations. We address these gaps through a unified framework that characterizes gauge-invariant parameter combinations across broad classes of quantum dynamics and reveals how coherent structure restricts gauge ambiguity, both in Hamiltonian dynamics and in weak perturbations around known coherent targets.

Further restrictions on gauge freedom arise when the noise is a small perturbation of a known ideal evolution. Our study of first-order learnability is inspired by Ref.~\cite{demendoza2026gaugeinvarianttheorysmallmarkovian}, which constructs linear spaces of first-order gauge-invariant properties for small Markovian errors in gate sets, and by Ref.~\cite{Learnability} which studies first-order learnable error rates for Pauli noise channels. In our framework, perfect single-qubit Clifford control further restricts the admissible gauges. Together with the weak-noise assumption, this allows us to characterize first-order learnability for noisy non-Clifford gates and Lindbladian evolution.

\section{Preliminaries}
\label{sec:preliminary}

\subsection{Pauli--Liouville representation}\label{subsec:Pauli-Liouville-preliminaries}

We begin by fixing the operator and superoperator notation used throughout this work. Let $D=2^n$, and let $\mathsf P_n=\{P_a:a\in\mathbb F_2^{2n}\}$ denote the $n$-qubit Pauli group after quotienting out its global phases. For $a=(x,z)\in\mathbb F_2^{2n}$, we choose the Hermitian representative
\begin{equation}\label{eq:prelim-Hermitian-Pauli-representative}
    P_a=i^{x\cdot z}X^xZ^z .
\end{equation}
Thus, every $P_a$ is a tensor product of $I,X,Y,$ and $Z$, and $P_0=I$. Binary addition labels multiplication up to a known phase:
\begin{equation}\label{eq:prelim-Pauli-product-phase}
    P_aP_b=\omega(a,b)P_{a\oplus b},
    \qquad
    \omega(a,b)\in\{\pm1,\pm i\}.
\end{equation}
We use $a,b,\ldots$ for binary Pauli labels and freely write $P_a$ or $a$ when the intended object is clear.

The Liouville representation regards an operator as a vector in a $D^2$-dimensional inner-product space. We use the normalized Pauli operators $P_a/\sqrt D$ as an orthonormal basis and abbreviate the corresponding basis vectors as $\dket{P_a}$. With this convention,
\begin{equation}\label{eq:prelim-Liouville-inner-product}
    \dbraket{P_a}{P_b}
    =D^{-1}\tr(P_aP_b)
    =\delta_{a,b}.
\end{equation}
A state $\rho$ is represented by a column vector $\dket{\rho}$, while a measurement effect or observable $M$ is represented by a row vector $\dbra{M}$. Their pairing reproduces the Born rule, $\dbraket{M}{\rho}=\tr(M\rho)$, with the normalization absorbed into their respective Liouville coordinates. This notation is useful because the probability generated by a circuit becomes an ordinary matrix product.

We distinguish a physical operator from the channel that it generates typographically. Roman letters such as $U$, $P$, and $C$ denote unitary gates, whereas the corresponding calligraphic letters denote their conjugation channels:
\begin{equation}\label{eq:prelim-unitary-channel-notation}
    \mathcal U(A)=UAU^\dagger,\qquad
    \mathcal P(A)=PAP^\dagger,\qquad
    \mathcal C(A)=CAC^\dagger.
\end{equation}
More generally, $\mathcal G$ denotes a linear quantum map. Once a map is written in the Pauli--Liouville representation, we use the same calligraphic symbol for its matrix; composition of channels is then matrix multiplication. Accordingly, expressions such as $\mathcal U\dket{P_a}$ and $\mathcal G_{a,b}$ always refer to the Liouville action of a channel, while $UP_aU^\dagger$ refers to ordinary operator multiplication.

The Pauli transfer matrix (PTM) of a linear map $\mathcal G$ is defined by
\begin{equation}\label{eq:prelim-PTM-definition}
    \mathcal G_{a,b}
    :=\dbra{P_a}\mathcal G\dket{P_b}
    =D^{-1}\tr\!\left[P_a\mathcal G(P_b)\right].
\end{equation}
Equivalently,
\begin{equation}\label{eq:prelim-PTM-expansion}
    \mathcal G
    =\sum_{a,b\in\mathbb F_2^{2n}}
    \mathcal G_{a,b}\dketbra{P_a}{P_b}.
\end{equation}
A Hermiticity-preserving map has a real PTM. In particular, all quantum channels, unitary channels, and Lindbladians considered below are represented by real matrices in this basis. If a circuit applies $\mathcal A_1,\ldots,\mathcal A_m$ in chronological order, its outcome probability is
\begin{equation}\label{eq:prelim-Liouville-circuit-probability}
    p=\dbra{M}\mathcal A_m\cdots\mathcal A_1\dket{\rho}.
\end{equation}
This expression will make both gauge transformations and the cancellation of SPAM-dependent factors transparent.

In many of our results, only the support of a Pauli operator matters. For $P=P_1\otimes\cdots\otimes P_n$, define its pattern $\pt(P)\in\{0,1\}^n$ by
\begin{equation}\label{eq:prelim-pattern-definition}
    \pt(P)_q=
    \begin{cases}
        0,&P_q=I,\\
        1,&P_q\in\{X,Y,Z\}.
    \end{cases}
\end{equation}
The Hamming weight of $\pt(P)$ is the Pauli weight $\operatorname{wt}(P)$. Paulis with the same pattern act nontrivially on the same qubits, although their local Pauli factors may differ. A single-qubit Clifford gate permutes $X,Y,$ and $Z$ up to signs and therefore preserves this pattern. This observation underlies the pattern-block structure of the gauge freedom derived later.

\subsection{Pauli characters and local Clifford symmetry}\label{subsec:characters-Schur-preliminaries}

Pauli characters encode commutation relations and provide the projection tool used in our learning protocols. For $P_a,P_b\in\mathsf P_n$, define
\begin{equation}\label{eq:prelim-Pauli-character}
    P_aP_b=\chi_{a,b}P_bP_a,
    \qquad
    \chi_{a,b}:=(-1)^{x_a\cdot z_b+z_a\cdot x_b}.
\end{equation}
Thus, $\chi_{a,b}=+1$ if $P_a$ and $P_b$ commute and $\chi_{a,b}=-1$ if they anticommute. For fixed $a$, the function $b\mapsto\chi_{a,b}$ is a group homomorphism from $\mathbb F_2^{2n}$ to $\{+1,-1\}$ and is therefore a character of the phase-free Pauli group.

The $4^n$ Pauli characters form an orthogonal basis for complex-valued functions on $\mathbb F_2^{2n}$:
\begin{equation}\label{eq:Pauli-character-orthogonality}
    \frac{1}{4^n}\sum_{b\in\mathbb F_2^{2n}}
    \chi_{a,b}\chi_{c,b}
    =\delta_{a,c}.
\end{equation}
Accordingly, the Pauli Fourier transform and its inverse are
\begin{equation}\label{eq:prelim-Pauli-Fourier-transform}
    \widehat f(a)
    =\frac{1}{4^n}\sum_b\chi_{a,b}f(b),
    \qquad
    f(b)=\sum_a\chi_{a,b}\widehat f(a).
\end{equation}
We use this transform later to pass between Lindbladian transfer functions and the conventional Hamiltonian and dissipative coefficients.

The same orthogonality relation gives a projector onto any chosen Pauli component. Let $\mathcal P_b$ be the unitary channel generated by the Pauli gate $P_b$. Since $P_b=P_b^\dagger$,
\begin{equation}\label{eq:prelim-Pauli-channel-eigenvalue}
    \mathcal P_b\dket{P_a}
    =\dket{P_bP_aP_b}
    =\chi_{a,b}\dket{P_a}.
\end{equation}
Averaging these channels with the character of a target Pauli $P_a$ gives
\begin{equation}\label{eq:prelim-character-projector}
    \frac{1}{4^n}\sum_{b\in\mathbb F_2^{2n}}
    \chi_{a,b}\mathcal P_b
    =\dketbra{P_a}{P_a}.
\end{equation}
Equation~\eqref{eq:prelim-character-projector} is the character-projection identity. The operator on the right is a linear projector rather than a physical quantum channel. Nevertheless, its action is experimentally accessible in expectation: sample $P_b$ uniformly, apply the physical Pauli gate, and multiply the measured outcome by the known sign $\chi_{a,b}$. Inserting such projections between target channels selects a prescribed sequence of PTM elements and therefore isolates a path through the Pauli transfer graph.

We also recall the representation-theoretic result responsible for the form of the admissible gauge. Let $\rho:G\rightarrow\operatorname{GL}(V)$ be a finite-dimensional complex representation. Schur's lemma states that if $V$ is irreducible, every linear map $T$ satisfying
\begin{equation}
    T\rho(g)=\rho(g)T
    \qquad\text{for all }g\in G
\end{equation}
is a scalar multiple of the identity on $V$. More generally, if $V=\bigoplus_\alpha V_\alpha$ is a multiplicity-free direct sum of pairwise inequivalent irreducible representations, then every such $T$ acts as an independent scalar on each $V_\alpha$. We will apply precisely this multiplicity-free form.

For one qubit, conjugation by the Clifford group leaves invariant the decomposition
\begin{equation}\label{eq:prelim-one-qubit-Clifford-decomposition}
    \operatorname{span}\{\dket{I}\}
    \oplus
    \operatorname{span}\{\dket{X},\dket{Y},\dket{Z}\}.
\end{equation}
The identity sector carries the trivial representation, while the traceless three-dimensional sector is irreducible. With independently controlled Clifford gates on $n$ qubits, tensor products of these two sectors give the subspaces
\begin{equation}\label{eq:prelim-pattern-subspaces}
    V_u
    =\operatorname{span}\{\dket{P}:\pt(P)=u\},
    \qquad
    u\in\{0,1\}^n.
\end{equation}
Each $V_u$ is irreducible under the local Clifford group, and $V_u$ and $V_v$ are inequivalent for $u\neq v$. The complete Pauli--Liouville space therefore decomposes as
\begin{equation}\label{eq:prelim-local-Clifford-decomposition}
    \mathbb C^{4^n}
    =\bigoplus_{u\in\{0,1\}^n}V_u.
\end{equation}

It follows from Schur's lemma that the commutant of all independently controlled single-qubit Clifford channels consists exactly of the pattern-block scalings
\begin{equation}\label{eq:prelim-pattern-block-commutant}
    T
    =\sum_{u\in\{0,1\}^n}\lambda_u\Pi_u
    =\sum_{P\in\mathsf P_n}
    \lambda_{\pt(P)}\dketbra{P}{P},
\end{equation}
where $\Pi_u$ is the projector onto $V_u$. Thus, a commuting map cannot distinguish $X$, $Y$, and $Z$ on a fixed support, but it may assign different scalars to different support patterns. When $T$ is an invertible real gauge transformation, every $\lambda_u$ is a nonzero real number. This is the representation-theoretic content of Lemma~\ref{lemma:gauge_schur}; the remaining conditions for a gauge to be valid come from preservation of the physical channel model, states, and measurements.

\subsection{Lindbladians and their Pauli parametrization}\label{subsec:Lindblad-preliminaries}

A time-independent Markovian evolution is described by a family of quantum channels $\mathcal G_t=e^{t\mathcal L}$, where the generator $\mathcal L$ determines the differential equation $\mathrm d\rho/\mathrm dt=\mathcal L(\rho)$. Here $\rho$ is the density matrix of the system, and the Markovian assumption means that its instantaneous rate of change depends only on its present value. The evolution is completely positive and trace preserving for every $t\geq0$ precisely when $\mathcal L$ can be written in Lindblad form,
\begin{equation}\label{eq:prelim-general-Lindblad-form}
    \mathcal L(\rho)=-i[H,\rho]+\sum_\mu\left(L_\mu\rho L_\mu^\dagger-\frac12\{L_\mu^\dagger L_\mu,\rho\}\right).
\end{equation}
Here $[A,B]=AB-BA$ and $\{A,B\}=AB+BA$ denote the commutator and anticommutator. The Hermitian operator $H$ generates coherent unitary evolution, whereas each $L_\mu$ describes a dissipative process, or quantum jump. Positive rates multiplying the dissipative terms may be absorbed into the normalization of the corresponding jump operators.

The two terms associated with a jump operator should be viewed together. If
\begin{equation}
    \mathcal D_L(\rho):=L\rho L^\dagger-\frac12\{L^\dagger L,\rho\},
\end{equation}
then cyclicity of the trace gives $\tr[\mathcal D_L(\rho)]=\tr(L^\dagger L\rho)-\tr(L^\dagger L\rho)=0$. Thus the anticommutator term compensates the trace change produced by $L\rho L^\dagger$. Complete positivity is a property of the full evolution $e^{t\mathcal L}$ and follows from the Lindblad form rather than from either contribution in $\mathcal D_L$ separately.

The representation in Eq.~\eqref{eq:prelim-general-Lindblad-form} is not unique. Let $D$ be the Hilbert-space dimension and decompose a jump operator as
\begin{equation}
    L=\alpha I+\widetilde L,
    \qquad
    \alpha=\frac{\tr(L)}{D},
    \qquad
    \tr(\widetilde L)=0.
\end{equation}
Substitution into the dissipator shows that the term proportional to $|\alpha|^2$ cancels and the two cross terms form a commutator:
\begin{equation}\label{eq:prelim-jump-identity-shift}
    \mathcal D_{\alpha I+\widetilde L}(\rho)
    =\mathcal D_{\widetilde L}(\rho)-i[H_\alpha,\rho],
    \qquad
    H_\alpha=\frac{i}{2}\left(\alpha^*\widetilde L-\alpha\widetilde L^\dagger\right).
\end{equation}
The operator $H_\alpha$ is Hermitian, so the identity component of every jump operator can be removed by replacing $L$ with $\widetilde L$ and shifting $H\mapsto H+H_\alpha$. We may therefore choose all jump operators to be traceless. We also choose $\tr(H)=0$, since adding a multiple of the identity to $H$ does not change $[H,\rho]$. These choices only fix a nonunique representation of the same generator; they are distinct from the experimental gauge freedom studied in this work.

Recall that $D=2^n$, and enumerate the Hermitian Pauli basis as $P_0=I,P_1,\ldots,P_{D^2-1}$. Expanding $H=\sum_{a\neq0}c_aP_a$ and $\widetilde L_\mu=\sum_{i\neq0}\ell_{\mu,i}P_i$ gives the canonical Pauli form
\begin{equation}\label{eq:prelim-Pauli-Lindblad-form}
    \mathcal L(\rho)=-i\sum_{a\neq0}c_a[P_a,\rho]+\sum_{i,j\neq0}\gamma_{i,j}\left(P_i\rho P_j-\frac12\{P_jP_i,\rho\}\right),
    \qquad
    \gamma_{i,j}=\sum_\mu\ell_{\mu,i}\ell_{\mu,j}^*.
\end{equation}
The coefficients $c_a$ are real. The Kossakowski matrix $\gamma=(\gamma_{i,j})$ is a Gram matrix and is therefore Hermitian and positive semidefinite. The condition $\gamma\succeq0$ concerns the matrix as a whole, not its entries individually: diagonal entries are real and nonnegative, whereas off-diagonal entries may be negative or complex and satisfy $\gamma_{j,i}=\gamma_{i,j}^*$.

Set $N=D^2-1=4^n-1$, and denote the set of $N\times N$ Hermitian matrices by
\begin{equation}
    \operatorname{Herm}(N):=\{A\in\mathbb C^{N\times N}:A=A^\dagger\}.
\end{equation}
This Hermitian matrix space has a real parametrization, write $A=X+iY$, where $X=\operatorname{Re}A$ and $Y=\operatorname{Im}A$ are real matrices. The condition $A=A^\dagger$ is equivalent to
\begin{equation}
    X^{\mathsf T}=X,
    \qquad
    Y^{\mathsf T}=-Y.
\end{equation}
Thus the real part is symmetric and the imaginary part is skew-symmetric. Conversely, every such pair $(X,Y)$ determines a unique Hermitian matrix $A=X+iY$. This gives the real-linear identification
\begin{equation}\label{eq:Herm-real-coordinate-identification}
    \operatorname{Herm}(N)
    \cong \operatorname{Sym}(N,\mathbb R)\oplus\operatorname{Skew}(N,\mathbb R)
    \cong \mathbb R^{N(N+1)/2}\oplus\mathbb R^{N(N-1)/2}
    \cong \mathbb R^{N^2}.
\end{equation}
Concretely, the corresponding $N^2$ real coordinates are the $N$ diagonal entries $A_{ii}$, together with $\operatorname{Re}A_{ij}$ and $\operatorname{Im}A_{ij}$ for every $i<j$. In this sense, $\operatorname{Herm}(N)$ is a real parameter space even though its elements are conveniently arranged as complex matrices. Multiplication by a real scalar preserves Hermiticity, whereas multiplication by a general complex scalar does not. Before imposing positivity or unitality, the coefficient space is therefore
\begin{equation}\label{eq:prelim-Lindblad-parameter-space}
    \Theta=\mathbb R^N\oplus\operatorname{Herm}(N)\cong\mathbb R^{N+N^2},
    \qquad
    \dim_{\mathbb R}\Theta=N+N^2=4^n(4^n-1).
\end{equation}
Physical generators occupy the cone $\gamma\succeq0$ inside this real vector space. Since a cone is not closed under subtraction, linear learnability is formulated on the surrounding space $\Theta$, with positivity imposed separately as a physical constraint.

Trace preservation, which is built into Eq.~\eqref{eq:prelim-general-Lindblad-form}, should not be confused with unitality. A generator is unital when
\begin{equation}\label{eq:prelim-Lindblad-unitality}
    \mathcal L(I)=\sum_{i,j\neq0}\gamma_{i,j}[P_i,P_j]=0.
\end{equation}
The unital coefficients form the real linear subspace $\Theta_{\mathrm u}=\{(c,\gamma)\in\Theta:\mathcal L_{c,\gamma}(I)=0\}$. Equivalently, an element of $\Theta_{\mathrm u}$ is specified by its real $N\times N$ transfer block on the traceless Pauli subspace, so
\begin{equation}\label{eq:prelim-unital-Lindblad-dimension}
    \Theta_{\mathrm u}\cong\mathbb R^{N\times N},
    \qquad
    \dim_{\mathbb R}\Theta_{\mathrm u}=N^2=(4^n-1)^2.
\end{equation}

\subsection{Robust phase estimation}\label{subsec:robust-phase-estimation}

Robust phase estimation converts estimates of an oscillation's sine and cosine into an estimate of its frequency~\cite{Kimmel2015RobustPhaseEstimation}. We summarize the procedure needed for our Hamiltonian protocols, without proofs. Suppose an unknown frequency $\omega$ can be probed through signals $\cos(\omega\tau)$ and $\sin(\omega\tau)$ at chosen evolution times $\tau$. A coarse bound $|\omega|\leq\Omega$ is assumed known. In our application, $\omega=2c_a$, and the signals are obtained by fitting the character-projected cycle data.

\begin{enumerate}
    \item \textbf{Choose the evolution times.} Select $\tau_0$ so that $\Omega\tau_0\leq\pi/2$, and use $\tau_k=2^k\tau_0$ for $k=0,\ldots,K$. The initial time fixes the phase branch, while the longer times provide increasing frequency resolution.

    \item \textbf{Estimate both signals.} At each time $\tau_k$, obtain estimates $\widehat C_k$ and $\widehat S_k$ of $\cos(\omega\tau_k)$ and $\sin(\omega\tau_k)$. A sufficient accuracy requirement is
    \begin{equation}\label{eq:prelim-RPE-signal-accuracy}
        |\widehat C_k-\cos(\omega\tau_k)|\leq\frac14,
        \qquad
        |\widehat S_k-\sin(\omega\tau_k)|\leq\frac14.
    \end{equation}
    Require these bounds to hold jointly with probability at least $1-\delta_k$. The accuracy is fixed: it need not improve as the target frequency precision increases. Any systematic error must fit within this same error budget.

    \item \textbf{Recover and unwrap the phase.} Compute
    \begin{equation}
        \widehat\phi_k=\operatorname{atan2}(\widehat S_k,\widehat C_k)
        \in(-\pi,\pi],
    \end{equation}
    where $\operatorname{atan2}$ returns the angle of the point $(\widehat C_k,\widehat S_k)$, taking both signs into account. Set $\widehat\omega_0=\widehat\phi_0/\tau_0$. For each $k\geq1$, choose the integer $j_k$ for which
    \begin{equation}\label{eq:prelim-RPE-unwrapping}
        \widehat\omega_k=\frac{\widehat\phi_k+2\pi j_k}{\tau_k}
    \end{equation}
    is closest to $\widehat\omega_{k-1}$. This uses the previous estimate to select the appropriate phase branch.

    \item \textbf{Return the frequency estimate.} Choose $K$ so that $\tau_K=\Theta(1/\varepsilon)$ and return $\widehat\omega_K$. Under Eq.~\eqref{eq:prelim-RPE-signal-accuracy}, the procedure gives $|\widehat\omega_K-\omega|\leq\varepsilon$ for a suitable constant in the choice of $\tau_K$, except with probability at most $\sum_{k=0}^K\delta_k$.
\end{enumerate}

For the resource scaling, assume that obtaining the fixed-accuracy signal estimates at scale $k$ with failure probability $\delta_k$ requires $O(\log(1/\delta_k))$ samples, each using evolution time $O(\tau_k)$. This holds for bounded measurement outcomes with a fixed margin for systematic error; for signals extracted by fitting, it additionally requires the fit to remain well conditioned uniformly across the evolution times. Taking $0<\delta<1/2$ and allocating $\delta_k=\delta/2^{K-k+1}$ yields success probability at least $1-\delta$ and total evolution time
\begin{equation}\label{eq:prelim-RPE-scaling}
    T_{\mathrm{tot}}=O\!\left(\frac{1}{\varepsilon}\log\frac{1}{\delta}\right).
\end{equation}
At fixed success probability, the precision therefore scales as $\varepsilon=O(1/T_{\mathrm{tot}})$, which is Heisenberg scaling. The number of time scales is $O(\log[1/(\varepsilon\tau_0)])$. For the Hamiltonian coefficient, return $\widehat c_a=\widehat\omega_K/2$; the same scaling applies.

\subsection{Signed logarithms and canonical real powers}\label{subsec:complex-logarithm-preliminaries}

Experimental protocols often determine combinations of channel parameters rather than the parameters individually. Recovering further quantities by classical postprocessing requires care when the parameters may have either sign. If an experiment determines $x^2$, both $x$ and $-x$ remain compatible with the result: the positive square root recovers $|x|$, but does not determine $x$. In contrast, knowledge of $x^3$ determines $x$ uniquely through its real cube root. We therefore consider postprocessing functions whose real values are uniquely determined by their inputs, without introducing a sign choice unsupported by the data.

To organize products and powers of such quantities, it is useful to represent them additively. For positive real numbers, the ordinary logarithm gives
\begin{equation}
    \log(xy)=\log x+\log y,\qquad x,y>0.
\end{equation}
This identity allows us to study multiplicative quantities using additive methods. Channel parameters, however, may also be negative, and the ordinary real logarithm is defined only for positive inputs. Taking the logarithm of the absolute value gives part of the desired information, since
\begin{equation}
    \log|xy|=\log|x|+\log|y|.
\end{equation}
It does not retain the sign: for example, $2$ and $-2$ both give $\log 2$. To recover the original number and determine the signs of products, we must record the sign separately.

We encode a positive sign by $0$ and a negative sign by $1$:
\begin{equation}
    s(x)=\begin{cases}
        0,&x>0,\\
        1,&x<0.
    \end{cases}
\end{equation}
With this encoding, the sign of a product is obtained by adding the two sign bits modulo $2$. In particular, two negative factors give $1+1=0$ modulo $2$, corresponding to a positive product. We write $\mathbb F_2=\{0,1\}$ for these two values with addition and multiplication modulo $2$. Thus
\begin{equation}
    s(xy)=s(x)+s(y)\quad\text{in }\mathbb F_2.
\end{equation}

Both the logarithmic magnitude and the sign now obey an addition rule. We collect them into a pair, called the \emph{signed logarithm}:
\begin{equation}\label{eq:signed-log}
    \operatorname{slog}(x):=(\log|x|,s(x)),
    \qquad x\in\mathbb R^\times:=\mathbb R\setminus\{0\}.
\end{equation}
The set of these pairs is denoted by $\mathbb R\oplus\mathbb F_2$. Pairs are added coordinate by coordinate, using ordinary addition for the first coordinate and addition modulo $2$ for the second. Consequently,
\begin{equation}\label{eq:slog-multiplication}
    \operatorname{slog}(xy)
    =\operatorname{slog}(x)+\operatorname{slog}(y).
\end{equation}
For example,
\begin{equation}
    \operatorname{slog}(-2)+\operatorname{slog}(-3)
    =(\log 2+\log 3,1+1)
    =(\log 6,0)=\operatorname{slog}(6).
\end{equation}
No information is lost in this representation: from a pair $(u,s)$, we recover the unique nonzero real number $x=(-1)^s e^u$.

We also need to describe powers of parameters. For a positive number, the familiar identity $\log(x^m)=m\log x$ turns a power into multiplication by its exponent. We seek the same rule for signed logarithms, while requiring the power itself to have a well-defined real value for either sign of the input.

Integer powers, including inverse powers of nonzero inputs, are well defined for either sign. The cube-root observation above extends to every odd root: each real number has a unique real root of odd order. Even roots do not provide the same operation on arbitrary signed inputs, since they do not exist as real numbers for negative inputs and do not invert an even power without losing the original sign.

More generally, write a rational exponent as $p/q$ in lowest terms, with $q>0$. If $q$ is odd, every nonzero real $x$ has a unique real $q$th root, and we define
\begin{equation}\label{eq:canonical-real-power}
    x^{p/q}:=\left(\sqrt[q]{x}\right)^p.
\end{equation}
We call this a \emph{canonical real power}: its value is determined without making an additional sign choice. If the reduced denominator is even, the required root does not exist for negative inputs. The rational exponents available for arbitrary nonzero real inputs are therefore precisely those with odd reduced denominators.

To describe this class of powers concisely, we collect their exponents into
\begin{equation}\label{eq:Z-localized-2}
    \mathbb Z_{(2)}:=\left\{\frac pq\in\mathbb Q:
        p,q\in\mathbb Z,\ q\text{ odd}\right\}.
\end{equation}
For example, $1/3,-2/5$, and every integer belong to $\mathbb Z_{(2)}$, whereas $1/2$ and $3/4$ do not. Adding, subtracting, or multiplying fractions with odd denominators again gives a fraction with an odd denominator. This closure under addition, subtraction, and multiplication makes $\mathbb Z_{(2)}$ a ring. Division by odd integers is allowed, but division by $2$ is not. The standard name for this ring is the \emph{localization of the integers at $2$}; we will use only the explicit description in Eq.~\eqref{eq:Z-localized-2}.

We can now determine how a power changes the signed logarithm. Taking an odd root preserves the sign. Raising the result to an even numerator removes a negative sign, while raising it to an odd numerator preserves that sign. Thus, for $\alpha=p/q\in\mathbb Z_{(2)}$ with $q$ odd,
\begin{equation}
    \log|x^\alpha|=\alpha\log|x|,
    \qquad s(x^\alpha)=(p\bmod 2)s(x).
\end{equation}
Only the parity of the numerator enters the second equation. We denote it by
\begin{equation}\label{eq:Z2-parity-map}
    \overline\alpha:=p\pmod 2\in\mathbb F_2.
\end{equation}
This value is independent of how the fraction is written with an odd denominator. Indeed, $p/q=p'/q'$ implies $pq'=p'q$; since $q$ and $q'$ are odd, $p$ and $p'$ have the same parity.

For a signed logarithm $(u,s)$, we therefore define multiplication by an allowed exponent to mean
\begin{equation}\label{eq:Z2-action-slog}
    \alpha(u,s):=(\alpha u,\overline\alpha s).
\end{equation}
For instance, multiplication by $1/3$ divides the logarithmic magnitude by three and leaves the sign unchanged, whereas multiplication by $2$ doubles the logarithmic magnitude and sets the sign bit to zero. With this definition, the familiar logarithmic power rule becomes
\begin{equation}\label{eq:slog-power}
    \operatorname{slog}(x^\alpha)
    =\alpha\operatorname{slog}(x),
    \qquad \alpha\in\mathbb Z_{(2)}.
\end{equation}
Equations~\eqref{eq:slog-multiplication} and~\eqref{eq:slog-power} express the two operations needed below: products become sums, and powers become multiplication by exponents. They also give the usual power identities
\begin{equation}
    x^{\alpha+\beta}=x^\alpha x^\beta,
    \qquad (xy)^\alpha=x^\alpha y^\alpha,
    \qquad (x^\alpha)^\beta=x^{\alpha\beta},
\end{equation}
for $x,y\in\mathbb R^\times$ and $\alpha,\beta\in\mathbb Z_{(2)}$.

If a quantity is known to be positive, arbitrary real powers can instead be defined by $x^a=e^{a\log x}$. Such prior information may concern either a parameter or a combination of parameters. We will distinguish operations available for arbitrary signed quantities from those that use a known-positive quantity.

\subsection{Multiplicative parameter functions as a free module}\label{subsec:module}

Once experiments determine a collection of nonzero quantities, their products, inverses, and canonical real powers can also be determined by classical postprocessing. We want to describe this family of functions systematically and identify independent quantities from which it can be generated. The signed logarithm makes this possible by turning multiplicative combinations into linear combinations of functions.

To define the ambient function space, we initially regard $\theta_1,\ldots,\theta_r$ as independent nonzero real variables, with $\theta\in\Theta=(\mathbb R^\times)^r$. Constraints imposed by a particular physical model will be incorporated later by restricting these functions to the corresponding parameter domain. We consider the multiplicative functions
\begin{equation}
    f_m(\theta)=\prod_{i=1}^r\theta_i^{m_i},
    \qquad m\in\mathbb Z_{(2)}^r,
\end{equation}
with powers defined by Eq.~\eqref{eq:canonical-real-power}. These functions are real and single-valued throughout $\Theta$. Integer exponents give Laurent monomials, that is, products of parameters and their inverses; the allowed fractional exponents also include their real odd roots. The parameters need not all be experimentally accessible. Determining which functions of them are learnable will be the subject of the following sections.

For each parameter, define the signed-logarithm function
\begin{equation}
    L_i:\Theta\longrightarrow\mathbb R\oplus\mathbb F_2,
    \qquad L_i(\theta)=\operatorname{slog}(\theta_i).
\end{equation}
Here $L_i$ is a function of the variable parameters, rather than the numerical signed logarithm obtained at one particular parameter value. The identities of the preceding subsection give
\begin{equation}\label{eq:laurent-log-linearization}
    \operatorname{slog}f_m(\theta)
    =\sum_{i=1}^r m_i L_i(\theta).
\end{equation}
Thus every multiplicative function in the chosen class is represented by a linear combination of the parameter functions $L_i$.

We collect these combinations into
\begin{equation}\label{eq:parameter-function-module}
    \mathcal F=\left\{\sum_{i=1}^r m_iL_i:
        m_i\in\mathbb Z_{(2)}\right\}.
\end{equation}
Addition and scalar multiplication of functions are defined by applying the operations in Eqs.~\eqref{eq:slog-multiplication} and~\eqref{eq:Z2-action-slog} at every parameter value. The set $\mathcal F$ is closed under both operations. Such an additive structure with scalars drawn from a ring is called a \emph{module}, just as a vector space has scalars drawn from a field. In the present case, the scalar ring is $\mathbb Z_{(2)}$: division by odd integers is allowed, while division by $2$ is not generally available.

Since the parameters vary independently, each element of $\mathcal F$ has a unique expression as a linear combination of $L_1,\ldots,L_r$. These parameter functions therefore provide a basis, leading to the following definition.

\begin{definition}[Free module of parameter functions]
    For the independent parameter variables specified above, the set $\mathcal F$ in Eq.~\eqref{eq:parameter-function-module}, with the function operations specified above, is the free $\mathbb Z_{(2)}$-module generated by $L_1,\ldots,L_r$. These functions form a basis of $\mathcal F$, and its rank is $r$.
\end{definition}

The term \emph{free} refers to the unique basis expansion, and \emph{rank} is the number of basis elements, analogous to dimension for a vector space. The exponent vector $m$ provides the coordinates of a function in this basis:
\begin{equation}\label{eq:free-module-expansion}
    \operatorname{slog}f_m=\sum_{i=1}^r m_iL_i
    \quad\longleftrightarrow\quad
    m=\sum_{i=1}^r m_ie_i\in\mathbb Z_{(2)}^r,
\end{equation}
where $e_i$ is the standard vector with a single $1$ in position $i$. Thus the free module of functions and the free module of exponent vectors are two representations of the same object. Restricting to a physical parameter domain may introduce relations among the basis functions and may make further parameter combinations learnable. We treat these model-specific restrictions separately in later discussion.

The correspondence also explains how experimentally determined functions generate further functions. For $m,n\in\mathbb Z_{(2)}^r$ and $\alpha\in\mathbb Z_{(2)}$,
\begin{equation}\label{eq:laurent-exponent-arithmetic}
    f_mf_n=f_{m+n},\qquad
    f_m^{-1}=f_{-m},\qquad
    (f_m)^\alpha=f_{\alpha m}.
\end{equation}
More generally, if experiments determine $f_1,\ldots,f_D$ in this class, then
\begin{equation}
    \operatorname{slog}\!\left(\prod_{j=1}^D f_j^{\alpha_j}\right)
    =\sum_{j=1}^D\alpha_j\operatorname{slog}f_j,
    \qquad \alpha_j\in\mathbb Z_{(2)}.
\end{equation}
The set of all such combinations is the submodule generated by $\operatorname{slog}f_1,\ldots,\operatorname{slog}f_D$; a submodule is a subset closed under the same addition and scalar multiplication. If $\operatorname{slog}f_1,\ldots,\operatorname{slog}f_D$ are independent over $\mathbb Z_{(2)}$, they form a basis of that submodule.

For example, if experiments determine $f_1=\theta_1\theta_2$ and $f_2=\theta_2^3$, then $f_1f_2^{-1/3}=\theta_1$. The same reconstruction is expressed additively as
\begin{equation}
    \operatorname{slog}f_1-\frac13\operatorname{slog}f_2=L_1,
    \qquad
    (1,1)-\frac13(0,3)=(1,0).
\end{equation}
This is the practical role of the module representation: it turns the construction of multiplicative postprocessing functions into the combination of their basis coordinates.

The signed logarithms of all learnable functions within the chosen class likewise form a submodule of $\mathcal F$, which we call the \emph{learnable submodule}. A basis of this submodule consists of independent signed-logarithm functions that generate it over $\mathbb Z_{(2)}$. A particular collection of measured functions need not generate the entire learnable space. Additional information about positive quantities can permit further inference; for example, knowledge of $\theta_i^4$ determines $\theta_i^2$ by the positive square root, although $2$ is not a $\mathbb Z_{(2)}$ multiple of $4$.

If some parameters themselves are known to be positive, we may enlarge the function class by allowing arbitrary real exponents on those parameters. For $r_+$ positive parameters and $r_{\pm}$ parameters of unrestricted sign, this gives
\begin{equation}\label{eq:mixed-laurent-monomial}
    f_{a,m}(\theta)
    =\prod_{i=1}^{r_+}\theta_i^{a_i}
     \prod_{j=1}^{r_{\pm}}\theta_{r_++j}^{m_j},
    \qquad a\in\mathbb R^{r_+},\quad
    m\in\mathbb Z_{(2)}^{r_{\pm}}.
\end{equation}
The corresponding exponent domain is
\begin{equation}\label{eq:mixed-exponent-space}
    \mathbb R^{r_+}\oplus\mathbb Z_{(2)}^{r_{\pm}}.
\end{equation}
This extension explicitly records where positivity permits arbitrary real powers. And the left summand $\mathbb{R}^{r+}$ is indeed a vector space. Unless stated otherwise, we use $\mathbb Z_{(2)}$ as the scalar ring.

In the main text, we nevertheless use the more familiar term ``space'' for such collections of functions. For example, ``function space'' may refer to the free $\mathbb Z_{(2)}$-module of parameter functions, while ``learnable space'' refers to its submodule of learnable functions. This terminology reflects the additive structure obtained after logarithmic linearization, together with the familiar notions of basis, dimension, and unique coordinates. Unless positivity is explicitly guaranteed, the allowed coefficients should be understood as in $\mathbb{Z}_{(2)}$ rather than arbitrary real numbers.

\section{Gauge freedom and notion of learnability}\label{sec:gauge-learnability}

A quantum calibration experiment probes an unknown implemented channel $\mathcal G$ using a collection of state-preparation, control, and measurement settings. The objective is to characterize $\mathcal G$, often by comparing it with a prescribed ideal evolution. For a circuit containing repeated applications of $\mathcal G$ interleaved with control gates  $\mathcal C_1,\ldots,\mathcal C_L$, an outcome probability takes the form
\begin{equation}\label{eq:exp_outcome}
    p=\dbra{M}\mathcal C_L\mathcal G\mathcal C_{L-1}
    \cdots\mathcal G\mathcal C_1\dket{\rho}.
\end{equation}
Here $\dket{\rho}$ and $\dbra{M}$ denote the effective prepared state and measurement.

In practice, state preparation and measurement are subject to non-trivial unknown errors. For a target initial state $\rho$, preparation error can produce the implemented state $\widetilde\rho=\Lambda^\rho(\rho)$, with $\Lambda^{\rho}$ being the preparation error channel. Similarly, if $\{M_i\}$ is the target POVM, we can model measurement error by a channel $\Lambda^M$ acting immediately before the ideal measurement. The implemented POVM is then $\{\widetilde M_i=(\Lambda^M)^\dagger(M_i)\}$. These error channels $\Lambda^{\rho},\Lambda^{M}$ may depend on the target preparation or measurement setting, but are assumed to remain fixed whenever the same SPAM setting is used.

In many experimental platforms, single-qubit gates can be calibrated to substantially higher fidelity than the multiqubit evolution of interest. Motivated by this observation, we consider an idealized model where all single-qubit Clifford gates are treated as noiseless. Throughout this work, the control and SPAM error model is therefore specified by the following assumptions:
\begin{itemize}
    \item \textbf{Assumption 1.} All single qubit Clifford gates can be perfectly implemented. 
    \item \textbf{Assumption 2.} The unknown state preparation and measurement error $\Lambda^{\rho},\Lambda^{M}$ depend only on their respective target SPAM setting, namely initial state $\rho$ and POVM $M$, and remain fixed and non-vanishing across circuits using those SPAM settings.
\end{itemize}
By non-vanishing we mean that the SPAM error will not severely erase the signal. For example, the fully depolarizing channel $\Lambda(\rho)=I/d$ is excluded by this property. More specifically, let $P$ be an intended Pauli observable, $\widetilde{\rho}$ be the actual prepared state whose target is a $+1$ eigenstate of $P$, and $ \widetilde{M}$ be the effective measurement observable whose target is the observable $P$. We assume
\begin{equation}
\Bigl|\dbraket{\widetilde{M}}{P}\dbraket{P}{\widetilde{\rho}}\Bigr|\geq\frac{1}{\operatorname{poly}(n)}.
\end{equation}
Thus, the relevant preparation and measurement components retain an inverse-polynomial signal.

Unknown SPAM errors give rise to an ambiguity in the description of the circuit elements. To see this, let $T$ be an invertible linear map on operator space. Inserting $T^{-1}T$ between successive factors in Eq.~\eqref{eq:exp_outcome} gives
\begin{equation}\label{eq:identical-outcome}
\begin{split}
    p&=\dbra{M}T^{-1}T\mathcal{C}_{L}T^{-1}T\mathcal{G}T^{-1}T\mathcal{C}_{L-1}T^{-1}T\cdots T^{-1}T\mathcal{G}T^{-1}T\mathcal{C}_{1}T^{-1}T\dket{\rho}\\
    &=(\dbra{M}T^{-1})(T\mathcal{C}_{L}T^{-1})(T\mathcal{G}T^{-1})(T\mathcal{C}_{L-1}T^{-1})\cdots (T\mathcal{G}T^{-1})(T\mathcal{C}_{1}T^{-1})(T\dket{\rho}).
\end{split}
\end{equation}
We call the invertible map $T$ a gauge. The corresponding gauge transformation leaves the outcome probability invariant and transforms the circuit elements as follows:
\begin{equation}
\begin{split}
    \mathcal{G}&~\rightarrow~T\mathcal{G}T^{-1},\qquad
    \mathcal{C}_{i}~\rightarrow~T\mathcal{C}_iT^{-1}\\
    \dket{\rho}&~\rightarrow~T\dket{\rho},\qquad
    \dbra{M}~\rightarrow~\dbra{M}T^{-1}.
\end{split}
\end{equation}
A gauge transformation is valid if it preserves the physical and structural assumptions of the model. Specifically, $T\mathcal G T^{-1}$ must remain CPTP, the transformed preparations must remain normalized quantum states, and the transformed measurement effects must form valid POVMs. In addition, every trusted control operation must remain unchanged. More specifically, $T\mathcal C_iT^{-1}=\mathcal C_i$. This requirement turns out to imposes great constraint to valid gauges. 

\begin{lemma}\label{lemma:gauge_schur}
    Under perfect single qubit Clifford assumption, any valid gauge $T$ has the form,
    \begin{equation}\label{eq:gauge_schur}
    \begin{split}
        T&=\sum_{i}\lambda_{\pt(P_i)}\dketbra{P_i}{P_i},
    \end{split}
    \end{equation}
    where $\pt(P)$ is an $n$-bit binary string such that $\pt_i=1$ if the $i$-th Pauli operator of $P$ is not identity and $\pt_i=0$ otherwise. Each support pattern $u$ has nonzero real scaling factor $\lambda_{u}$. State normalization fixes the identity-pattern factor to $\lambda_{0}=1$.
\end{lemma}

\begin{proof}
    For each pattern $a\in\{0,1\}^{n}$, define the subspace
    \begin{equation}
        V_a=\operatorname{span}\{\dket{P}:\pt(P)=a\}.
    \end{equation}
    On each qubit, the Clifford representation decomposes the Pauli space into the trivial subspace $\operatorname{span}\{I\}$ and the irreducible subspace $\operatorname{span}\{X,Y,Z\}$. Therefore, under the local Clifford group, each $V_a$ is an irreducible representation: the $i$-th tensor factor is the trivial representation when $a_i=0$ and the three-dimensional irreducible representation when $a_i=1$. Moreover, $V_a$ and $V_b$ are inequivalent whenever $a\neq b$.
    
    Since the single-qubit Clifford gates are noiseless, a valid gauge must satisfy $T\mathcal{C}=\mathcal{C}T$ for every local Clifford $\mathcal{C}$. By Schur's lemma, any map commuting with all local Clifford gates must act as a scalar $\lambda_a$ on each $V_a$. Hence
    \begin{equation}
        T=\sum_{a\in\{0,1\}^{n}}\lambda_a\Pi_a
        =\sum_i\lambda_{\pt(P_i)}\dketbra{P_i}{P_i},
    \end{equation}
    where $\Pi_a$ is the projector onto $V_a$. Since $T$ is invertible and Pauli transfer matrix is a real matrix, $\lambda_a\neq0\in\mathbb{R}$ for every $a$. Only the identity Pauli operator contributes to the trace, so preserving state trace requires $\lambda_{0}=1$.
\end{proof}

The additional requirements that the gauge-transformed state, channel, and measurement remain physical are generally less restrictive. For example, for a finite collection of strictly positive prepared states and POVM effects, sufficiently small changes of the gauge preserve positivity by continuity. Therefore unless otherwise stated, we assume that the SPAM operators admit a neighborhood of the identity gauge such that for any diagonal gauge $T$ within this neighborhood,
\begin{equation}
    T(\rho)\geq0,\qquad (T^{-1})^{\dagger}(M_j)\geq 0,
\end{equation}
always hold for every initial state $\rho$ and POVM measurement $\{M_j\}$ under consideration. 

To formulate the calibration target, let $\mathcal{G}_{\theta}$ be a channel model with parameter vector $\theta=(\theta_1,\cdots,\theta_{d})\in\Theta$. Here $\Theta$ is the parameter space and $d$ is the dimension of it. For example, for a known ideal gate $\mathcal{U}$ followed by a Pauli noise, one can take the Pauli eigenvalues of the noise channel as $\theta$ and writes
\begin{equation}
\begin{split}
    \mathcal{G}_{\theta}&=\left(\sum_{i}\theta_{i}\dket{P_i}\dbra{P_i}\right)\mathcal{U}.
\end{split}
\end{equation}
Here since $\theta_0$ always equals to one, the parameter space is $\Theta=\mathbb{R}^{4^n-1}$ with further constraints including $|\theta_{i}|\leq 1$ and positivity of the Pauli channel. For a general channel, the independent Pauli transfer matrix (PTM) entries provide a direct parametrization. More specifically, for an $n$-qubit channel,
\begin{equation}\label{eq:parametrize}
    \mathcal{G}_{\theta}=\begin{pmatrix}
        1 & 0 & \cdots & 0\\
        \theta_{1,0} & \theta_{1,1} & \cdots & \theta_{1,4^{n}-1}\\
        \vdots & \vdots & \ddots & \vdots\\
        \theta_{4^{n}-1,0} & \theta_{4^{n}-1,1} & \cdots & \theta_{4^{n}-1,4^{n}-1}
    \end{pmatrix}.
\end{equation}
Trace preservation fixes the first row, while complete positivity gives restrictions to the remaining entries. 
We use this entry-wise parametrization in the discussion for general quantum channels. Other model-specific parametrization will be introduced as needed later.

Two channels related by a valid gauge transformation produce identical outcome probabilities when their initial state and measurement are also transformed by the same gauge, as shown in Eq.~\eqref{eq:identical-outcome}. The associated parameter vectors therefore cannot be distinguished by any allowed experiments. This motivates the following definition of gauge-equivalent parameter vectors.

\begin{definition}[Gauge-equivalent parameter vectors]
    Two parameter vectors $\theta$ and $\theta'$ are gauge equivalent, denoted by $\theta\sim_{\mathrm{g}}\theta'$, if there exists a valid gauge $T$ such that
    \begin{equation}
        \mathcal{G}_{\theta}=T\mathcal{G}_{\theta'}T^{-1}.
    \end{equation}
\end{definition}

A quantity of interest can be represented by a function $f:\Theta\rightarrow\mathbb{R}$ of parameters. Since gauge-equivalent parameters are experimentally indistinguishable, such a quantity can be determined from the observed probability only if $f(\theta)$ takes the same value for all gauge-equivalent $\theta$. This leads to the definition of learnable function.

\begin{definition}[Learnable function]
    A function $f:\Theta\rightarrow\mathbb{R}$ is learnable if it is constant on every gauge-equivalent parameter,  namely,
    \begin{equation}\label{eq:gauge_invariant_function}
        \theta\sim_{\mathrm{g}}\theta'
        \quad\Longrightarrow\quad
        f(\theta)=f(\theta').
    \end{equation}
\end{definition}
Under the entry-wise parametrization in Eq.~\eqref{eq:parametrize}, Lemma~\ref{lemma:gauge_schur} ensures a simple class of learnable functions.

\begin{example}
    Let $\mathcal{G}$ be the Pauli transfer matrix of a quantum channel. Each entry $\mathcal{G}_{ij}$ satisfying $\pt(P_i)=\pt(P_j)$ is learnable. More specifically,
    \begin{equation}
        (T\mathcal GT^{-1})_{i,j}
        = \frac{\lambda_{\pt(P_i)}}{\lambda_{\pt(P_j)}}\mathcal G_{i,j}
        = \mathcal G_{i,j}.
    \end{equation}
\end{example}

The example suggests looking for products in which input and output gauge factors cancel to one. This cancellation has a multiplicative structure, we therefore focus on multiplicative functions of nonzero parameters, written as
\begin{equation}
    f_m(\theta) = \prod_i\theta_i^{m_i},
\end{equation}
There are two additional advantages to considering functions of this form. First, taking logarithm yields a linear combination of logarithms of the parameters $\theta_i$, enabling a clean characterization of the learnable space. Second, gauge-invariant functions of this form can be estimated through experimental protocols developed in Sec~\ref{subsec: CPTP-protocol}, making them of both theoretical and experimental interest.

\section{Learning general quantum channels}
\label{sec:CPTP}

\subsection{Learnability of quantum channels}\label{subsec:CPTP}

For a general quantum channel, each PTM entry describes the transfer of one Pauli component into another. The gauge in Eq.~\eqref{eq:gauge_schur} multiplies each PTM entry by the ratio of the gauge factors associated with its output and input patterns, which could leave certain multiplicative combinations unchanged. Our aim is to characterize all such multiplicative combinations. To this end, we represent Pauli transfers as edges of a graph and identify the gauge-invariant multiplicative functions through its cycle structure. This construction extends the pattern transfer approach developed for Clifford gates with Pauli noise~\cite{Learnability,Chen2026} to arbitrary quantum channels.
 
\begin{definition}[Pauli transfer graph]
    The Pauli transfer graph $\mathsf{P}(\mathcal G)=(V,E)$ has one vertex for each Pauli operator. Its directed edges are
    \begin{equation}
        E = \{(P_j,P_i):\mathcal G_{i,j}\neq0\}.
    \end{equation}
    The edge $(P_j,P_i)$ runs from the input Pauli $P_j$ to the output Pauli $P_i$ and carries weight $\mathcal G_{i,j}$.
\end{definition}

The edge weights are precisely the nonzero PTM entries, so multiplicative functions can be expressed as products of these weights. Writing $g_e$ for the weight of edge $e$, we consider
\begin{equation}\label{eq:CPTP-multiplicative-function}
    f_x(\mathcal G)=\prod_{e\in E}g_e^{x_e},
    \qquad x\in X=\mathbb Z_{(2)}^{|E|}.
\end{equation}
As explained in Sec.~\ref{subsec:complex-logarithm-preliminaries}, exponents in $\mathbb Z_{(2)}$ are rational numbers with odd reduced denominators, so these powers have unique real values even when the edge weights are negative. We consider channels with a fixed set of nonzero PTM entries, keeping $E$ fixed and every inverse factor well defined. Each function is specified by its exponent vector $x$, allowing us to study learnability through the corresponding vector.

To study how gauge transformations affect these products, we express them additively while retaining the signs of the edge weights. Recall the signed logarithm
\begin{equation}
    \operatorname{slog}(z)=(\log|z|,s(z))
    \in\mathbb R\oplus\mathbb F_2,
\end{equation}
where $s(z)=0$ for $z>0$ and $s(z)=1$ for $z<0$. The sign coordinate is evaluated modulo $2$. The identities
\begin{equation}
    \operatorname{slog}(ab)=\operatorname{slog}(a)+\operatorname{slog}(b),
    \qquad
    \operatorname{slog}(a^\alpha)=\alpha\operatorname{slog}(a)
\end{equation}
convert products and powers into sums and scalar multiplication. Since a signed logarithm is a pair, multiplication by an exponent acts separately on its two coordinates. Specifically, for $(u,s)\in\mathbb{R}\oplus \mathbb{F}_{2}$ and $\alpha=p/q\in\mathbb Z_{(2)}$ with $q$ odd, we use
\begin{equation}\label{eq:CPTP-scalar-action}
    \alpha(u,s):=(\alpha u,(p\bmod 2)s).
\end{equation}
The logarithmic magnitude is multiplied by $\alpha$, while the sign bit is multiplied by the parity of $p$ in $\mathbb F_2$. 

With these rules, we can express a product of edge weights as a sum of their signed logarithms. Collecting the signed logarithms of the edge weights into the vector
\begin{equation}
    w=(\operatorname{slog}(g_e))_{e\in E},
\end{equation}
we obtain
\begin{equation}\label{eq:CPTP-logarithmic-function}
    \operatorname{slog}f_x(\mathcal G)
    =\sum_{e\in E}x_e w_e
    =:\langle x,w\rangle.
\end{equation}
The notation $\langle x,w\rangle$ follows the familiar sum-of-products form of an inner product, with each multiplication $x_e w_e$ defined by Eq.~\eqref{eq:CPTP-scalar-action} since each $x_{e}\in\mathbb{Z}_{(2)}$ while each $w_{e}\in\mathbb{R}\oplus\mathbb{F}_{2}$. The result has two components: the logarithmic magnitude and the sign bit of $f_x$. We can therefore determine whether $f_x$ is gauge invariant by examining how a gauge changes $w$.

For an edge $e=(P,Q)$, a valid diagonal gauge $T$ transforms its weight as
\begin{equation}\label{eq:CPTP-edge-gauge}
    g'_e=\frac{\lambda_{\pt(Q)}}{\lambda_{\pt(P)}}g_e.
\end{equation}
Taking signed logarithms gives the additive relation
\begin{equation}\label{eq:CPTP-logarithmic-gauge-shift}
    w'=w+\Delta_T,
\end{equation}
where $w'$ collects the transformed signed logarithms and
$(\Delta_T)_e=\operatorname{slog}(\lambda_{\pt(Q)}/\lambda_{\pt(P)})$.
The corresponding change in $\operatorname{slog}f_x$ is $\langle x,\Delta_T\rangle$. Thus $f_x$ is learnable precisely when this change vanishes in both components for every valid gauge, giving the following criterion.

\begin{lemma}\label{lemma:vector_learnable}
    For each valid diagonal gauge $T$, define the shift vector $\Delta_T$ by
    \begin{equation}
        (\Delta_T)_e=\operatorname{slog}\!\left(
        \frac{\lambda_{\pt(Q)}}{\lambda_{\pt(P)}}\right),
        \qquad e=(P,Q).
    \end{equation}
    A multiplicative function $f_x$ with exponent vector $x\in\mathbb Z_{(2)}^{|E|}$ is learnable if and only if
    \begin{equation}\label{eq:vector_invariant}
        \langle x,\Delta_T\rangle=(0,0)
        \quad\text{in }\mathbb R\oplus\mathbb F_2
    \end{equation}
    for every valid gauge $T$.
\end{lemma}

\begin{proof}
    To recover a real number from its signed logarithm, define
    \begin{equation}
        \operatorname{sexp}(u,s):=(-1)^s e^u.
    \end{equation}
    This is the inverse of $\operatorname{slog}$, and it equals one precisely when $(u,s)=(0,0)$. Applying it to the change in the signed logarithm of $f_x$ gives
    \begin{equation}
        \frac{f_x(T\mathcal GT^{-1})}{f_x(\mathcal G)}
        =\operatorname{sexp}\!\bigl(\langle x,\Delta_T\rangle\bigr).
    \end{equation}
    The function is therefore invariant under every valid gauge exactly when the stated sum vanishes.
\end{proof}

Lemma~\ref{lemma:vector_learnable} expresses learnability in terms of the gauge-induced shifts $\Delta_{T}$. Each entry of this vector depends only on support patterns at the two ends of an edge, not on the particular Pauli axes. To describe the gauge action, we can therefore simplify the Pauli transfer graph by merging vertices with the same pattern, while retaining the individual edges and weights. We call the resulting graph the \textit{pattern transfer graph}. Grouping vertices in this way also reflects the available control operations: any two Pauli operators with the same pattern can be transformed into each other by noiseless single-qubit Clifford gates. 

\begin{definition}[Pattern transfer graph]
    The pattern transfer graph $\Gamma_{\pt}(\mathcal{G})$ is a weighted directed multigraph with vertex set $V_{\pt}=\{0,1\}^n$ consisting of the support patterns. For each nonzero PTM entry $\mathcal{G}_{i,j}\neq 0$, it contains a distinct edge
    \begin{equation}
        e_{ij}:\pt(P_j)\longrightarrow\pt(P_i).
    \end{equation}
    Entries with the same input and output patterns give parallel edges, while entries satisfying $\pt(P_i)=\pt(P_j)$ give self-loops.
\end{definition}

Compared to the Pauli transfer graph $\mathsf{P}(\mathcal{G})$, this pattern transfer graph $\Gamma_{\mathrm{pt}}(\mathcal{G})$ retains every edge while combining vertices sharing the same pattern. A Pauli transfer between equal patterns becomes a self-loop, recovering the individually learnable entries discussed in Sec.~\ref{sec:gauge-learnability}. More general invariants arise by joining several transfers into a closed route.

\begin{definition}[Cycle space]
    Let $c$ be a cycle in the pattern transfer graph $\Gamma_{\pt}(\mathcal{G})$, where edges may be traversed in either direction. Choose a direction to traverse around $c$. Its cycle vector $x_c$ is defined by
    \begin{equation}
        (x_{c})_{e}=\begin{cases}
            +1, & \text{if $c$ traverses $e$ in its direction},\\
            -1, & \text{if $c$ traverses $e$ in the opposite direction},\\
            ~0, & \text{if $e$ does not belong to $c$}.
        \end{cases}
    \end{equation}
    Self-loops and cycles formed by two distinct parallel edges are included. 
    The cycle space $Z\bigl(\Gamma_{\pt}(\mathcal{G})\bigr)$ is the set of all finite linear combinations of these cycle vectors with coefficients in $\mathbb{Z}_{(2)}$,
    \begin{equation}
        Z\bigl(\Gamma_{\pt}(\mathcal{G})\bigr)=\operatorname{span}_{\mathbb{Z}_{(2)}}\{x_c:c\text{ is a cycle in }\Gamma_{\pt}(\mathcal{G})\}.
    \end{equation}
\end{definition}

Each cycle vector also serves as the exponent vector of a
multiplicative function.The sign in a cycle vector specifies whether an edge weight is multiplied or divided. For example, select three edges $u\to v$, $v\to w$, and $u\to w$ in the pattern transfer graph, and denote their weights by $g_{v,u}$, $g_{w,v}$, and $g_{w,u}$, respectively. These weights refer to the selected edges; other edges with the same endpoint patterns may also be present. Traversing the cycle in the order $u\to v\to w\to u$ gives
\begin{equation}
    f_c(\mathcal G)
    =\frac{g_{v,u}g_{w,v}}{g_{w,u}}.
\end{equation}
Here $g_{w,u}$ is in the denominator since the cycle traverses the opposite ($w\to u$) as its original direction ($u\to w$).
Under a valid gauge transformation, this function transforms as
\begin{equation}
    f_c'(\mathcal G)
    =\frac{\left(\frac{\lambda_{v}}{\lambda_{u}}g_{ v,u}\right)\left(\frac{\lambda_{w}}{\lambda_{v}}g_{w,v}\right)}{\frac{\lambda_{w}}{\lambda_{u}}g_{w,u}}=f_{c}(\mathcal{G}).
\end{equation}
Thus, the cycle defines a gauge-invariant multiplicative function. The same gauge invariance occurs for every cycle vector. To see this, we use a characterization of the cycle space in terms of the directed incidence matrix.

Let $B\in\mathbb Z^{|V_{\pt}|\times|E|}$ be the directed incidence matrix of $\Gamma_{\pt}(\mathcal G)$. For an edge $e$ directed from $u$ to $v$, with $u\neq v$, the corresponding $e$-th column in $B$ is defined as
\begin{equation}
    B_{u,e}=-1,\qquad B_{v,e}=+1,
\end{equation}
with all other entries in that column equal to zero. For a self loop, the entire column is zero. The cycle space is then precisely the kernel of $B:\mathbb Z_{(2)}^{E}\to\mathbb Z_{(2)}^{V_{\pt}}$:
\begin{equation}
    Z\bigl(\Gamma_{\pt}(\mathcal{G})\bigr)=\ker B.
\end{equation}
Indeed, each entry of $Bx$ measures the difference between the sums of exponents on incoming and outgoing edges:
\begin{equation}\label{eq:CPTP-vertex-balance}
    (Bx)_v
    =\sum_{e:\,t(e)=v}x_e-\sum_{e:\,s(e)=v}x_e.
\end{equation}
Here each directed edge is written as $e=(s(e),t(e))$, therefore $s(e)$ and $t(e)$ denote the start and end vertices of edge $e$. Thus, $(Bx)_v$ is the sum of the exponents on edges entering $v$ minus the sum on edges leaving $v$. For a cycle vector, these contributions cancel at every vertex, so $Bx_c=0$. 

Writing the signed logarithmic gauge factors as $\ell_v=\operatorname{slog}(\lambda_v)$, the gauge shift is then $\Delta_{T}=B^{\mathsf T}\ell$. The invariance condition of $f_x$ in Lemma~\ref{lemma:vector_learnable} then becomes: 
\begin{equation}\label{eq:CPTP-logarithmic-boundary}
    \langle x,\Delta_T\rangle
    =\langle Bx,\ell\rangle=(0,0),
\end{equation}
In particular, $Bx=0$ guarantees Eq.~\eqref{eq:CPTP-logarithmic-boundary} to hold for arbitrary gauge $T$, so every finite linear combination of cycle vectors with coefficients in $\mathbb{Z}_{(2)}$ defines a learnable function.
Conversely, if the nonidentity-pattern gauge factors can vary independently near the identity, this condition is also necessary. The following theorem therefore identifies the cycle space as the full set of learnable exponent vectors.

\begin{theorem}[Cycle-space characterization of channel learnability]\label{thm:general_cptp_learnability}
    Consider the nonzero PTM entries of a general quantum channel, with exponent space $X=\mathbb Z_{(2)}^{|E|}$. Suppose that the valid diagonal gauges contain a neighborhood of the identity in which the nonidentity-pattern scalings can vary independently, with $\lambda_{0^n}=1$. The learnable exponent vectors are exactly
    \begin{equation}\label{eq:CPTP-learnable-space}
        \mathfrak L
        =Z\bigl(\Gamma_{\pt}(\mathcal G)\bigr).
    \end{equation}
    The remaining gauge-dependent classes are described by the quotient
    \begin{equation}\label{eq:decomposition}
        \mathfrak U
        =X/Z\bigl(\Gamma_{\pt}(\mathcal G)\bigr).
    \end{equation}
    Every representative of a nonzero class in $\mathfrak U$ defines an unlearnable multiplicative function.
\end{theorem}

\begin{proof}
    We first show that $f_x$ is learnable if and only if $Bx=0$. If $Bx=0$, then
    \begin{equation}
        \langle x,\Delta_T\rangle=\langle Bx,\ell\rangle=(0,0)
    \end{equation}
    for every valid gauge $T$. Lemma~\ref{lemma:vector_learnable} therefore proves learnability.

    Conversely, suppose $Bx\neq0$. Every column of $B$ sums to zero, so $\sum_v(Bx)_v=0$. Thus at least one nonidentity pattern $v$ satisfies $(Bx)_v\neq0$. Choose $\lambda_v=e^t$ and all other gauge factors equal to one. By the neighborhood assumption, this gives a valid gauge $T$ for sufficiently small real $t$. All its sign bits vanish, and hence
    \begin{equation}
        \langle x,\Delta_T\rangle=\bigl(t(Bx)_v,0\bigr),
        \qquad
        \frac{f_x(T\mathcal GT^{-1})}{f_x(\mathcal G)}
        =e^{t(Bx)_v}.
    \end{equation}
    For small nonzero $t$, this ratio differs from one, so $f_x$ is not learnable. This establishes $\mathfrak L=\ker B$ over $\mathbb Z_{(2)}$.

    To identify this kernel with the cycle space, choose a spanning forest $F$ of the graph, disregarding edge directions. Each edge $e$ outside $F$ forms a cycle with the unique forest path connecting its endpoints; a self-loop forms a cycle by itself. Choose the traversal so that its cycle vector $x_{c_e}$ has coefficient $+1$ on $e$. Every cycle vector lies in $\ker B$. Conversely, for $x\in\ker B$, form
    \begin{equation}
        y=x-\sum_{e\notin F}x_e x_{c_e}.
    \end{equation}
    The coefficients $x_e$ belong to $\mathbb Z_{(2)}$, so this is an allowed linear combination. Since each $c_e$ contains only one edge outside the forest, $y$ is supported on $F$ and satisfies $By=0$. At a leaf, this condition forces the coefficient on its only incident forest edge to vanish. Repeating after removing that edge gives $y=0$. Thus the cycle vectors generate the full kernel over $\mathbb Z_{(2)}$.

    Finally, $x$ and $y$ represent the same quotient class precisely when $x-y$ lies in the cycle space, or equivalently when $Bx=By$. Every representative of a nonzero class has $Bx\neq0$ and therefore defines an unlearnable function.
\end{proof}

Theorem~\ref{thm:general_cptp_learnability} also allows us to count the independent learnable combinations and the remaining gauge degrees of freedom. Let $\Gamma_{\mathrm{pt}}(\mathcal{G})$ have $|V|$ vertices,
$|E|$ edges, and $c(\Gamma_{\mathrm{pt}}(\mathcal{G}))$ connected components, where
connectivity is considered without regard to edge directions. Then the cycle space has rank
\begin{equation}\label{eq:CPTP-cycle-dimension}
    \operatorname{rank}_{\mathbb Z_{(2)}} Z(\Gamma_{\mathrm{pt}}(\mathcal{G}))=|E|-|V|+c(\Gamma_{\mathrm{pt}}(\mathcal{G})),
\end{equation}
and the quotient has rank
\begin{equation}\label{eq:CPTP-quotient-dimension}
    \operatorname{rank}_{\mathbb Z_{(2)}}\mathfrak U=|V|-c(\Gamma_{\mathrm{pt}}(\mathcal{G})).
\end{equation}
The cycle space rank counts the independent generators of the learnable multiplicative functions, while the quotient rank counts the gauge degree of freedom.

For an $n$-qubit channel $\mathcal{G}$ whose pattern transfer graph is connected, there are $2^n$ pattern vertices, giving $|E|-2^n+1$ independent cycle generators, among which $\mathcal{G}_{0,0}$ is a self-loop with known value one. There are $2^n-1$ remaining gauge degrees of freedom. This count agrees with the diagonal gauge $T=\sum_i\lambda_{\pt(P_i)}\dketbra{P_i}{P_i}$ : fixing $\lambda_{0^n}=1$ leaves one free factor for each of the $2^n-1$ nonidentity patterns. 

\subsection{Learning protocols for quantum channels}
\label{subsec: CPTP-protocol}

The preceding subsection characterizes the space of learnable multiplicative functions of a quantum channel. We now construct explicit protocols for estimating these functions in the presence of SPAM errors. Our first step is to isolate a product of PTM entries along a sequence of Pauli transfers. This can be achieved using the Pauli character projection technique, which selects one Pauli component at each stage of the circuit.

For Pauli operators $P,Q$, define the character $\chi(P,Q)$ such that it equals $+1$ if $P$ and $Q$ commute and $-1$ when they anticommute. The character projection identity gives
\begin{equation}
    \dketbra{Q}{Q}
    =\mathbb{E}_{P\sim\mathsf{P}_n}\bigl[\chi(Q,P)\mathcal{P}\bigr]
    =\frac{1}{|\mathsf{P}_n|}\sum_{P\in\mathsf{P}_n}\chi(Q,P)\mathcal{P}.
\end{equation}
Here $\mathcal{P}$ denotes the channel of conjugation by $P$ in the Pauli-Liouville representation. Although this projector is not itself a physical channel, its action can be realized in expectation by sampling Pauli gates and weighting the measurement outcomes by their characters~\cite{Helsen2019}.

Consider a directed path $(Q_0,Q_1,\cdots,Q_{L})$ in the Pauli transfer graph for a $n$-qubit channel $\mathcal{G}$. To isolate its weight $\prod_{i=0}^{L-1}\mathcal{G}_{Q_{i+1},Q_{i}}$, sample $P_0,\cdots,P_{L}$ independently and uniformly from the $n$-qubit Pauli group $\mathsf{P}_{n}$ and implement the circuit
\begin{equation}
    \dbra{\widetilde{M}}\mathcal{P}_{L}\mathcal{G}\mathcal{P}_{L-1}\cdots\mathcal{P}_1\mathcal{G}\mathcal{P}_0\dket{\widetilde{\rho}}.
\end{equation}
Here $\dket{\widetilde{\rho}}$ denotes the noisy initial state and $\widetilde{M}$ is the effective noisy measurement observable. Multiplying each outcome by $\prod_{i=0}^{L}\chi(Q_i,P_i)$ and averaging over all independently sampled Pauli $P_0,\cdots,P_{L}$ gives
\begin{equation}
\begin{split}
    &\quad\mathbb{E}_{P_i\sim\mathsf{P}_{n}}\left[\prod_{i=0}^{L}\chi(Q_i,P_i)\dbra{\widetilde{M}}\mathcal{P}_{L}\mathcal{G}\mathcal{P}_{L-1}\cdots\mathcal{P}_1\mathcal{G}\mathcal{P}_0\dket{\widetilde{\rho}}\right]\\
    &=\dbra{\widetilde{M}}\mathbb{E}_{P_L\sim\mathsf{P}_n}\Bigl(\chi(P_L,Q_L)\mathcal{P}_{L}\Bigr)\mathcal{G}\mathbb{E}_{P_{L-1}\sim\mathsf{P}_n}\Bigl(\chi(P_{L-1},Q_{L-1})\mathcal{P}_{L-1}\Bigr)\cdots \mathbb{E}_{P_1\sim\mathsf{P}_n}\Bigl(\chi(P_1,Q_1)\mathcal{P}_1\Bigr)\mathcal{G}\mathbb{E}_{P_0\sim\mathsf{P}_n}\Bigl(\chi(P_0,Q_0)\mathcal{P}_0\Bigr)\dket{\widetilde{\rho}}\\
    &=\dbraket{\widetilde{M}}{Q_L}\dbra{Q_L}\mathcal{G}\dketbra{Q_{L-1}}{Q_{L-1}}\cdots\dketbra{Q_{1}}{Q_{1}}\mathcal{G}\dket{Q_0}\dbraket{Q_0}{\widetilde{\rho}}\\
    &=\dbraket{\widetilde{M}}{Q_L}\dbraket{Q_0}{\widetilde{\rho}}\prod_{i=0}^{L-1}\mathcal{G}_{Q_{i+1},Q_{i}}.
\end{split}
\end{equation}
Thus, character projection isolates the desired path product $\prod_{i=0}^{L-1}\mathcal{G}_{Q_{i+1},Q_{i}}$, while the preparation and measurement enter only through two endpoint overlaps $\dbraket{\widetilde{M}}{Q_L}\dbraket{Q_0}{\widetilde{\rho}}$. These overlaps are unknown in the presence of SPAM errors, so the path product cannot generally be inferred from this signal alone.

To remove this dependence on the unknown SPAM errors, we use the cycle structure identifies in the preceding subsection. Since individual cycles form a basis for the cycle space, it suffices to estimate the function associated with these cycles. We first consider an oriented cycle, meaning that all edges are traversed in their own directions, which can be written as
\begin{equation}\label{eq:cycle_pauli}
    c=(e_0,e_1,\ldots,e_{L-1}),\qquad e_i=(P_i,Q_i),
\end{equation}
Here, consecutive edges of a pattern cycle need not share the same Pauli operators. They need only to have the same support pattern, namely $\pt(Q_i)=\pt(P_{i+1})$ with $P_{L}=P_{0}$. Then there exists local Clifford gates to connect these operators, allowing the transfers to be implemented sequentially. Specifically, for each $i$, choose a local Clifford transformation $C_i$ such that
\begin{equation}\label{eq:cycle_clifford}
    \mathcal{C}_i\dket{Q_i}=\dket{P_{i+1}}.
\end{equation}
Combining these controls with character projection produces a cycle block whose character-weighted average is
\begin{equation}
\begin{split}
    \mathcal{B}&=\mathbb{E}_{R_i\sim\mathsf{P}_{n}}\left[\chi(P_0,R_0)\prod_{i=0}^{L-1}\chi(Q_i,R_{i+1})\mathcal{C}_{L-1}\mathcal{R}_{L}\mathcal{G}\mathcal{R}_{L-1}\mathcal{C}_{L-2}\cdots\mathcal{C}_0\mathcal{R}_1\mathcal{G}\mathcal{R}_0\right]\\
    &=\Bigl(\prod_{i=0}^{L-1}\mathcal{G}_{Q_i,P_i}\Bigr)\dket{P_0}\dbra{P_0}.
\end{split}
\end{equation}
Repeating this block $m$ times in a circuit with Pauli operators $R_i$ drawn uniformly and independently on each block therefore yields,
\begin{equation}
\begin{split}
    F(m)&=\dbra{\widetilde{M}}\mathcal{B}^{m}\dket{\widetilde{\rho}}\\
    &=\dbraket{\widetilde{M}}{P_0}\dbraket{P_0}{\widetilde{\rho}}\Bigl(\prod_{i=0}^{L-1}\mathcal{G}_{Q_i,P_i}\Bigr)^{m}.
\end{split}
\end{equation}
For fixed preparation and measurement settings, the SPAM errors only affect the depth-independent amplitude $\dbraket{\widetilde{M}}{P_0}\dbraket{P_0}{\widetilde{\rho}}$. By fitting the signals at several depth $m$ to an exponential function $\hat{F}(m)=Ar^{m}$, the cycle product can be extracted robustly from the fitted $r$.

For a cycle beginning and ending at $P_0$, ideally preparing an $+1$ eigenstate $\rho$ of $P_0$ and measuring the observable $M=P_0$ gives the maximal overlap $\dbraket{M}{P_0}\dbraket{P_0}{\rho}$. Due to SPAM error, the actual implementation can be noisy and we denote them as $\widetilde{\rho}$ and $\widetilde{M}$. We assume that the SPAM error does not exponentially suppress the measured signal and the resulting amplitude $\dbraket{\widetilde{M}}{P_0}\dbraket{P_0}{\widetilde{\rho}}$ is at least inverse-polynomial in the number of qubits. With these preparation and measurement settings, Box~\ref{box:oriented_cycle} then gives the complete protocol.
\begin{mybox}[label=box:oriented_cycle]{Learning an oriented cycle}
    \begin{enumerate}
        \item Given the oriented cycle $c$, determine the Pauli operators $P_i,Q_i$ and the local Clifford transformations $C_i$ according to Eqs.~\eqref{eq:cycle_pauli} and~\eqref{eq:cycle_clifford}.

        \item Ideally prepare a $+1$ eigenstate of $P_0$ and measure the observable $P_0$. Denote the actually prepared state by $\widetilde{\rho}$ and the actually measured observable $\widetilde{M}$.

        \item Fix a positive integer depth $m$. For every repetition $\ell\in\{1,\ldots,m\}$, sample independently and uniformly $L+1$ Pauli operators from the Pauli group. Use $\mathbf{R}_{\ell}$ to denote these operators and $\mathcal{S}(\mathbf{R}_{\ell})$ to be a cycle block with inserting $\mathbf{R}_{\ell}$: 
        \begin{equation}\label{eq:random_oriented_cycle_block}
        \begin{split}
            \mathbf{R}_{\ell}&=(R_{\ell,0},\ldots,R_{\ell,L})
            \in\mathsf{P}_n^{L+1}\\
            \mathcal{S}(\mathbf{R}_{\ell})&=\left(\prod_{i=0}^{L-1}\mathcal{C}_{i}\mathcal{R}_{\ell,i+1}\mathcal{G}\right)\mathcal{R}_{\ell, 0}.
        \end{split}
        \end{equation}
        Writing $\mathbf{R}=(\mathbf{R}_1,\ldots,\mathbf{R}_m)$ for the complete random sequence, implement the depth-$m$ circuit
        \begin{equation}
            \mathcal{S}_{m}(\mathbf{R})=\prod_{\ell=1}^{m}\mathcal{S}(\mathbf{R}_{\ell}).
        \end{equation}
        Measure the expectation of noisy observable $\widetilde{M}$ as: 
        \begin{equation}\label{eq:oriented_outcome_probability}
            f(m,\mathbf{R})=\dbra{\widetilde{M}}\mathcal{S}_{m}\dket{\widetilde{\rho}}.
        \end{equation}

        \item For each depth $m$, sample different Pauli operators $\mathbf{R}$ and estimate the character weighted expectation:
        \begin{equation}\label{eq:RB_data}
            F(m)=\mathbb{E}_{\mathbf{R}\sim\mathsf{P}_{n}^{(L+1)\times m}}\left[\left(\prod_{\ell=1}^{m}\chi(P_{0},\mathcal{R}_{\ell,0})\Bigl(\prod_{i=0}^{L-1}\chi(Q_{i},\mathcal{R}_{\ell, i+1})\Bigr)\right)f(m,\mathbf{R})\right].
        \end{equation}

        \item Repeat the estimation for several positive depths $m$, including both odd and even depths, and fit $F(m)$ to the single-exponential model
        \begin{equation}
            \widehat{F}(m)=A\cdot r^m.
        \end{equation}
        Return the fitted decay parameter $r$ as the oriented-cycle product.
    \end{enumerate}
\end{mybox}

We next consider an unoriented cycle where the traversal of the cycle follows some edges in their assigned direction and others in the opposite direction, which can be written as a path: 
\begin{equation}
    c'=(e_0,e_1,\cdots,e_{L-1}),\qquad e_{i}=(P_i,Q_i) \text{ or } (Q_i,P_i),
\end{equation}
where the condition $\pt(Q_i)=\pt(P_{i+1})$ with $P_{L}=P_{0}$ still holds. Compare this path to the cycle in Eq.~\eqref{eq:cycle_pauli}. The difference is that in an unoriented cycle, certain edges can be traversed in the reversed direction. The corresponding exponent vector of $c'$ is therefore:
\begin{equation}
    (x_{c'})_{e_i}=\left\{\begin{aligned}
        1,&\qquad e_i=(P_i,Q_i)\\
        -1,&\qquad e_{i}=(Q_i,P_i)\\
        0,&\qquad e_{i}\not\in c'
    \end{aligned}\right..
\end{equation}
The gauge-invariant function associated to cycle $c'$ is: 
\begin{equation}
    f_{x_{c'}}(\mathcal{G})=\frac{\prod_{e:\,(x_{c'})_e=+1}g_e}{\prod_{e:\,(x_{c'})_e=-1}g_e}.
\end{equation}

\begin{figure}[htbp]
    \centering
    \begin{tikzpicture}[
    x=1cm,
    y=1cm,
    source/.style={
        circle,
        draw=green!45!black,
        fill=green!10,
        line width=1pt,
        minimum size=9mm,
        inner sep=0pt,
        font=\small\bfseries
    },
    sink/.style={
        circle,
        draw=violet!65!black,
        fill=violet!9,
        line width=1pt,
        minimum size=9mm,
        inner sep=0pt,
        font=\small\bfseries
    },
    internal/.style={
        circle,
        draw=black!55,
        fill=white,
        line width=0.75pt,
        minimum size=4.5mm,
        inner sep=0pt
    },
    plus chain/.style={
        -{Stealth[length=2.4mm,width=1.7mm]},
        draw=blue!70!black,
        line width=1.25pt
    },
    minus chain/.style={
        -{Stealth[length=2.4mm,width=1.7mm]},
        draw=orange!85!black,
        line width=1.25pt
    },
    chain label/.style={
        font=\small,
        fill=white,
        inner sep=1.5pt
    }
]
    \node[source] (s1) at (-2.75, 1.45) {$s_1$};
    \node[sink]   (t1) at ( 2.75, 1.45) {$t_1$};
    \node[source] (s2) at ( 2.75,-1.45) {$s_2$};
    \node[sink]   (t2) at (-2.75,-1.45) {$t_2$};

    \node[internal] (p1) at ( 0,    1.45) {};
    \node[internal] (m1) at ( 2.75, 0) {};
    \node[internal] (p2) at ( 0,   -1.45) {};
    \node[internal] (m2) at (-2.75, 0) {};

    \draw[plus chain] (s1) -- (p1);
    \draw[plus chain] (p1) -- (t1);
    \node[chain label] at (0,1.93) {$\gamma_1^+$};

    \draw[plus chain] (s2) -- (p2);
    \draw[plus chain] (p2) -- (t2);
    \node[chain label] at (0,-1.93) {$\gamma_2^+$};

    \draw[minus chain] (s2) -- (m1);
    \draw[minus chain] (m1) -- (t1);
    \node[chain label] at (3.25,0) {$\gamma_1^-$};

    \draw[minus chain] (s1) -- (m2);
    \draw[minus chain] (m2) -- (t2);
    \node[chain label] at (-3.25,0) {$\gamma_2^-$};

    \draw[gray!75,dashed,line width=0.8pt,-{Stealth[length=2mm]}]
        (0.58,0) arc[start angle=0,end angle=-300,radius=0.58];
\end{tikzpicture}
    \caption{Decomposition of an unoriented cycle into maximal consistently oriented chains; the dashed central arrow indicates the chosen clockwise traversal. Green vertices are sources and violet vertices are sinks. Blue chains $\gamma_k^+$ carry coefficient $+1$ and their physical direction agrees with the traversal. Orange chains $\gamma_k^-$ carry coefficient $-1$: the cycle traverses them from $t_k$ to $s_{k+1}$, opposite to their physical direction $s_{k+1}\to t_k$. Thus two physical chains leave every source and two enter every sink.}
    \label{fig:unoriented-cycle-chains}
\end{figure}
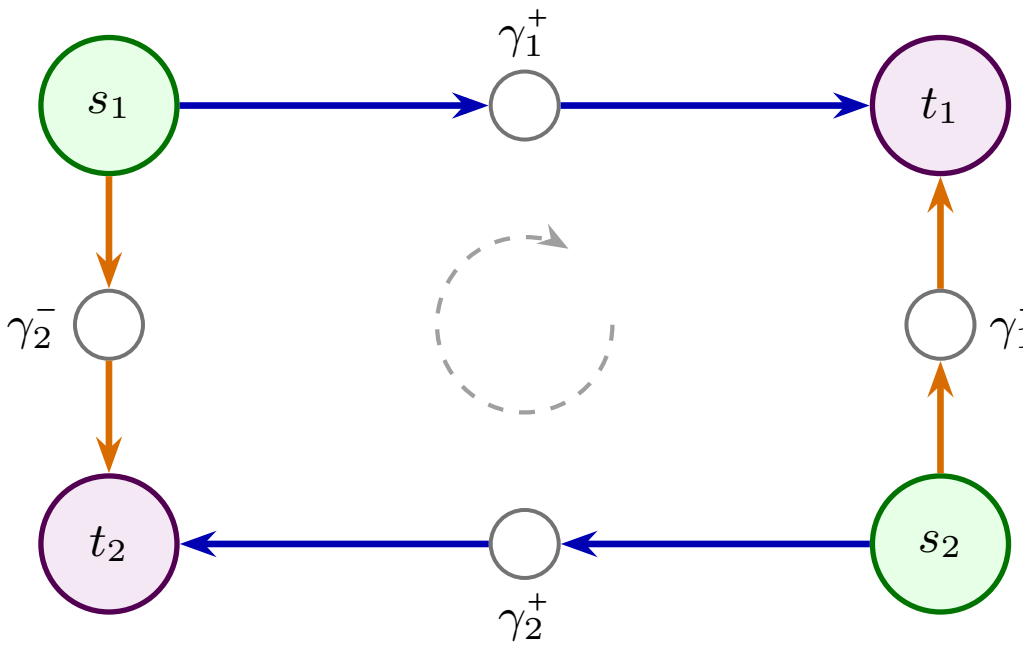

Divide the cycle whenever the sign changes. This decomposes its edges into maximal chains that are consistently oriented. At every division point, the two adjacent edges either both point outward or both point inward. We call the corresponding vertices sources and sinks, respectively. Sources and sinks alternate around the cycle, as illustrated in Figure~\ref{fig:unoriented-cycle-chains}.

Label the sources by $s_1,\ldots,s_r$ and the sinks by $t_1,\ldots,t_r$ in cyclic order. The chains whose edges carry coefficient $+1$ form directed paths $\gamma_k^+:s_k\rightarrow t_k$. A chain carrying coefficient $-1$ is traversed backward when defining the cycle vector, but its physical edges form a directed path $\gamma_k^-:s_{k+1}\rightarrow t_k$, where $s_{r+1}=s_1$. Thus, each sink is reached by one forward chain and one reversed chain:
\begin{equation}\label{eq:unoriented_chain_decomposition}
    s_k\xrightarrow{\gamma_k^+}t_k
    \xleftarrow{\gamma_k^-}s_{k+1},
    \qquad k=1,\ldots,r,
    \qquad s_{r+1}=s_1.
\end{equation}
Although the negative chains are traversed from $t_k$ to $s_{k+1}$ around the chosen cycle, their experiments are always performed in the physical direction $s_{k+1}\rightarrow t_k$. Their weights will later appear in the denominator.

To implement these path experiements, choose a fixed Pauli representative $P_{s}$ for each source pattern $s$ and $P_{t}$ for each sink pattern $t$. At each source, choose a preparation setting that ideally produces a $+1$ eigenstate $\rho_{s}$ of $P_{s}$. At each sink, choose a measurement setting for the observable $M_{t}=P_{t}$. The actual initial state and the effective measurement observable after SPAM error is written as $\widetilde{\rho}_{s}$ and $\widetilde{M}_{t}$ respectively. Use the same preparation setting for both chains leaving a given source and the same measurement setting for both chains entering a given sink. The associated SPAM errors are therefore the same in the paired experiments.

For a directed chain $\gamma:s\rightarrow t$, local Clifford gates connect $P_{s}$ to the initial Pauli operator of the chain and its terminating Pauli operator to $P_{t}$. Apply the character projection construction then gives:
\begin{equation}\label{eq:chain_path_quantity}
    W_{\gamma}=\dbraket{\widetilde{M}_{t}}{P_{t}}\dbraket{P_{s}}{\widetilde{\rho}_{s}}\prod_{e\in\gamma}g_e,
\end{equation}
where $\dbraket{P_s}{\widetilde{\rho}_s}$ is the unknown but fixed state-preparation overlap at the source, $\dbraket{\widetilde M_{t}}{P_{t}}$ is the unknown but fixed measurement overlap at the sink. The two endpoint factors depend only on the chosen source and sink settings, allowing them to cancel when the path signals are combined. We formulate the above protocol in Box~\ref{box:unoriented_cycle}.

\begin{mybox}[label=box:unoriented_cycle]{Learning an unoriented cycle}
    \begin{enumerate}
        \item Given an unoriented cycle $c$, assign $\pm 1$ signs to its edges, and split the cycle at every change of sign. Label the resulting chains as in Eq.~\eqref{eq:unoriented_chain_decomposition}, with $\gamma_k^+:s_k\rightarrow t_k$ and $\gamma_k^-:s_{k+1}\rightarrow t_k$.       
        \item At a source $s$, ideally prepare a $+1$ eigenstate of $P_s$. At a sink $t$, ideally measure the observable $P_{t}$. Denote the actually prepared state by $\widetilde{\rho_{s}}$ and the actually measured observable $\widetilde{M_{t}}$. Use the same noisy preparation or measurement setting whenever that vertex occurs as an endpoint.
        \item Perform the following character-projected path experiment for each of the $2r$ directed chains. For a fixed chain $\gamma:s\rightarrow t$ containing $d$ edges, relabel its edges in their physical direction as
        \begin{equation}
            \gamma=\bigl((P_0,Q_0),(P_1,Q_1),\ldots,(P_{d-1},Q_{d-1})\bigr).
        \end{equation}
        Note that $\pt(P_0)=\pt(P_s)$ and $\pt(Q_{d-1})=\pt(P_{t})$ are always be satisfied. Therefore there exists local Clifford transformations $C_{\mathrm{in}}$, $C_0,\ldots,C_{d-2}$, and $C_{\mathrm{out}}$ such that
        \begin{equation}
        \begin{split}
            \mathcal{C}_{\mathrm{in}}\dket{P_s}&=\dket{P_0},\\
            \mathcal{C}_{i}\dket{Q_i}&=\dket{P_{i+1}},\qquad 0\leq i\leq d-2,\\
            \mathcal{C}_{\mathrm{out}}\dket{Q_{d-1}}&=\dket{P_t}.
        \end{split}
        \end{equation}
        Sample independently and uniformly $d+1$ Pauli operators
        \begin{equation}
            \mathbf R=(R_0,R_1,\ldots,R_d)\in\mathsf P_n^{d+1},
        \end{equation}
        and implement the random path circuit
        \begin{equation}\label{eq:random_chain_circuit}
            \mathcal S_{\gamma}(\mathbf R)=\mathcal{C}_{\mathrm{out}}\mathcal{R}_{d}\mathcal{G}\left(\prod_{i=0}^{d-2}\mathcal{C}_i\mathcal{R}_{i+1}\mathcal{G}\right)\mathcal{C}_{\mathrm{in}}\mathcal{R}_0.
        \end{equation}
        Using the source state $\widetilde\rho_s$ and the sink measurement $\widetilde{M_t}$, estimate
        \begin{equation}
            f_{\gamma}(\mathbf R)
            =\dbra{\widetilde{M_t}}\mathcal S_{\gamma}(\mathbf R)\dket{\widetilde{\rho_s}}.
        \end{equation}
        Finally, average over the sampled Pauli operators with the corresponding character weights:
        \begin{equation}\label{eq:chain_character_average}
        \begin{split}
            W_{\gamma}
            &=\mathbb E_{\mathbf R\sim\mathsf P_n^{d+1}}
            \left[
                \chi(P_s,R_0)
                \left(\prod_{i=0}^{d-1}\chi(Q_i,R_{i+1})\right)
                f_{\gamma}(\mathbf R)
            \right]\\
            &=\dbraket{\widetilde{M_t}}{P_t}
              \dbraket{P_s}{\widetilde\rho_s}
              \prod_{i=0}^{d-1}\mathcal G_{Q_i,P_i}.
        \end{split}
        \end{equation}
        \item Combine the estimates as
        \begin{equation}\label{eq:unoriented_cycle_estimator}
            \widehat{f}_{x_{c'}}(\mathcal{G})=\frac{\prod_{k=1}^{r}W_{\gamma_k^+}}{\prod_{k=1}^{r}W_{\gamma_k^-}}.
        \end{equation}
    \end{enumerate}
\end{mybox}

To verify Eq.~\eqref{eq:unoriented_cycle_estimator}, substitute Eq.~\eqref{eq:chain_path_quantity} to obtain
\begin{equation}
\begin{split}
    \frac{\prod_{k=1}^{r}W_{\gamma_k^+}}{\prod_{k=1}^{r}W_{\gamma_k^-}}&=\frac{\prod_{k=1}^{r}\dbraket{\widetilde{M}_{t_k}}{P_{t_k}}\dbraket{P_{s_k}}{\widetilde{\rho_{s_k}}}}{\prod_{k=1}^{r}\dbraket{\widetilde{M}_{t_k}}{P_{t_k}}\dbraket{P_{s_{k+1}}}{\widetilde{\rho_{s_{k+1}}}}}\cdot\frac{\prod_{e:(x_{c'})_{e}=+1}g_e}{\prod_{e:(x_{c'})_{e}=-1}g_e}\\
    &=\frac{\prod_{e:(x_{c'})_{e}=+1}g_e}{\prod_{e:(x_{c'})_{e}=-1}g_e}.
\end{split}
\end{equation}
The second line is because the denominator contains the same sources as the numerator, only cyclically relabeled, and each sink occurs once in both, therefore returns the desired unoriented-cycle function $f_{x_{c'}}(\mathcal{G})$. Each reversed chain is accompanied by one forward chain, so a cycle with $r$ reversed chains requires $2r$ path experiments.

\subsection{Other parametrization}

The preceding subsections characterize learnable products of PTM entries and provide protocols for estimating them. In experiments, however, the quantities of interest are often parameters of a specified noise model. For an implemented channel $\mathcal G_\theta=\Lambda_\theta\mathcal U$, where $\mathcal U$ is the prescribed ideal unitary channel, we therefore seek learnable functions of the noise parameters $\theta$. We consider models in which each PTM entry is a known constant times a product of parameter powers. This form allows us to translate PTM cycle quantities into a family of learnable parameter functions.

\begin{definition}[Product-entry channel model]\label{def:product-entry}
    Suppose a noisy quantum channel is in the form of:
    \begin{equation}
        \mathcal{G}_{\theta}=\Lambda_{\theta}\mathcal{U},
    \end{equation}
    where $U$ is the target unitary, $\Lambda_{\theta}$ is a noise channel parametrized by $\theta=(\theta_1,\cdots,\theta_{k})\in\mathbb{R}^{k}$. The channel $\mathcal{G}_{\theta}$ is said to satisfy the product-entry channel model if the Pauli transfer matrix of $\mathcal{G}$ satisfies:  
    \begin{equation}
        \mathcal{G}_{i,j}=C_{ij}\prod_{l=1}^{k}\theta_{l}^{m_{ij,l}},
    \end{equation}
    where $C_{ij}\in\mathbb{R},m_{ij,l}\in\mathbb{Z}_{(2)}$ are constants. 
\end{definition}

The advantage of the product-entry model is that the dependence of the PTM entries on the model parameters becomes linear after taking signed logarithms. We restrict attention to a parameter region on which every $\theta_l$ is nonzero and use the signed logarithm. Let $e_1,\ldots,e_{|E|}$ be the ordered edges of the Pauli transfer graph. Write $C_a\neq0$ and $m_{a,l}$ for the constants associated with the PTM entry on edge $e_a$. Define the signed-logarithmic parameter and PTM-entry vectors by
\begin{equation}
    \vartheta=(\operatorname{slog}(\theta_1),\ldots,\operatorname{slog}(\theta_k))^{\mathsf T},
    \qquad
    \widetilde g_a=\operatorname{slog}\left(\frac{g_a}{C_a}\right).
\end{equation}
The product-entry relation then gives
\begin{equation}
    \widetilde g_a=\sum_{l=1}^{k}m_{a,l}\vartheta_l.
\end{equation}

\begin{definition}[Logarithmic parameter map]
    The logarithmic parameter map of a product-entry channel model is the $\mathbb Z_{(2)}$-linear map
    \begin{equation}
        \mathcal Q:(\mathbb R\oplus\mathbb F_2)^k\longrightarrow(\mathbb R\oplus\mathbb F_2)^{|E|},
        \qquad
        \widetilde g=\mathcal Q\vartheta,
    \end{equation}
    whose matrix elements are
    \begin{equation}
        \mathcal Q_{a,l}=m_{a,l}.
    \end{equation}
\end{definition}
Matrix multiplication here uses the scalar action in Eq.~\eqref{eq:Z2-action-slog}. The matrix entries act on logarithmic magnitudes in the usual way and on sign bits through their reduction modulo $2$. The sufficient learnability result below does not require this map to be injective.

The same coefficient matrix also relates the exponents of PTM products to those of parameter products. Its transpose acts on exponent vectors as
\begin{equation}
    \mathcal Q^{\mathsf T}:\mathbb Z_{(2)}^{|E|}
    \longrightarrow\mathbb Z_{(2)}^k,
    \qquad y=\mathcal Q^{\mathsf T}x.
\end{equation}
Here the matrix entries multiply exponents within $\mathbb Z_{(2)}$, whereas $\mathcal Q\vartheta$ uses the scalar action on signed-logarithm pairs. These two operations are related by
\begin{equation}\label{eq:parameter_pullback}
    \langle x,\widetilde g\rangle
    =\langle x,\mathcal Q\vartheta\rangle
    =\langle\mathcal Q^{\mathsf T}x,\vartheta\rangle.
\end{equation}
Thus, a PTM function represented by $x$ induces a parameter function represented by $\mathcal Q^{\mathsf T}x$.

\begin{theorem}[Learnable parameter functions]
    Under the assumptions of Theorem~\ref{thm:general_cptp_learnability}, a parameter product $f_y(\theta)=\prod_{l=1}^k\theta_l^{y_l}$ is learnable if
    \begin{equation}\label{eq:parametric_learnable_space}
        y=\mathcal Q^{\mathsf T}x
        \quad\text{for some}\quad
        x\in Z\bigl(\Gamma_{\pt}(\mathcal G)\bigr).
    \end{equation}
\end{theorem}

\begin{proof}
    Choose $x\in Z\bigl(\Gamma_{\pt}(\mathcal G)\bigr)$ and set $y=\mathcal Q^{\mathsf T}x$. Both $x_a$ and $m_{a,l}$ lie in $\mathbb Z_{(2)}$, so $y_l=\sum_a m_{a,l}x_a$ also lies in this ring. The canonical real powers obey the product and power rules established in Sec.~\ref{subsec:complex-logarithm-preliminaries}, including for negative parameter values. Consequently,
    \begin{equation}
    \begin{split}
        \prod_{a=1}^{|E|}\left(\frac{g_a}{C_a}\right)^{x_a}
        &=\prod_{a=1}^{|E|}\left(\prod_{l=1}^k\theta_l^{m_{a,l}}\right)^{x_a}\\
        &=\prod_{l=1}^k\theta_l^{\sum_a m_{a,l}x_a}
        =\prod_{l=1}^k\theta_l^{y_l}.
    \end{split}
    \end{equation}
    Since $x$ belongs to the cycle space over $\mathbb Z_{(2)}$, the PTM product $\prod_a g_a^{x_a}$ is gauge invariant by Theorem~\ref{thm:general_cptp_learnability}. The factors $C_a$ are known and nonzero, and their powers have unique real values under the same convention. Dividing by $\prod_a C_a^{x_a}$ therefore gives a learnable parameter function with exponent vector $y$.
\end{proof}

The parameter product can be estimated by measuring the corresponding PTM product and dividing by the known factor $\prod_a C_a^{x_a}$. This follows the reduced-model approach of Ref.~\cite{Chen2026}. In practice, if $\{x_1,\ldots,x_D\}$ is a basis of the PTM cycle space, then the vectors
\begin{equation}
    \mathcal Q^{\mathsf T}x_1,\ldots,\mathcal Q^{\mathsf T}x_D,
\end{equation}
generate $\mathcal Q^{\mathsf T}Z\bigl(\Gamma_{\pt}(\mathcal G)\bigr)$ over $\mathbb Z_{(2)}$. Products, inverses, and real odd roots of the corresponding cycle quantities therefore give this family of learnable parameter functions.

\section{Learning Hamiltonian evolution}
\label{sec:Hamiltonian}

The preceding section treats the implemented evolution as a general quantum channel. On mainstream quantum platforms, including trapped ions, neutral atoms, and superconducting circuits, gates are implemented through controlled evolution over prescribed durations, possibly in conjunction with additional control operations~\cite{Leibfried2003GeometricGate,Isenhower2010NeutralAtomCNOT,DiCarlo2009SuperconductingProcessor,Sung2021TunableCouplerGates}. Calibration experiments can vary this duration and thereby probe a family of channels generated by the same underlying dynamics. In addition, some quantum platforms are designed for analog operation, evolving continuously rather than through sequences of digital gates~\cite{Bernien2017QuantumSimulator}, and offer opportunities for analog quantum applications~\cite{xu2026AEDP,Liu2024HShadow} that likewise require calibration. These observations raise a natural question: what additional information becomes learnable when the evolution time can be varied and the structure of the generator is taken into account?

We address this question for Hamiltonian evolution in this section and for Lindbladian evolution in the next. A valid gauge must preserve the assumed form of the dynamics throughout the family of evolution times. We first determine the restrictions this requirement places on the gauge, and then use the remaining gauge freedom to characterize the learnable properties of these dynamics.

\subsection{Hamiltonian gauge freedom}

We consider evolution under a time-independent Hamiltonian, allowing coherent errors such as an incorrect interaction strength or an unwanted Hamiltonian term, but no dissipation. After a gauge transformation, a single Hamiltonian $H'$ must still describe the experiments at all evolution times. This requirement places additional restrictions on the gauge.

Let $H$ be the implemented Hamiltonian and define the associated Liouvillian by $\mathcal L_H(\rho)=-i[H,\rho]$. The state evolves according to
\begin{equation}
    \frac{\mathrm d\rho}{\mathrm dt}=\mathcal L_H(\rho).
\end{equation}
Denote the Pauli--Liouville matrix of this evolution channel by $\mathcal U_t^{(H)}$. Using $\mathcal L_H$ also for the matrix of the Liouvillian in the same representation, we have
\begin{equation}
    \dket{\rho(t)}=\mathcal U_t^{(H)}\dket{\rho(0)},\qquad
    \mathcal U_t^{(H)}=e^{t\mathcal L_H}.
\end{equation}
Since conjugation on a matrix exponentiation equals the exponentiation of the conjugated matrix, 
the gauge transformation of the evolution channel induces the same conjugation on its Liouvillian,
\begin{equation}
    T\mathcal U_t^{(H)}T^{-1}
    =T e^{t\mathcal L_H}T^{-1}
    =e^{t(T\mathcal L_HT^{-1})}.
\end{equation}
The transformed Liouvillian $T\mathcal{L}_{H}T^{-1}$ describes Hamiltonian evolution for all times if and only if there exists a single time-independent Hamiltonian $H'$ such that
\begin{equation}\label{eq:Hamiltonian-gauge-equivalency}
    T\mathcal L_HT^{-1}=\mathcal L_{H'}.
\end{equation}
We call $H$ and $H'$ gauge equivalent if their Liouvillians satisfy Eq.~\eqref{eq:Hamiltonian-gauge-equivalency}. 

To determine the restrictions imposed by Eq.~\eqref{eq:Hamiltonian-gauge-equivalency}, expand $H=\sum_{a}c_aP_a$ with real coefficients $c_a$. We omit the identity term, which does not affect the evolution. The Liouvillian acts on a Pauli operator $P_b$ as
\begin{equation}\label{eq:original-Hamiltonian}
\begin{split}
    \mathcal L_H(P_b)&=-i\sum_a c_a[P_a,P_b]\\
    &=-2i\sum_{a:\{P_a,P_b\}=0}c_aP_aP_b.
\end{split}
\end{equation}
For each Hamiltonian term $c_aP_a$, the coefficient of $P_aP_b$ is $-2ic_a$, independent of the anticommuting input $P_b$. This independence must be preserved for the transformed Liouvillian to describe another Hamiltonian. A diagonal gauge of the form in Eq.~\eqref{eq:gauge_schur} gives
\begin{equation}\label{eq:gauge-Hamiltonian}
    (T\mathcal L_HT^{-1})(P_b)
    =-2i\sum_{a:\{P_a,P_b\}=0}
    c_a\frac{\lambda_{\pt(P_{a\oplus b})}}{\lambda_{\pt(P_b)}}P_aP_b.
\end{equation}
For fixed $b$, the operators $P_aP_b$ are linearly independent as $a$ varies. Comparing their coefficients in Eqs.~\eqref{eq:original-Hamiltonian} and~\eqref{eq:gauge-Hamiltonian}, a single transformed coefficient $c_a'$ exists only if the gauge ratio $\lambda_{\pt(P_{a\oplus b})}/\lambda_{\pt(P_b)}$ multiplying $c_a$ is independent of $b$. This yields the following condition.

\begin{lemma}[Gauge restriction for Hamiltonian evolution]\label{lemma:Hamiltonian_gauge}
    Let $H=\sum_{a\neq0}c_aP_a$ be a Hamiltonian with Liouvillian $\mathcal L_H(\rho)=-i[H,\rho]$. A diagonal gauge $T$ transforms $\mathcal L_H$ into another Hamiltonian Liouvillian if and only if, for every $a$ with $c_a\neq0$, there exists a constant $\mu_a$ such that
    \begin{equation}\label{eq:Hamiltonian_gauge_ratio}
        \frac{\lambda_{\pt(P_{a\oplus b})}}{\lambda_{\pt(P_b)}}=\mu_a
        \qquad\text{for every $b$ satisfying }\{P_a,P_b\}=0.
    \end{equation}
    In this case, $T\mathcal L_HT^{-1}=\mathcal L_{H'}$, where, up to an additive multiple of the identity,
    \begin{equation}
        H'=\sum_{a:c_a\neq0}\mu_a c_aP_a.
    \end{equation}
\end{lemma}

Suppose $P_a$ anticommutes with $P_b$, then it also anticommutes with $P_{a\oplus b}$, applying Lemma~\ref{lemma:Hamiltonian_gauge} to these two inputs gives
\begin{equation}
    \mu_a=\frac{\lambda_{\pt(P_{a\oplus b})}}{\lambda_{\pt(P_b)}}=\frac{\lambda_{\pt(P_b)}}{\lambda_{\pt(P_{a\oplus b})}}.
\end{equation}
Here the second equal sign is because $P_{a\oplus (a\oplus b)}=P_{b}$. Multiplying the two ratios gives $\mu_a^2=1$. Thus a Hamiltonian-preserving gauge can change a nonzero coefficient only by a sign. The gauge factors themselves need not have unit magnitude, but their magnitudes agree in each ratio and cancel. Separating each gauge factor into its magnitude and sign therefore distinguishes the freedom that leaves the Hamiltonian unchanged from the possible changes in its coefficients.

\begin{theorem}[Hamiltonian-preserving gauges]\label{thm:Hamiltonian_gauge_space}
    Suppose Hamiltonian $H$ is in the form of $H=\sum_{a}c_aP_a$, with $\mathcal L_H(\rho)=-i[H,\rho]$ being its Liouvillian. For a diagonal gauge $T$, write $T=DS$ with 
    \begin{equation}\label{eq:Hamiltonian_gauge_decomposition}
        D=\sum_b d_{\pt(P_b)}\dketbra{P_b}{P_b},
        \qquad
        S=\sum_b(-1)^{s_{\pt(P_b)}}\dketbra{P_b}{P_b}.
    \end{equation}
    where $d_{u}>0$ and $s_{u}\in\mathbb{F}_2$ for each pattern $u$. 
    Then $T\mathcal L_HT^{-1}$ is a Hamiltonian Liouvillian if and only if $[D,\mathcal L_H]=0$ and, for each $a$ with $c_a\neq0$, there exists $t_a\in\mathbb F_2$ such that
    \begin{equation}\label{eq:Hamiltonian_sign_constraint}
        s_{\pt(P_{a\oplus b})}+s_{\pt(P_b)}=t_a
        \quad\text{whenever }\{P_a,P_b\}=0.
    \end{equation}
    In this case the gauge transformed Hamiltonian $H'$ satisfying $\mathcal{L}_{H'}=T\mathcal{L}_{H}T^{-1}$ is,
    \begin{equation}
    \begin{split}
        H'&=\sum_{a:c_a\neq0}(-1)^{t_a}c_aP_a,
    \end{split}
    \end{equation}
    up to an additive multiple of the identity in $H'$.
\end{theorem}

\begin{proof}
    If $T$ preserves Hamiltonian evolution, applying Lemma~\ref{lemma:Hamiltonian_gauge} to the inputs $P_b$ and $P_{a\oplus b}$ gives $\mu_a=\pm1$ for every nonzero coefficient. Taking absolute values in Eq.~\eqref{eq:Hamiltonian_gauge_ratio} gives $d_{\pt(P_{a\oplus b})}=d_{\pt(P_b)}$ on every nonzero matrix entry of $\mathcal L_H$, which is equivalent to $[D,\mathcal L_H]=0$. Writing $\mu_a=(-1)^{t_a}$ then gives Eq.~\eqref{eq:Hamiltonian_sign_constraint}. Conversely, these two conditions make the gauge ratio equal to $(-1)^{t_a}$ for every input anticommuting with $P_a$. Lemma~\ref{lemma:Hamiltonian_gauge} yields the stated Hamiltonian $H'$. Since $D$ and $S$ commute, the factor $D$ cancels from the transformed Liouvillian.
\end{proof}

Different Hamiltonian terms can involve the same gauge factors, so their coefficient signs cannot in general change independently. We next examine these relations to determine which signs are fixed and which can change together.

\subsection{Learnability of Hamiltonian coefficients}

To characterize the remaining sign freedom, we adapt the pattern transfer graph to Hamiltonian evolution. Each nonzero term $c_aP_a$ gives a transfer from $P_b$ to $P_{a\oplus b}$ whenever $P_a$ and $P_b$ anticommute. The following graph records these transfers together with their associated Hamiltonian terms.

\begin{definition}[Hamiltonian pattern transfer graph]\label{def:Hamiltonian_pattern_graph}
    Consider a Hamiltonian $H$ in the form $H=\sum_{a}c_{a}P_a$. Let $\mathcal A_H=\{a\neq0:c_a\neq0\}$ denote its Pauli support. The Hamiltonian pattern transfer graph $\Gamma_H=(V_H,E_H)$ is a directed multigraph with vertex set $V_H=\{0,1\}^n$. For every $a\in\mathcal A_H$ and every $b$ satisfying $\{P_a,P_b\}=0$, it contains an edge
    \begin{equation}
        e_{a,b}:\pt(P_b)\longrightarrow\pt(P_{a\oplus b})
    \end{equation}
    labeled by $a$.
\end{definition}

For an odd-weight term $P_a$, choose $P_b$ with the same support and a different nonidentity Pauli on each supported qubit. The two operators anticommute, while $P_b$ and $P_{a\oplus b}$ have the same support. This gives a self-loop labeled by $a$ in $\Gamma_H$. Combined with Lemma~\ref{lemma:Hamiltonian_gauge}, the self-loop enforces the factor $\mu_a=1$. Therefore $c_a$ is gauge invariant. This establishes the following corollary.

\begin{corollary}\label{cor:odd-Hamiltonian}
    For $H=\sum_{a}c_aP_a$, every coefficient $c_a$ with $\operatorname{wt}(P_a)$ odd is learnable.
\end{corollary}

For even-weight terms, the individual coefficient signs can remain ambiguous. These signs cannot always change independently, however. This is because the gauge assigns the same value to all Pauli operators with the same support, while a Hamiltonian coefficient must acquire the same sign along every edge carrying its label. Edges with different labels can share support patterns, so a sign choice for one coefficient can also constrain the signs of other coefficients. To describe how overlapping Hamiltonian supports give rise to these relations, we introduce the support hypergraph.

\begin{definition}[Hamiltonian support hypergraph]
\label{def:Hamiltonian_support_hypergraph}
    Let $H=\sum_a c_aP_a$ and let $\mathcal A_H$ denote the labels
    of its nonzero nonidentity terms. The support hypergraph
    $\mathscr H_H$ has vertex set $[n]$ denotes the $n$ qubits and one hyperedge
    \begin{equation}
        A_a=\operatorname{supp}(P_a)
    \end{equation}
    for each $a\in\mathcal A_H$. Two qubits belong to the same
    connected component if they can be joined by a sequence
    of overlapping hyperedges.
\end{definition}

Consider a group of qubits connected through overlapping Hamiltonian supports, and let $C\subseteq[n]$ be the corresponding connected component of $\mathscr H_H$. Define $U$ to be the Pauli support pattern that is nonzero on every qubit in $C$ and zero outside $C$. 

For an even-weight term $P_a$ with support $A_a\subseteq C$, choose $j\in A_a$ and an input Pauli $P_b$ supported on $U\backslash\{j\}$such that it anticommutes with $P_a$ on every qubit in $A_a\backslash{j}$. Operators $P_a$ and $P_b$ then anticommute on $|A_a|-1$ qubits, an odd number, and hence anticommute globally. Their product is nonidentity at $j$ and remains nonidentity on $A_a\backslash\{j\}$, while its support outside $A_a$ is unchanged. Thus the output pattern is $U$, giving an $a$-labeled edge from $U\backslash\{j\}$ to $U$. By Lemma~\ref{lemma:Hamiltonian_gauge}, the sign acquired by $c_a$ is therefore
\begin{equation}\label{eq:r}
    \mu_a=r_j,\qquad r_j:=\frac{\lambda_U}{\lambda_{U\backslash\{j\}}}.
\end{equation}
This construction applies to every $j\in A_a$, so all ratios $r_j$ on the support of an even-weight term are equal. In particular, two even-weight terms containing the same qubit $j$ must acquire the same sign, since both signs equal $r_j$.

Odd-weight terms also relate these ratios, although their own coefficients are gauge invariant. Let $P_a$ have odd weight and choose distinct $j,k\in A_a$. Take $P_b$ supported on $U\backslash\{j\}$ such that $P_b$ anticommutes with $P_a$ on qubits $i\in A_a\backslash\{j,k\}$ but satisfies $(P_b)_k=(P_a)_k$ on qubit $k$. The operators $P_a,P_b$ anticommute on $|A_a|-2$ qubits, again an odd number. In their product, the factor at $j$ becomes nonidentity and the factor at $k$ becomes the identity, while all other qubits retain their support. This gives an $a$-labeled edge from $U\backslash\{j\}$ to $U\backslash\{k\}$. Corollary~\ref{cor:odd-Hamiltonian} gives $\mu_a=1$, so Lemma~\ref{lemma:Hamiltonian_gauge} requires $\lambda_{U\backslash\{j\}}=\lambda_{U\backslash\{k\}}$. By Eq.~\eqref{eq:r}, the ratios $r_j$ and $r_k$ have these equal values as denominators and the same numerator $\lambda_U$, yielding $r_j=r_k$.

Thus the ratios $r_j$ agree on each Hamiltonian support, including odd-weight supports that connect different even-weight terms. Following a sequence of overlapping supports extends this equality throughout $C$. All even-weight coefficients in the component must therefore acquire the same sign. The following corollary also establishes that the signs for different components can be chosen independently.

\begin{corollary}\label{cor:even-Hamiltonian}
    Let $C(a)$ be the connected component of $\mathscr H_H$ containing $A_a$. The action of Hamiltonian-preserving gauges on the nonzero even-weight coefficients is
    \begin{equation}\label{eq:Hamiltonian_component_sign}
        c_a'=\sigma_{C(a)}c_a,
        \qquad \operatorname{wt}(P_a)\text{ even}.
    \end{equation}
    The signs $\sigma_C\in\{+1,-1\}$ can be chosen independently for each component containing even-weight terms, assuming the corresponding gauges preserve SPAM physicality. In particular, the magnitudes and the relative signs of even-weight coefficients within each component are learnable.
\end{corollary}

\begin{proof}
    The preceding argument establishes the common sign within each component. To show that these signs are independent, choose $\sigma_C\in\{+1,-1\}$ for every component containing even-weight terms, set $\sigma_C=1$ for the remaining components, and define
    \begin{equation}
        \lambda_u=\prod_C\sigma_C^{|u\cap C|}.
    \end{equation}
    Along an $a$-labeled edge from $u$ to $v$, the patterns agree outside $A_a$. Within $A_a$, a site belongs to both patterns when the input Pauli anticommutes locally with $P_a$, and to exactly one pattern otherwise. Anticommutation therefore requires an odd number of sites in $A_a\cap u\cap v$. Consequently,
    \begin{equation}
        |v\cap C(a)|-|u\cap C(a)|
        \equiv \operatorname{wt}(P_a)-1\pmod 2.
    \end{equation}
    The gauge ratio is therefore $1$ for every odd-weight label and $\sigma_{C(a)}$ for every even-weight label. By Lemma~\ref{lemma:Hamiltonian_gauge}, this gauge preserves Hamiltonian evolution and realizes the prescribed component signs.

    Note that the SPAM physicality assumption ensures that every componentwise sign choice is physically admissible. Without it, SPAM physicality may exclude some choices, but the invariance of coefficient magnitudes and relative signs within each component still holds.
\end{proof}

The common sign ambiguity leaves both $c_ac_{a'}$ and $c_a/c_{a'}$ invariant for nonzero even-weight coefficients in the same component. Thus relative signs remain learnable even when individual signs are not fixed.

For example, consider a Hamiltonian with nonzero coefficients for every nonidentity Pauli operator, on $n\geq2$ qubits. Its support hypergraph is globally connected. Writing $H=H_{\mathrm{odd}}+H_{\mathrm{even}}$, where the two sums contain the odd- and even-weight terms respectively, the only two candidates consistent with the gauge ambiguity are
\begin{equation}
    H_{\mathrm{odd}}+H_{\mathrm{even}},
    \qquad H_{\mathrm{odd}}-H_{\mathrm{even}}.
\end{equation}
For a chain Hamiltonian $H=\sum_iJ_iZ_iZ_{i+1}+\sum_ih_iX_i$ with all $J_i\neq0$, the coefficients $h_i$, magnitudes $|J_i|$, and ratios $J_i/J_j$ are learnable. If some $J_i$ vanish, the relative signs remain learnable within each connected segment.

The odd–even distinction has a physical interpretation through spin reversal. For a single spin, reversal sends the Bloch vector to its opposite and commutes with every spin rotation. Its extension to $n$ qubits is
\begin{equation}
\mathcal R(A)=Y^{\otimes n}A^T Y^{\otimes n},
\end{equation}
where transposition is taken in the computational basis. On Hermitian operators, this is the action of spin-$\tfrac12$ time reversal \cite{Rungta2001Universal}. Each non-identity Pauli factor changes sign, giving $\mathcal R(P)=(-1)^{\mathrm{wt}(P)}P$. Transposition also reverses the order of products in a commutator, so the transformed dynamics is generated by
\begin{equation}
H'=-\mathcal R(H)=H_{\mathrm{odd}}-H_{\mathrm{even}}.
\end{equation}
The two sign changes therefore cancel for odd-weight terms and reinforce for even-weight terms.

The remaining ambiguity is highly restricted, consisting only of a common sign change of the even-weight coefficients in each connected component. Additional assumptions can therefore remove this ambiguity and make all Hamiltonian coefficients learnable. For example, existing Hamiltonian-learning protocols achieve efficient estimation of all coefficients under suitable constant bounds on SPAM errors~\cite{Huang_2023,Ma_2026}. Our first-order analysis in Sec~\ref{subsec:first-order-Lindbladian-errors} provides another instance, showing all Hamiltonian coefficients are learnable in a pertubasive model. These results show that Hamiltonian structure sharply restricts the gauge freedom, while additional information of SPAM can remove the remaining ambiguity and make individual coefficients learnable.

\subsection{Heisenberg scaling learning protocols}
\label{subsec:Hamiltonian-protocol}

We next construct protocols for estimating the gauge-invariant quantities identified in the preceding subsection. These protocols can address SPAM errors beyond the bounds assumed in previous Hamiltonian-learning work~\cite{Huang_2023,Ma_2026}. As in Sec.~\ref{subsec: CPTP-protocol}, Pauli character projection isolates a closed path in the pattern transfer graph. Repeating the corresponding circuit produces an exponential signal whose depth dependence determines the path weight, while SPAM enters through a depth-independent amplitude. Exponential fitting and robust phase estimation then recover the learnable quantities with Heisenberg scaling.

We first estimate the coefficients magnitude of Pauli terms one at a time. For a target term $c_aP_a$, we can first use Hamiltonian reshaping  to isolate the effective Hamiltonian $H_{\mathrm{eff}}=c_aP_a$ by fast single-qubit Pauli control, removing all other unwanted Pauli terms. Appendix~\ref{app:Hamiltonian-reshaping} discusses how to efficiently implement this Hamiltonian reshaping in detail.

The effective evolution under $H_{\mathrm{eff}}=c_aP_a$ allows us to estimate $|c_a|$ through a cosine signal. We first apply character projection to select a Pauli operator $P_b$ that anticommutes with $P_a$, then evolve it for time $\tau$. Writing $Q=-iP_aP_b$, which is Hermitian, and $\mathcal L_{\mathrm{eff}}(\rho)=-ic_a[P_a,\rho]$, we obtain
\begin{equation}\label{eq:Hamiltonian-protocol-Pauli-evolution}
    e^{\tau\mathcal L_{\mathrm{eff}}}\dket{P_b}
    =\cos(2c_a\tau)\dket{P_b}+\sin(2c_a\tau)\dket{Q}.
\end{equation}
Applying the $P_b$ character projector again selects the unchanged Pauli component, forming a cycle with weight $\cos(2c_a\tau)$. Repeating this cycle to depth $m$ gives
\begin{equation}\label{eq:Hamiltonian-protocol-cosine-signal}
    F_{\cos}(\tau,m)=A\cos^m(2c_a\tau),
\end{equation}
where $A=\dbraket{\widetilde M}{P_b}\dbraket{P_b}{\rho}$, with $\rho$ and $\widetilde M$ denoting the actual prepared state and measured observable. The unknown SPAM errors enter through the depth-independent amplitude $A$, so fitting the depth dependence determines the cosine signal. This construction applies to both odd- and even-weight terms and allows $|c_a|$ to be estimated by varying $\tau$, as described in Appendix~\ref{app:cosine-only-estimation}.

For odd-weight $P_a$, the sine signal can also be extracted. Choose $P_b$ with the same support as $P_a$ and anticommuting with it on every supported qubit, as in the self-loop construction preceding Corollary~\ref{cor:odd-Hamiltonian}. Then $Q$ has the same support as $P_b$. A local Clifford $C$ can therefore satisfy $CQC^\dagger=P_b$, including the known sign of $Q$. Selecting $Q$ after the evolution and applying $C$ closes a cycle with weight $\sin(2c_a\tau)$, giving
\begin{equation}\label{eq:Hamiltonian-protocol-sine-signal}
    F_{\sin}(\tau,m)=A\sin^m(2c_a\tau).
\end{equation}
Together, the sine and cosine estimates determine the signed coefficient $c_a$ through robust phase estimation from Sec.~\ref{subsec:robust-phase-estimation}. The corresponding experiments are summarized in Box~\ref{box:odd-Pauli}.

\begin{mybox}[label=box:odd-Pauli]{Learning odd weight Pauli operators}
    \begin{enumerate}
        \item Given the target Hamiltonian $P_{a}$, select a Hamiltonian $P_{b}$ such that $\pt(P_a)=\pt(P_b)=\pt(P_aP_b)$ and $\{P_a,P_b\}=0$. This can always be achieved for odd weight $P_{a}$. Let $Q=-iP_aP_b$ and choose a local Clifford $C$ such that
        \begin{equation}
            \mathcal{C}\dket{Q}=\dket{P_b}.
        \end{equation}

        \item Ideally prepare a $+1$ eigenstate of $P_b$ and measure the observable $P_b$. Denote the actually prepared state by $\rho$ and the actually measured observable $\widetilde{M}$. Assume that these SPAM operations are fixed across all circuits and depths, and that $\dbraket{P_b}{\rho}$ and $\dbraket{\widetilde M}{P_b}$ are non-negligible.

        \item Suppose $|2c_a|\leq\Omega$. Choose $\tau_0$ such that $\Omega\tau_0\leq\pi/2$ and use evolution times $\tau_k=2^k\tau_0$ for $k=0,\ldots,K$, with $\tau_K=\Theta(1/\varepsilon)$ for target coefficient precision $\varepsilon$.

        \item Fix a time scale $k$ and positive integer depth $m$. For every repetition $\ell\in\{1,\ldots,m\}$, sample independently and uniformly Pauli operators $R_{\ell,0},R_{\ell,1}\in \mathsf{P}_{n}$. Denote $\mathbf{R}_{\ell}=(R_{\ell,0},R_{\ell,1})$. Implement the following block circuit
        \begin{equation}\label{eq:Hamiltonian_RB_circuit}
            \mathcal{S}^{(k)}(\mathbf{R}_{\ell})=\mathcal{C}\mathcal{R}_{\ell,1}e^{\mathcal{L}_{\mathrm{eff}}\tau_k}\mathcal{R}_{\ell,0}.
        \end{equation}
        Here $\mathcal{L}_{\mathrm{eff}}(\rho)=-ic_a[P_{a},\rho]$ is the Lindbladian operator for the effective Hamiltonian. $e^{\mathcal{L}_{H}t}$ is the Pauli Liouville representation of evolving Hamiltonian $H$ for time $t$. Writing $\mathbf{R}=(\mathbf{R}_1,\ldots,\mathbf{R}_m)$ for the complete random sequence, implement the depth-$m$ circuit
        \begin{equation}
            \mathcal{S}_{m}^{(k)}(\mathbf{R})=\prod_{\ell=1}^{m}\mathcal{S}^{(k)}(\mathbf{R}_{\ell}).
        \end{equation}
        Measure the expectation of noisy observable $\widetilde{M}$ as: 
        \begin{equation}
            f(k,m,\mathbf{R})=\dbra{\widetilde{M}}\mathcal{S}_{m}^{(k)}\dket{\rho}.
        \end{equation}

        \item For each depth $m$, sample different Pauli operators $\mathbf{R}$ and estimate the character weighted expectation:
        \begin{equation}\label{eq:Hamiltonian_RB_data}
            F(k,m)=\mathbb{E}_{\mathbf{R}\sim\mathsf{P}_{n}^{2m}}\left[\left(\prod_{\ell=1}^{m}\chi(P_{b},\mathcal{R}_{\ell,0})\chi(P_{a}P_{b},\mathcal{R}_{\ell,1})\right)f(k,m,\mathbf{R})\right].
        \end{equation}

        \item At each time $\tau_k$, repeat the estimation for several depths $m$, including both even and odd depths, and fit $F(k,m)$ to the single-exponential model
        \begin{equation}
            F(k,m)=A_k r_k^m.
        \end{equation}
        The fitted weight estimates $\sin(2c_a\tau_k)$. To estimate $\cos(2c_a\tau_k)$, omit $\mathcal C$ in Eq.~\eqref{eq:Hamiltonian_RB_circuit} and replace the output character in Eq.~\eqref{eq:Hamiltonian_RB_data} by $\chi(P_b,\mathcal R_{\ell,1})$. Estimate both signals to the fixed additive accuracy specified in Eq.~\eqref{eq:prelim-RPE-signal-accuracy}.

        \item Apply robust phase estimation as described in Sec.~\ref{subsec:robust-phase-estimation} to the fitted sine and cosine signals across different time scale $\tau_{k}$ to estimate $c_a$ with precision $\varepsilon$ in the Heisenberg scaling.
    \end{enumerate}
\end{mybox}

It remains to determine the relative signs of even-weight coefficients within each connected component of the support hypergraph. We call two even-weight terms neighbors if their supports overlap or can be joined through a sequence of odd-weight Hamiltonian supports. Any two even-weight terms in a connected component can be linked by a sequence of neighboring pairs. It therefore suffices to determine the relative sign within each such pair.

Consider neighboring terms $c_aP_a$ and $c_{a'}P_{a'}$. We construct a cycle containing an edge of weight $\sin(2c_a\tau)$ and an edge of weight $\sin(2c_{a'}\tau)$. If the supports of $P_a$ and $P_{a'}$ overlap, these two edges alone form the cycle. Otherwise, the cycle also includes one edge of weight $\sin(2c_{o_\ell}\tau)$ for each odd-weight term $P_{o_\ell}$ used to connect their supports. Since the odd-weight coefficients are already learnable, their sine factors can be determined separately. Dividing these factors out of the cycle weight gives $\sin(2c_a\tau)\sin(2c_{a'}\tau)$, which, together with the individual cosine signals, determines the relative sign of the two coefficients. Box~\ref{box:even-Pauli-cycle} gives the cycle construction.

\begin{mybox}[label=box:even-Pauli-cycle]{Constructing a cycle for two neighboring even-weight terms}
    \begin{enumerate}
        \item Let $P_a$ and $P_{a'}$ be the two even-weight terms, and let
        $S$ be the union of all supports in their connected
        component of $\mathscr H_H$. Choose a reduced sequence of qubits
        $i_0,i_1,\ldots,i_k$ and odd-weight Hamiltonian terms
        $P_{o_1},\ldots,P_{o_k}$ such that
        \begin{equation}\label{eq:odd-support-path}
        \begin{split}
            i_0&\in\operatorname{supp}(P_a),\\
            i_k&\in\operatorname{supp}(P_{a'}),\\
            i_{\ell-1},i_\ell
            &\in\operatorname{supp}(P_{o_\ell})
            \quad\text{for every }\ell\in\{1,\ldots,k\}.
        \end{split}
        \end{equation}
        Require $i_{\ell-1}\neq i_\ell$ for every $\ell$. Such a sequence is
        obtained from any support-hypergraph path from $P_a$ to $P_{a'}$ whose
        internal hyperedges have odd weight: choose a qubit from every
        intersection of consecutive supports and remove steps for which the
        chosen qubit does not change. If $P_a$ and $P_{a'}$ intersect directly,
        one may take $k=0$ and choose
        $i_0\in\operatorname{supp}(P_a)\cap\operatorname{supp}(P_{a'})$.

        \item For each $\ell\in\{0,\ldots,k\}$, define the pattern vertex
        $v_\ell=S\backslash\{i_\ell\}$. Since $P_a$ has even weight and
        $i_0\in\operatorname{supp}(P_a)$, the Hamiltonian pattern transfer
        graph contains an $a$-labeled edge
        \begin{equation}
            S\longrightarrow v_0.
        \end{equation}
        We use here the reverse of the edge
        $v_0\longrightarrow S$ constructed in the discussion preceding
        Corollary~\ref{cor:even-Hamiltonian}.

        \item For every $\ell\in\{1,\ldots,k\}$, the term $P_{o_\ell}$ has odd
        weight and its support contains the distinct qubits
        $i_{\ell-1}$ and $i_\ell$. The discussion preceding
        Corollary~\ref{cor:even-Hamiltonian} therefore supplies an
        $o_\ell$-labeled edge
        \begin{equation}
            v_{\ell-1}\longrightarrow v_\ell.
        \end{equation}

        \item Since $P_{a'}$ has even weight and
        $i_k\in\operatorname{supp}(P_{a'})$, add the $a'$-labeled edge
        \begin{equation}
            v_k\longrightarrow S.
        \end{equation}
        The resulting closed path is
        \begin{equation}\label{eq:even-Pauli-cycle}
            S\xrightarrow{a}v_0
            \xrightarrow{o_1}\cdots
            \xrightarrow{o_k}v_k
            \xrightarrow{a'}S,
        \end{equation}
        where $v_\ell=S\backslash\{i_\ell\}$. When $k=0$, the middle path is
        absent and the two even-labeled edges form a two-edge cycle between
        $S$ and $v_0$.

        \item For an edge labeled by $h$, choose an input Pauli $P$ with the required support that anticommutes with $P_h$. Under the isolated Hamiltonian $c_hP_h$, character projection selects the Hermitian output Pauli $Q_h=-iP_hP$ with coefficient $\sin(2c_h\tau)$. Its support matches the input support of the next edge, so a local Clifford can take $Q_h$, including its known sign, to that input Pauli. Include the final Clifford returning to the initial Pauli to close the cycle.
    \end{enumerate}
\end{mybox}

Once the cycle is specified, isolate the Hamiltonian term associated with each edge and apply the same character-projection and repetition procedure as in Box~\ref{box:odd-Pauli}. The character-weighted signal at depth $m$ has the form $F(\tau,m)=A_{\mathrm{cyc}}W(\tau)^m$, where $A_{\mathrm{cyc}}$ is a depth-independent SPAM amplitude and
\begin{equation}\label{eq:Hamiltonian-relative-sign-cycle-weight}
    W(\tau)=\sin(2c_a\tau)\sin(2c_{a'}\tau)\prod_{\ell=1}^k\sin(2c_{o_\ell}\tau).
\end{equation}
Fitting the depth dependence estimates this signed cycle weight. For each neighboring pair used in the reconstruction, assume that $\tau_0$ can be chosen so that the product of the auxiliary odd-weight sine factors remains bounded away from zero by a constant at all sampled evolution times. This condition holds automatically, for example, for directly overlapping terms, which require no auxiliary factors. Dividing out these known factors gives
\begin{equation}\label{eq:Hamiltonian-pair-sine-signal}
    G_{a,a'}(\tau)=\sin(2c_a\tau)\sin(2c_{a'}\tau).
\end{equation}
For directly overlapping terms, the cycle has only two edges and no division is needed.

Combining this product with the individual cosine estimates yields
\begin{equation}\label{eq:Hamiltonian-relative-sign-cosines}
\begin{split}
    \cos\bigl(2(c_a+c_{a'})\tau\bigr)&=\cos(2c_a\tau)\cos(2c_{a'}\tau)-G_{a,a'}(\tau),\\
    \cos\bigl(2(c_a-c_{a'})\tau\bigr)&=\cos(2c_a\tau)\cos(2c_{a'}\tau)+G_{a,a'}(\tau).
\end{split}
\end{equation}
The cosine-only estimation procedure in Appendix~\ref{app:cosine-only-estimation} then determines $|c_a+c_{a'}|$ and $|c_a-c_{a'}|$ in Heisenberg scaling. Consequently,
\begin{equation}
    c_ac_{a'}=\frac{|c_a+c_{a'}|^2-|c_a-c_{a'}|^2}{4},
\end{equation}
which determines the relative sign of the two coefficients. Applying this construction to neighboring pairs recovers the relative signs throughout each connected component.

\subsection{Parallel structural learning}\label{subsec:Hamiltonian-structural-learning}

A general Hamiltonian can have exponentially many Pauli terms, However many physical Hamiltonians contain relatively few nonzero terms. If the support of these terms are known, the preceding protocols can estimate their coefficients individually. When their locations and Pauli labels are unknown, however, testing candidate operators one by one does not exploit the small number of terms property actually present. We therefore seek a collective procedure that identifies the nonzero terms before estimating their coefficients. Consider a sparse Hamiltonian of the form
\begin{equation}\label{eq:sparse_k_local_Hamiltonian}
    H=\sum_{a\in\mathcal A_H}c_aP_a
\end{equation}
with at most $M=|A_{H}|$ nonzero terms. We further assume that each nonzero term acts on at most $k$ qubits, as is typical of Hamiltonians with few-body interactions. We normalize the Hamiltonian so that $|c_a|\leq1$ and assume a known lower bound $|c_a|\geq c_{\min}>0$ for every nonzero coefficient. This lower bound may depend on system size and sets the precision needed to distinguish a present term from an absent one. Our goal is to develop a protocol to efficiently  identify the support $\mathcal A_H$. Then apply the preceding protocols only to the terms naturally learns all desired information.

The idea is to probe many commuting terms in the same experiment. For example, Pauli averaging over products of single-qubit $Z$ gates reshapes the Hamiltonian into one containing only $Z$-type terms~\cite{Ma_2026}. To probe this Hamiltonian, choose an $X$-type Pauli operator $X^b=\bigotimes_i X_i^{b_i}$, where the binary string $b\in\{0,1\}^n$ specifies the qubits on which it acts. Character projection selects the input component $\dket{X^b}$, evolve it under the reshaped $Z$-type Hamiltonian, and apply character projection before measurement to extract the coefficient of $X^b$ in the output. At short times, the change in this coefficient reveals the sum of squared coefficients of the Hamiltonian terms that anticommute with $X^b$. Different choices of $b$ select different sums, allowing the individual squared coefficients to be recovered from shared measurements. Repeating this construction with other choices of local Pauli axes reveals terms of other Pauli types.

To describe these measurements, choose a local Pauli frame $\beta=(\beta_1,\ldots,\beta_n)\in\{X,Y,Z\}^n$ and write $P_\beta(u)=\bigotimes_i\beta_i^{u_i}$ for a binary string $u$. Averaging over the commuting group $\mathcal K_\beta=\{P_\beta(u):u\in\mathbb F_2^n\}$ gives
\begin{equation}\label{eq:commuting_sector_Hamiltonian}
    H_\beta=\frac{1}{|\mathcal K_\beta|}\sum_{R\in\mathcal K_\beta}RHR^\dagger
    =\sum_u c_{\beta,u}P_\beta(u).
\end{equation}
Here $c_{\beta,u}$ is the coefficient of $P_\beta(u)$ in $H$, or zero if that term is absent. This average can be implemented by fast Pauli control as in Appendix~\ref{app:Hamiltonian-reshaping} with the control Paulis sampled from $\mathcal K_\beta$. A local Clifford then takes the surviving terms to $\widetilde H_\beta=\sum_u c_{\beta,u}Z^u$.

For a uniformly chosen $b\in\mathbb F_2^n$, let $\rho_b$ and $M_b$ denote the actual prepared state and measured observable for the $X^b$ experiment. Applying the character projector to $X^{b}$ before and after evolution under $\widetilde H_\beta$ gives the measured signal
\begin{equation}\label{eq:structural-projected-measurement}
    F_\beta(b;t)=\dbraket{M_b}{X^b}\dbra{X^{b}} e^{t\mathcal L_{\widetilde H_\beta}}\dket{X^b}\dbraket{X^{b}}{\rho_b}
    =A_bf_\beta(b;t).
\end{equation}
Here $A_b=\dbraket{M_b}{X^b}\dbraket{X^b}{\rho_b}$ is the SPAM amplitude and $f_\beta(b;t)=\dbra{X^b}e^{t\mathcal L_{\widetilde H_\beta}}\dket{X^b}$ is the coefficient of $X^b$ after evolution. The zero-depth reference uses the same preparation, measurement, and character projection without evolution, and therefore measures $A_b$. Comparing the two signals determines $f_\beta(b;t)$ independently of this unknown amplitude. To improve sensitivity when $f_\beta(b;t)$ is close to one, we repeat the projected cycle to depth $m$, obtaining $A_bf_\beta(b;t)^m$, and take the $m$th root of its ratio to the reference. Appendix~\ref{app:structural-learning} chooses $m$ so that this signal remains bounded away from zero relative to $A_b$. The single-cycle short-time change satisfies
\begin{equation}\label{eq:structural_linear_measurement}
    y_\beta(b;t):=\frac{1-f_\beta(b;t)}{2t^2}
    =\sum_{u:u\cdot b=1}c_{\beta,u}^2+O(\kappa M^2t^2).
\end{equation}
Here $\kappa=\min\{M,9^k\}$ combines the sparsity and hypercontractivity bounds on the fourth-order remainder, derived in Appendix~\ref{app:structural-learning}. The condition $u\cdot b=1$ selects the terms that anticommute with $X^b$ in the leading order.

These sums can be viewed as linear measurements of a sparse vector. For a fixed frame, set $d_{\beta,u}=c_{\beta,u}^2$ for the nonzero candidate strings with $\operatorname{wt}(u)\leq k$. At most $M$ entries are nonzero. Writing $y_\beta^0(b)=\sum_{u:u\cdot b=1}d_{\beta,u}$ for the ideal leading term in Eq.~\eqref{eq:structural_linear_measurement}, we have
\begin{equation}\label{eq:structural-Walsh-model}
    2y_\beta^0(b)=\sum_u d_{\beta,u}
                  -\sum_u d_{\beta,u}(-1)^{u\cdot b}.
\end{equation}
Thus the measured function has a sparse expansion in the binary characters $(-1)^{u\cdot b}$, together with an unknown constant term. Compressed sensing reconstructs this expansion jointly from uniformly sampled patterns by minimizing the sum of absolute values of its coefficients subject to consistency with the data~\cite{Rauhut2010,CandesRombergTao2006}. A number of patterns linear in $M$ up to logarithmic factors suffices, provided the error in each estimated $y_\beta(b;t)$ is at most a sufficiently small constant times $c_{\min}^2$. Thresholding the recovered squared coefficients then distinguishes present terms from absent ones. Appendix~\ref{app:structural-learning} defines the sensing matrix and reconstruction problem and proves the error and resource bounds. A weight-$w$ term survives a uniformly chosen frame with probability $3^{-w}$. Repeating the search over $O(3^k\log(M/\delta))$ independent frames therefore includes every nonzero term in at least one frame with probability at least $1-\delta$.

Let $N_k=\sum_{j=1}^{k}\binom{n}{j}$ denote the number of candidate supports in each frame. Assume the preparation and measurement overlaps satisfy $|A_b|=|\dbraket{M_b}{X^b}\dbraket{X^b}{\rho_b}|\geq A_{\mathrm{SPAM}}>0$ for every input pattern used. The following theorem gives the resulting support-recovery guarantee.

\begin{theorem}[Structural learning of sparse local Hamiltonians]\label{thm:sparse-Hamiltonian-support-learning}
    Let $H=\sum_{a\in\mathcal A_H}c_aP_a$ be an $n$-qubit Hamiltonian with at most $M\geq1$ nonzero terms of weight at most $k\leq n$, where $c_{\min}\leq|c_a|\leq1$ for a known lower bound $c_{\min}>0$. Assume ideal single-qubit control, access to fast Hamiltonian reshaping, and preparation and measurement satisfying $|\dbraket{M_b}{X^b}\dbraket{X^b}{\rho_b}|\geq A_{\mathrm{SPAM}}>0$ for every input pattern used. For $0<\delta<1/2$, there is a protocol that identifies $\mathcal A_H$ exactly and estimates its squared coefficients to additive error at most $c_{\min}^2/4$, with joint success probability at least $1-\delta$. Writing $\kappa=\min\{M,9^k\}$, its circuit-sample count $N_{\mathrm{supp}}$ and total Hamiltonian evolution time $T_{\mathrm{supp}}$ satisfy
    \begin{equation}\label{eq:structural-polynomial-cost}
        N_{\mathrm{supp}}=\widetilde O\!\left(
            \frac{3^kM^3}{A_{\mathrm{SPAM}}^2c_{\min}^4}\right),
        \qquad
        T_{\mathrm{supp}}=\widetilde O\!\left(
            \frac{3^k\sqrt\kappa\,M^3}{A_{\mathrm{SPAM}}^2c_{\min}^5}\right).
    \end{equation}
    Here $\widetilde O$ contains only logarithmic factors in $M$, $N_k$, $3^k$, $1/c_{\min}$, and $1/\delta$. Classical processing is polynomial in $n$, $M$, $1/c_{\min}$, and the number of measurement records for fixed $k$. Evolution time includes every repeated Hamiltonian evolution in each circuit and excludes the duration of the ideal control pulses.
\end{theorem}

\begin{proof}[Proof sketch]
    Choose $B=O(3^k\log(M/\delta))$ independent random frames so that every nonzero term is retained with high probability. Within each frame, Eq.~\eqref{eq:structural-Walsh-model} expresses the collective signal as a sparse expansion in bounded orthonormal characters. Compressed sensing reconstructs its squared coefficients from $L=\widetilde O(M)$ uniformly sampled patterns, provided each curvature estimate has error at most a sufficiently small constant times $c_{\min}^2$.

    To attain this accuracy, choose the single-cycle time $t=\Theta(c_{\min}/(M\sqrt\kappa))$ and repeat the projected cycle to depth $m=\Theta(\kappa M/c_{\min}^2)$. The bias in Eq.~\eqref{eq:structural_linear_measurement} is then a sufficiently small fraction of $c_{\min}^2$, while $f_\beta(b;t)^m$ remains bounded away from zero. Taking the $m$th root of the measured ratio between depths $m$ and zero yields the required curvature accuracy with $\widetilde O(M^2/(A_{\mathrm{SPAM}}^2c_{\min}^4))$ samples per pattern. Multiplying by $BL$ gives the sample bound in Eq.~\eqref{eq:structural-polynomial-cost}. Each nonreference sample uses time $mt=\Theta(\sqrt\kappa/c_{\min})$, giving the stated total evolution time.

    Stable recovery bounds all squared-coefficient errors by $c_{\min}^2/4$. Thresholding at $c_{\min}^2/2$ identifies the nonzero terms. Appendix~\ref{app:structural-learning} gives the constants, reconstruction guarantee, and allocation of failure probabilities that ensure joint success probability at least $1-\delta$.
\end{proof}

Testing candidate Pauli terms individually would require separate experiments for the $N_k$ candidate supports, which grow in number as $n^k$ for fixed $k$. Here, all candidate squared coefficients are reconstructed jointly from the same measurement data, so the experimental cost depends only polylogarithmically on $N_k$, with a factor $3^k$ for sampling the local Pauli frames. For constant $M$, $c_{\min}$, $A_{\mathrm{SPAM}}$, and fixed failure probability, the experimental cost is therefore $3^k\operatorname{polylog}(N_k)$, which is polylogarithmic in system size.

Once the nonzero terms are identified, the protocols of Sec.~\ref{subsec:Hamiltonian-protocol} estimate their coefficients to the desired precision and recover the even-weight relative signs within each component. Thus the precision required for support recovery is set by $c_{\min}$, while finer coefficient estimates are obtained only for the terms that are present.

\section{Learning Lindbladian evolution}
\label{sec:Lindbladian}

We now extend the learnability analysis to time-independent Lindbladian evolution, which includes both coherent dynamics and Markovian dissipation. This setting describes, for example, a target Hamiltonian evolution accompanied by coherent control errors, dephasing, or amplitude damping. As reviewed in Sec.~\ref{subsec:Lindblad-preliminaries}, the Hamiltonian part of the generator describes the coherent dynamics, while the jump operators describe the dissipative processes.

At any fixed time $t\geq0$, the evolution $\mathcal G_t=e^{t\mathcal L}$ is a quantum channel, so the channel learnability results apply. Access to different evolution times, however, imposes an additional requirement: the channels must all arise from the same physical generator. The question is whether this requirement restricts the gauge freedom and makes additional quantities learnable, as the restriction to Hamiltonian evolution did in the preceding section. A similarity transformation acts consistently on the entire family, since $T e^{t\mathcal L}T^{-1}=e^{tT\mathcal L T^{-1}}$. The issue is therefore whether the transformed generator $T\mathcal L T^{-1}$ still has Lindblad form.

For unrestricted Lindbladian generators, we find that the learnability characterization is analogous to that for general quantum channels: under the corresponding neighborhood assumption on admissible gauges, learnable products and ratios of the generator's Pauli transfer entries are described by the same cycle-space construction. To establish this result, we first derive how the Hamiltonian coefficients and dissipative matrix change under the gauge transformations allowed by the trusted controls, and examine when the transformed generator remains physical. We then use this gauge structure to characterize the learnable functions and develop protocols for extracting them from short-time evolution.

\subsection{Reconstruction of gauge-transformed Lindbladians}

Consider an $n$-qubit system whose density operator $\rho(t)$ evolves according to $d\rho(t)/dt=\mathcal L(\rho(t))$, where $\mathcal L$ is a time-independent Lindblad generator in the form
\begin{equation}\label{eq:Lindbladian-definition}
    \mathcal L(\rho)=-i\sum_{a}c_a[P_a,\rho]
    +\sum_{i,j\neq0}\gamma_{i,j}
    \left(P_i\rho P_j-\frac12\{P_jP_i,\rho\}\right).
\end{equation}
The real coefficients $c_a$ describe the Hamiltonian contribution, while the positive semidefinite Kossakowski matrix $\gamma=(\gamma_{i,j})$ describes dissipation. We set $c_0=0$, since the identity component of the Hamiltonian does not affect the evolution. With this convention, the coefficients $c$ and $\gamma$ are unique.

The evolution generated by $\mathcal L$ over a duration $t\geq0$ defines the quantum channel $\mathcal{G}_{t}^{\mathcal{L}}=e^{\mathcal{L}t}$. Under a valid gauge $T$, this channel transforms as
\begin{equation}
    T\mathcal{G}_{t}^{\mathcal{L}}T^{-1}=e^{(T\mathcal{L}T^{-1})t}.
\end{equation}
The transformed channels describe Lindbladian evolution for all $t\geq0$ if and only if
\begin{equation}\label{eq:Lindbladian-gauge-equivalency}
    \mathcal{L}'=T\mathcal{L}T^{-1}
\end{equation}
is also a Lindblad generator. Two Lindblad generators are \emph{gauge equivalent} if they are related by such a transformation. When the transformed preparations and measurements are also physical, the two descriptions give identical experimental probabilities at every evolution time.

To examine the restriction imposed by Eq.~\eqref{eq:Lindbladian-gauge-equivalency}, we first express $\mathcal L$ in coordinates in which the gauge action is simple. View a Lindblad generator $\mathcal{L}$ in the Pauli--Liouville representation. Its matrix entries are $\mathcal{L}_{i,j}=2^{-n}\tr[P_i\mathcal{L}(P_j)]$. For each transfer label $a\in\{0,1\}^{2n}$, define the \emph{Pauli transfer function} $F_a:\{0,1\}^{2n}\to\mathbb C$ by
\begin{equation}\label{eq:Lindbladian-transfer-entry}
    F_a(b):=\omega(a,b)^*\mathcal L_{a\oplus b,b}.
\end{equation}
Here the phase $\omega(a,b)\in\{\pm1,\pm i\}$ is defined by $P_aP_b=\omega(a,b)P_{a\oplus b}$ and $*$ denotes complex conjugation. Thus $F_a$ collects the transfers from $P_b$ to $P_{a\oplus b}$ as the input label $b$ varies, with a known phase included in its definition.  Under the same label $a$, the function described all transfers obtained through multiplication of Pauli operator $P_{a}$. 

Substituting Eq.~\eqref{eq:Lindbladian-definition} gives
\begin{equation}\label{eq:Fab-c-gamma}
    F_{a}(b)=-ic_a(1-\chi_{a,b})+\sum_{j\notin\{0,a\}}\omega(j\oplus a,j)\gamma_{j\oplus a,j}\Bigl(\chi_{b,j}-\frac{\chi_{a,j}}{2}(1+\chi_{a,b})\Bigr),
\end{equation}
where the Pauli character $\chi_{a,b}$ equals $+1$ when $P_a$ and $P_b$ commute and $-1$ otherwise. The dependence on $b$ is expressed entirely through these characters. Their orthogonality allows us to recover $c$ and $\gamma$ by a Fourier transform.

For each fixed $a$, define the Fourier transformation of $F_a$ by
\begin{equation}\label{eq:Lindbladian-Fourier-transform}
    \widehat F_a(k)=\frac{1}{4^n}
    \sum_{b\in\{0,1\}^{2n}}\chi_{k,b}F_a(b),
\end{equation}
where $k\in\{0,1\}^{2n}$ labels the Fourier modes. By character orthogonality, Eq.~\eqref{eq:Pauli-character-orthogonality}, the inverse transform is
\begin{equation}
    F_{a}(b)=\sum_{k\in\{0,1\}^{2n}}\chi_{k,b}\widehat{F}_{a}(k).
\end{equation}
We also define the Fourier support of $F_a$ as
\begin{equation}\label{eq:Fourier-support-definition}
    \operatorname{supp}(\widehat F_a)=
    \{k\in\{0,1\}^{2n}:\widehat F_a(k)\neq0\}.
\end{equation}
In these coordinates, the coefficients of $\mathcal L$ can be reconstructed explicitly. The reconstruction is an algebraic identity such that it applies to any linear map of the form \eqref{eq:Lindbladian-definition}, even when its coefficients do not describe a physical Lindblad generator.

\begin{lemma}[Fourier reconstruction]\label{lemma:Lindbladian-Fourier-reconstruction}
    Let $\mathcal L$ be a linear map of the form \eqref{eq:Lindbladian-definition}, with $c_0=0$ and arbitrary complex coefficients $c_a$ and $\gamma_{i,j}$. Define $F_a$ and $\widehat F_a$ by Eqs.~\eqref{eq:Lindbladian-transfer-entry} and~\eqref{eq:Lindbladian-Fourier-transform}. Then
    \begin{equation}\label{eq:Lindbladian-Fourier-inverse}
    \begin{split}
        c_a&=\frac{i}{2}\bigl(\widehat F_a(0)-\widehat F_a(a)\bigr),
        \qquad a\neq0,\\
        \gamma_{i,j}&=\omega(i,j)^*\widehat F_{i\oplus j}(j),
        \qquad i,j\neq0.
    \end{split}
    \end{equation}
\end{lemma}

Each coefficient $\gamma_{i,j}$ is therefore a single Fourier coefficient, up to a known phase. For $a\neq0$, the Hamiltonian coefficient $c_a$ is a difference of two Fourier coefficients. Appendix~\ref{app:Lindbladian-reconstruction} gives the more detailed proof. In the proof, neither reality of $c$ nor Hermiticity or positivity of $\gamma$ is needed, that is why these identities can be established for nonphysical coefficients setting. 

The Pauli transfer functions transform particularly simply under a valid gauge $T$. Writing $F'_a$ for the transfer functions of $\mathcal L'=T\mathcal LT^{-1}$ and using the gauge factors defined in Eq.~\eqref{eq:gauge_schur}, we obtain
\begin{equation}\label{eq:Lindbladian-transfer-gauge}
    F'_a(b)=\frac{\lambda_{\pt(P_{a\oplus b})}}
    {\lambda_{\pt(P_b)}}F_a(b).
\end{equation}
Combining this transformation with the coefficient relations in Lemma~\ref{lemma:Lindbladian-Fourier-reconstruction} gives explicit expressions for $c'$ and $\gamma'$ whenever $\mathcal L'$ admits the form \eqref{eq:Lindbladian-definition}, and allows us to test whether these coefficients describe a physical Lindblad generator. The following theorem establishes that, for a real diagonal gauge, such a representation always exists with real $c'$ and Hermitian $\gamma'$. Thus $\mathcal L'$ is a Lindblad generator precisely when $\gamma'$ is positive semidefinite.

\begin{theorem}[Gauge-transformed generator]\label{thm:Lindbladian-reconstruction}
\label{thm:gauge-transformed-Lindbladian-reconstruction}
    Let $\mathcal L$ be a Lindblad generator, and let $T$ be a real diagonal gauge of the form \eqref{eq:gauge_schur}. Then $\mathcal L'=T\mathcal LT^{-1}$ admits a unique representation of the form \eqref{eq:Lindbladian-definition} with $c'_0=0$, real coefficients $c'_a$, and a Hermitian matrix $\gamma'$. In terms of the Pauli transfer functions of $\mathcal L$, these coefficients are
    \begin{equation}\label{eq:gauge-transformed-coefficients-through-F}
    \begin{split}
        c'_a&=\frac{i}{2\cdot4^n}\sum_b(1-\chi_{a,b})
        \frac{\lambda_{\pt(P_{a\oplus b})}}{\lambda_{\pt(P_b)}}F_a(b),\\
        \gamma'_{i,j}&=\frac{\omega(i,j)^*}{4^n}\sum_b\chi_{j,b}
        \frac{\lambda_{\pt(P_{i\oplus j\oplus b})}}{\lambda_{\pt(P_b)}}F_{i\oplus j}(b),
    \end{split}
    \end{equation}
    Here $a\neq0$, $i,j\neq0$, and both sums run over $b\in\{0,1\}^{2n}$. The map $\mathcal L'$ is a Lindblad generator if and only if $\gamma'\succeq0$.
\end{theorem}

The theorem reduces the physicality of the transformed generator to a single matrix condition, $\gamma'\succeq0$. Reality of $c'$ and Hermiticity of $\gamma'$ follow automatically because a real diagonal gauge preserves Hermiticity and the condition $\tr[\mathcal L(X)]=0$ for every operator $X$. Every linear map with these two properties admits a unique canonical representation with real Hamiltonian coefficients and a Hermitian Kossakowski matrix. Appendix~\ref{app:Lindbladian-reconstruction} proves this assertion and gives the transformed coefficients explicitly in terms of the original $c$ and $\gamma$.

We next examine whether this positivity condition permits an open neighborhood of gauges around the identity. Suppose $\gamma$ is positive definite, with smallest eigenvalue $\mu>0$. The transformed matrix $\gamma'$ depends continuously on the gauge factors in Eq.~\eqref{eq:gauge-transformed-coefficients-through-F} and equals $\gamma$ at the identity gauge. Therefore, for all sufficiently small changes of the gauge factors,
\begin{equation}\label{eq:Lindbladian-gauge-positivity}
    \|\gamma'-\gamma\|_{\mathrm{op}}<\mu,
    \qquad
    \lambda_{\min}(\gamma')\geq
    \mu-\|\gamma'-\gamma\|_{\mathrm{op}}>0.
\end{equation}
Here $\|\cdot\|_{\mathrm{op}}$ denotes the operator norm, and $\lambda_{\min}$ denotes the smallest eigenvalue. Thus, when $\gamma\succ0$, every sufficiently small diagonal gauge transformation preserves the Lindblad form of the generator.

The gauge-transformed generator must satisfy $\gamma'\succeq0$, similarly in the general channel case the corresponding transformed channels $T\mathcal G_tT^{-1}$ must remain CPTP. In the unstructured setting, when the generator and channels lie in the interior of their respective sets of physical models, continuity ensures that sufficiently small gauge changes preserve both conditions. Thus Lindbladian evolution imposes no additional local restriction on the gauge freedom already present in the quantum channel setting.

When $\gamma$ is singular, or a preparation state or measurement effect lies on the boundary of its positive cone, the existence of such a neighborhood requires a separate analysis or assumption. Purely Hamiltonian evolution, for which $\gamma=0$, is one such case. Additional structural restrictions on the dissipator must also be preserved and may further constrain the gauges. The neighborhood argument therefore explains why unrestricted dissipative models with $\gamma\succ0$ can retain the local gauge freedom of general channels, while Hamiltonian and other structured models may admit fewer gauges.

\subsection{Learnability of Lindbladian functions}
The cycle-space characterization from Sec.~\ref{subsec:CPTP} applies to Lindbladian generators once their transfer-matrix entries are written in terms of the functions $F_a$. By definition,
\begin{equation}
    (\mathcal{L})_{i,j}=\omega(i\oplus j,j)F_{i\oplus j}(j).
\end{equation}
Thus each directed transfer from $P_j$ to $P_i$ is represented, up to a known Pauli phase, by $F_{i\oplus j}(j)$. Since the gauge affects of these entries is the same as in the general channel case, Theorem~\ref{thm:general_cptp_learnability} gives the learnable cycle products directly. The corollary states this correspondence explicitly in terms of the functions $F_a$.

\begin{corollary}\label{cor:Lindbladian-function-cycle}
    Construct the pattern transfer graph directly from the nonzero PTM entries of $\mathcal L$, as in Sec.~\ref{subsec:CPTP}. Let
    \begin{equation}
    \begin{split}
        c&=(e_0,e_1,\ldots,e_{L-1}),\\
        e_\ell&=(P_{a_\ell},P_{b_\ell})
        \text{ or }(P_{b_\ell},P_{a_\ell}).
    \end{split}
    \end{equation}
    be a cycle in its underlying undirected multigraph, traversed from the pattern of $P_{a_\ell}$ to the pattern of $P_{b_\ell}$ on the $\ell$-th step, so that $\pt(P_{b_\ell})=\pt(P_{a_{\ell+1}})$ with $\ell+1$ interpreted modulo $L$. If the physical direction of $e_\ell$ agrees with the traversal, its Lindbladian Pauli-function label is $F_{a_\ell\oplus b_\ell}(a_\ell)$. If its physical direction is opposite to the traversal, its label is $F_{a_\ell\oplus b_\ell}(b_\ell)$ and it enters with exponent $-1$. Consequently, the function
    \begin{equation}\label{eq:learnable-Lindbladian-cycle-function}
        f_c(F)
        =\frac{\displaystyle\prod_{\ell:\,e_\ell=(P_{a_\ell},P_{b_\ell})}
        F_{a_\ell\oplus b_\ell}(a_\ell)}
        {\displaystyle\prod_{\ell:\,e_\ell=(P_{b_\ell},P_{a_\ell})}
        F_{a_\ell\oplus b_\ell}(b_\ell)}
    \end{equation}
    is learnable. Under the neighborhood assumption of Theorem~\ref{thm:general_cptp_learnability}, choosing a basis of undirected cycles gives a multiplicative basis for all learnable product-of-power functions of the nonzero Lindbladian Pauli functions.
\end{corollary}

\begin{proof}
    If $e_\ell=(P_{a_\ell},P_{b_\ell})$, its PTM weight is $(\mathcal L)_{b_\ell,a_\ell}=\omega(a_\ell\oplus b_\ell,a_\ell)F_{a_\ell\oplus b_\ell}(a_\ell)$ and appears in the numerator of the PTM cycle function. If $e_\ell=(P_{b_\ell},P_{a_\ell})$, its PTM weight is $(\mathcal L)_{a_\ell,b_\ell}=\omega(a_\ell\oplus b_\ell,b_\ell)F_{a_\ell\oplus b_\ell}(b_\ell)$ and appears in the denominator. Therefore the PTM cycle function from Theorem~\ref{thm:general_cptp_learnability} differs from $f_c(F)$ only by a known product of Pauli phases $\omega$. Removing this known nonzero factor preserves learnability, and the cycle-basis statement follows from the same theorem.
\end{proof}

The coefficients $c_a$ and $\gamma_{i,j}$ are often the parameters of physical interest. In the unrestricted Lindbladian model, however, each coefficient $c_{a}$ or $\gamma_{i,j}$ is related to $4^{n}$ different $F_{a}(b)$ coefficients, which may include large gauge degree of freedom. And the gauge transformed value of $c'_{a}$ and $\gamma'_{i,j}$ can depend on gauge parameters in a highly nonlinear way. Although particular nontrivial combinations may remain gauge invariant, characterizing them in full generality is complicated and beyond the scope of this subsection. we therefore focus here on the cycle products of the transfer functions $F_a(b)$. Section~\ref{sec:first-order} revisits the coefficient-level learnability problem under an additional first-order assumption and provides much cleaner results.

\subsection{Protocols for learning Lindbladians}

The learnability characterization in Corollary~\ref{cor:Lindbladian-function-cycle} can be turned into a protocol by evolving the target Lindbladian operator for a short time.  For every $t\geq0$, the physical evolution $\mathcal{G}_{t}= e^{t\mathcal L}$ is a quantum channel, and its short-time expansion is
\begin{equation}\label{eq:short-time-Lindbladian-expansion}
    \mathcal{G}_{t}=I+t\mathcal L+O(t^2).
\end{equation}
We may therefore apply the channel protocols in Boxes~\ref{box:oriented_cycle} and~\ref{box:unoriented_cycle} to $\mathcal{G}_{t}$ and extract the leading coefficient as $t\to0$. For a PTM edge $e=(P_j,P_i)$ with $i\neq j$, define
\begin{equation}\label{eq:Lindbladian-edge-weight}
    \ell_e=(\mathcal L)_{i,j}=\omega(i\oplus j,j)F_{i\oplus j}(j).
\end{equation}
Equation~\eqref{eq:short-time-Lindbladian-expansion} gives
\begin{equation}\label{eq:short-time-off-diagonal-edge}
    (\mathcal G_t)_{i,j}=t\ell_e+O(t^2).
\end{equation}
Let $x$ be a signed cycle vector supported on such off-diagonal PTM edges, and write $d_x=\sum_e x_e$. The corresponding channel cycle function satisfies
\begin{equation}\label{eq:short-time-cycle-limit}
    \prod_e(\mathcal G_t)_{e}^{x_e}
    =t^{d_x}\prod_e\ell_e^{x_e}\bigl(1+O(t)\bigr),
    \qquad
    \prod_e\ell_e^{x_e}
    =\lim_{t\to0}t^{-d_x}\prod_e(\mathcal G_t)_{e}^{x_e}.
\end{equation}
The product on the right differs from the Lindbladian cycle function in Eq.~\eqref{eq:learnable-Lindbladian-cycle-function} only by known Pauli phases.  Thus every cycle-space function of the off-diagonal Lindbladian transfer entries can be obtained from short-time channel cycle functions.

PTM-diagonal entries require a slightly different normalization because the identity channel contributes at zeroth order.  For every Pauli $P_j$,
\begin{equation}\label{eq:short-time-diagonal-edge}
    (\mathcal G_t)_{j,j}=1+t(\mathcal L)_{j,j}+O(t^2),
    \qquad
    (\mathcal L)_{j,j}=\lim_{t\to0}\frac{(\mathcal G_t)_{j,j}-1}{t}.
\end{equation}
Since $(\mathcal G_t)_{j,j}$ is a self-loop of the channel pattern-transfer graph, it is individually learnable by the channel self-loop protocol.  Every diagonal Lindbladian entry is consequently learnable from its initial slope.  A general Lindbladian cycle vector decomposes into its PTM-diagonal self-loops and an off-diagonal circulation, because removing self-loops does not change the incidence constraint; the latter can in turn be decomposed into off-diagonal cycles.  Equations~\eqref{eq:short-time-cycle-limit} and~\eqref{eq:short-time-diagonal-edge} therefore suffice for an arbitrary learnable cycle function.

\begin{mybox}[label=box:short-time-Lindbladian]{Learning Lindbladian cycle functions from short-time evolution}
    \begin{enumerate}
        \item Choose a cycle vector $x$ in the pattern-transfer graph of $\mathcal L$.  Separate its PTM-diagonal edges $e=(P_j,P_j)$ from its off-diagonal edges $e=(P_j,P_i)$ with $i\neq j$, and record the known Pauli phases relating the latter to their functions $F_a(b)$.

        \item Choose several sufficiently short evolution times $t_1,\ldots,t_R$.  At each time $t_r$, implement the exact channel $\mathcal G_{t_r}=e^{t_r\mathcal L}$.  Decompose the off-diagonal part of $x$ into cycles and apply Box~\ref{box:oriented_cycle} or Box~\ref{box:unoriented_cycle} to each one, according to its orientation, to estimate their product
        \begin{equation}
            R_x(t_r)=\prod_{e=(P_j,P_i):\,i\neq j}(\mathcal G_{t_r})_e^{x_e}.
        \end{equation}

        \item For the off-diagonal part, rescale the estimate by $t_r^{-d_x}$ and extrapolate to $t=0$.  By Eq.~\eqref{eq:short-time-cycle-limit}, the intercept is $\prod_{e=(P_j,P_i):\,i\neq j}\ell_e^{x_e}$.  Divide by the known product of Pauli phases to obtain the corresponding product of powers of $F_a(b)$.

        \item For every PTM-diagonal edge appearing in $x$, estimate the self-loop $(\mathcal G_{t_r})_{j,j}$ at the same short times and fit its initial slope according to Eq.~\eqref{eq:short-time-diagonal-edge}.  Combine these diagonal entries with the off-diagonal cycle function to return the complete Lindbladian cycle function.
    \end{enumerate}
\end{mybox}

The character projections and cycle constructions are exactly those of the channel protocols, so their cancellation of fixed preparation and measurement errors is inherited unchanged.  Using several evolution times also permits the $O(t)$ bias remaining after the rescaling in Eq.~\eqref{eq:short-time-cycle-limit} to be removed by extrapolation.  Smaller times reduce this truncation bias but also suppress a product containing several off-diagonal edges, giving the usual tradeoff between short-time bias and statistical precision. 

\section{First-order learnability}\label{sec:first-order}

In calibration, the intended operation is usually known and the unknown quantities are small deviations from it. And recent advances in quantum control have made this regime experimentally relevant across different platforms. For example, two-qubit gate fidelities around $99.9\%$ have been demonstrated with trapped-ion hyperfine qubits~\cite{Ballance2016HighFidelity} and superconducting fluxonium qubits~\cite{Ding2023FluxoniumGates}. Such high-fidelity implementations motivate characterizing the leading deviations from a known target. This prior information changes the learnability problem. A gauge may transform errors between the dynamics and SPAM, but it must leave the prescribed ideal operation unchanged at leading order. Transfers already present in the ideal dynamics therefore constrain the relative preparation and measurement scales, and these constraints can make additional noise parameters accessible.

\subsection{First-order learnability of quantum channels}

To formulate the problem, we first describe this mechanism for a channel close to a known unitary operation. Let $\mathcal U$ denote the known ideal channel and write the implemented channel in terms of its physical noise parameters $\vartheta=(\theta_1,\cdots,\theta_{d})\in\mathbb{R}^{d}$:
\begin{equation}
    \mathcal{G}_{\vartheta}=\mathcal{U}+\Lambda_{\vartheta},
    \label{eq:first-order-channel-model}
\end{equation}
We assume that each component of $\vartheta$ and each entry of the induced channel perturbation $\Lambda_{\vartheta}$ is in $O(\varepsilon)$, with $\varepsilon$ being a known first order small quantity. Note that for $\Lambda_{\vartheta}$ to be a first-order small perturbation, the components of the parameter vector $\vartheta$ need not themselves be first-order small. However, we wish to capture the physical setting in which small deviations in the experimental implementation produce small errors. We therefore additionally require all components of $\vartheta$, which represent implementation imperfections or environmental effects, to be first-order small.

A linear function of the $O(\varepsilon)$ noise parameters is first-order learnable if its leading value is unaffected by gauge ambiguity. Since gauge-equivalent parameter families are experimentally indistinguishable, their function values must agree up to $O(\varepsilon^2)$, as formalized below.

\begin{definition}[First-order learnable linear function]\label{def:first-order-learnable-function}
    Let $\vartheta$ be the parameter vector for channel $\mathcal{G}_{\vartheta}=\mathcal{U}+\Lambda_{\vartheta}$. A linear function of parameters $f_{x}(\vartheta)=\langle x,\vartheta\rangle$ is called first-order learnable if, whenever two parameter vector $\vartheta$ and $\vartheta'$ are related by a valid gauge $T$ satisfying
    \begin{equation}
        T(\mathcal{U}+\Lambda_{\vartheta})T^{-1}
        =\mathcal{U}+\Lambda_{\vartheta'},
    \end{equation}
    their values agree up to second order, in the sense that
    \begin{equation}\label{eq:first-order-invariant}
        \langle x,\vartheta'-\vartheta\rangle=O(\varepsilon^2).
    \end{equation}
\end{definition}
Here we consider regular gauge families, meaning each entry of $T$ is of constant order. This scaling ensures that conjugation does not itself change the order in $\varepsilon$ of a transfer-matrix element. To characterize the space for these learnabel linear functions, we first identify the relevant pattern transfer graphs and the leading-order action of the gauge.

\begin{definition}[Ideal pattern transfer graph]\label{def:ideal-pattern-transfer-graph}
    Let $\mathcal{G}_{\vartheta}=\mathcal{U}+\Lambda_{\vartheta}$ be a noisy implementation of the known ideal channel $\mathcal{U}$. The ideal pattern-transfer graph associated with $\mathcal{G}_{\vartheta}$, denoted by $\Gamma_{\pt}^{\mathrm{id}}(\mathcal{G}_{\vartheta})$, is the pattern-transfer graph of $\mathcal{U}$:
    \begin{equation}
        \Gamma_{\pt}^{\mathrm{id}}(\mathcal{G}_{\vartheta})=\Gamma_{\pt}(\mathcal{U}).
    \end{equation}
    Thus, an edge is present precisely when the corresponding Pauli transfer matrix element of $\mathcal{U}$ is nonzero, independently of the additional edges or weight corrections introduced by $\Lambda_{\vartheta}$.
\end{definition}

The ideal pattern-transfer graph records the nonzero entries of the ideal channel. The corresponding edges in the implemented channel have weights $e_{i,j}=\mathcal{U}_{i,j}+O(\varepsilon)$ and therefore remain nonzero for sufficiently small $\varepsilon$. Any edge absent from the ideal graph is induced by the noise and has weight $O(\varepsilon)$. To describe these edges, we first contract the graph by connected commponents. Here, we define connected components by ignoring edge directions, such that two vertices belong to the same component if a path connects them when edges may be traversed in either direction. To describe how the noise-induced edges connect these components, we next contract each connected component of the ideal graph to a single vertex.
 
\begin{definition}[Noise-induced component transfer graph]\label{def:noise-induced-component-transfer-graph}
    Let $C_1,\ldots,C_M$ be the connected components of $\Gamma_{\pt}^{\mathrm{id}}(\mathcal{G}_{\vartheta})$. The noise-induced component transfer graph, denoted by $\Gamma_{\pt}^{\mathrm{noise}}(\mathcal{G}_{\vartheta})$, is obtained from the full pattern-transfer graph $\Gamma_{\pt}(\mathcal{G}_{\vartheta})$ by contracting each $C_m$ to a single vertex and retaining only edges absent from $\Gamma_{\pt}^{\mathrm{id}}(\mathcal{G}_{\vartheta})$.
\end{definition}

More explicitly, the edge associated with the Pauli transfer matrix element $(\mathcal{G}_{\vartheta})_{i,j}$ is retained precisely when $\mathcal{U}_{i,j}=0$ and $(\mathcal{G}_{\vartheta})_{i,j}\neq 0$. Its weight is then
\begin{equation}
    (\mathcal{G}_{\vartheta})_{i,j}=(\Lambda_{\vartheta})_{i,j}=O(\varepsilon).
\end{equation}
Distinct noise-induced edges are retained separately, so the contracted graph may contain parallel edges and self-loops. The contraction is natural because, as the next lemma shows, the gauge parameters on vertices in the same ideal connected component agree at leading order.

\begin{lemma}[Gauge restriction for first-order channels]\label{lemma:first-order-component-gauge}
    Let $\mathcal{G}_{\vartheta}=\mathcal{U}+\Lambda_{\vartheta}$ and let $T$ be a diagonal gauge of the form in Eq.~\eqref{eq:gauge_schur}. Suppose $T$ is a valid gauge satisfying
    \begin{equation}
        T(\mathcal{U}+\Lambda_{\vartheta})T^{-1}=\mathcal{U}+\Lambda_{\vartheta'}.
    \end{equation}
    Then, for every connected component $C$ of the ideal pattern transfer graph $\Gamma_{\pt}^{\mathrm{id}}(\mathcal{G})$, there exists a nonzero $\kappa_C$ of constant order of magnitude such that the gauge parameters satisfy
    \begin{equation}
        \lambda_v=\kappa_C+O(\varepsilon)
    \end{equation}
    for every pattern vertex $v\in C$. The constants $\kappa_C$ may differ between connected components.
\end{lemma}

\begin{proof}
    Consider an ideal edge from the pattern $u=\pt(P_j)$ to the pattern $v=\pt(P_i)$. By definition, the corresponding ideal Pauli transfer matrix element satisfies $\mathcal{U}_{i,j}\neq0$. Under the diagonal gauge $T$, this matrix element transforms as
    \begin{equation}
        \mathcal{G}'_{i,j}=\frac{\lambda_v}{\lambda_u}\left(\mathcal{U}_{i,j}+(\Lambda_{\vartheta})_{i,j}\right).
    \end{equation}
    Because the transformed channel has the same ideal part $\mathcal{U}$, it must also satisfy $\mathcal{G}'_{i,j}=\mathcal{U}_{i,j}+O(\varepsilon)$. Since $(\Lambda_{\vartheta})_{i,j}=O(\varepsilon)$ and $\mathcal{U}_{i,j}\neq0$, comparison of the zeroth-order terms gives
    \begin{equation}
        \frac{\lambda_v}{\lambda_u}=1+O(\varepsilon).
    \end{equation}
    Now fix a reference vertex $v_C$ in a connected component $C$ and let $\kappa_C$ be the zeroth-order value of $\lambda_{v_C}$. Every vertex $v\in C$ is joined to $v_C$ by an undirected path of ideal edges. Applying the preceding ratio relation along this path, with the ratio inverted whenever an edge is traversed against its orientation, yields $\lambda_v/\lambda_{v_C}=1+O(\varepsilon)$. Regularity of the gauge therefore implies $\lambda_v=\kappa_C+O(\varepsilon)$. For a component consisting of a single vertex, the same conclusion holds by taking $\kappa_C$ to be the zeroth-order value of its unrestricted gauge parameter.
\end{proof}

Lemma~\ref{lemma:first-order-component-gauge} distinguishes the gauge action within an ideal connected component from that between different components. To express these actions in terms of noise parameters, we assign one coordinate to each edge of the two graphs.

For an ideal edge $e$, its known nonzero weight $\mathcal U_e$ provides a reference, so we define the relative correction by
\begin{equation}
    \theta_e^{\mathrm{id}}=\mathcal G_e/\mathcal U_e-1.
\end{equation}
Or equivalently $\mathcal G_e=\mathcal U_e(1+\theta_e^{\mathrm{id}})$. This allows the first order small $\theta_{e}^{\mathrm{id}}$ to induce a small deviation in the ideal edges.

For a noise-induced edge, the ideal weight vanishes, so there is no nonzero ideal reference. We restrict to edges with a nonzero first-order weight, meaning that $\mathcal G_e/\varepsilon$ has a finite, nonzero limit as $\varepsilon\to0$. Using the signed logarithm from Sec.~\ref{subsec:complex-logarithm-preliminaries}, define
\begin{equation}
    \operatorname{slog}(\mathcal G_e/\varepsilon)=(\theta_e^{\mathrm{noise}}/\varepsilon,\sigma_e),
\end{equation}
where $\theta_e^{\mathrm{noise}}=\varepsilon\log|\mathcal G_e/\varepsilon|$ and $\sigma_e=s(\mathcal G_e)\in\mathbb F_2$ denotes the sign of $\mathcal{G}_{e}$. Here $s(x)=0$ if $x>0$ and equals $1$ otherwise. Equivalently, $\mathcal G_e=(-1)^{\sigma_e}\varepsilon\exp(\theta_e^{\mathrm{noise}}/\varepsilon)$. The normalization makes $\theta_e^{\mathrm{noise}}=O(\varepsilon)$, while $\sigma_e$ retains the sign in $\mathbb{F}_{2}$. We retain both the logarithmic magnitudes and the signs in the parameter collection:
\begin{equation}
    \vartheta=\theta^{\mathrm{id}}\oplus(\theta^{\mathrm{noise}},\sigma),
\end{equation}
with $\sigma=(\sigma_e)_{e\in E_{\mathrm{noise}}}$. To define a function $f_{x}=\langle x,\vartheta\rangle$, we consider the exponent vector to be $x=x_{\mathrm{id}}\oplus x_{\mathrm{noise}}\in\mathbb R^{|E_{\mathrm{id}}|}\oplus\mathbb Z_{(2)}^{|E_{\mathrm{noise}}|}$, with $E_{\mathrm{id}}$ and $E_{\mathrm{noise}}$ denoting the edges in the ideal graph and noise graph respectively. Then the function can be defined as
\begin{equation}
    f_{x}(\vartheta)=\langle x_{\mathrm{id}},\theta^{\mathrm{id}}\rangle+\langle x_{\mathrm{noise}},(\theta^{\mathrm{noise}},\sigma)\rangle.
\end{equation}
For the noise-induced edges, we combine the parameters using the signed-logarithm rules of Sec.~\ref{subsec:complex-logarithm-preliminaries}. We sum the real coordinates with coefficients $x_e$, giving $\sum_e x_e\theta_e^{\mathrm{noise}}$, and sum the sign bits with coefficients $\overline{x_e}$ modulo $2$, giving $\sum_e\overline{x_e}\sigma_e\in\mathbb F_2$. Here $\overline{x_e}$ is the parity of the numerator of $x_e$ when written as a fraction with an odd denominator. 

Rigorously speaking, $f_{x}$ should be understood as $f_{x}:\mathbb{R}^{|E_{\mathrm{id}}|+|E_{\mathrm{noise}}|}\oplus\mathbb{F}_{2}^{|E_{\mathrm{noise}}|}\rightarrow \mathbb{R}\oplus\mathbb{F}_2$, where $\mathbb{R}$ records the the product $\langle x_{\mathrm{id}},\theta^{\mathrm{id}}\rangle$ and real part of $\langle x_{\mathrm{noise}},(\theta^{\mathrm{noise}},\sigma)\rangle$ and $\mathbb{F}_2$ records the sign of $\langle x_{\mathrm{noise}},(\theta^{\mathrm{noise}},\sigma)\rangle$. For these functions, agreement up to $O(\varepsilon^2)$ means that the real components differ by $O(\varepsilon^2)$ and the sign components agree exactly in $\mathbb{F}_2$ for sufficiently small $\varepsilon$.

To describe the gauge action, allow the component factors $\kappa_C$ to have either sign and expand the logarithmic magnitudes:
\begin{equation}\label{eq:first-order-log-gauge-parameters}
    \log|\lambda_v|=k_{C(v)}+\varepsilon s_v+O(\varepsilon^2),
\end{equation}
where $C(v)$ is the ideal connected component containing $v$, $k_C=\log|\kappa_C|$ describes its leading gauge magnitude, and $s_v$ describes the first-order variation within that component. In the following, we will consider an unstructured noise model in which the gauge parameters $s_v$ and $k_C$ can vary independently. By Lemma~\ref{lemma:first-order-component-gauge}, the gauge signs are constant within each ideal component for sufficiently small $\varepsilon$. Writing $\tau_C=s(\kappa_C)\in\mathbb F_2$, a noise-induced edge from $C$ to $D$ has sign transformation $\sigma'_e=\sigma_e+\tau_D-\tau_C$ in $\mathbb F_2$. Thus a negative component factor flips the signs of edges connecting it to components with positive factors, while signs on edges within a component remain unchanged.

Since edge weights transform by ratios of endpoint gauge factors, the leading changes of these coordinates are expressed by the incidence matrices of the two graphs. The resulting first-order learnable subspace is characterized in the following theorem.

\begin{theorem}[Decomposition of first-order linear learnability]\label{thm:first-order-learnable-decomposition}
    Let $B_{\mathrm{id}}$ and $B_{\mathrm{noise}}$ be the incidence matrices of $\Gamma_{\pt}^{\mathrm{id}}(\mathcal{G})$ and $\Gamma_{\pt}^{\mathrm{noise}}(\mathcal{G})$, respectively. A linear function
    \begin{equation}
        f_x(\vartheta)=\left\langle x_{\mathrm{id}},\theta^{\mathrm{id}}\right\rangle+\left\langle x_{\mathrm{noise}},(\theta^{\mathrm{noise}},\sigma)\right\rangle
    \end{equation}
    with $x_{\mathrm{id}}\in\mathbb R^{E_{\mathrm{id}}}$ and $x_{\mathrm{noise}}\in\mathbb Z_{(2)}^{E_{\mathrm{noise}}}$ is first-order learnable if and only if
    \begin{equation}
        \begin{split}
            x_{\mathrm{id}}&\in Z\left(\Gamma_{\pt}^{\mathrm{id}}(\mathcal{G})\right),\\
            x_{\mathrm{noise}}&\in Z\left(\Gamma_{\pt}^{\mathrm{noise}}(\mathcal{G})\right).
        \end{split}
    \end{equation}
    Here the ideal cycle space is taken over $\mathbb R$ and the noise cycle space over $\mathbb Z_{(2)}$. Hence the first-order learnable coefficient set decomposes as
    \begin{equation}\label{eq:first-order-learnable-decomposition}
        \mathfrak{L}^{(1)}
        =Z\left(\Gamma_{\pt}^{\mathrm{id}}(\mathcal{G})\right)
        \oplus Z\left(\Gamma_{\pt}^{\mathrm{noise}}(\mathcal{G})\right).
    \end{equation}
\end{theorem}

\begin{proof}
    If $e:u\to v$ is an ideal edge, then $C(u)=C(v)$. Using Eq.~\eqref{eq:first-order-log-gauge-parameters} in $\mathcal{G}'_e=\lambda_v\mathcal{G}_e/\lambda_u$ gives
    \begin{equation}
        (\theta_e^{\mathrm{id}})'=\theta_e^{\mathrm{id}}+\varepsilon(s_v-s_u)+O(\varepsilon^2).
    \end{equation}
    If $e$ is a noise-induced edge directed from the ideal component $C$ to the ideal component $D$, taking the logarithmic magnitude of the transformed edge weight and multiplying by $\varepsilon$ gives
    \begin{equation}
        (\theta_e^{\mathrm{noise}})'=\theta_e^{\mathrm{noise}}+\varepsilon(k_D-k_C)+O(\varepsilon^2).
    \end{equation}
    Therefore the complete leading gauge action is
    \begin{equation}\label{eq:first-order-gauge-decomposition}
        \begin{split}
            (\theta^{\mathrm{id}})'&=\theta^{\mathrm{id}}+\varepsilon B_{\mathrm{id}}^{\mathsf T}s+O(\varepsilon^2),\\
            (\theta^{\mathrm{noise}})'&=\theta^{\mathrm{noise}}+\varepsilon B_{\mathrm{noise}}^{\mathsf T}k+O(\varepsilon^2),\\
            \sigma'&=\sigma+B_{\mathrm{noise}}^{\mathsf T}\tau\quad\text{in }\mathbb F_2.
        \end{split}
    \end{equation}
    It follows that the gauge-induced change in the real component of $f_x$ is
    \begin{equation}
        \varepsilon\left\langle B_{\mathrm{id}}x_{\mathrm{id}},s\right\rangle
        +\varepsilon\left\langle B_{\mathrm{noise}}x_{\mathrm{noise}},k\right\rangle
        +O(\varepsilon^2).
    \end{equation}
    Since $s$ and $k$ can be varied independently, this change is $O(\varepsilon^2)$ for every valid gauge if and only if $B_{\mathrm{id}}x_{\mathrm{id}}=0$ and $B_{\mathrm{noise}}x_{\mathrm{noise}}=0$. The change in the sign component is $\langle\overline{x}_{\mathrm{noise}},B_{\mathrm{noise}}^{\mathsf T}\tau\rangle=\langle\overline{B_{\mathrm{noise}}x_{\mathrm{noise}}},\tau\rangle$ in $\mathbb F_2$, where the bar denotes the parity map on $\mathbb Z_{(2)}$. Thus $B_{\mathrm{noise}}x_{\mathrm{noise}}=0$ also guarantees exact sign invariance, without any positivity assumption on $\kappa_C$. The two kernel conditions therefore ensure both required forms of agreement and give the cycle spaces in Eq.~\eqref{eq:first-order-learnable-decomposition}. In particular, a non-learnable linear function from one sector cannot become learnable by adding a function from the other sector.
\end{proof}

The two cycle conditions have a simple interpretation. On the ideal graph, cycle combinations cancel the gauge changes in the relative edge corrections. On the noise-induced graph, a cycle can be turned into a closed path in the original graph by joining its noise-induced edges with paths inside the ideal connected components. Although the weights on these connecting paths also contain unknown errors, their ideal values suffice to recover the first-order noise information.

The reason is that a correction to an ideal edge is multiplied by an already small noise contribution. For example, let a noise-induced edge have weight $\varepsilon a$ and a connecting ideal edge have weight $u+\varepsilon b$, where $u\neq0$ is known and $a,b=O(1)$. Their product is
\begin{equation}
    (\varepsilon a)(u+\varepsilon b)
    =\varepsilon au+\varepsilon^2ab
    =\varepsilon au+O(\varepsilon^2).
\end{equation}
Thus replacing the connecting edge by $u$ leaves the first-order term unchanged. More generally, replacing the ideal-edge weights along a fixed connecting path by their known ideal values changes its multiplicative contribution by a relative factor $1+O(\varepsilon)$. In the noise coordinates, which are $\varepsilon$ times logarithmic magnitudes, this produces only an $O(\varepsilon^2)$ change. The signs of the ideal-edge weights also remain equal to their known ideal signs for sufficiently small $\varepsilon$. Consequently, their unknown corrections do not affect the noise-induced cycle function to first order.

At first sight, a theoretically learnable cycle function may appear inaccessible to a first-order experiment. Consider two noise-induced edges $e_1=(u,v)$ and $e_2=(v,u)$ in the full pattern-transfer graph. Their weights transform as $\mathcal G'_{e_1}=(\lambda_v/\lambda_u)\mathcal G_{e_1}$ and $\mathcal G'_{e_2}=(\lambda_u/\lambda_v)\mathcal G_{e_2}$. The gauge factors therefore cancel in their product, including when either factor is negative:
\begin{equation}
    \mathcal G'_{e_1}\mathcal G'_{e_2}
    =\frac{\lambda_v}{\lambda_u}\frac{\lambda_u}{\lambda_v}
    \mathcal G_{e_1}\mathcal G_{e_2}
    =\mathcal G_{e_1}\mathcal G_{e_2}.
\end{equation}
Its normalized signed logarithm gives
\begin{equation}
    \operatorname{slog}\!\left(\frac{\mathcal G_{e_1}\mathcal G_{e_2}}{\varepsilon^2}\right)
    =\left(
        \frac{\theta_{e_1}^{\mathrm{noise}}+\theta_{e_2}^{\mathrm{noise}}}{\varepsilon},
        \sigma_{e_1}+\sigma_{e_2}
    \right),
\end{equation}
where the sign bits are added in $\mathbb F_2$. Thus both the real parameter sum and the combined sign are gauge invariant. However, directly measuring their underlying cycle product gives only an $O(\varepsilon^2)$ signal. More generally, the direct path experiment for a chain containing $k$ noise-induced edges produces an $O(\varepsilon^k)$ signal. Nevertheless, it is unnecessary to estimate this higher-order signal directly. We can instead estimate each edge through a separate $O(\varepsilon)$ experiment and cancel the additional SPAM factors using zero-depth reference experiments at the intermediate vertices. Then every theoretically first-order learnable linear function in $\mathfrak L^{(1)}$ from Eq.~\eqref{eq:first-order-learnable-decomposition} can be reconstructed using only character-projected signals whose leading unknown contributions are of order $O(\varepsilon)$ or larger.

Consider a cycle $c_{\mathrm{noise}}$ in the noise-induced graph for $\mathcal{G}_{\vartheta}$. Let $c$ be a cycle in the full pattern transfer graph of $\mathcal{G}_{\vartheta}$ such that, when restricted on the noise induced graph, it equals $c_{\mathrm{noise}}$. We call such $c$ a lifted cycle which can be obtained by joining successive noise-induced edges with paths in the corresponding ideal connected components. Denote the cycle vector of the cycle $c$ by $x\in\mathbb{Z}_{(2)}^{|E_{\mathrm{id}}|+|E_{\mathrm{noise}}|}$. The protocol below measures each edge separately and combines the results classically. Zero-depth reference measurements remove the SPAM factors, including their signs, without requiring a circuit whose signal contains the product of all noise-induced edges.

\begin{mybox}[label=box:first_order_cycle]{First-order learning of a cycle edge by edge}
    Each circuit contains at most one use of the unknown channel. Products along the cycle are reconstructed in classical post-processing.
    \begin{enumerate}
        \item For every pattern vertex $v$ occurring in the lifted cycle $c$, choose a Pauli representative $P_v$. Ideally prepare a $+1$ eigenstate of $P_v$ and measure the observable $P_v$. Denote the actual state and observable by $\rho_v$ and $\widetilde M_v$, and define their Pauli overlaps as $r_v=\dbraket{P_v}{\rho_v}$ and $m_v=\dbraket{\widetilde M_v}{P_v}$. Use the same SPAM settings every time the vertex $v$ occurs and assume $r_v$ and $m_v$ are nonzero.

        \item For every such vertex $v$, perform the zero-depth character-projected experiment
        \begin{equation}\label{eq:first-order-cycle-anchor}
            A_v=\mathbb E_{R\sim\mathsf P_n}\left[\chi(P_v,R)\dbra{\widetilde M_v}\mathcal R\dket{\rho_v}\right]=m_vr_v.
        \end{equation}

        \item For each physical directed edge $e:u\to v$ in the lifted cycle, let $P_e$ and $Q_e$ be the input and output Pauli operators defining that PTM edge, so its weight is $g_e=\mathcal G_{Q_e,P_e}$. Choose local Clifford transformations satisfying
        \begin{equation}
            \begin{split}
                \mathcal C_e^{\mathrm{in}}\dket{P_u}&=\dket{P_e},\\
                \mathcal C_e^{\mathrm{out}}\dket{Q_e}&=\dket{P_v}.
            \end{split}
        \end{equation}
        Independently sample $R_0,R_1\in\mathsf P_n$, implement
        \begin{equation}
            \mathcal S_e(R_0,R_1)=\mathcal C_e^{\mathrm{out}}\mathcal R_1\mathcal G\mathcal C_e^{\mathrm{in}}\mathcal R_0,
        \end{equation}
        and estimate the character-weighted expectation
        \begin{equation}\label{eq:first-order-single-edge-signal}
            \begin{split}
                W_e
                &=\mathbb E_{R_0,R_1\sim\mathsf P_n}\left[\chi(P_u,R_0)\chi(Q_e,R_1)\dbra{\widetilde M_v}\mathcal S_e(R_0,R_1)\dket{\rho_u}\right]\\
                &=m_vr_u g_e.
            \end{split}
        \end{equation}
        For a noise-induced edge this is an $O(\varepsilon)$ signal, while for an ideal edge it has a nonzero zeroth-order value.

        \item\label{step:cancellation} For every vertex $v$, compute
        \begin{equation}\label{eq:first-order-cycle-vertex-exponent}
            q_v=\sum_{e:\,s(e)=v}x_e=\sum_{e:\,t(e)=v}x_e,
        \end{equation}
        where the equality follows from the cycle condition $Bx=0$. To combine the measurements without losing signs, compute the logarithmic magnitude and sign bit separately:
        \begin{equation}\label{eq:first-order-edgewise-cycle-estimator}
            \begin{split}
                L_x&=\sum_e x_e\log|W_e|-\sum_v q_v\log|A_v|,\\
                S_x&=\sum_e\overline{x_e}s(W_e)-\sum_v\overline{q_v}s(A_v)
                \quad\text{in }\mathbb F_2.
            \end{split}
        \end{equation}
        Here the bar denotes the parity map defined in Sec.~\ref{subsec:complex-logarithm-preliminaries}. At each vertex, the incoming and outgoing exponents agree, so the reference $A_v=m_vr_v$ cancels both SPAM factors. Thus $(L_x,S_x)$ is the signed logarithm of the lifted cycle product $\prod_e g_e^{x_e}$.

        \item Remove the known ideal-edge contributions and the noise normalization. Let $E_{\mathrm{id}}(x)$ and $E_{\mathrm{noise}}(x)$ denote the ideal and noise-induced edges used by the lifted cycle, and return
        \begin{equation}\label{eq:first-order-cycle-parameter-estimator}
            \begin{split}
                a_x&=\varepsilon\left[L_x
                -\left(\sum_{e\in E_{\mathrm{noise}}(x)}x_e\right)\log\varepsilon
                -\sum_{e\in E_{\mathrm{id}}(x)}x_e\log|\mathcal U_e|\right],\\
                b_x&=S_x-\sum_{e\in E_{\mathrm{id}}(x)}\overline{x_e}s(\mathcal U_e)
                \quad\text{in }\mathbb F_2.
            \end{split}
        \end{equation}
        On an ideal edge, $\log|g_e|-\log|\mathcal U_e|=O(\varepsilon)$ and $s(g_e)=s(\mathcal U_e)$ for sufficiently small $\varepsilon$. Consequently,
        \begin{equation}
            a_x=\sum_{e\in E_{\mathrm{noise}}(x)}x_e\theta_e^{\mathrm{noise}}+O(\varepsilon^2),
            \qquad
            b_x=\sum_{e\in E_{\mathrm{noise}}(x)}\overline{x_e}\sigma_e.
        \end{equation}
        These are the real and sign components of the function associated with the original noise-induced cycle.
    \end{enumerate}
\end{mybox}

The cancellation in Step~\ref{step:cancellation} can also be seen directly from the product. Since $Bx=0$, each preparation factor $r_v$ and measurement factor $m_v$ occurs with the same total exponent $q_v$, giving
\begin{equation}
    \frac{\prod_e W_e^{x_e}}{\prod_v A_v^{q_v}}
    =\frac{\left(\prod_e g_e^{x_e}\right)\prod_v(r_vm_v)^{q_v}}
    {\prod_v(r_vm_v)^{q_v}}
    =\prod_e g_e^{x_e}.
\end{equation}
The canonical powers for exponents in $\mathbb Z_{(2)}$ preserve this identity for weights and SPAM overlaps of either sign. Step~\ref{step:cancellation} performs the same cancellation using sums of logarithmic magnitudes and sign bits.

For the two-edge example, the reconstruction is simply
\begin{equation}
    \mathcal G_{e_1}\mathcal G_{e_2}=\frac{W_{e_1}W_{e_2}}{A_uA_v}.
\end{equation}
Both $W_{e_1}$ and $W_{e_2}$ are measured separately as $O(\varepsilon)$ signals. Their $O(\varepsilon^2)$ product is formed only in classical post-processing. The same principle applies to a cycle containing $k$ noise-induced edges: the experiment measures the individual $O(\varepsilon)$ edge signals and the reference signals, rather than an $O(\varepsilon^k)$ product signal. This is the advantage of learning the cycle edge by edge.

\subsection{First-order learnability of Lindbladian errors}\label{subsec:first-order-Lindbladian-errors}

We now apply the preceding first-order framework to study a Lindbladian with a known ideal Hamiltonian and small coherent and dissipative error. Consider
\begin{equation}\label{eq:first-order-Lindbladian-model}
    \mathcal L=\mathcal L_0+\mathcal E,
    \qquad
    \mathcal E=O(\varepsilon),
\end{equation}
where there is a known ideal Hamiltonian $H_0$ such that $\mathcal{L}_0(\rho)=-i[H_0,\rho]$ and
\begin{equation}\label{eq:first-order-Lindbladian-error}
\begin{split}
    \mathcal E(\rho)=-i\sum_a\widetilde c_a[P_a,\rho]+\sum_{i,j\neq0}\gamma_{i,j}\left(P_i\rho P_j-\frac12\{P_jP_i,\rho\}\right),
\end{split}
\end{equation}
where the coherent error coefficients $\widetilde c_a$ and dissipative error coefficients $\gamma_{i,j}$ are themselves $O(\varepsilon)$. These are the physical error coefficients, together ensuring the generator error $\mathcal E=O(\varepsilon)$ is first order small.

To simplify the problem, we assume that the error generator is unital, i.e. $\mathcal E(I)=0$. Together with trace preservation, the $0$ node is isolated in the pattern transfer graph. It therefore suffices to consider the remaining $2^{n}-1$ nodes. The ideal graph $\Gamma_0^{\mathrm{id}}(\mathcal L)$ is formed from the nonzero PTM entries of $\mathcal L_0$. Every pair $P_b,P_c$ with $\tr(P_c[H_0,P_b])\neq0$ gives an edge $\pt(P_b)\to\pt(P_c)$, and distinct PTM entries remain distinct edges. Equivalently, it is the Hamiltonian pattern transfer graph of the ideal Hamiltonian $H_0$.

Assume further that $\Gamma_0^{\mathrm{id}}(\mathcal L)$ is connected on all nonzero patterns. Its contracted noise-induced graph, denoted by $\Gamma^{\mathrm{noise}}(\mathcal L)$, then has a single vertex, so every noise-induced edge becomes a self-loop. Consequently, every noise induced transfer has endpoints whose gauge ratio is $1+O(\varepsilon)$. Multiplying this ratio by an $O(\varepsilon)$ transfer changes it only at second order. Such transfers are therefore individually learnable to first order, even though the two endpoint patterns need not coincide.

Define the error transfer functions $F_{a}^{\mathrm{err}}$ by
\begin{equation}\label{eq:first-order-error-transfer-function}
    \mathcal E\dket{P_b}=\sum_aF_a^{\mathrm{err}}(b)\omega(a,b)\dket{P_{a\oplus b}}.
\end{equation}
Similarly, let $F_a^0(b)$ denote the transfer function of $\mathcal L_0$. Then the transfer function of the physical generator $\mathcal L$ is
\begin{equation}\label{eq:first-order-total-Lindbladian-transfer-function}
    F_a^{\mathcal L}(b)=F_a^0(b)+F_a^{\mathrm{err}}(b),
\end{equation}
where $F_a^{\mathrm{err}}(b)=O(\varepsilon)$ is the physical error transfer coefficient and carries the same $O(\varepsilon)$ scaling as $(\widetilde c,\gamma)$. Equation~\eqref{eq:Fab-c-gamma} shows that it is linear in these coefficients, while unitality is equivalent to $F_a^{\mathrm{err}}(0)=0$ for every $a$. Since every noise-induced transfer weight is individually first-order learnable, any real linear combination of these weights is also first-order learnable. We can therefore work directly with the weights themselves, without forming cycle products or taking logarithms.

On the ideal edges, the relative corrections $\theta_e^{\mathrm{id}}=g_e^{\mathrm{err}}/g_e^0$ are already linear in the physical error weights, with known denominators. Thus the learnability problem in both sectors can be expressed in terms of real linear combinations of $g^{\mathrm{err}}$ and described using real vector spaces.

To describe these real linear functions, we use the following graph spaces over $\mathbb R$. For a directed graph $\Gamma=(V,E)$, the \emph{edge space} $C_1(\Gamma)=\mathbb R^E$ consists of vectors assigning one real coefficient to each edge. Let $B$ be its incidence matrix, with $-1$ at an edge's source and $+1$ at its target. And a self-loop has a zero column. The cycle space is $Z(\Gamma)=\ker B$, whose vectors have equal total incoming and outgoing coefficients at each vertex. The \emph{cut space} is $U(\Gamma)=\operatorname{im}B^{\mathsf T}$. With the usual inner product on edge coefficients,
\begin{equation}
    C_1(\Gamma)=Z(\Gamma)\oplus U(\Gamma),
    \qquad U(\Gamma)=Z(\Gamma)^\perp.
\end{equation}
In particular, when every edge is a self-loop, $B=0$ and the cycle space is the entire edge space. With these definition, we can develop the following learnability theorem. 

\begin{proposition}[First-order Lindbladian learnability]\label{prop:first-order-connected-Lindbladian}
    Let $E_{\mathrm{id}}$ and $E_{\mathrm{noise}}$ be the edge sets of $\Gamma_0^{\mathrm{id}}(\mathcal L)$ and $\Gamma^{\mathrm{noise}}(\mathcal L)$, respectively, and write $m_{\mathrm{id}}=|E_{\mathrm{id}}|$ and $m_{\mathrm{noise}}=|E_{\mathrm{noise}}|$. On an ideal edge $e$, write the physical weight as $g_e=g_e^0+g_e^{\mathrm{err}}$, where $g_e^{\mathrm{err}}=O(\varepsilon)$, and define $D_0=\operatorname{diag}(g_e^0)_{e\in E_{\mathrm{id}}}$. On a noise-induced edge $P_b\to P_{a\oplus b}$, the physical weight is $g_e=g_e^{\mathrm{err}}=\omega(a,b)F_a^{\mathrm{err}}(b)=O(\varepsilon)$. Denote by $\mathfrak L_F^{(1)}$ the real coefficient vectors $x$ for which $\langle x,g^{\mathrm{err}}\rangle$ is first-order learnable, and by $\mathfrak U_F^{(1)}$ the complementary coefficient space obtained from the ideal cut space. All edge, cycle, and cut spaces in this proposition are taken over $\mathbb R$. Then
    \begin{equation}\label{eq:first-order-Lindbladian-F-space}
        \begin{split}
            \mathfrak L_F^{(1)}&=D_0^{-1}Z\bigl(\Gamma_0^{\mathrm{id}}(\mathcal L)\bigr)\oplus C_1\bigl(\Gamma^{\mathrm{noise}}(\mathcal L)\bigr),\\
            \mathfrak U_F^{(1)}&=D_0^{-1}U\bigl(\Gamma_0^{\mathrm{id}}(\mathcal L)\bigr).
        \end{split}
    \end{equation}
    In particular, every noise-induced transfer function $F_a^{\mathrm{err}}(b)$ is individually first-order learnable. Moreover,
    \begin{equation}\label{eq:first-order-Lindbladian-dimensions}
        \dim\mathfrak L_F^{(1)}=m_{\mathrm{id}}+m_{\mathrm{noise}}-2^n+2,
        \qquad
        \dim\mathfrak U_F^{(1)}=2^n-2.
    \end{equation}
\end{proposition}

\begin{proof}
    On the ideal edges, $\theta^{\mathrm{id}}=D_0^{-1}g_{\mathrm{id}}^{\mathrm{err}}$. Thus a learnable cycle combination $\langle z,\theta^{\mathrm{id}}\rangle$ equals $\langle D_0^{-1}z,g_{\mathrm{id}}^{\mathrm{err}}\rangle$, giving the coefficient space $D_0^{-1}Z(\Gamma_0^{\mathrm{id}}(\mathcal L))$. The same change of coordinates maps the ideal cut space to $D_0^{-1}U(\Gamma_0^{\mathrm{id}}(\mathcal L))$, which is a direct-sum complement, not necessarily an orthogonal complement.

    For a noise-induced edge $e:u\to v$, connectedness gives $\lambda_v/\lambda_u=1+O(\varepsilon)$. Since $g_e^{\mathrm{err}}=O(\varepsilon)$,
    \begin{equation}
        (g_e^{\mathrm{err}})'=\frac{\lambda_v}{\lambda_u}g_e^{\mathrm{err}}
        =g_e^{\mathrm{err}}+O(\varepsilon^2).
    \end{equation}
    Every such error weight is therefore individually first-order learnable, and arbitrary real linear combinations give the entire noise edge space. For nonzero first-order weights, this also preserves the sign and changes $\theta_e^{\mathrm{noise}}$ only by $O(\varepsilon^2)$, in agreement with the signed-log formulation. The direct argument remains valid for weights whose first-order term vanishes. Dividing by the known nonzero phase $\omega(a,b)$ gives the same first-order invariance for $F_a^{\mathrm{err}}(b)$.

    Finally, the connected ideal graph has $2^n-1$ vertices, so its cycle and cut spaces have dimensions $m_{\mathrm{id}}-2^n+2$ and $2^n-2$, respectively. Each of the $m_{\mathrm{noise}}$ noise-induced edges becomes a self-loop after contraction and contributes one further learnable coordinate. This proves Eqs.~\eqref{eq:first-order-Lindbladian-F-space} and~\eqref{eq:first-order-Lindbladian-dimensions}.
\end{proof}

For the full unital error model, all $(4^n-1)^2$ transfers between nonidentity Pauli operators are present. Proposition~\ref{prop:first-order-connected-Lindbladian} therefore gives
\begin{equation}\label{eq:first-order-general-unital-Lindbladian-dimensions}
    \dim\mathfrak L_F^{(1)}=(4^n-1)^2-2^n+2,
    \qquad
    \dim\mathfrak U_F^{(1)}=2^n-2.
\end{equation}

It remains to express this result in terms of the conventional Lindbladian coefficients. Let $d=4^n$. Under the traceless-jump convention reviewed in Sec.~\ref{subsec:Lindblad-preliminaries}, their real parameter space is
\begin{equation*}
    \Theta=\mathbb R^{d-1}\oplus\operatorname{Herm}(d-1).
\end{equation*}
We restrict to the unital subspace
\begin{equation*}
    \Theta_{\mathrm u}=\left\{(\widetilde c,\gamma)\in\Theta:\mathcal E_{\widetilde c,\gamma}(I)=0\right\}.
\end{equation*}
For $(\widetilde c,\gamma)\in\Theta_{\mathrm u}$, the error generator preserves the traceless Pauli subspace. We therefore define $A$ to be the linear map from $(\widetilde c,\gamma)$ to the $(d-1)\times(d-1)$ Pauli transfer block of $\mathcal E$, excluding only line and column involving $0$, written as $g^{\mathrm{err}}=A(\widetilde c,\gamma)$. A learnable linear functional $x$ of the error transfer weights pulls back to the coefficient functional $A^{\mathsf T}x$. Hence the learnable coefficient space is
\begin{equation}\label{eq:first-order-Lindbladian-coefficient-space}
    \mathfrak L_{\widetilde c,\gamma}^{(1)}=A^{\mathsf T}\mathfrak L_F^{(1)}=\left\{A^{\mathsf T}x:x\in\mathfrak L_F^{(1)}\right\}.
\end{equation}
Trace preservation and unitality set the identity-output row and identity-input column of the error generator to zero. The remaining $(d-1)\times(d-1)$ block therefore specifies the entire generator, and Lemma~\ref{lemma:Lindbladian-Fourier-reconstruction} recovers its coefficients $(\widetilde c,\gamma)$ uniquely and linearly. Conversely, any real block of this size, completed with a zero identity row and column, defines a Hermiticity-preserving, trace-annihilating, unital map and hence a unique coefficient vector in $\Theta_{\mathrm u}$. Thus $A$ gives a one-to-one linear correspondence between $\Theta_{\mathrm u}$ and $\mathbb R^{(d-1)\times(d-1)}$. Consequently, $\dim\Theta_{\mathrm u}=(d-1)^2=(4^n-1)^2$, and the change of parametrization preserves both the learnable and unlearnable dimensions:
\begin{equation}\label{eq:first-order-Lindbladian-coefficient-dimensions}
    \dim\mathfrak L_{\widetilde c,\gamma}^{(1)}=(4^n-1)^2-2^n+2,
    \qquad
    \dim\mathfrak U_{\widetilde c,\gamma}^{(1)}=2^n-2.
\end{equation}
Thus the pattern-transfer graph determines the learnable combinations of $F_a^{\mathrm{err}}(b)$, and the Pauli Fourier transformation carries them directly to learnable combinations of $\widetilde c_a$ and $\gamma_{i,j}$.

The dimension count describes the total amount of learnable information, but it does not identify which conventional Lindbladian coefficients can be recovered individually. The simplest case is the diagonal Fourier block $a=0$, for which
\begin{equation}\label{eq:first-order-Pauli-rate-transform}
    \gamma_{j,j}=\frac{1}{4^n}\sum_b\chi_{j,b}F_0^{\mathrm{err}}(b).
\end{equation}
For $b\neq0$, $F_0^{\mathrm{err}}(b)$ is a PTM self-loop and is therefore individually learnable. The remaining value $F_0^{\mathrm{err}}(0)=0$ is fixed by unitality. Equation~\eqref{eq:first-order-Pauli-rate-transform} then shows that every diagonal rate $\gamma_{j,j}$ is individually learnable. This is the continuous-time analogue of the Fourier relation between Pauli decay parameters and Pauli error rates~\cite{Learnability,vanDenBerg2023}.

The same idea applies to the off-diagonal coefficients, but now some Fourier modes overlap with transfers generated by the ideal Hamiltonian. Lemma~\ref{lemma:Lindbladian-Fourier-reconstruction} associates $\gamma_{i,j}$ with the Fourier mode $\widehat F_{i\oplus j}^{\mathrm{err}}(j)$, while $\widetilde c_a$ is determined by the two modes at $k=0$ and $k=a$. It is therefore enough to understand when an individual Fourier mode is invariant under the first-order gauge freedom. The following proposition gives a complete criterion.

\begin{proposition}[Individual first-order learnability of Lindbladian Fourier modes]\label{prop:first-order-individual-Lindbladian-Fourier}
    Write $H_0=\sum_{a}c_a^0P_a$ and let $\mathcal A_0=\{a:c_a^0\neq0\}$ denote its Pauli support. Under the assumptions of Proposition~\ref{prop:first-order-connected-Lindbladian},
    \begin{equation}\label{eq:first-order-individual-Fourier-characterization}
        \widehat F_a^{\mathrm{err}}(k)\text{ is individually first-order learnable}
        \quad\Longleftrightarrow\quad
        a\notin\mathcal A_0\quad\text{or}\quad
        \chi_{a,k}=1\quad\text{or}\quad
        \pt(P_k)=\pt(P_{a\oplus k}).
    \end{equation}
\end{proposition}

\begin{proof}[Proof sketch]
    A first-order gauge changes an error transfer only through its ideal contribution: $\Delta F_a^{\mathrm{err}}(b)=\varepsilon(s_{\pt(P_{a\oplus b})}-s_{\pt(P_b)})F_a^0(b)+O(\varepsilon^2)$. Thus every Fourier coefficient in a block with $a\notin\mathcal A_0$ is invariant to first order. When $a\in\mathcal A_0$ and $\chi_{a,k}=1$, the terms indexed by $b$ and $a\oplus b$ cancel in the Fourier sum. In the remaining case, collecting the terms by Pauli pattern shows that the variation vanishes precisely when $\pt(P_k)=\pt(P_{a\oplus k})$. If the patterns differ, a gauge potential on a single-qubit pattern produces a nonzero first-order change. The full calculation is given in Appendix~\ref{app:first-order-Lindbladian-Fourier-proof}.
\end{proof}

The three conditions in Proposition~\ref{prop:first-order-individual-Lindbladian-Fourier} have simple interpretations. The first means that ideal Hamiltonian does not have Pauli operator $P_{a}$ as a term. In the second, $P_a$ and $P_k$ commute, so the Fourier character is unchanged under the pairing $b\mapsto a\oplus b$. The third means that the transfer associated with the Fourier mode has the same input and output pattern. Translating these conditions through the Pauli Fourier transformation gives the following coefficient-level statement.

\begin{corollary}[Individual first-order learnability of Lindbladian coefficients]\label{cor:first-order-individual-Lindbladian-coefficients}
    Consider an $n$-qubit Lindbladian $\mathcal L(\rho)=-i[H_0,\rho]+\mathcal E(\rho)$ with known ideal Hamiltonian $H_0=\sum_{a}c_a^0P_a$ and error generator
    \begin{equation}
        \mathcal E(\rho)=-i\sum_{a\neq0}\widetilde c_a[P_a,\rho]
        +\sum_{i,j\neq0}\gamma_{i,j}\left(P_i\rho P_j-\frac12\{P_jP_i,\rho\}\right),
    \end{equation}
    where all $\widetilde c_a$ and $\gamma_{i,j}$ are $O(\varepsilon)$. Assume that $\mathcal E$ is unital, $\mathcal E(I)=0$, and that the ideal pattern-transfer graph is connected on all nonzero patterns when edge directions are ignored. 

    Let $\mathcal A_0=\{a:c_a^0\neq0\}$. Then every coherent error coefficient $\widetilde c_a$ is individually first-order learnable. For $i,j\neq0$, the coefficient $\gamma_{i,j}$ is individually first-order learnable whenever at least one of the following three condition holds
    \begin{equation}\label{eq:first-order-individual-gamma-conditions}
        i\oplus j\notin\mathcal A_0,
        \qquad
        \chi_{i,j}=1,
        \qquad\text{or}\qquad
        \pt(P_i)=\pt(P_j).
    \end{equation}
    If none of these conditions holds, $\operatorname{Im}\gamma_{i,j}$ remains individually first-order learnable, whereas $\operatorname{Re}\gamma_{i,j}$ need not be. Hence every first-order unlearnable direction lies in this exceptional real off-diagonal sector of the Kossakowski matrix.
\end{corollary}
To see the corollary directly, Lemma~\ref{lemma:Lindbladian-Fourier-reconstruction} gives
\begin{equation}\label{eq:first-order-coefficient-Fourier-identities}
    \widetilde c_a=\frac{i}{2}\left[\widehat F_a^{\mathrm{err}}(0)-\widehat F_a^{\mathrm{err}}(a)\right],
    \qquad
    \gamma_{i,j}=\omega(i,j)^*\widehat F_{i\oplus j}^{\mathrm{err}}(j).
\end{equation}
The two Fourier modes determining $\widetilde c_a$ satisfy $\chi_{a,0}=\chi_{a,a}=1$, so Proposition~\ref{prop:first-order-individual-Lindbladian-Fourier} makes every coherent coefficient learnable. Applying the same proposition to the second identity gives the three alternatives in Eq.~\eqref{eq:first-order-individual-gamma-conditions}. In the remaining case, $P_i$ and $P_j$ anticommute and hence $\omega(i,j)=\pm i$. Equation~\eqref{eq:first-order-Fourier-pattern-gauge-change} then shows that the first-order gauge variation of $\gamma_{i,j}$ is real. It can therefore affect $\operatorname{Re}\gamma_{i,j}$ but not $\operatorname{Im}\gamma_{i,j}$.

Although the exceptional real coefficients need not be individually learnable, further information remains available through linear combinations whose gauge changes cancel. Appendix~\ref{app:collective-Lindbladian-learnability} characterizes all such combinations and gives a direct procedure for finding them.

An immediate consequence is a fully individually learnable and experimentally relevant error model.  Let $\mathsf W_1$ denote the set of weight-one Pauli indices and suppose the dissipative error satisfies $\gamma_{i,j}=0$ unless $i,j\in\mathsf W_1$.  If $P_i$ and $P_j$ act on the same qubit, then they have the same pattern; if they act on different qubits, then they commute.  Thus one of the last two conditions in Eq.~\eqref{eq:first-order-individual-gamma-conditions} always holds.  We obtain the following corollary.

\begin{corollary}[Unital weight-one dissipation]\label{cor:first-order-weight-one-complete-learnability}
    Suppose the ideal Hamiltonian pattern-transfer graph is connected on all nonzero patterns and the first-order error generator is unital, with an arbitrary coherent error and a Kossakowski matrix supported on $\mathsf W_1\times\mathsf W_1$.  Then every coherent coefficient $\widetilde c_a$ and every allowed Kossakowski coefficient $\gamma_{i,j}$ in this model is individually first-order learnable, subject only to the known linear relations imposed by unitality and Hermiticity.
\end{corollary}

This corollary illustrates how model structure can eliminate the $2^n-2$ cut directions present in the unrestricted unital Lindbladian model: none of those gauge directions remains inside the weight-one dissipative parameter space.

\section{Case study}\label{sec:case}

\subsection{Non-Clifford gates with Pauli noise}

We begin with a deliberately simple example: a known non-Clifford gate followed by an unknown Pauli channel.  Unlike the Clifford setting, Pauli noise after a non-Clifford gate is not in general justified by randomized compiling, so this model should not be viewed as a generic description of experimental noise.  Its purpose here is instead to illustrate a basic consequence of the pattern-transfer graph and to isolate the role played by the ideal gate.

In the usual Clifford-plus-Pauli-noise setting, the ideal Clifford channel permutes the Pauli operators.  The resulting graph therefore contains relatively few self-loops, and SPAM-robust experiments generally determine cycle products of Pauli-noise parameters rather than the individual parameters~\cite{Learnability}.  A non-Clifford gate can mix Pauli operators and, in particular, can have a nonzero diagonal overlap with many of them.  Each such overlap creates a self-loop and makes the corresponding Pauli-noise parameter individually learnable.

Let $G$ be a known $n$-qubit unitary channel, not necessarily Clifford, and let its noisy implementation be
\begin{equation}\label{eq:nonclifford-Pauli-noise-model}
    \widetilde{\mathcal G}=\Lambda\mathcal G,
    \qquad
    \Lambda=\sum_{P\in\mathsf P_n}\lambda_P\dketbra{P}{P},
\end{equation}
where $\mathcal G$ denotes the Pauli transfer matrix of $G$ and $\Lambda$ is an unknown Pauli channel.  We take $\lambda_I=1$ and regard the remaining $\lambda_P$ as the noise parameters.

\begin{proposition}[Individual learnability from a diagonal overlap]
\label{prop:nonclifford-diagonal-overlap}
    For every nonidentity Pauli operator $P$, if
    \begin{equation}\label{eq:nonclifford-diagonal-overlap-condition}
        \tr\!\bigl(PG(P)\bigr)\neq0,
    \end{equation}
    then the Pauli-noise parameter $\lambda_P$ is individually learnable.  In particular, if Eq.~\eqref{eq:nonclifford-diagonal-overlap-condition} holds for every $P\neq I$, then the entire Pauli channel $\Lambda$ is learnable.
\end{proposition}

\begin{proof}
    The diagonal Pauli-transfer entry of the implemented channel is
    \begin{equation}\label{eq:nonclifford-Pauli-diagonal-entry}
        \widetilde{\mathcal G}_{P,P}
        =\lambda_P\mathcal G_{P,P}
        =2^{-n}\lambda_P\tr\!\bigl(PG(P)\bigr).
    \end{equation}
    This entry is represented by a self-loop at the pattern vertex $\pt(P)$. A self-loop lies in the cycle space of the pattern transfer graph and is therefore learnable by Theorem~\ref{thm:general_cptp_learnability}.  Since $G$ is known and the overlap in Eq.~\eqref{eq:nonclifford-diagonal-overlap-condition} is nonzero, we can divide the learned entry by its known ideal value to obtain
    \begin{equation}
        \lambda_P
        =\frac{\widetilde{\mathcal G}_{P,P}}{\mathcal G_{P,P}}
        =\frac{2^n\widetilde{\mathcal G}_{P,P}}
        {\tr\!\bigl(PG(P)\bigr)}.
    \end{equation}
\end{proof}

For comparison, if $G$ is Clifford, then $G(P)=\pm P'$ for a Pauli $P'$, and $\tr(PG(P))$ is nonzero only when $P'=P$.  The proposition then recovers individual learnability only for Pauli operators fixed by the Clifford action. A non-Clifford gate need not permute the Pauli basis, and its diagonal overlaps can be nonzero for substantially more Pauli operators.  It can consequently turn parameters that occur only in nontrivial cycle products in the Clifford setting into individually learnable quantities.

The diagonal-overlap condition is sufficient but not necessary.  Even when $\mathcal G_{P,P}=0$, off-diagonal PTM entries may further restrict the gauge and allow $\lambda_P$ to be isolated from learnable cycle products.  This motivates the following complete characterization.

For each output Pauli $P$, define its set of input patterns by
\begin{equation}\label{eq:nonclifford-input-pattern-set}
    \mathcal I(P)=\left\{\pt(Q):\mathcal G_{P,Q}\neq0\right\}
    \subseteq\{0,1\}^n.
\end{equation}
Since $G$ is a unitary channel, every row is nonzero and $\mathcal I(P)$ is nonempty.

\begin{definition}[Pauli-input constraint graph]
    The Pauli-input constraint graph $K_G$ has vertex set $\{0,1\}^n$.  Two distinct patterns $u$ and $v$ are connected if, for some Pauli $R$,
    \begin{equation}
        u,v\in\mathcal I(R).
    \end{equation}
    Equivalently, each set $\mathcal I(R)$ forms a clique in $K_G$.
\end{definition}

All patterns in $\mathcal I(P)$ lie in one connected component, denoted by $C_{\mathrm{in}}(P)$.  Let $C_{\mathrm{out}}(P)$ be the component containing $\pt(P)$.

\begin{proposition}[Individual learnability of Pauli-noise parameters]
\label{prop:nonclifford-Pauli-complete-characterization}
    Suppose every $\lambda_P$ is nonzero and every sufficiently small diagonal gauge preserving the model in Eq.~\eqref{eq:nonclifford-Pauli-noise-model} remains physical.  Then, for every nonidentity Pauli $P$,
    \begin{equation}\label{eq:nonclifford-individual-characterization}
        \lambda_P\text{ is individually learnable}
        \quad\Longleftrightarrow\quad
        C_{\mathrm{out}}(P)=C_{\mathrm{in}}(P).
    \end{equation}
\end{proposition}

\begin{proof}
    Write a diagonal gauge compatible with the trusted local Clifford gates as
    \begin{equation}
        T=\sum_R\tau_{\pt(R)}\dketbra{R}{R}.
    \end{equation}
    On a nonzero entry in the $P$-th output row, it acts as
    \begin{equation}\label{eq:nonclifford-model-gauge-action}
        \widetilde{\mathcal G}_{P,Q}=\lambda_P\mathcal G_{P,Q}
        \quad\longmapsto\quad
        \frac{\tau_{\pt(P)}}{\tau_{\pt(Q)}}
        \lambda_P\mathcal G_{P,Q}.
    \end{equation}
    For the transformed channel to retain the form $\Lambda'\mathcal G$, one parameter $\lambda'_P$ must multiply the entire $P$-th row.  Hence, for every $\pt(Q)\in\mathcal I(P)$,
    \begin{equation}\label{eq:nonclifford-transformed-Pauli-parameter}
        \lambda'_P
        =\frac{\tau_{\pt(P)}}{\tau_{\pt(Q)}}\lambda_P.
    \end{equation}
    Since $\lambda_P\neq0$, this expression is independent of $Q$ only if $\tau_u=\tau_v$ for all $u,v\in\mathcal I(P)$.  Over all output Paulis, these are precisely the constraints that $\tau$ be constant on each connected component of $K_G$. If $C_{\mathrm{out}}(P)=C_{\mathrm{in}}(P)$, the ratio in Eq.~\eqref{eq:nonclifford-transformed-Pauli-parameter} is one, so $\lambda'_P=\lambda_P$ under every valid gauge.

    Conversely, if the two components differ, assign them distinct gauge values sufficiently close to one while keeping $\tau$ constant within every component.  This gauge preserves the model and remains physical by assumption, but it changes $\lambda_P$.  Thus $\lambda_P$ is unlearnable.
\end{proof}

The constraint graph also captures how individual learnability propagates beyond self-loops.  For example, suppose $P$ has no diagonal overlap but there is a Pauli $Q$ satisfying
\begin{equation}\label{eq:nonclifford-two-cycle-condition}
    \mathcal G_{Q,Q}\neq0,
    \qquad
    \mathcal G_{P,Q}\mathcal G_{Q,P}\neq0.
\end{equation}
The nonzero entries $\mathcal G_{Q,Q}$ and $\mathcal G_{Q,P}$ place $\pt(Q)$ and $\pt(P)$ in the same component of $K_G$, while $\pt(Q)\in\mathcal I(P)$.  Proposition~\ref{prop:nonclifford-Pauli-complete-characterization} therefore implies that $\lambda_P$ is individually learnable.  The same fact can be seen directly from the learnable two-cycle product
\begin{equation}
    \widetilde{\mathcal G}_{P,Q}\widetilde{\mathcal G}_{Q,P}
    =\lambda_P\lambda_Q\mathcal G_{P,Q}\mathcal G_{Q,P}.
\end{equation}
Here $\lambda_Q$ is learned from its self-loop, after which the displayed product determines $\lambda_P$.  Longer chains of constraints similarly propagate learnability even when no two-edge cycle of this form is present.

As a concrete multiqubit example for which the entire Pauli channel is individually learnable, consider the three-qubit Toffoli gate.  To evaluate its diagonal PTM overlaps, first consider the locally Clifford-equivalent gate
\begin{equation}
    V=CCZ=\sum_{x\in\mathbb F_2^3}(-1)^{f(x)}\ket{x}\!\bra{x},
    \qquad f(x)=x_1x_2x_3.
\end{equation}
Using the Hermitian phase convention, write a three-qubit Pauli operator as $P=i^{a\cdot b}X^aZ^b$, where $a,b\in\mathbb F_2^3$.  A direct calculation in the computational basis gives
\begin{equation}\label{eq:CCZ-diagonal-Pauli-overlap}
    2^{-3}\tr\!\bigl(PVPV^{\dagger}\bigr)
    =\frac{1}{8}\sum_{x\in\mathbb F_2^3}
    (-1)^{f(x)+f(x\oplus a)}
    =
    \begin{cases}
        1, & a=000,\\
        \frac{1}{2}, & a\neq000.
    \end{cases}
\end{equation}
In particular, the overlap is nonzero for every Pauli operator.  The Toffoli gate is related to $CCZ$ by
\begin{equation}
    U_{\mathrm{Toffoli}}
    =(I\otimes I\otimes H)V(I\otimes I\otimes H).
\end{equation}
Conjugation by the local Clifford $I\otimes I\otimes H$ merely permutes the Pauli basis, so the diagonal PTM overlaps of the Toffoli channel are likewise all nonzero.  Proposition~\ref{prop:nonclifford-diagonal-overlap} therefore implies that, for a Toffoli gate followed by an arbitrary three-qubit Pauli channel, all $63$ nontrivial Pauli-noise parameters $\lambda_P$ are individually learnable.

\subsection{Finite-time $ZZ$ rotation with coherent, $T_1$, and $T_2$ noise}

We next consider a more realistic two-qubit example.  Rotations generated by a $ZZ$ interaction are experimentally important entangling operations: they arise naturally from Ising-type couplings, have been implemented on several quantum-computing platforms, and form basic building blocks for digital quantum simulation~\cite{Wu2019,Bao2024,Layden2026}.  Since a generic-angle $R_{ZZ}(\theta)$ gate is non-Clifford, understanding which features of its coherent and incoherent errors admit SPAM-robust characterization is therefore of direct practical interest.

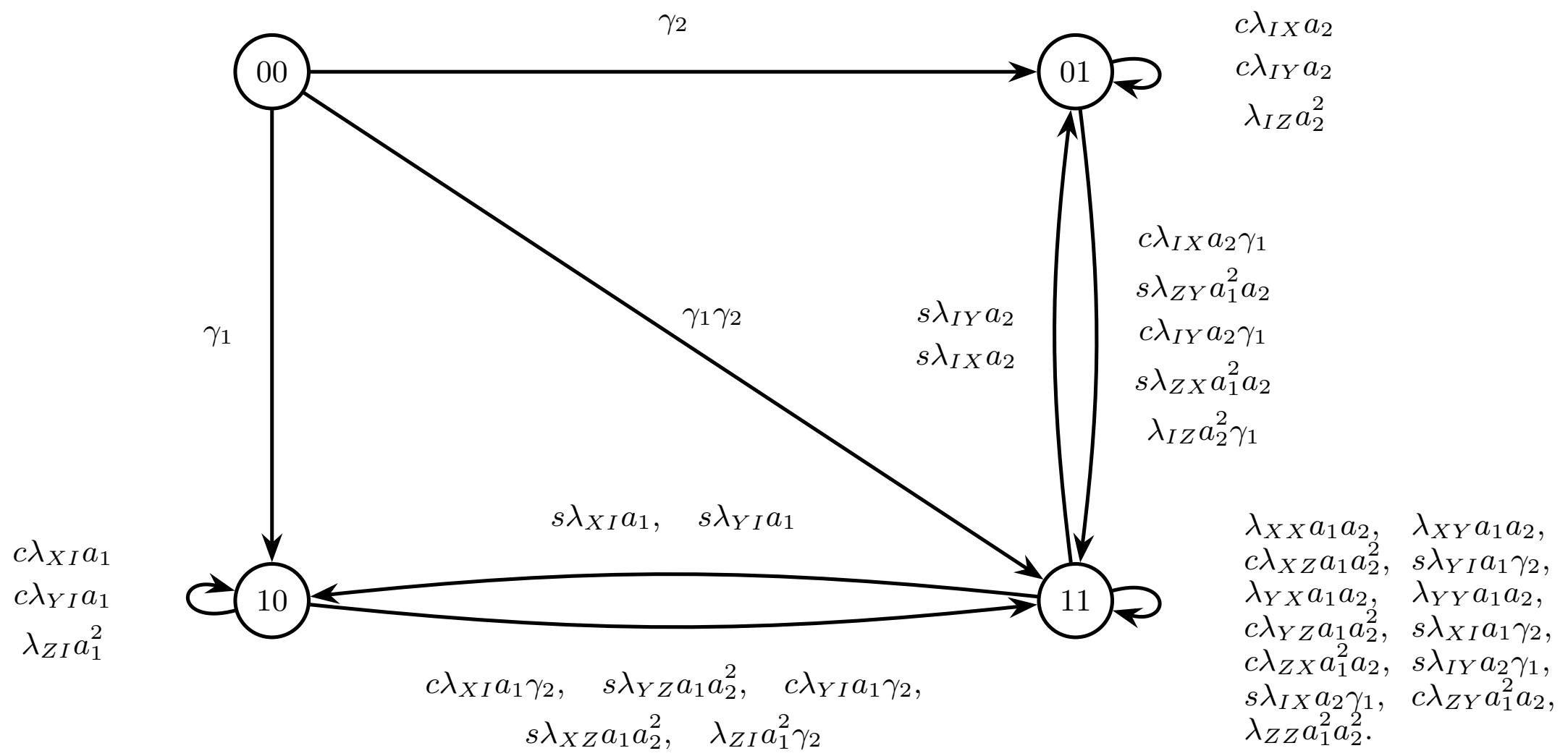
\begin{figure}[htbp]
    \centering
    \resizebox{0.8\textwidth}{!}{    \begin{tikzpicture}[
      scale=0.85,transform shape,
      weight/.append style={font=\small,inner sep=2pt},
      edge/.append style={line width=0.9pt}
    ]
      \node[vertex] (v00) at (0,5.4) {$00$};
      \node[vertex] (v01) at (8.2,5.4) {$01$};
      \node[vertex] (v10) at (0,0)   {$10$};
      \node[vertex] (v11) at (8.2,0) {$11$};
    
      \draw[edge] (v00) -- node[weight,above=3mm] {$\gamma_2$} (v01);
      \draw[edge] (v00) -- node[weight,left=3mm]  {$\gamma_1$} (v10);
      \draw[edge] (v00) -- node[weight,above right] {$\gamma_1\gamma_2$} (v11);
    
      \draw[edge] (v01) to[loop right]
        node[weight,right=7mm] {$\begin{gathered}
          c\lambda_{IX}a_2\\[-1pt]
          c\lambda_{IY}a_2\\[-1pt]
          \lambda_{IZ}a_2^2
        \end{gathered}$} (v01);
    
      \draw[edge] (v11) to[bend left=7]
        node[weight,left=3mm] {$\begin{gathered}
          s\lambda_{IY}a_2\\[-1pt]
          s\lambda_{IX}a_2
        \end{gathered}$} (v01);
    
      \draw[edge] (v01) to[bend left=7]
        node[weight,right=3mm] {$\begin{gathered}
          c\lambda_{IX}a_2\gamma_1\\[-1pt]
          s\lambda_{ZY}a_1^2a_2\\[-1pt]
          c\lambda_{IY}a_2\gamma_1\\[-1pt]
          s\lambda_{ZX}a_1^2a_2\\[-1pt]
          \lambda_{IZ}a_2^2\gamma_1
        \end{gathered}$} (v11);
    
      \draw[edge] (v11) to[bend right=6]
        node[weight,above=3mm] {$\begin{gathered}
          s\lambda_{XI}a_1,\quad s\lambda_{YI}a_1
        \end{gathered}$} (v10);
    
      \draw[edge] (v10) to[bend right=6]
        node[weight,below=3mm] {$\begin{gathered}
          c\lambda_{XI}a_1\gamma_2,\quad s\lambda_{YZ}a_1a_2^2,\quad
          c\lambda_{YI}a_1\gamma_2,\\[-1pt]
          s\lambda_{XZ}a_1a_2^2,\quad \lambda_{ZI}a_1^2\gamma_2
        \end{gathered}$} (v11);
    
      \draw[edge] (v10) to[loop left]
        node[weight,left=7mm] {$\begin{gathered}
          c\lambda_{XI}a_1\\[-1pt]
          c\lambda_{YI}a_1\\[-1pt]
          \lambda_{ZI}a_1^2
        \end{gathered}$} (v10);
    
      \draw[edge] (v11) to[loop right]
        node[weight,right=8mm,yshift=-3mm] {$\begin{array}{@{}ll@{}}
          \lambda_{XX}a_1a_2,       & \lambda_{XY}a_1a_2,\\[-1pt]
          c\lambda_{XZ}a_1a_2^2,      & s\lambda_{YI}a_1\gamma_2,\\[-1pt]
          \lambda_{YX}a_1a_2,       & \lambda_{YY}a_1a_2,\\[-1pt]
          c\lambda_{YZ}a_1a_2^2,      & s\lambda_{XI}a_1\gamma_2,\\[-1pt]
          c\lambda_{ZX}a_1^2a_2,      & s\lambda_{IY}a_2\gamma_1,\\[-1pt]
          s\lambda_{IX}a_2\gamma_1,& c\lambda_{ZY}a_1^2a_2,\\[-1pt]
          \lambda_{ZZ}a_1^2a_2^2.       &
        \end{array}$} (v11);
    \end{tikzpicture}}
    \caption{Pattern transfer graph of the noisy $ZZ$ rotation in
    Eq.~\eqref{eq:ZZ-rotation-channel}. Each displayed expression is the
    weight $g_e/C_e$ of a distinct PTM edge in the product-entry
    parametrization of Eq.~\eqref{eq:ZZ-product-entry-form}.}
    \label{fig:ZZ-rotation}
\end{figure}

We model three error mechanisms: a coherent overrotation by an unknown angle $\delta$, a two-qubit Pauli channel describing stochastic dephasing and decoherence, and independent amplitude damping on the two qubits.  Define
\begin{equation}
    R_{ZZ}(\alpha)=e^{-i\alpha ZZ/2},
    \qquad
    \mathcal R_{ZZ}(\alpha)(\rho)
    =R_{ZZ}(\alpha)\rho R_{ZZ}(\alpha)^{\dagger}.
\end{equation}
The noisy operation is
\begin{equation}\label{eq:ZZ-rotation-channel}
    \widetilde{\mathcal R}_{ZZ}
    =\bigl(\mathcal A_1(\gamma_1)\otimes\mathcal A_2(\gamma_2)\bigr)
    \circ\Lambda_P\circ\mathcal R_{ZZ}(\delta)
    \circ\mathcal R_{ZZ}(\theta),
\end{equation}
where $\theta$ is known and
\begin{equation}
    \Lambda_P=\sum_{P\in\mathsf P_2}\lambda_P\dketbra{P}{P},
    \qquad \lambda_{II}=1.
\end{equation}
Here $\gamma_i\in[0,1]$ is the amplitude-damping probability on qubit $i$. Writing $a_i=\sqrt{1-\gamma_i}$, the action of amplitude damping noise on the single-qubit Pauli basis is
\begin{equation}\label{eq:amplitude-damping-Pauli-action}
    I\mapsto I+\gamma_i Z,\qquad
    X\mapsto a_iX,\qquad
    Y\mapsto a_iY,\qquad
    Z\mapsto a_i^2Z.
\end{equation}
Since rotations about the same Pauli axis add, it is convenient to define
\begin{equation}\label{eq:ZZ-auxiliary-parameters}
    \phi=\theta+\delta,\qquad
    c=\cos\phi,\qquad s=\sin\phi.
\end{equation}
To place this noise model in the product-entry framework of Definition~\ref{def:product-entry}, we use the auxiliary parameter vector
\begin{equation}\label{eq:ZZ-product-entry-parameters}
    \eta=
    \left(c,s,a_1,a_2,\gamma_1,\gamma_2,
    (\lambda_P)_{P\in\mathsf P_2\backslash\{II\}}\right).
\end{equation}
These parameters obey the known physical constraints
\begin{equation}\label{eq:ZZ-product-entry-constraints}
    c^2+s^2=1,
    \qquad
    a_i^2=1-\gamma_i,\qquad i=1,2.
\end{equation}
The purpose of this change of variables is to expose the product-entry structure: every nonzero PTM edge weight can be written as
\begin{equation}\label{eq:ZZ-product-entry-form}
    g_e=C_e\prod_{\ell}\eta_{\ell}^{m_{e,\ell}},
\end{equation}
with known constants $C_e$ and $m_{e,\ell}$, exactly as in Definition~\ref{def:product-entry}.  The components of $\eta$ are therefore not regarded as independent physical parameters; the relations in Eq.~\eqref{eq:ZZ-product-entry-constraints} are retained when recovering $\delta$ and the damping probabilities. Figure~\ref{fig:ZZ-rotation} shows the resulting pattern transfer graph. Multiple labels beside the same arrow represent distinct parallel PTM edges, and the known factors $C_e$ are omitted because they do not affect learnability.

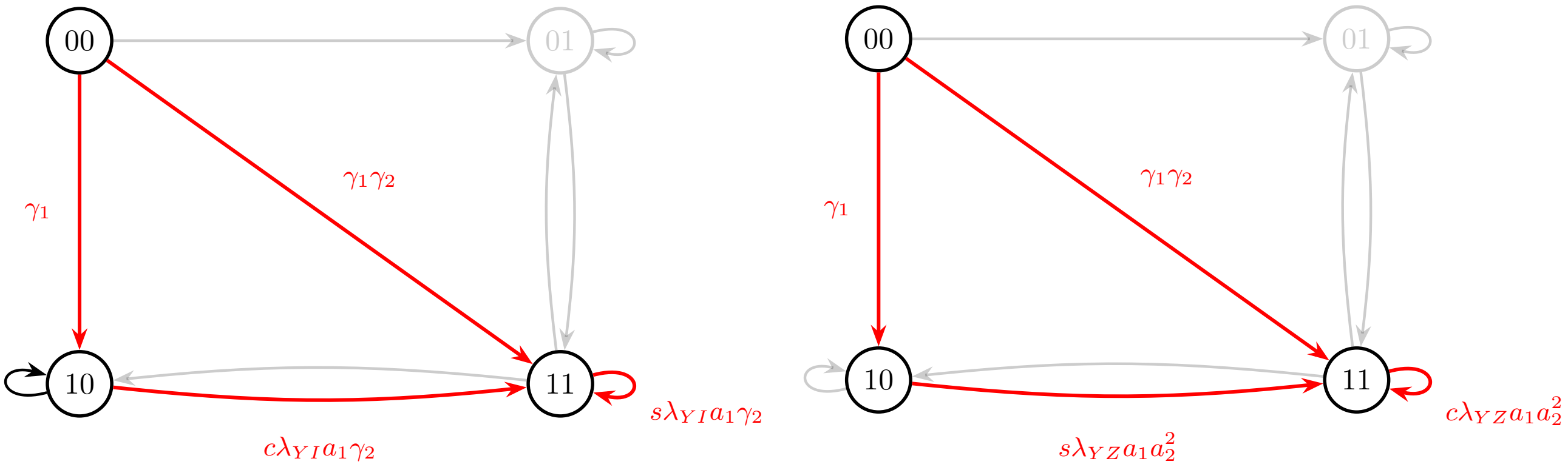
\begin{figure}[htbp]
    \centering
    \resizebox{0.48\textwidth}{!}{
\begin{tikzpicture}[
  vertex/.append style={font=\normalsize,line width=1.1pt},
  weight/.append style={font=\small,inner sep=3pt},
  highlight/.style={edge,red,line width=1.2pt},
  background edge/.style={edge,opacity=0.2}
]
  \node[vertex] (v00) at (0,4) {$00$};
  \node[vertex,opacity=0.2] (v01) at (5.6,4) {$01$};
  \node[vertex] (v10) at (0,0) {$10$};
  \node[vertex] (v11) at (5.6,0) {$11$};

  \draw[background edge] (v00) -- (v01);
  \draw[background edge] (v01) to[loop right] (v01);
  \draw[background edge] (v11) to[bend left=7] (v01);
  \draw[background edge] (v01) to[bend left=7] (v11);
  \draw[background edge] (v11) to[bend right=6] (v10);
  \draw[edge] (v10) to[loop left] (v10);

  \draw[highlight] (v00) -- node[weight,left=2mm] {$\gamma_1$} (v10);
  \draw[highlight] (v00) -- node[weight,above right=2mm] {$\gamma_1\gamma_2$} (v11);
  \draw[highlight] (v10) to[bend right=6]
    node[weight,below=3mm] {$c\lambda_{YI}a_1\gamma_2$} (v11);
  \draw[highlight] (v11) to[loop right]
    node[weight,below right=1mm] {$s\lambda_{YI}a_1\gamma_2$} (v11);
\end{tikzpicture}}
    \resizebox{0.48\textwidth}{!}{
\begin{tikzpicture}[
  vertex/.append style={font=\normalsize,line width=1.1pt},
  weight/.append style={font=\small,inner sep=3pt},
  highlight/.style={edge,red,line width=1.2pt},
  background edge/.style={edge,opacity=0.2}
]
  \node[vertex] (v00) at (0,4) {$00$};
  \node[vertex,opacity=0.2] (v01) at (5.6,4) {$01$};
  \node[vertex] (v10) at (0,0) {$10$};
  \node[vertex] (v11) at (5.6,0) {$11$};

  \draw[background edge] (v00) -- (v01);
  \draw[background edge] (v01) to[loop right] (v01);
  \draw[background edge] (v11) to[bend left=7] (v01);
  \draw[background edge] (v01) to[bend left=7] (v11);
  \draw[background edge] (v11) to[bend right=6] (v10);
  \draw[background edge] (v10) to[loop left] (v10);

  \draw[highlight] (v00) -- node[weight,left=2mm] {$\gamma_1$} (v10);
  \draw[highlight] (v00) -- node[weight,above right=2mm] {$\gamma_1\gamma_2$} (v11);
  \draw[highlight] (v10) to[bend right=6]
    node[weight,below=3mm] {$s\lambda_{YZ}a_1a_2^2$} (v11);
  \draw[highlight] (v11) to[loop right]
    node[weight,below right=1mm] {$c\lambda_{YZ}a_1a_2^2$} (v11);
\end{tikzpicture}}
    \caption{Two signed cycle invariants used to determine $\gamma_2$.  Red
    edges enter the corresponding cycle product, while faded edges are not
    used.}
    \label{fig:learning gamma}
\end{figure}

We first show that the two amplitude-damping probabilities are individually learnable.  Consider the two signed cycles highlighted in Figure~\ref{fig:learning gamma}.  The left cycle gives the invariant
\begin{equation}\label{eq:gamma_cycle_1}
    r_{2,c}
    =\frac{\gamma_1(c\lambda_{YI}a_1\gamma_2)}
    {(s\lambda_{YI}a_1\gamma_2)(\gamma_1\gamma_2)}
    =\frac{c}{s\gamma_2},
\end{equation}
whereas the right cycle gives
\begin{equation}\label{eq:gamma_cycle_2}
    r_{2,s}
    =\frac{\gamma_1(s\lambda_{YZ}a_1a_2^2)}
    {(c\lambda_{YZ}a_1a_2^2)(\gamma_1\gamma_2)}
    =\frac{s}{c\gamma_2}.
\end{equation}
We work here in the generic regime in which the displayed factors are nonzero. Thus multiplying the two invariants cancels both the rotation angle and the Pauli eigenvalues:
\begin{equation}\label{eq:learn-gamma2}
    r_{2,c}r_{2,s}=\frac{1}{\gamma_2^2}.
\end{equation}
Because $\gamma_2\geq0$, this determines $\gamma_2$ uniquely.  Exchanging the two qubits and using the analogous cycles through the pattern vertex $01$ determines $\gamma_1$. Consequently, $a_1$ and $a_2$ are also known through the relation $a_i=\sqrt{1-\gamma_i}$. Once the damping parameters are known, several self-loop weights immediately determine the seven Pauli eigenvalues
\begin{equation}\label{eq:ZZ-individually-learnable-Pauli}
    \lambda_{IZ},\ \lambda_{ZI},\ \lambda_{XX},\ \lambda_{XY},\
    \lambda_{YX},\ \lambda_{YY},\ \lambda_{ZZ}.
\end{equation}
For example, the $IZ$ self-loop has weight $\lambda_{IZ}a_2^2$, and the $XX$ self-loop has weight $\lambda_{XX}a_1a_2$; all accompanying damping factors are now known and can be divided out. The same graph determines the remaining eight eigenvalues, viewed as separate parameters in the decomposition, up to one common binary ambiguity.  Let
\begin{equation}
    \mathcal S_{\mathrm{ac}}
    =\{IX,IY,XI,YI,XZ,YZ,ZX,ZY\},
\end{equation}
which is precisely the set of two-qubit Paulis that anticommute with $ZZ$. For every $P\in\mathcal S_{\mathrm{ac}}$, a self-loop in Figure~\ref{fig:ZZ-rotation} determines $c\lambda_P$.  Moreover, Eq.~\eqref{eq:gamma_cycle_2} and the learned value of $\gamma_2$ give
\begin{equation}\label{eq:ZZ-learn-tangent}
    \tan\phi=\frac{s}{c}=\gamma_2r_{2,s}.
\end{equation}
Thus $\phi=\theta+\delta$ is determined modulo $\pi$, and the magnitude and relative sign information of every $\lambda_P$ is fixed.  The only remaining transformation is
\begin{equation}\label{eq:ZZ-binary-ambiguity}
    \phi\longmapsto\phi+\pi,\qquad
    \lambda_P\longmapsto-\lambda_P
    \quad\text{for every }P\in\mathcal S_{\mathrm{ac}},
\end{equation}
with all other parameters unchanged.  Indeed, $\mathcal R_{ZZ}(\phi+\pi)$ differs from $\mathcal R_{ZZ}(\phi)$ by conjugation with $ZZ$, whose Pauli eigenvalues are $-1$ exactly on $\mathcal S_{\mathrm{ac}}$.  The sign changes in Eq.~\eqref{eq:ZZ-binary-ambiguity} absorb this conjugation into $\Lambda_P$, leaving the complete channel in Eq.~\eqref{eq:ZZ-rotation-channel} unchanged.

Therefore the parameterization is identifiable up to a single $\mathbb Z_2$ ambiguity.  Crucially, this does not represent missing physical information about the implemented operation.  The two parameter assignments in Eq.~\eqref{eq:ZZ-binary-ambiguity} produce exactly the same composed channel, so every operationally meaningful property of $\widetilde{\mathcal R}_{ZZ}$ is learned.  What remains undetermined is only how that channel is decomposed into a coherent $ZZ$ overrotation and a Pauli channel.

Channel data cannot select between two exactly identical decompositions, even with perfect SPAM.  In a calibration setting, however, the conventional weak-error assumption fixes the parameterization: one may require $\delta$ to lie in an interval of width less than $\pi$ around zero, or equivalently select the Pauli channel close to the identity rather than the branch close to a deterministic $ZZ$ error.  Under this standard branch convention, the overrotation, both amplitude-damping probabilities, and all fifteen nontrivial Pauli eigenvalues are individually determined.  Thus the noisy $ZZ$-rotation model is fully learnable in the operational sense, and fully parameter-learnable once the physically natural weak-error branch is chosen.

\subsection{$ZZ$ rotation with first-order unital Lindbladian noise}

The preceding analysis exploits the product-entry structure of a concrete finite-time noise model. We now use the same $ZZ$ interaction to illustrate a different strength of the first-order framework: it can give a complete learnability characterization without specifying the individual coherent and dissipative error mechanisms. To keep this characterization closed within the nonidentity Pauli sector, we restrict the perturbation to be unital. Thus the present model and the preceding one make complementary assumptions. Here the error parametrization is much more general, but the errors are required to be weak and unital.

We therefore take the ideal generator to be the $ZZ$ interaction and write the implemented generator as
\begin{equation}\label{eq:first-order-ZZ-Lindbladian-model}
    \mathcal L=\mathcal L_0+\mathcal E,
    \qquad
    \mathcal L_0(\rho)=-i[JZZ,\rho],
    \qquad
    \mathcal E=O(\varepsilon),
    \qquad
    \mathcal E(I)=0,
\end{equation}
Apart from its magnitude and unitality, the perturbation $\mathcal E$ is unrestricted: it may contain an arbitrary coherent Hamiltonian error and arbitrary dissipative coefficients compatible with a unital Lindbladian, including off-diagonal coefficients and correlated multiqubit dissipation. The condition $\mathcal E(I)=0$ eliminates transfers from the identity to traceless Pauli operators, so the identity pattern decouples and the learnability problem is completely described by the graph on the three nonzero patterns. This is precisely the setting in which the general coefficient-level results of the preceding section yield a clean answer. The restriction does exclude nonunital processes such as ordinary amplitude damping; its role here is therefore not to claim the most general experimental noise model, but to isolate a broad class for which every learnable and unlearnable direction can be identified explicitly.

The ideal Hamiltonian pattern-transfer graph is connected on the three nonzero two-qubit patterns. Indeed, $XI\leftrightarrow YZ$ connects $10$ with $11$, while $IX\leftrightarrow ZY$ connects $01$ with $11$. The ideal Pauli support is simply $\mathcal A_0=\{ZZ\}$, so Proposition~\ref{prop:first-order-individual-Lindbladian-Fourier} and Corollary~\ref{cor:first-order-individual-Lindbladian-coefficients} apply. In particular, every coherent error coefficient $\widetilde c_a$ is individually first-order learnable.

We next apply the three conditions in Eq.~\eqref{eq:first-order-individual-gamma-conditions} to the $\gamma$ matrix. If $P_iP_j$ is not proportional to $ZZ$, equivalently $i\oplus j\neq ZZ$, then the first condition applies. If $i\oplus j=ZZ$ but $P_i$ and $P_j$ commute, the second condition applies. The only remaining possibility is therefore
\begin{equation}\label{eq:first-order-ZZ-exceptional-conditions}
    i\oplus j=ZZ,
    \qquad
    \chi_{i,j}=-1,
    \qquad
    \pt(P_i)\neq\pt(P_j).
\end{equation}
Multiplication by $ZZ$ exchanges $I\leftrightarrow Z$ and $X\leftrightarrow Y$ on each qubit. Condition~\eqref{eq:first-order-ZZ-exceptional-conditions} therefore requires one qubit to carry an $X/Y$ pair, which produces the anticommutation, and the other to carry an $I/Z$ pair, which changes the pattern. Up to Hermitian conjugation, this leaves exactly four exceptional entries:
\begin{equation}\label{eq:first-order-ZZ-exceptional-gamma}
    \gamma_{ZX,IY},\qquad \gamma_{ZY,IX},\qquad
    \gamma_{XZ,YI},\qquad \gamma_{XI,YZ}.
\end{equation}
Every other $\gamma$ coefficient is individually learnable. The imaginary parts of the four entries in Eq.~\eqref{eq:first-order-ZZ-exceptional-gamma} are also individually learnable; only their real parts can vary under a first-order gauge. The remaining ambiguity can be obtained explicitly from Eq.~\eqref{eq:first-order-Fourier-pattern-gauge-change}. Let $s_{10},s_{01},s_{11}$ be the gauge potentials of the three nonzero patterns. Using $c_{ZZ}^0=J$ and the Pauli phases of the four pairs gives
\begin{equation}\label{eq:first-order-ZZ-exceptional-variations}
\begin{split}
    \Delta\operatorname{Re}\gamma_{ZX,IY}
    &=\frac{\varepsilon J}{2}(s_{10}-s_{11})+O(\varepsilon^2),\\
    \Delta\operatorname{Re}\gamma_{ZY,IX}
    &=-\frac{\varepsilon J}{2}(s_{10}-s_{11})+O(\varepsilon^2),\\
    \Delta\operatorname{Re}\gamma_{XZ,YI}
    &=\frac{\varepsilon J}{2}(s_{01}-s_{11})+O(\varepsilon^2),\\
    \Delta\operatorname{Re}\gamma_{XI,YZ}
    &=-\frac{\varepsilon J}{2}(s_{01}-s_{11})+O(\varepsilon^2).
\end{split}
\end{equation}
The opposite signs within each pair show directly that
\begin{equation}\label{eq:first-order-ZZ-learnable-gamma-sums}
    \operatorname{Re}(\gamma_{ZX,IY}+\gamma_{ZY,IX}),
    \qquad
    \operatorname{Re}(\gamma_{XZ,YI}+\gamma_{XI,YZ})
\end{equation}
are learnable. By contrast, the corresponding differences respond independently to $s_{10}-s_{11}$ and $s_{01}-s_{11}$ and span the two unlearnable directions. Thus the number $223$ has a direct interpretation: among the $(4^2-1)^2=225$ independent real parameters of a general two-qubit unital error generator, only these two real combinations are lost to gauge freedom. All remaining $223$ independent combinations are first-order learnable.

\subsection{$Z$-type Lindbladian jumps}

Dephasing provides a natural intermediate case between purely Hamiltonian evolution and a completely general Lindbladian.  In many platforms the computational basis is physically preferred, and the dominant dissipative couplings are diagonal in that basis.  Motivated by this regime, we consider the simplified model in which the dissipative noise is generated entirely by $Z$-type operators, while retaining an otherwise arbitrary Hamiltonian term.  Such noise includes independent single-qubit dephasing, correlated dephasing, and coherent linear combinations of multiqubit $Z$ strings.  The Abelian structure of such model imposes additional constraints on the gauge and makes several physically meaningful coefficients individually learnable. Let $\mathsf Z_n=\{Z^u:u\in\mathbb F_2^n\}$ be the subgroup of $Z$-type Pauli operators.  We consider a Lindbladian of the form
\begin{equation}\label{eq:Z-type-Lindbladian}
\begin{split}
    \mathcal L_Z(\rho)
    &=-i[H,\rho]+\sum_{u,v\in\mathbb F_2^n\backslash\{0\}}
    \gamma_{u,v}
    \left(
        Z^u\rho Z^v-\frac{1}{2}\{Z^{u\oplus v},\rho\}
    \right),
\end{split}
\end{equation}
where the support of $H$ is unrestricted and $\gamma_Z=(\gamma_{u,v})_{u,v\neq0}$ is Hermitian and positive semidefinite. Equivalently, all Lindblad jump operators belong to the linear span of $\mathsf Z_n\backslash\{I\}$.  The diagonal entries $\gamma_{u,u}$ are the rates of the Pauli dephasing jumps $Z^u$, while off-diagonal entries describe correlations between different $Z$ strings.

We first identify an immediate and useful consequence of this structure.  For a Pauli $P_b$, the diagonal PTM element of the generator is
\begin{equation}\label{eq:Z-type-diagonal-generator-entry}
    (\mathcal L_Z)_{b,b}
    =F_0(b)
    =\sum_{u\neq0}\gamma_{u,u}
    \bigl(\chi(P_b,Z^u)-1\bigr).
\end{equation}
Neither the Hamiltonian nor the off-diagonal entries of $\gamma_Z$ contribute to this expression.  Moreover, $(\mathcal L_Z)_{b,b}$ is a self-loop at $\pt(P_b)$ and is therefore individually gauge invariant.

\begin{proposition}[Learnability of the Pauli dephasing rates]
\label{prop:Z-type-diagonal-rates}
    For the $Z$-type Lindbladian in Eq.~\eqref{eq:Z-type-Lindbladian}, every diagonal Kossakowski coefficient $\gamma_{u,u}$ is individually learnable.  It is reconstructed from the diagonal PTM entries by
    \begin{equation}\label{eq:Z-type-rate-reconstruction}
        \gamma_{u,u}
        =\frac{1}{4^n}\sum_{b\in\mathsf P_n}
        \chi(Z^u,P_b)(\mathcal L_Z)_{b,b},
        \qquad u\neq0.
    \end{equation}
    Consequently, if $\gamma_Z$ is diagonal, then every dissipative parameter of the dephasing Lindbladian is individually learnable, even in the presence of an arbitrary Hamiltonian.
\end{proposition}

\begin{proof}
    Equation~\eqref{eq:Z-type-diagonal-generator-entry} is the $a=0$ case of Eq.~\eqref{eq:Lindbladian-PT-function}.  Since the input and output Pauli are identical, every $(\mathcal L_Z)_{b,b}$ is unchanged by the diagonal gauges in Eq.~\eqref{eq:gauge_schur}.  Taking the Pauli Fourier transform of Eq.~\eqref{eq:Z-type-diagonal-generator-entry} and using character orthogonality gives Eq.~\eqref{eq:Z-type-rate-reconstruction}, in agreement with Eq.~\eqref{eq:Lindbladian-diagonal-Fourier-coefficients}.  Thus each $\gamma_{u,u}$ is a known linear combination of individually learnable self-loop entries.
\end{proof}

This reconstruction also has a direct experimental interpretation.  For a short evolution time $t$,
\begin{equation}
    \bigl(e^{t\mathcal L_Z}\bigr)_{b,b}
    =1+t(\mathcal L_Z)_{b,b}+O(t^2).
\end{equation}
A character-projected self-loop experiment estimates the left-hand side in a SPAM-robust manner.  Repeating it at several short times extracts the linear coefficient $(\mathcal L_Z)_{b,b}$, after which the Fourier transform in Eq.~\eqref{eq:Z-type-rate-reconstruction} returns all Pauli dephasing rates. The proposition therefore covers, for example, simultaneous local and correlated dephasing generated by
\begin{equation}
    \sum_{u\neq0}\gamma_u
    \left(Z^u\rho Z^u-\rho\right).
\end{equation}
Naively, the Fourier reconstruction uses all $4^n$ diagonal generator entries; the proposition is therefore an identifiability statement rather than an efficiency claim.  When the dephasing rates are local or sparse, the structural learning techniques discussed earlier can be used to reduce this cost.

The Abelian $Z$-type structure also constrains the off-diagonal transfer functions, following from a general Fourier duality between support and translation invariance. For a $Z$-type transfer label $a$, Lemma~\ref{lemma:Lindbladian-Fourier-reconstruction} and the restriction of both Kossakowski indices to $\mathsf Z_n$ imply
\begin{equation}\label{eq:Z-type-Fourier-support}
    \widehat F_a(k)=0
    \qquad\text{whenever }P_k\notin\mathsf Z_n.
\end{equation}
Indeed, the modes $k=0,a$ are $Z$-type, while every other mode is proportional to $\gamma_{k\oplus a,k}$ and can be nonzero only when both $P_k$ and $P_{k\oplus a}$ are $Z$-type. Since all elements of $\mathsf Z_n$ commute, $\chi_{k,z}=1$ whenever $P_k,P_z\in\mathsf Z_n$. Fourier inversion therefore gives
\begin{equation}\label{eq:Z-type-transfer-translation-symmetry}
\begin{split}
    F_a(b\oplus z)
    &=\sum_{k:\,P_k\in\mathsf Z_n}\chi_{k,b\oplus z}\widehat F_a(k)\\
    &=\sum_{k:\,P_k\in\mathsf Z_n}\chi_{k,b}\chi_{k,z}\widehat F_a(k)
    =F_a(b),
    \qquad P_z\in\mathsf Z_n.
\end{split}
\end{equation}
For a non-$Z$-type $a$, on the other hand, no pair of $Z$-type jump indices can satisfy $u\oplus v=a$.  Hence the dissipative contribution vanishes and
\begin{equation}\label{eq:non-Z-transfer-Hamiltonian-only}
    F_a(b)=-ic_a\bigl(1-\chi(P_a,P_b)\bigr),
    \qquad P_a\notin\mathsf Z_n.
\end{equation}

\begin{proposition}[Gauge restriction from $Z$-type jumps]
\label{prop:Z-type-gauge-restriction}
    Let
    \begin{equation}
        T=\sum_b\lambda_{\pt(P_b)}\dketbra{P_b}{P_b}
    \end{equation}
    be a diagonal gauge that maps the Lindbladian in Eq.~\eqref{eq:Z-type-Lindbladian} to another Lindbladian of the same form. Define
    \begin{equation}
        q_a(b)=
        \frac{\lambda_{\pt(P_{a\oplus b})}}
        {\lambda_{\pt(P_b)}}.
    \end{equation}
    Then the following statements hold wherever the corresponding transfer functions are nonzero:
    \begin{enumerate}
        \item If $P_a\notin\mathsf Z_n$ and $c_a\neq0$, then $q_a(b)$ is independent of $b$ over all $P_b$ anticommuting with $P_a$.  Thus the gauge acts on $c_a$ exactly as in the purely Hamiltonian case.

        \item If $P_a\in\mathsf Z_n$, then
        \begin{equation}\label{eq:Z-type-gauge-translation}
            q_a(b\oplus z)=q_a(b)
            \qquad\text{for every }z\in\mathsf Z_n,
        \end{equation}
        and, in particular,
        \begin{equation}\label{eq:Z-type-transfer-sign}
            q_a(b)\in\{+1,-1\},
            \qquad
            F'_a(b)=\pm F_a(b).
        \end{equation}
    \end{enumerate}
\end{proposition}

\begin{proof}
    If $P_a\notin\mathsf Z_n$, Eq.~\eqref{eq:non-Z-transfer-Hamiltonian-only} must retain the same Hamiltonian form after the gauge transformation $F'_a(b)=q_a(b)F_a(b)$.  Therefore $q_a(b)$ must be constant on the anticommuting inputs, proving the first statement.

    If $P_a\in\mathsf Z_n$, both the original and transformed transfer functions obey Eq.~\eqref{eq:Z-type-transfer-translation-symmetry}. Comparing
    \begin{equation}
        F'_a(b\oplus z)=q_a(b\oplus z)F_a(b)
        \quad\text{and}\quad
        F'_a(b)=q_a(b)F_a(b)
    \end{equation}
    gives Eq.~\eqref{eq:Z-type-gauge-translation}.  Taking $z=a$ yields
    \begin{equation}
        q_a(b\oplus a)
        =\frac{\lambda_{\pt(P_b)}}
        {\lambda_{\pt(P_{a\oplus b})}}
        =q_a(b)^{-1}.
    \end{equation}
    Equation~\eqref{eq:Z-type-gauge-translation} then implies $q_a(b)^2=1$, proving Eq.~\eqref{eq:Z-type-transfer-sign}.
\end{proof}

The first part of Proposition~\ref{prop:Z-type-gauge-restriction} imports the basic Hamiltonian conclusions into every non-$Z$ sector: odd-weight non-$Z$ Hamiltonian coefficients are individually learnable, while the magnitudes of even-weight non-$Z$ coefficients are learnable.  Relations among their remaining signs can be read from the Hamiltonian pattern graph using the non-$Z$ labels and may be tightened further by the $Z$-sector constraints. The second part shows that a $Z$-sector transfer function cannot be continuously rescaled by a valid structure-preserving gauge; at most, it can change sign.  Thus $|F_a(b)|$ is learnable for $Z$-type $a$, and cycle relations determine relative signs among many such functions.  Additional non-$Z$ Hamiltonian terms can connect these sign sectors and remove further ambiguities.

\subsection{Weight-one Lindbladian jumps}

Weight-one jump operators describe one of the most natural nontrivial Lindbladian models for quantum devices.  The standard Markovian mechanisms attached to an individual qubit all have this form: pure dephasing is generated by $Z_q$, relaxation and thermal excitation by $\sigma_q^{\mp}=(X_q\mp iY_q)/2$, and fluctuations of local control fields by linear combinations of $X_q,Y_q,Z_q$.  More broadly, the empirical success of sparse low-weight Pauli--Lindblad descriptions on multiqubit processors indicates that locality is an effective organizing principle for experimentally relevant noise~\cite{vanDenBerg2023}.  It is therefore useful to determine what can be learned when the dissipative Pauli support is restricted to weight one.

Restricting the dissipative support to weight-one Pauli operators does not require the noise on different qubits to be independent. A jump operator may be any linear combination $L_\mu=\sum_{i:\operatorname{wt}(P_i)=1}\ell_{\mu,i}P_i$, including collective decay such as $L=\sum_q\alpha_q\sigma_q^{-}$ induced by a shared environment. Accordingly, the Kossakowski matrix may have off-diagonal entries that couple Pauli operators on different qubits. The model therefore includes independent $T_1$ and $T_2$ processes, anisotropic single-qubit noise, and spatially correlated or collective decay. What it  excludes are jump operators containing Pauli terms of weight two or higher. Since there are $3n$ weight-one Pauli operators, the Kossakowski matrix has size $3n\times3n$, and the number of dissipative parameters grows quadratically with the number of qubits, making it far more structured than a general Lindbladian. Let
\begin{equation}
    \mathsf W_1=\{a\in\mathsf P_n:\operatorname{wt}(P_a)=1\}.
\end{equation}
We consider
\begin{equation}\label{eq:weight-one-Lindbladian}
\begin{split}
    \mathcal L_{(1)}(\rho)
    &=-i\sum_{a\neq0}c_a[P_a,\rho]+\sum_{i,j\in\mathsf W_1}\gamma_{i,j}
    \left(P_i\rho P_j-\frac{1}{2}\{P_jP_i,\rho\}\right),
\end{split}
\end{equation}
where $\gamma=(\gamma_{i,j})_{i,j\in\mathsf W_1}$ is Hermitian and positive semidefinite.  Equivalently, every jump operator can be chosen in the form $L_\mu=\sum_{i\in\mathsf W_1}\ell_{\mu,i}P_i$.  Off-diagonal entries of $\gamma$ between different qubits therefore retain collective and spatially correlated weight-one noise.

The locality assumption immediately separates the Pauli-transfer function into sectors.  A dissipative term contributing to $F_a$ must have $a=i\oplus j$ for some $i,j\in\mathsf W_1$, and hence $\operatorname{wt}(P_a)\leq2$.  Consequently,
\begin{equation}\label{eq:weight-one-high-weight-sector}
    F_a(b)=-ic_a\bigl(1-\chi_{a,b}\bigr)
    \qquad\text{whenever }\operatorname{wt}(P_a)\geq3.
\end{equation}
All sectors of weight greater than two are therefore exactly Hamiltonian, with no mixing whatsoever between $c_a$ and the dissipative coefficients.  Every learnability conclusion from the Hamiltonian analysis consequently remains valid: each odd-weight $c_a$ with $\operatorname{wt}(P_a)\geq3$ is individually learnable, while for every even-weight $c_a$ with $\operatorname{wt}(P_a)\geq4$, the magnitude $|c_a|$ are learnable as well as all relative signs and ratios within each Hamiltonian support component formed by purely Hamiltonian sectors by Corollary~\ref{cor:even-Hamiltonian}. The remaining sectors contain additional information about the dissipator. The following proposition summarizes the coefficients that can be isolated without further assumptions.

\begin{proposition}[Learnability under weight-one jumps]\label{prop:weight-one-Lindbladian}
    For the Lindbladian in Eq.~\eqref{eq:weight-one-Lindbladian}:
    \begin{enumerate}
        \item every Hamiltonian coefficient with $\operatorname{wt}(P_a)>2$ retains the similar  learnable information of the purely Hamiltonian model. Odd-weight coefficients are individually learnable.  Magnitudes of coefficients are learnable. Two even-weight Hamiltonian terms have learnable relative signs whenever they belong to the same connected component of the support hypergraph formed by terms whose transfer sectors are purely Hamiltonian.
        \item every entry of the real symmetric matrix $\operatorname{Re}\gamma$ is individually learnable;
        \item every weight-one Hamiltonian coefficient $c_a$ is individually learnable;
        \item for weight-two $a=u\oplus v$, where $u,v\in\mathsf W_1$ act on different qubits, the quantities
        \begin{equation}\label{eq:weight-one-mixed-invariants}
            \bigl(c_a-\operatorname{Im}\gamma_{u,v}\bigr)^2,\qquad
            \bigl(c_a+\operatorname{Im}\gamma_{u,v}\bigr)^2
        \end{equation}
        are learnable. Equivalently, $c_a^2+(\operatorname{Im}\gamma_{u,v})^2$ and $c_a\operatorname{Im}\gamma_{u,v}$ are learnable, although $c_a$ and $\operatorname{Im}\gamma_{u,v}$ need not be individually identifiable without an additional prior.
    \end{enumerate}
\end{proposition}

\begin{proof}
    Equation~\eqref{eq:weight-one-high-weight-sector} proves the first statement: the PTM entries in every sector of weight greater than two coincide with those of the purely Hamiltonian model, so all invariants from Corollaries~\ref{cor:odd-Hamiltonian} and~\ref{cor:even-Hamiltonian} remain available. The diagonal entries $\gamma_{i,i}$ are reconstructed from the diagonal generator entries by Eq.~\eqref{eq:Lindbladian-diagonal-Fourier-coefficients}, and hence are individually learnable. Now let $r,s,a\in\mathsf W_1$ act on the same qubit and choose their orientation so that $P_rP_s=iP_a$. Direct substitution into Eq.~\eqref{eq:Lindbladian-PT-function} gives
    \begin{equation}\label{eq:weight-one-same-qubit-self-loops}
        F_a(r)=-2i\bigl(c_a+\operatorname{Re}\gamma_{r,s}\bigr),
        \qquad
        F_a(s)=-2i\bigl(c_a-\operatorname{Re}\gamma_{r,s}\bigr).
    \end{equation}
    Both entries preserve the support pattern of the input Pauli and are therefore self-loops.  They are individually gauge invariant, and their sum and difference determine $c_a$ and $\operatorname{Re}\gamma_{r,s}$ separately.

    It remains to consider $u,v\in\mathsf W_1$ on different qubits and $a=u\oplus v$.  Choose single-qubit Paulis $\bar u$ and $\bar v$ that anticommute with $u$ and $v$, respectively.  The input $\bar u\oplus\bar v$ and its output under the $a$-sector have the same weight pattern, so the corresponding self-loop gives
    \begin{equation}\label{eq:weight-one-cross-qubit-real-part}
        F_a(\bar u\oplus\bar v)=-4\operatorname{Re}\gamma_{u,v}.
    \end{equation}
    This learns every cross-qubit entry of $\operatorname{Re}\gamma$.  On the other hand,
    \begin{equation}\label{eq:weight-one-cross-qubit-mixed-edges}
        F_a(\bar u)=-2i\bigl(c_a-\operatorname{Im}\gamma_{u,v}\bigr),
        \qquad
        F_a(\bar v)=-2i\bigl(c_a+\operatorname{Im}\gamma_{u,v}\bigr).
    \end{equation}
    Translation by $v$ or $u$, respectively, produces the reverse edge with the same Pauli-transfer function. Each pair forms a two-edge cycle, which determines the square of the corresponding expression in Eq.~\eqref{eq:weight-one-cross-qubit-mixed-edges}. This proves Eq.~\eqref{eq:weight-one-mixed-invariants} and the proposition.
\end{proof}

The proposition gives a useful physical summary. All diagonal dissipative rates and all real correlations among weight-one Pauli components are accessible, as are arbitrary local coherent fields and all higher odd-weight Hamiltonian terms. An imaginary correlation between different qubits can produce the same transfer sector as a two-qubit Hamiltonian term. Prior knowledge of the intended Hamiltonian or of a specific relaxation model can resolve this ambiguity. Without such a prior, the combinations in Eq.~\eqref{eq:weight-one-mixed-invariants} are the identifiable quantities in the weight-two sector.

We have not included $\operatorname{Im}\gamma_{r,s}$ for $P_r$ and $P_s$ on the same qubit in Proposition~\ref{prop:weight-one-Lindbladian}, because these coefficients are not individually learnable in general. Their physical meaning is nevertheless explicit: they generate transfers from the identity to weight-one Pauli operators and hence describe the nonunital part of the noise. To see this, choose the orientation $P_rP_s=iP_a$. The two terms related by Hermiticity contribute
\begin{equation}\label{eq:weight-one-nonunital-drift-pair}
\begin{split}
    \gamma_{r,s}[P_r,P_s]+\gamma_{s,r}[P_s,P_r]=(\gamma_{r,s}-\gamma_{s,r})[P_r,P_s]=-4\operatorname{Im}\gamma_{r,s}P_a
\end{split}
\end{equation}
to $\mathcal L_{(1)}(I)$. Hamiltonian terms vanish on $I$, and weight-one Paulis on different qubits commute. Therefore
\begin{equation}\label{eq:weight-one-nonunital-drift}
    \mathcal L_{(1)}(I)
    =-4\sum_{q=1}^n\left(
    \operatorname{Im}\gamma_{Y_q,Z_q}X_q
    +\operatorname{Im}\gamma_{Z_q,X_q}Y_q
    +\operatorname{Im}\gamma_{X_q,Y_q}Z_q
    \right).
\end{equation}
There are three independent same-qubit entries of $\operatorname{Im}\gamma$ per qubit and three weight-one coefficients in the nonunital part of $\mathcal L_{(1)}(I)$ per qubit. Equation~\eqref{eq:weight-one-nonunital-drift} therefore gives a one-to-one correspondence between these two $3n$-dimensional real spaces. For a fixed qubit, the three terms in Eq.~\eqref{eq:weight-one-nonunital-drift} are parallel PTM edges from the identity pattern to the same weight-one pattern and acquire the same factor under a diagonal gauge. The ratio of any two nonzero edge weights is consequently a signed cycle product. Thus, for a cyclic triple satisfying $P_rP_s=iP_a$, quantities such as $\operatorname{Im}\gamma_{r,s}/{\operatorname{Im}\gamma_{s,a}}$
are learnable whenever the denominator is nonzero. The standard cycle-product protocol therefore determines the relative values of the three local nonunital coefficients.

\subsection{Weight-one plus $Z$-type Lindbladian jumps}

The two preceding models capture complementary experimental mechanisms.  Weight-one jumps describe local relaxation, excitation, dephasing, and collective decay mediated by linear combinations of single-qubit operators, whereas multiqubit $Z$-type jumps describe correlated phase noise generated by common-mode frequency fluctuations or longitudinal coupling to an environment.  It is therefore natural to allow both structures simultaneously.  The resulting model retains the most common local dissipative processes together with potentially long-range correlated dephasing, while still excluding generic multiqubit transverse jumps.

Define the allowed Pauli support $\mathsf S_{1Z}=\mathsf W_1\cup\bigl(\mathsf Z_n\backslash\{I\}\bigr)$,
and consider
\begin{equation}\label{eq:weight-one-Z-Lindbladian}
\begin{split}
    \mathcal L_{1Z}(\rho)
    &=-i\sum_{a\neq0}c_a[P_a,\rho]+\sum_{i,j\in\mathsf S_{1Z}}\gamma_{i,j}
    \left(P_i\rho P_j-\frac{1}{2}\{P_jP_i,\rho\}\right),
\end{split}
\end{equation}
where $\gamma_{1Z}=(\gamma_{i,j})_{i,j\in\mathsf S_{1Z}}$ is Hermitian and positive semidefinite.  Equivalently, each jump operator belongs to the linear span of all weight-one Paulis and all nonidentity $Z$ strings.  In particular, Eq.~\eqref{eq:weight-one-Z-Lindbladian} permits coherent correlations between the two sectors, not merely a sum of an independent weight-one dissipator and an independent dephasing dissipator.

To expose the structure, define the transverse weight
\begin{equation}
    \operatorname{wt}_{\perp}(P)
    =\bigl|\{q:P_q\in\{X,Y\}\}\bigr|.
\end{equation}
If a dissipative coefficient contributes to $F_a$, then $a=i\oplus j$ for two indices $i,j\in\mathsf S_{1Z}$.  This observation partitions the transfer labels into four classes.  If $\operatorname{wt}_{\perp}(P_a)\geq2$ and $\operatorname{wt}(P_a)\geq3$, no such pair exists, so the sector is purely Hamiltonian and all conclusions of Corollaries~\ref{cor:odd-Hamiltonian} and~\ref{cor:even-Hamiltonian} remain valid.  If $P_a$ is $Z$-type and has weight at least two, only $Z$-type pairs contribute, and the translation symmetry of Eq.~\eqref{eq:Z-type-transfer-translation-symmetry} applies.  If $\operatorname{wt}_{\perp}(P_a)=\operatorname{wt}(P_a)\in\{1,2\}$, every contributing pair is weight one, reducing the sector to Proposition~\ref{prop:weight-one-Lindbladian}.  The only genuinely new class consists of labels with exactly one transverse factor and a nonempty $Z$-type tail.

There is one minor overlap at a single-qubit label $P_a=Z_q$: besides the $Z$-type pairs, the weight-one pair $X_q,Y_q$ can contribute. In Fourier space these contributions are nevertheless separated, since the former occur at $Z$-type indices and the latter at $X_q,Y_q$. The universally learnable diagonal rates from Proposition~\ref{prop:Z-type-diagonal-rates} remain unchanged, while the remaining combinations in this exceptional sector can be read directly from the Fourier reconstruction of Lemma~\ref{lemma:Lindbladian-Fourier-reconstruction}. We now focus on the genuinely mixed class, where the new gauge argument is required.

Let $P_a$ have a unique transverse factor on qubit $q$ and write
\begin{equation}\label{eq:mixed-one-Z-decomposition}
    a=x\oplus z,\qquad
    x\in\{X_q,Y_q\},\qquad
    z=Z_B,\qquad
    \varnothing\neq B\subseteq[n]\backslash\{q\}.
\end{equation}
Define the other transverse Pauli on qubit $q$ by $x^\sharp=x\oplus Z_q$ and set $\widetilde z=Z_q\oplus z$.  The only decompositions of $a$ into two allowed dissipative indices are
\begin{equation}
    a=x\oplus z=x^\sharp\oplus\widetilde z.
\end{equation}
With this notation, Lemma~\ref{lemma:Lindbladian-Fourier-reconstruction} gives the six-point support bound
\begin{equation}\label{eq:mixed-one-Z-Fourier-support}
    \operatorname{supp}(\widehat F_a)
    \subseteq
    \mathcal K_a
    :=\{0,a,x,x^\sharp,z,\widetilde z\}.
\end{equation}
The first two indices contain the Hamiltonian contribution and the anticommutator corrections, while the remaining four encode the two dissipative pairs and their Hermitian conjugates.  Thus the apparently complicated mixed sector is still controlled by only four dissipative Fourier modes.

Consider the subgroup
\begin{equation}\label{eq:mixed-one-Z-translation-group}
    \mathsf H_a
    =\{\beta\in\mathsf P_n:
      \chi_{\beta,x}=\chi_{\beta,x^\sharp}
      =\chi_{\beta,z}=1\}.
\end{equation}
Every $\beta\in\mathsf H_a$ commutes with all six indices in $\mathcal K_a$. By the same Fourier-inversion argument used to derive the $Z$-type translation symmetry in Eq.~\eqref{eq:Z-type-transfer-translation-symmetry}, the support bound in Eq.~\eqref{eq:mixed-one-Z-Fourier-support} gives
\begin{equation}\label{eq:mixed-one-Z-translation}
\begin{split}
    F_a(b\oplus\beta)
    &=F_a(b),\quad\forall\beta\in\mathsf{H}_{a}.
\end{split}
\end{equation}
Concretely, $\beta$ must be the identity on $q$, may be arbitrary outside $\{q\}\cup B$, and may contain transverse Paulis on an even number of sites in $B$; in particular, every $Z$ string supported on $B$ belongs to $\mathsf H_a$.  Any gauge-transformed generator that preserves the model in Eq.~\eqref{eq:weight-one-Z-Lindbladian} must obey the same translation symmetry.

\begin{proposition}[Mixed transverse--dephasing sector]\label{prop:mixed-one-Z-sector}
    Let $a=x\oplus Z_B$ be as in Eq.~\eqref{eq:mixed-one-Z-decomposition}, and define
    \begin{equation}
        q_a(b)=
        \frac{\lambda_{\pt(P_{a\oplus b})}}
        {\lambda_{\pt(P_b)}}.
    \end{equation}
    Suppose $\beta\in\mathsf H_a$ exchanges the endpoint patterns of the $a$-labeled transfer,
    \begin{equation}\label{eq:mixed-one-Z-pattern-exchange}
        \pt(P_{b\oplus\beta})=\pt(P_{a\oplus b}),
        \qquad
        \pt(P_{a\oplus b\oplus\beta})=\pt(P_b).
    \end{equation}
    Wherever $F_a(b)\neq0$, one then has $q_a(b)\in\{+1,-1\}$.
    And consequently, $|F_a(b)|$ is individually learnable for every input admitting such a pattern-exchanging translation.  If the two endpoint patterns already coincide, then $F_a(b)$ itself is individually learnable.
\end{proposition}

\begin{proof}
    Let $F_{a}'$ be a transfer function obtained from $F_{a}$ by a valid gauge transformation. Since both $F_a$ and $F'_a$ obey Eq.~\eqref{eq:mixed-one-Z-translation}, the gauge relation $F'_a(b)=q_a(b)F_a(b)$ implies
    \begin{equation}
        q_a(b\oplus\beta)=q_a(b).
    \end{equation}
    On the other hand, Eq.~\eqref{eq:mixed-one-Z-pattern-exchange} reverses the numerator and denominator of the gauge ratio, so
    \begin{equation}
        q_a(b\oplus\beta)=q_a(b)^{-1}.
    \end{equation}
    Hence $q_a(b)^2=1$.  The final statement is the self-loop case $\pt(P_b)=\pt(P_{a\oplus b})$, for which the gauge ratio equals one directly.
\end{proof}

We now give an explicit construction of the pattern-exchanging translations in Proposition~\ref{prop:mixed-one-Z-sector}. Choose a subset $J\subseteq B$, and let $T_J$ be any Pauli string supported exactly on $J$, with an $X$ or $Y$ on each site in $J$. The $Z$ string $\beta_J=Z_{B\backslash J}$ belongs to $\mathsf H_a$. Consider the four inputs
\begin{equation}\label{eq:mixed-one-Z-paired-inputs}
\begin{split}
    b_1&=x^\sharp\oplus T_J,
    \quad b'_1=x^\sharp\oplus T_J\oplus Z_{B\backslash J},\\
    b_2&=Z_q\oplus T_J,
    \quad b'_2=Z_q\oplus T_J\oplus Z_{B\backslash J}.
\end{split}
\end{equation}
In both pairs, $b'_i=b_i\oplus\beta_J$. Identifying a binary pattern with its support set, their endpoint patterns are
\begin{equation}\label{eq:mixed-one-Z-paired-patterns}
\begin{split}
    \pt(P_{b_i})&=\{q\}\cup J,
    \quad\pt(P_{a\oplus b_i})=\{q\}\cup B,\\
    \pt(P_{b'_i})&=\{q\}\cup B,
    \quad\pt(P_{a\oplus b'_i})=\{q\}\cup J,
\end{split}
\end{equation}
for $i=1,2$. Thus the two transfers in each pair connect the same pattern vertices in opposite directions. Translation invariance gives $F_a(b_i)=F_a(b_i\oplus\beta_{J})=F_{a}(b'_{i})$.
For $J\neq B$, each equality therefore produces a two-edge cycle whose product is $F_a(b_i)^2$, up to known Pauli phases. Consequently, $|F_a(b_1)|$ and $|F_a(b_2)|$ are learnable. For example, $J=\varnothing$ gives cycles between the patterns $\{q\}$ and $\{q\}\cup B$, while $J=\{j\}$ gives cycles between $\{q,j\}$ and $\{q\}\cup B$.

The coefficient combinations contained in these functions can also be written explicitly. Let $p_J=(-1)^{|J|}$. Fourier inversion over the six indices in $\mathcal K_a$ uses the following characters:
\begin{equation}\label{eq:mixed-one-Z-character-table}
\begin{array}{c|rrrrrr}
    &0&a&x&x^\sharp&z&\widetilde z\\ \hline
    b_1&1&-p_J&-1&1&p_J&-p_J\\
    b_2&1&-p_J&-1&-1&p_J&p_J
\end{array}
\end{equation}
It follows that
\begin{equation}\label{eq:mixed-one-Z-explicit-functions}
\begin{split}
    F_a(b_1)
    &=\widehat F_a(0)-p_J\widehat F_a(a)-\widehat F_a(x)
      +\widehat F_a(x^\sharp)+p_J\widehat F_a(z)
      -p_J\widehat F_a(\widetilde z),\\
    F_a(b_2)
    &=\widehat F_a(0)-p_J\widehat F_a(a)-\widehat F_a(x)
      -\widehat F_a(x^\sharp)+p_J\widehat F_a(z)
      +p_J\widehat F_a(\widetilde z).
\end{split}
\end{equation}
The four dissipative modes are, explicitly,
\begin{equation}\label{eq:mixed-one-Z-Fourier-gamma-map}
\begin{split}
    \widehat F_a(x)&=\omega(z,x)\gamma_{z,x},
    \quad \widehat F_a(z)=\omega(x,z)\gamma_{x,z},\\
    \widehat F_a(x^\sharp)&=\omega(\widetilde z,x^\sharp)
    \gamma_{\widetilde z,x^\sharp},
    \quad \widehat F_a(\widetilde z)=\omega(x^\sharp,\widetilde z)
    \gamma_{x^\sharp,\widetilde z}.
\end{split}
\end{equation}
The first row corresponds to the dissipative pair $(x,z)$ and the second to $(x^\sharp,\widetilde z)$; within each row, the two Kossakowski coefficients are related by Hermiticity. Equation~\eqref{eq:mixed-one-Z-explicit-functions} therefore gives the precise dissipative combinations probed by the two cycle families.

The two remaining Fourier modes satisfy $\widehat F_a(a)-\widehat F_a(0)=2ic_a$, while their common dissipative part contributes to both modes. The parity of $J$ determines whether the Hamiltonian part survives in Eq.~\eqref{eq:mixed-one-Z-explicit-functions}. For both inputs in Eq.~\eqref{eq:mixed-one-Z-paired-inputs},
\begin{equation}\label{eq:mixed-one-Z-Hamiltonian-parity}
    \chi_{a,b_1}=\chi_{a,b_2}=(-1)^{|J|+1}=-p_J.
\end{equation}
The Hamiltonian part $-ic_a(1-\chi_{a,b_i})$ is therefore $-2ic_a$ when $|J|$ is even and zero when $|J|$ is odd. Finally, take $J=B$. Then $\beta_J=I$, all four endpoints in Eq.~\eqref{eq:mixed-one-Z-paired-patterns} have the same pattern $\{q\}\cup B$, and $F_a(b_1)$ and $F_a(b_2)$ are self-loops rather than two-edge-cycle quantities. They are consequently learnable with their signs. If $\operatorname{wt}(P_a)=1+|B|$ is odd, then $|B|$ is even and these self-loops contain $c_a$ together with the dissipative coefficients. If $\operatorname{wt}(P_a)$ is even, then $|B|$ is odd and the self-loops contain only dissipative combinations; choosing any even-cardinality $J$ instead gives two-edge cycles whose learnable absolute values contain $c_a$. In this way, the two input families and the two parities of $J$ give concrete access to four different combinations of the dissipative modes, with either signed or absolute-value information determined by whether the corresponding transfer is a self-loop or a two-edge cycle.

This example shows that restrictions on Lindblad jump operators can identify explicit learnable combinations even when local and correlated dephasing processes occur together. It also illustrates the increasing complexity of the coefficient-level problem: several dissipative pairs can contribute to the same transfer sector, and the resulting learnable functions mix Hamiltonian and dissipative coefficients. The constructions above provide concrete guarantees for selected combinations, but do not constitute a complete characterization of the learnable information in the combined model. We leave a systematic analysis of the relations among these combinations, the remaining gauge ambiguities, and the additional simplifications available under further assumptions on the jump operators to future work.

\section{Discussion and outlook}
\label{sec:outlook}

In this work, we developed a unified framework for determining which properties of quantum evolution remain learnable in the presence of unknown SPAM errors. Assuming noiseless single-qubit Clifford control, we first identified the admissible gauge transformations and then characterized the information invariant under them. For general quantum channels, this characterization is encoded by cycles of the pattern transfer graph, while Pauli character projection provides explicit protocols for isolating and learning the corresponding cycle quantities. The framework therefore answers two complementary questions: what information is fundamentally accessible, and how that information can be extracted experimentally.

Additional structure in the dynamics leads to substantially sharper conclusions. For Hamiltonian evolution, coherence strongly restricts the gauge freedom: odd-weight coefficients are individually learnable, while the magnitudes and appropriate relative signs of even-weight coefficients remain accessible. For general Lindbladian evolution, we reconstructed the Hamiltonian and dissipative coefficients after a gauge transformation and related their learnability to cycle invariants of the Lindbladian transfer functions. When the error generator is a first-order perturbation of a known ideal evolution, the ideal dynamics further restricts the gauge. Under unital noise and a connected nonzero ideal pattern graph, this yields explicit conditions for the individual learnability of Hamiltonian and dissipative coefficients. The case studies illustrate how these general results specialize to experimentally motivated coherent, relaxation, dephasing, and local Lindbladian noise models.

Several questions remain open. Our results suggest that stronger constraints toward coherent evolution leave less gauge ambiguity and make more information learnable. Although this intuition is supported by the models studied here, a general quantitative description remains to be developed. Developing such a description for both quantum channels and continuous-time dynamics would require identifying which features of coherent structure constrain gauge freedom and quantifying how those constraints change as more general noise is admitted. This would turn the qualitative hierarchy observed here into a systematic account of how physical structure determines learnability.

A more specific problem is to characterize learnability for a general $k$-local Lindbladian model, in which the Pauli expansions of the Hamiltonian and jump operators contain only terms of weight at most $k$. The precise problem is to determine all diagonal pattern gauges that preserve this locality constraint and then identify the combinations of Hamiltonian and dissipative coefficients invariant under those gauges. Our examples with weight-one and Z-type jumps provide initial instances of this analysis. Extending these results to general few-body dissipative dynamics would give a more complete account of how locality constrains SPAM ambiguity.

Another direction is to revisit our assumption of perfect, independently addressable single-qubit Clifford control, which can be relaxed in two distinct ways. First, some devices permit only collective control on groups of qubits, so the same single-qubit gate must be applied throughout a region. Given such a control architecture, one may ask which gauges commute with the available operations and how the resulting learnable information depends on the degree of addressability. Second, the control gates themselves may be imperfect. Treating their errors as unknown, or as first-order perturbations of the intended controls, couples their characterization to that of the target evolution and introduces additional gauge freedom. Understanding these two settings would clarify the spatial addressability and control accuracy required for SPAM-robust calibration of quantum dynamics of interest.

Finally, learnability identifies the information available in principle but does not by itself guarantee an optimal learning protocol. The character-projection protocols developed here directly estimate individual cycle quantities, while our structural Hamiltonian protocol uses sparsity and locality to learn many coefficients in parallel. Developing more efficient protocols for structured Hamiltonians, extending the comparatively limited protocol results for Lindbladian learning, and integrating the resulting estimators into closed-loop calibration for practical quantum gates are promising steps toward scalable characterization.

\acknowledgements

We thank Zhi-Yuan Wei for helpful discussions. Z.X.~acknowledges support from QuICS Lanczos Graduate Fellowship. S.C. acknowledges funding provided by the Institute for Quantum Information and Matter, an NSF Physics Frontiers Center (NSF Grant PHY-2317110). G.L. acknowledges funding support from the Turing AI Institute of Nanjing (Grant No.~TR-IIIS 002), the National Natural Science Foundation of China (Grant No.~12575023), the Quantum Science and Technology-National Science and Technology Major Project (Grants No.~2021ZD0300804 and No.~2021ZD0300702), and the CCF-QuantumCtek Superconducting Quantum Computing Special Cooperation Program (Grant No.~CCF-QC2025005). The original ideas and primary proofs were developed by the authors. Generative artificial intelligence tools were subsequently used to assist with secondary proofs, proof simplification, clarification of mathematical exposition, language editing, manuscript organization, and figure preparation. Specifically, the statements and proofs of Propositions 2 and 4 were generated using these tools. And the complexity for structural learning in Theorem 4 was improved by AI tools. For the remaining results, the authors first established the proofs and then used AI assistance to simplify the arguments or improve their presentation. The authors checked all AI-assisted content and take full responsibility for the manuscript.

\nocite{forceTitles}
\bibliographystyle{apsrev4-2}

\bibliography{ref}
\clearpage

\appendix
\section{Hamiltonian learning}\label{app:Hamiltonian-learning}

This appendix section develops technical tools and detailed proofs for the Hamiltonian learning section.

\subsection{Hamiltonian reshaping with fast Pauli control}\label{app:Hamiltonian-reshaping}

To estimate a coefficient $c_a$, we first isolate its term in $H=\sum_b c_bP_b$. As in the main text, the identity contribution is omitted. For an $n$-qubit Pauli operator, define the Pauli centralizer of $P_a$ by
\begin{equation}
    \operatorname{Cent}(P_a)=\{R\in\mathsf P_n:[R,P_a]=0\}.
\end{equation}
Averaging over these operators preserves $P_a$ and removes every other nonidentity Pauli term,
\begin{equation}\label{eq:Hamiltonian_reshaping}
    \frac{1}{|\operatorname{Cent}(P_a)|}\sum_{R\in\operatorname{Cent}(P_a)}RHR^\dagger=c_aP_a.
\end{equation}
Indeed, $P_a$ commutes with every member of its centralizer. For any other nonidentity $P_b$, there is a member $R_0$ that anticommutes with $P_b$. Pairing $R$ with $R_0R$, with irrelevant Pauli phases omitted, shows that their contributions to the average of $P_b$ have opposite signs.

The average can be implemented by fast Pauli control, following the Hamiltonian-reshaping approach of Ref.~\cite{Huang_2023}. Divide an evolution time $\tau$ into $N$ intervals of length $\Delta=\tau/N$. In each interval, independently sample $R_q$ uniformly from $\operatorname{Cent}(P_a)$ and implement
\begin{equation}
    R_q e^{-iH\Delta}R_q^\dagger=e^{-i(R_qHR_q^\dagger)\Delta}.
\end{equation}
Every $R_q$ is a product of single-qubit Pauli gates. For sufficiently short intervals, averaging the resulting experiments approximates evolution under $c_aP_a$. More precisely, let $\Phi_\Delta$ be the channel averaged over one interval. Expanding to first order gives $\Phi_\Delta=\mathcal I+\Delta\mathcal L_{c_aP_a}+O(\Delta^2\|H\|^2)$. Independence between intervals makes the averaged channel $\Phi_\Delta^N$, with
\begin{equation}\label{eq:appendix-reshaping-error}
    \bigl\|\Phi_\Delta^N-e^{\tau\mathcal L_{c_aP_a}}\bigr\|_\diamond
    =O\!\left(\frac{\tau^2\|H\|^2}{N}\right).
\end{equation}
Here $\|H\|$ is the operator norm. The estimate follows by bounding the second-order error in each interval and summing over the $N$ channel compositions. Thus the reshaping error can be kept within a fixed signal-accuracy budget by increasing the number of Pauli pulses with $\tau$. The total Hamiltonian evolution time remains $\tau$. The evolution-time scaling in this paper assumes ideal fast control and does not include pulse count or pulse duration in that cost.

Uniform sampling from $\operatorname{Cent}(P_a)$ requires no enumeration of the group. First suppose $P_a$ is $Z$-type with support $A$. Write $R=X^xZ^z$, ignoring its overall phase, with $x,z\in\mathbb F_2^n$. Commutation with $P_a$ imposes the single parity condition
\begin{equation}
    \sum_{i\in A}x_i=0\pmod 2.
\end{equation}
Sample all bits of $z$ and all bits of $x$ outside $A$ independently and uniformly. Within $A$, sample all but one bit of $x$ independently and choose the final bit to satisfy the parity condition. This gives the uniform distribution on the centralizer. For a general $P_a$, choose a local Clifford $V$ such that $VP_aV^\dagger$ is $Z$-type, sample a commuting Pauli $R$ by this rule, and use $V^\dagger RV$. The resulting operator is still a product of single-qubit Paulis and can be applied directly.

\subsection{Heisenberg scaling from cosine measurements}\label{app:cosine-only-estimation}

The signal $C(t)=\cos(2c_at)$ determines the magnitude of $c_a$. We give a direct procedure that estimates this magnitude with Heisenberg scaling using only estimates of $C(t)$. The idea is to maintain an interval containing the nonnegative frequency and choose each evolution time so that cosine is monotone and has a slope bounded away from zero throughout that interval. Inverting cosine then reduces the interval width by a constant factor. Inversion on a known monotone branch is also used in iterative quantum amplitude estimation~\cite{Grinko2021IterativeAmplitudeEstimation}; here we give a self-contained construction for continuously selectable evolution times.

Write $\nu=2|c_a|$, so that $C(t)=\cos(\nu t)$, and assume a known bound $0\leq\nu\leq\Omega$. Fix $\epsilon>0$ and $0<\delta<1/2$. We assume that, at any chosen time $t>0$, a fresh estimate satisfies
\begin{equation}\label{eq:appendix-cosine-accuracy}
    \Pr\!\left[|\widehat C(t)-C(t)|>\eta\right]\leq p,
    \qquad \eta=10^{-3},
\end{equation}
using $O(\log(1/p))$ samples, each costing evolution time $O(t)$. This guarantee is required conditional on the preceding observations when $t$ is chosen adaptively. For independent outcomes in $[-1,1]$ with mean $C(t)$, the sample mean of $N$ outcomes has error probability at most $2\exp(-N\eta^2/2)$, which gives the stated sample bound. For the signals extracted by exponential fitting in our protocol, the assumption requires the SPAM amplitude to be bounded away from zero and the fit to remain uniformly well conditioned. Systematic errors, including reshaping errors, must fit within the same fixed additive-error budget. The numerical constants below are chosen for simplicity rather than optimized.

Under this measurement assumption, we will construct an estimate satisfying
\begin{equation}\label{eq:appendix-cosine-guarantee}
    \Pr[|\widehat\nu-\nu|\leq\epsilon]\geq1-\delta,
    \qquad
    T_{\mathrm{tot}}=O\!\left(\frac{1}{\epsilon}\log\frac{1}{\delta}\right).
\end{equation}
We first prove correctness whenever all requested estimates obey the error bound in Eq.~\eqref{eq:appendix-cosine-accuracy}, and then assign their failure probabilities and bound the total cost.

\emph{Step 1: Locate the frequency, including the possibility that it is close to zero.}
If $\Omega\leq\epsilon$, return $\widehat\nu=0$. Otherwise, set
\begin{equation}
    t_j=2^jt_0,\qquad t_0=\frac{\pi}{4\Omega},
    \qquad J=\min\{j:t_j\geq2/\epsilon\}.
\end{equation}
Estimate $C(t_j)$ in increasing order, stopping at the first estimate with $\widehat C(t_j)\leq1/2$. Denote this time by $t_*$. Until this stop, the phase remains on the first decreasing branch of cosine. Indeed, initially $\nu t_0\leq\pi/4$. At a time without a stop, $C(t_j)>0.499$, so on this branch $\nu t_j<\arccos(0.499)$. The phase at the next time is therefore less than $2\arccos(0.499)<2.2<\pi$, proving the claim inductively. At a stopping time, $C(t_*)\leq0.501$, and consequently
\begin{equation}\label{eq:appendix-cosine-first-crossing}
    1<\arccos(0.501)\leq\nu t_*<2.2.
\end{equation}
If no stop occurs through $t_J$, then $\nu t_J<\arccos(0.499)<1.1$, implying $\nu<\epsilon$. Return $\widehat\nu=0$ in this case. Thus no positive lower bound on $\nu$ is needed.

When a stop occurs, obtain a fresh cosine estimate at $d=t_*/2$. Equation~\eqref{eq:appendix-cosine-first-crossing} places $\nu d$ in $[0.5,1.1]$. On this interval cosine is decreasing, with slope magnitude greater than $0.47$. Clip $\widehat C(d)$ to $[\cos(1.1),\cos(0.5)]$, meaning replace any value outside this interval by the nearest endpoint, and call the result $\overline C(d)$. Since $C(d)$ belongs to this interval, clipping cannot increase its error. The mean value theorem gives
\begin{equation}
    \widetilde\nu=\frac{\arccos\overline C(d)}{d},
    \qquad
    |\widetilde\nu-\nu|\leq\frac{\eta}{0.47d}
    <\frac{0.005}{t_*}.
\end{equation}
Hence the interval
\begin{equation}\label{eq:appendix-cosine-initial-interval}
    I_0=[L_0,U_0]
    =\left[\widetilde\nu-\frac{0.005}{t_*},
            \widetilde\nu+\frac{0.005}{t_*}\right]
\end{equation}
contains $\nu$. Its width is $w_0=0.01/t_*$. Moreover, clipping ensures $\widetilde\nu\geq1/t_*$, so $L_0\geq0.995/t_*>99w_0$. This gives a narrow positive interval from which to refine the estimate.

\emph{Step 2: Halve the interval with one cosine estimate.}
Suppose $I=[L,U]$ contains $\nu$, has width $w=U-L$, and satisfies $L\geq99w$. Write $m=(L+U)/2$. Choose an evolution time $s$ such that
\begin{equation}\label{eq:appendix-cosine-refinement-time}
    \frac{1}{4w}\leq s\leq\frac{1}{2w},
    \qquad ms=\frac{\pi}{2}+q\pi,
    \qquad q\in\mathbb Z_{\geq0}.
\end{equation}
Such a time exists: the times $(\pi/2+q\pi)/m$ are spaced by $\pi/m<1/(4w)$, the length of the allowed time interval, and the first such time lies below its lower endpoint. For every candidate frequency $x\in I$,
\begin{equation}
    |xs-ms|\leq\frac{ws}{2}\leq\frac14.
\end{equation}
Therefore $x\mapsto\cos(xs)$ is monotone on $I$, and its derivative has magnitude at least $s\cos(1/4)>0.96s$ throughout $I$. The previous frequency interval thus specifies a unique inverse branch even when the accumulated phase is large.

Estimate $C(s)$, clip the estimate to the range of $x\mapsto\cos(xs)$ on $I$, and invert this branch to obtain $z\in I$. Again clipping cannot increase the error, and the mean value theorem yields
\begin{equation}\label{eq:appendix-cosine-refinement-error}
    |z-\nu|\leq\frac{\eta}{0.96s}
    \leq\frac{4\eta}{0.96}w<\frac w4.
\end{equation}
The interval $I\cap[z-w/4,z+w/4]$ therefore contains $\nu$ and has width at most $w/2$. Enlarge it within $I$, if necessary, to an interval of width exactly $w/2$, and use that as the next interval. Its lower endpoint cannot decrease, so the condition $L\geq99w$ is preserved. Repeating this step gives widths $w_k=2^{-k}w_0$. Stop after
\begin{equation}
    K=\max\!\left\{0,\left\lceil\log_2\frac{w_0}{2\epsilon}\right\rceil\right\}
\end{equation}
steps and return the midpoint of the final interval. Its width is at most $2\epsilon$, so $|\widehat\nu-\nu|\leq\epsilon$.

\emph{Step 3: Bound the failure probability and total evolution time.}
To avoid a logarithmic overhead from assigning equal failure probabilities to all scales, we demand higher confidence for the inexpensive early measurements. A related nonuniform allocation is used in modified iterative quantum amplitude estimation~\cite{Fukuzawa2023ModifiedAmplitudeEstimation}. Assign the search estimate at $t_j$ failure probability
\begin{equation}
    p_j=\frac{\delta}{6}2^{j-J},\qquad j=0,\ldots,J,
\end{equation}
and the additional estimate at $d$ failure probability $\delta/3$. If refinement is needed, assign the estimate at step $k$ failure probability
\begin{equation}
    r_k=\frac{\delta}{6}2^{k-(K-1)},
    \qquad k=0,\ldots,K-1.
\end{equation}
The search probabilities sum to less than $\delta/3$. Conditional on any search history that reaches refinement, $K$ is fixed by $t_*$, and the refinement probabilities also sum to less than $\delta/3$. Applying the conditional measurement guarantee and a union bound, all required estimates are accurate with probability at least $1-\delta$. Steps 1 and 2 then prove the accuracy claim in Eq.~\eqref{eq:appendix-cosine-guarantee}.

For the search cost, $t_j=2^{j-J}t_J$ and $2/\epsilon\leq t_J<4/\epsilon$, giving
\begin{equation}
\begin{split}
    T_{\mathrm{search}}
    & =O\!\left(\sum_{j=0}^J t_j
       \left[\log\frac{6}{\delta}+(J-j)\log2\right]\right)\\
    & =O\!\left(t_J\sum_{r=0}^{\infty}2^{-r}
       \left[\log\frac{6}{\delta}+r\log2\right]\right)
      =O\!\left(\frac{1}{\epsilon}\log\frac{1}{\delta}\right).
\end{split}
\end{equation}
The additional estimate at $d\leq t_J/2$ costs $O(\epsilon^{-1}\log(1/\delta))$ as well. For refinement, let $s_k$ be the time chosen at width $w_k$. If $K\geq1$, the last width before a measurement obeys $w_{K-1}>2\epsilon$, and
\begin{equation}
    s_k\leq\frac{1}{2w_k}
    =\frac{2^{k-(K-1)}}{2w_{K-1}}
    <\frac{2^{k-(K-1)}}{4\epsilon}.
\end{equation}
The same convergent geometric sum therefore bounds the refinement cost:
\begin{equation}
    T_{\mathrm{refine}}
    =O\!\left(\sum_{k=0}^{K-1}s_k\log\frac1{r_k}\right)
    =O\!\left(\frac{1}{\epsilon}\log\frac1\delta\right).
\end{equation}
There is no refinement cost when $K=0$. Combining the three costs proves Eq.~\eqref{eq:appendix-cosine-guarantee}. At fixed success probability, the total evolution time is $O(1/\epsilon)$.

Returning $\widehat\nu/2$ estimates $|c_a|$ to error $\epsilon/2$ with the same scaling. Every measured signal is invariant under $c_a\mapsto-c_a$; the construction neither measures nor reconstructs a sine signal. The same procedure applies to the frequencies $2|c_a+c_{a'}|$ and $2|c_a-c_{a'}|$ in the relative-sign construction of Sec.~\ref{subsec:Hamiltonian-protocol}, provided the resulting cosine estimates satisfy the measurement assumption above.

\subsection{Support recovery for sparse local Hamiltonians}\label{app:structural-learning}

We prove Theorem~\ref{thm:sparse-Hamiltonian-support-learning} for an $n$-qubit Hamiltonian $H=\sum_{a\in\mathcal A_H}c_aP_a$ with at most $M\geq1$ nonzero Pauli terms. Each term acts on at most $k$ qubits, and its coefficient satisfies $c_{\min}\leq|c_a|\leq1$, where $c_{\min}>0$ is known. We assume ideal single-qubit control and retain the dependence on $c_{\min}$ and the SPAM amplitude explicitly. The proof derives the measured signal, recovers the squared coefficients from shared measurements, and bounds the resources needed to identify all nonzero terms.

\subsubsection{The collective signal.}
Choose a local Pauli frame $\beta=(\beta_1,\ldots,\beta_n)\in\{X,Y,Z\}^n$. For $u\in\mathbb F_2^n$, define $P_\beta(u)=\bigotimes_i\beta_i^{u_i}$ and the commuting Pauli group $\mathcal K_\beta=\{P_\beta(u):u\in\mathbb F_2^n\}$. Pauli averaging over this group gives
\begin{equation}
    H_\beta=\frac{1}{|\mathcal K_\beta|}\sum_{R\in\mathcal K_\beta}RHR^\dagger
    =\sum_u c_{\beta,u}P_\beta(u),
\end{equation}
where $c_{\beta,u}$ is the coefficient of $P_\beta(u)$ in $H$, or zero if that term is absent. Any other Pauli term anticommutes with a member of $\mathcal K_\beta$ and has zero average. Sampling a uniform control Pauli requires only independent uniform bits $u_i$, so the fast-control construction of Appendix~\ref{app:Hamiltonian-reshaping} implements this average. A local Clifford transformation takes each $\beta_i$ to $Z_i$, giving the effective Hamiltonian $\widetilde H_\beta=\sum_u c_{\beta,u}Z^u$, where $Z^u=\bigotimes_i Z_i^{u_i}$.

For a binary pattern $b\in\mathbb F_2^n$, let $X^b=\bigotimes_i X_i^{b_i}$. Character projection onto $X^b$ before and after evolution selects its coefficient in the evolved operator,
\begin{equation}
    f_\beta(b;t)=2^{-n}\operatorname{Tr}\!\left[X^b e^{-it\widetilde H_\beta}X^b e^{it\widetilde H_\beta}\right].
\end{equation}
Only terms with $u\cdot b=1$ anticommute with $X^b$, where the binary dot product is evaluated modulo two. Since all terms in $\widetilde H_\beta$ commute with one another,
\begin{equation}
    e^{-it\widetilde H_\beta}X^be^{it\widetilde H_\beta}
    =X^b\exp\!\left(2it\sum_{u:u\cdot b=1}c_{\beta,u}Z^u\right).
\end{equation}
Multiplication by $X^b$ leaves a diagonal operator inside the trace. Define $h_{\beta,b}(x)=\sum_{u:u\cdot b=1}c_{\beta,u}(-1)^{u\cdot x}$, evaluating the trace in this basis therefore yields
\begin{equation}\label{eq:structural_collective_signal}
\begin{split}
    f_{\beta}(b;t)&=2^{-n}\Tr\left[\exp\!\left(2it\sum_{u:u\cdot b=1}c_{\beta,u}Z^u\right)\right]\\
    &=2^{-n}\Tr\left[\exp\!\left(\sum_{x}2it\ketbra{x}{x}\sum_{u:u\cdot b=1}c_{\beta,u}\bra{x}Z^u\ket{x}\right)\right]\\
    &=2^{-n}\Tr\left[\exp\!\left(\sum_{x}2ith_{\beta,b}(x)\ketbra{x}{x}\right)\right]\\
    &=2^{-n}\sum_{x}\exp\bigl(2ith_{\beta,b}(x)\bigr)\\
    &=2^{-n}\sum_{x}\cos\bigl(2th_{\beta,b}(x)\bigr).
\end{split}
\end{equation}
Here the second line is obtained by multiplying two identity operators before and after $Z^{u}$ and using the condition $\bra{x}Z^{u}\ket{y}=\delta_{x,y}(-1)^{u\cdot x}$ for any $u\neq 0$. The fourth line is because exponentiating a diagonal matrix equals exponentiating the diagonal entries of it.  
The imaginary terms cancel because $h_{\beta,b}(x\oplus b)=-h_{\beta,b}(x)$. 

Expanding Eq.~\eqref{eq:structural_collective_signal} at short $t$ limit, one obtains
\begin{equation}
\begin{split}
    f_{\beta}(b;t)&=2^{-n}\sum_{x}\left(1-\frac{1}{2}\cdot 4t^2h^2_{\beta,b}(x)+O\bigl(t^4h_{\beta,b}^{4}(x)\bigr)\right)\\
    &=1-2t^2\left(2^{-n}\sum_{x}h^2_{\beta,b}(x)\right)+2^{-n}\sum_{x}O\bigl(t^4h_{\beta,b}^{4}(x)\bigr).
\end{split}
\end{equation}
We first evaluate the quadratic terms
\begin{equation}
\begin{split}
    2^{-n}\sum_{x}h^2_{\beta,b}(x)&=2^{-n}\sum_{x}\sum_{u:u\cdot b=1}\sum_{v:v\cdot b=1}c_{\beta,u}c_{\beta,v}(-1)^{(u\oplus v)\cdot x}\\
    &=2^{-n}\sum_{u:u\cdot b=1}\sum_{v:v\cdot b=1}c_{\beta,u}c_{\beta,v}\delta_{u,v}2^{n}\\
    &=\sum_{u:u\cdot b=1}c_{\beta,u}^2.
\end{split}
\end{equation}
This is the sum of the square magnitude of all nonzero Pauli terms, which would be the useful collective information for later post-processing. 

The leftover fourth order term in Eq.~\eqref{eq:structural_collective_signal} is the error that remains to be bounded. Denote $I_{b}=\{u:u\cdot b=1,c_{\beta,u}\neq 0\}$ to be the index set of nonzero Pauli operators anti-commuting with $X^{b}$, 
\begin{equation}
\begin{split}
    2^{-n}\sum_{x}O\bigl(t^4h_{\beta,b}^{4}(x)\bigr)&=O\left(t^42^{-n}\sum_{x}\sum_{u,v,w,y\in I_{b}}c_{\beta, u}c_{\beta, v}c_{\beta, w}c_{\beta, y}(-1)^{(u\oplus v\oplus w\oplus y)\cdot x}\right)\\
    &=O\left(t^4\sum_{u,v,w,y\in I_{b}}\delta_{u\oplus v\oplus w\oplus y,0}\right)\\
    &=O(M^3t^4).
\end{split}
\end{equation}
The second line utilizes the bound $|c_{u}|\leq 1$ for arbitrary $u$. The last line follows from the sparsity assumption that $|I_{b}|\leq M$.  

Locality gives a second bound on the fourth moment through hypercontractivity. For a real function $g:\mathbb F_2^n\to\mathbb R$, define its normalized $L^p$ norm by
\begin{equation}
    \|g\|_{L^p}=\left(2^{-n}\sum_{x\in\mathbb F_2^n}|g(x)|^p\right)^{1/p},
    \qquad 1\leq p<\infty.
\end{equation}
We say that $g$ has degree at most $k$ if its Walsh transformation $\widehat g(v)=2^{-n}\sum_xg(x)(-1)^{v\cdot x}$ is nonzero only for strings $v$ with $\operatorname{wt}(v)\leq k$. For function $h_{\beta,b}(x)=\sum_{u:u\cdot b=1}c_{\beta,u}(-1)^{u\cdot x}$, its Walsh transformation is 
\begin{equation}
\begin{split}
    \widehat{h}_{\beta, b}(v)&=2^{-n}\sum_{x}(-1)^{v\cdot x}h_{\beta, b}(x)\\
    &=2^{-n}\sum_{x}\sum_{u:u\cdot b=1}(-1)^{(v\oplus u)\cdot x}c_{\beta, u}\\
    &=c_{\beta, v}1_{v\cdot b=1}.
\end{split}
\end{equation}
By the locality assumption, $c_{\beta, v}$ is nonzero only if $\mathrm{wt}(v)\leq k$. Therefore indeed $h_{\beta, b}$ has degree at most $k$.


The Bonami--Beckner hypercontractive inequality implies that every such degree-$k$ function satisfies~\cite{Bonami1970,Beckner1975}
\begin{equation}\label{eq:structural-hypercontractivity}
    \|g\|_{L^4}\leq3^{k/2}\|g\|_{L^2},
    \qquad\text{equivalently}\qquad
    \mathbb E_x[g(x)^4]\leq9^k\bigl(\mathbb E_x[g(x)^2]\bigr)^2,
\end{equation}
where $\mathbb E_x$ denotes the uniform average over $\mathbb F_2^n$. Writing $S_{\beta,b}=\sum_{u:u\cdot b=1}c_{\beta,u}^2$, the quadratic-moment identity above gives
\begin{equation}
    \mathbb E_x[h_{\beta,b}(x)^4]
       \leq9^kS_{\beta,b}^2\leq9^kM^2.
\end{equation}
The sparsity bound can also be expressed in terms of $S_{\beta,b}$. Cauchy--Schwarz gives $|h_{\beta,b}(x)|^2\leq |I_b|S_{\beta,b}\leq MS_{\beta,b}$ pointwise, and hence $\mathbb E_x[h_{\beta,b}(x)^4]\leq MS_{\beta,b}^2$. Combining the two bounds, define
\begin{equation}\label{eq:structural-fourth-moment-bound}
    \kappa=\min\{M,9^k\},\qquad
    \mathbb E_x[h_{\beta,b}(x)^4]
       \leq\kappa S_{\beta,b}^2\leq\kappa M^2.
\end{equation}
This estimate holds uniformly over all frames $\beta$ and patterns $b$. The Taylor bound $|\cos z-1+z^2/2|\leq |z|^4/24$ therefore yields the explicit fourth-order remainder
\begin{equation}\label{eq:structural-fourth-order-remainder}
    \left|f_\beta(b;t)-1+2t^2S_{\beta,b}\right|
       \leq\frac23\kappa M^2t^4.
\end{equation}
Thus the short-time change in Eq.~\eqref{eq:structural_linear_measurement} obeys
\begin{equation}\label{eq:structural-curvature-bias}
    y_\beta(b;t)=S_{\beta,b}+O(\kappa M^2t^2),
    \qquad
    |y_\beta(b;t)-S_{\beta,b}|\leq\frac13\kappa M^2t^2.
\end{equation}
The remainder is $O(M^3t^4)$ when $M\leq9^k$, while the locality bound improves it to $O(9^kM^2t^4)$ when $M>9^k$. We next show how to estimate $S_{\beta,b}$ from experimental data with controlled error.

\subsubsection{Estimating the short-time signal in the presence of SPAM.}
The short-time signal $f_\beta(b;t)$ is close to one, so repeated projected cycles can amplify its departure from one while keeping the measured signal bounded away from zero. We use a reference at depth zero and a second experiment at a suitably chosen depth $m$ to exploit this sensitivity.

Fix a frame $\beta$ and pattern $b$, and abbreviate $f=f_\beta(b;t)$ and $S=S_{\beta,b}$. Let $\rho_b$ and $M_b$ denote the actual prepared state and measured observable, with SPAM amplitude $A_b=\dbraket{M_b}{X^b}\dbraket{X^b}{\rho_b}$ satisfying $|A_b|\geq A_{\mathrm{SPAM}}>0$. Character projection onto $X^b$ before and after each evolution of duration $t$ selects a one-dimensional transfer with weight $f$. Repeating this projected cycle $m$ times therefore gives
\begin{equation}\label{eq:structural-depth-signal}
    F_m=A_bf^m,\qquad F_0=A_b.
\end{equation}
The reference uses the same preparation, measurement, and character projection without evolution. Projection between successive evolutions is essential for the power $f^m$ in Eq.~\eqref{eq:structural-depth-signal}. Both experiments have character-weighted outcomes bounded by one in magnitude, and each depth-$m$ sample uses total Hamiltonian evolution time $mt$.

We choose the single-cycle time $t$ to control the short-time bias and the depth $m$ to amplify the resulting small change in $f$. For a desired  accuracy $0<\eta\leq1$, Eq.~\eqref{eq:structural-curvature-bias} allows us to set
\begin{equation}\label{eq:structural_precision_choice}
    t^2=\frac{\eta}{4\kappa M^2},
\end{equation}
which makes $|y_\beta(b;t)-S|\leq\eta/12$.

For the the depth choice, the first principle is to guarantee that the decay $f^{m}$ remains constantly large. For $f\in[1/2,1]$, Applying the Bernoulli's inequality
\begin{equation}\label{eq:structural-depth-contrast}
\begin{split}
    f^m&\geq1-m(1-f)\\
    &=1-m2^{-n}\sum_{x}\Bigl(1-\cos\bigl(2th_{\beta, b}(x)\bigr)\Bigr)\\
    &\geq 1-2mt^2\mathbb{E}_{x}[h^2_{\beta, b}(x)]\\
    &=1-2mt^2S\\
    &\geq 1-2mt^2M.
\end{split}
\end{equation}
The first line is by the Bernoulli's inequality. The second line is by the definition of $f$ in Eq.~\eqref{eq:structural_collective_signal}. The third line is because of the inequality $1-\cos z\leq z^2/2$. And the last line is because $c_{\beta,u}^2\leq 1$ and there are less than $M$ terms. A uniform choice is to require $m(1-f)\leq1/2$, which can be ensured by $m\leq1/(4Mt^2)$.


We take the largest integer satisfying this sufficient condition:
\begin{equation}\label{eq:structural-depth-bounds}
    m=\left\lfloor\frac{1}{4Mt^2}\right\rfloor
      =\left\lfloor\frac{\kappa M}{\eta}\right\rfloor.
\end{equation}
By the inequality $a/2\leq\lfloor a\rfloor\leq a$ for $a\geq1$, the bounds on $mt^2$ follow
\begin{equation}
    \frac{1}{8M}\leq mt^2\leq\frac{1}{4M}.
\end{equation}
The total evolution time per nonreference sample are therefore 
\begin{equation}\label{eq:structural-amplified-time}
    mt=\Theta\!\left(\sqrt{\frac\kappa\eta}\right).
\end{equation}

It remains to recover the single-cycle signal from the measured ratio. Write $r=F_m/F_0=f^m\in[1/2,1]$ and $\widehat r=\widehat F_m/\widehat F_0$. If the empirical means satisfy $\widehat F_m=F_m+e_m$ and $\widehat F_0=F_0+e_0$, with $|e_m|,|e_0|\leq\alpha\leq A_{\mathrm{SPAM}}/2$, then the denominator is nonzero and
\begin{equation}\label{eq:structural_ratio_error}
    |\widehat r-r|
    =\left|\frac{e_m-re_0}{A_b+e_0}\right|
    \leq\frac{4\alpha}{A_{\mathrm{SPAM}}}.
\end{equation}
Choose $\xi=\eta/(32M)$ and $\alpha=A_{\mathrm{SPAM}}\xi/4$. Hoeffding's inequality, applied to both empirical means, ensures this accuracy event, and hence $|\widehat r-r|\leq\xi$, with probability at least $1-\delta_0$ using
\begin{equation}\label{eq:structural_f_sample_complexity}
    N_{\mathrm{shot}}=O\!\left(
       \frac{1}{A_{\mathrm{SPAM}}^2\xi^2}\log\frac1{\delta_0}\right)
\end{equation}
samples in total across the two depths. Since $\xi\leq1/32$, the empirical ratio on this event lies in $[1/4,5/4]$ and is therefore positive. We can directly take its positive real $m$th root and estimate the curvature as
\begin{equation}\label{eq:structural_ratio_estimator}
    \widehat f=\left(\frac{\widehat F_m}{\widehat F_0}\right)^{1/m},
    \qquad \widehat y_\beta(b;t)=\frac{1-\widehat f}{2t^2}.
\end{equation}
A zero empirical reference or a nonpositive ratio is counted as an estimation failure; neither occurs on the accuracy event above. On $[1/4,5/4]$, the derivative of $r^{1/m}$ is at most $4/m$. The mean value theorem consequently gives
\begin{equation}
    |\widehat f-f|\leq\frac{4\xi}{m},\qquad
    |\widehat y_\beta(b;t)-y_\beta(b;t)|
       \leq\frac{2\xi}{mt^2}\leq16M\xi=\frac\eta2.
\end{equation}
Adding the short-time bias yields
\begin{equation}\label{eq:structural-amplified-curvature-error}
    |\widehat y_\beta(b;t)-S_{\beta,b}|
       \leq\frac\eta2+\frac{\kappa M^2t^2}{3}
       =\frac{7\eta}{12}<\eta.
\end{equation}
The $m$th root reduces the statistical error in $f$ by a factor of order $m$, while the bias remains governed by the single-cycle time $t$. Substituting $\xi=\eta/(32M)$ into Eq.~\eqref{eq:structural_f_sample_complexity} gives
\begin{equation}\label{eq:structural_shots_per_measurement}
    N_{\mathrm{shot}}=O\!\left(
       \frac{M^2}{A_{\mathrm{SPAM}}^2\eta^2}\log\frac1{\delta_0}\right).
\end{equation}
The evolution-time cost per depth-$m$ sample is given by Eq.~\eqref{eq:structural-amplified-time}.
These bounds assume exact reshaping. Finite-control errors can be accommodated by allocating a fixed fraction of the ratio-error budget $\xi$ to the bias of the entire depth-$m$ signal and reducing the statistical error accordingly. In particular, a uniform error of order $\xi/m$ in each projected cycle weight suffices, since each weight has magnitude at most one. The bound in Appendix~\ref{app:Hamiltonian-reshaping} permits this with sufficiently frequent ideal Pauli control; the Hamiltonian evolution time of a depth-$m$ sample remains $mt$.

\subsubsection{Recovering all candidate coefficients by compressed sensing.}
The estimates $\widehat y_\beta(b;t)$ provide noisy linear measurements of a sparse vector of squared coefficients. Compressed sensing reconstructs such a vector jointly from few measurements. We now formulate this reconstruction and state the two standard results needed for its guarantee~\cite{Rauhut2010,CandesRombergTao2006}.

For a fixed frame $\beta$, define the candidate set and squared coefficients by
\begin{equation}
    \mathcal U_k=\{u\in\mathbb F_2^n:1\leq\operatorname{wt}(u)\leq k\},
    \qquad d_{\beta,u}=c_{\beta,u}^2.
\end{equation}
The size of the candidate set is $|\mathcal U_k|=N_k=\sum_{j=1}^k\binom nj$. By the sparsity assumption, at most $M$ entries of $d_\beta=(d_{\beta,u})_{u\in\mathcal U_k}$ are nonzero. Let $\chi_u(b)=(-1)^{u\cdot b}$ denote a Walsh character, with $\chi_0(b)=1$. The identity $\mathbf1_{u\cdot b=1}=(1-\chi_u(b))/2$ gives
\begin{equation}\label{eq:structural-augmented-Walsh-model}
    2S_{\beta,b}=\sum_{u\in\mathcal U_k}d_{\beta,u}
       -\sum_{u\in\mathcal U_k}d_{\beta,u}\chi_u(b)
       =\sum_{u\in\{0\}\cup\mathcal U_k}\theta_{\beta,u}\chi_u(b),
\end{equation}
where
\begin{equation}\label{eq:structural-augmented-coefficients}
    \theta_{\beta,0}=\sum_{u\in\mathcal U_k}d_{\beta,u},
    \qquad \theta_{\beta,u}=-d_{\beta,u}\quad(u\in\mathcal U_k).
\end{equation}
The coordinate indexed by $0$ is an auxiliary constant term in the measured function, not the coefficient of the identity in the Hamiltonian. It allows us to treat the unknown sum of squared coefficients without a separate experiment. The augmented vector $\theta_\beta\in\mathbb R^D$, with $D=N_k+1$, has at most $s=\min\{M+1,D\}$ nonzero entries; such a vector is called $s$-sparse.

Choose $L$ independent uniform patterns $b_1,\ldots,b_L\in\mathbb F_2^n$. Define the sensing matrix $\Phi_\beta\in\mathbb R^{L\times D}$ and data vector $z_\beta\in\mathbb R^L$ by
\begin{equation}\label{eq:structural-sensing-matrix}
    (\Phi_\beta)_{\ell,u}=\frac{\chi_u(b_\ell)}{\sqrt L},
    \qquad (z_\beta)_\ell=\frac{2\widehat y_\beta(b_\ell;t)}{\sqrt L},
\end{equation}
with $u\in\{0\}\cup\mathcal U_k$. Equation~\eqref{eq:structural-augmented-Walsh-model} then reads 
\begin{equation}
    z_\beta=\Phi_\beta\theta_\beta+e_\beta.
\end{equation}
Here $e_\beta\in\mathbb R^L$ is the normalized estimation-error vector, with entries
\begin{equation}
    (e_\beta)_\ell=\frac{2}{\sqrt L}
       \bigl[\widehat y_\beta(b_\ell;t)-S_{\beta,b_\ell}\bigr],
    \qquad \ell=1,\ldots,L.
\end{equation}
It includes both the statistical error in estimating the curvature and the short-time bias relative to $S_{\beta,b_\ell}$.
For a real vector $v$, write $\|v\|_1=\sum_j|v_j|$, $\|v\|_2=(\sum_j|v_j|^2)^{1/2}$, and $\|v\|_\infty=\max_j|v_j|$. If each curvature estimate has error at most $\eta$, the normalized error vector obeys
\begin{equation}\label{eq:structural-CS-noise}
    \|e_\beta\|_2^2
    =\frac4L\sum_{\ell=1}^L
       |\widehat y_\beta(b_\ell;t)-S_{\beta,b_\ell}|^2
    \leq4\eta^2.
\end{equation}
The remaining task is to recover the sparse vector $\theta_\beta$ from the noisy linear equations $z_\beta=\Phi_\beta\theta_\beta+e_\beta$. Two ingredients are needed: the sampled equations must distinguish different sparse vectors, and the reconstruction procedure must tolerate the error $\|e_\beta\|_2\leq2\eta$ established above.

First consider distinguishability. Let $\theta_{1},\theta_{2}$ be two $s$-sparse vectors and denote $v=\theta_1-\theta_{2}$ to be the difference between the two $s$-sparse vectors, which has at most $2s$ nonzero entries. If $\Phi_\beta v$ were much smaller than $v$, those vectors would give nearly identical measurements and would be difficult to distinguish in the presence of error. We therefore ask that $\Phi_\beta$ preserve the squared Euclidean norm of every such difference to within a relative error of $1/3$. This requirement is called the \emph{restricted isometry property} of order $2s$, and takes the form
\begin{equation}\label{eq:structural-CS-RIP}
    \frac23\|v\|_2^2\leq\|\Phi_\beta v\|_2^2
        \leq\frac43\|v\|_2^2
    \quad\text{for every $v\in\mathbb R^D$ with at most $2s$ nonzero entries}.
\end{equation}
In particular, its lower bound rules out two distinct $s$-sparse vectors producing exactly the same data. The numerical tolerance $1/3$ is a fixed choice sufficient for the stability result used below.

Our random sensing matrix satisfies this requirement with sufficiently many sampled patterns because its columns are evaluations of Walsh characters. Under uniform sampling of $b$, distinct characters are orthogonal and each character has magnitude one. For indices $u,v\in\{0\}\cup\mathcal U_k$, let $\mathbf1_{u=v}$ equal one when $u=v$ and zero otherwise. Then
\begin{equation}
    \mathbb E_b[\chi_u(b)\chi_v(b)]=\mathbf1_{u=v},
    \qquad |\chi_u(b)|=1.
\end{equation}
A collection of functions with these properties is a \emph{bounded orthonormal system}. Orthogonality means that the normalized sensing matrix preserves squared norms on average over its random patterns. The sampling theorem for bounded orthonormal systems strengthens this average statement to the simultaneous guarantee in Eq.~\eqref{eq:structural-CS-RIP} for all sparse vectors~\cite[Theorem~8.4]{Rauhut2010}.

To apply that theorem, let $B$ be the total number of local Pauli frames, chosen in the next subsubsection. We require the sensing matrix in each frame to satisfy Eq.~\eqref{eq:structural-CS-RIP} with failure probability at most $\delta/(3B)$. Specializing the bounded-orthonormal-system sampling theorem of Ref.~\cite[Theorem~8.4]{Rauhut2010} to our Walsh characters gives the following sufficient number of independent uniform patterns per frame, for a sufficiently large universal constant $C_{\mathrm{samp}}>0$:
\begin{equation}\label{eq:structural-pattern-count}
    L=\left\lceil C_{\mathrm{samp}}s
          \log^4\!\left(\frac{6BD}{\delta}\right)\right\rceil.
\end{equation}
This nonoptimized choice grows linearly with the sparsity $s\leq M+1$, up to logarithmic factors, and only logarithmically with the number $D=N_k+1$ of candidate coefficients. The guarantee holds regardless of the locations or signs of the nonzero entries.

We next specify how to reconstruct the coefficients. For a candidate vector $x\in\mathbb R^D$, the residual $\Phi_\beta x-z_\beta$ measures its disagreement with the observations. We accept candidates with residual norm at most $2\eta$, the bound already obtained for the true vector. Among these candidates, we minimize $\|x\|_1$, the sum of the absolute values of its entries. This choice favors sparse solutions while giving a convex optimization problem that can be solved without enumerating possible supports. The procedure is known as \emph{basis pursuit denoising}. Defining $\widehat\theta_\beta$ to be any minimizer, it reads
\begin{equation}\label{eq:structural-CS-estimator}
    \widehat\theta_\beta\in\mathop{\rm arg\,min}_{x\in\mathbb R^D}\|x\|_1
    \quad\text{subject to}\quad
    \|\Phi_\beta x-z_\beta\|_2\leq2\eta.
\end{equation}
The true $\theta_\beta$ satisfies the constraint, so the optimization is feasible.

The connection between this optimization and sparse recovery is supplied by the stability theorem for basis pursuit denoising~\cite{CandesRombergTao2006,Rauhut2010}. It states that if the true vector is $s$-sparse and the sensing matrix satisfies Eq.~\eqref{eq:structural-CS-RIP}, then the error of every minimizer is bounded by a constant times the measurement-error tolerance. Let $C_{\mathrm{rec}}\geq1$ be a universal constant for our fixed relative tolerance $1/3$. Applying the theorem with measurement-error tolerance $2\eta$ gives
\begin{equation}\label{eq:structural-CS-stability}
    \|\widehat\theta_\beta-\theta_\beta\|_2
       \leq2C_{\mathrm{rec}}\eta.
\end{equation}
The reconstructed vector need not be exactly sparse for this bound to hold. The bound also allows arbitrary errors within the prescribed norm, including the short-time bias in $e_\beta$, and is independent of the magnitudes of the true coefficients. Thus the auxiliary constant coordinate requires no separate error analysis.

To obtain the squared coefficients, we now apply this vector error bound to each nonzero candidate index. For each $u\in\mathcal U_k$, Eq.~\eqref{eq:structural-augmented-coefficients} gives $d_{\beta,u}=-\theta_{\beta,u}$. We therefore use $-\widehat\theta_{\beta,u}$ as its estimate, replacing a negative result by zero because the true squared coefficient is nonnegative. This replacement cannot increase the error. Choose the curvature accuracy $\eta$ small enough that the vector error is below the support-detection threshold, and denote the resulting coefficient estimates by $\widehat d_{\beta,u}$:
\begin{equation}\label{eq:structural-CS-precision}
    \eta=\frac{c_{\min}^2}{16C_{\mathrm{rec}}},
    \qquad
    \widehat d_{\beta,u}=\max\{0,-\widehat\theta_{\beta,u}\},
    \quad u\in\mathcal U_k.
\end{equation}
Each coordinate error is at most the Euclidean norm of the full error vector, so Eq.~\eqref{eq:structural-CS-stability} now yields, simultaneously for every candidate,
\begin{equation}\label{eq:structural-CS-coefficient-error}
    \max_{u\in\mathcal U_k}|\widehat d_{\beta,u}-c_{\beta,u}^2|
       \leq2C_{\mathrm{rec}}\eta
       =\frac{c_{\min}^2}{8}<\frac{c_{\min}^2}{4}.
\end{equation}
The spare error margin permits solving the convex problem to finite numerical accuracy. Thresholding at $c_{\min}^2/2$ distinguishes absent terms, whose squared coefficients vanish, from present terms, whose squared coefficients are at least $c_{\min}^2$. If a term is retained by several frames, any one of its estimates may be used. Thus $L$ grows linearly with the sparsity up to logarithmic factors; $c_{\min}$ determines the required accuracy of each measured curvature rather than an additional inverse-power factor in the number of patterns.

\subsubsection{Frame coverage and total resources.}
A nonzero term of weight $w\leq k$ survives in a uniform frame when each of its $w$ nonidentity Pauli operators matches the corresponding local axis. This has probability $3^{-w}\geq3^{-k}$. Its probability of being missed by $B$ independent frames is at most $\exp(-B3^{-k})$. Taking
\begin{equation}\label{eq:structural-frame-count}
    B=\left\lceil3^k\log\frac{3M}{\delta}\right\rceil
\end{equation}
ensures that all terms are covered with probability at least $1-\delta/3$. For each frame, choose the number of patterns in Eq.~\eqref{eq:structural-pattern-count}; a union bound makes Eq.~\eqref{eq:structural-CS-RIP} hold in every frame except with probability at most $\delta/3$. Allocate failure probability $\delta_0=\delta/(3BL)$ to each curvature estimate. The measurement bounds then hold for all frames and patterns except with probability at most $\delta/3$. On the intersection of these three events, the reconstruction above identifies every nonzero term and estimates its squared coefficient to the accuracy required in Theorem~\ref{thm:sparse-Hamiltonian-support-learning}. The joint failure probability is at most $\delta$.

Since $C_{\mathrm{rec}}$ is a universal constant, Eq.~\eqref{eq:structural-CS-precision} gives $\eta=\Theta(c_{\min}^2)$. Equations~\eqref{eq:structural_precision_choice},~\eqref{eq:structural_shots_per_measurement}, and~\eqref{eq:structural-amplified-time} then give
\begin{equation}
    t=\Theta\!\left(\frac{c_{\min}}{M\sqrt\kappa}\right),
    \qquad m=\Theta\!\left(\frac{\kappa M}{c_{\min}^2}\right),
    \qquad mt=\Theta\!\left(\frac{\sqrt\kappa}{c_{\min}}\right),
\end{equation}
and $\widetilde O(M^2/(A_{\mathrm{SPAM}}^2c_{\min}^4))$ samples per pattern, including the reference measurements. Here $\widetilde O$ suppresses only logarithmic factors in $M$, $N_k$, $3^k$, $1/c_{\min}$, and $1/\delta$. Since $s\leq M+1\leq2M$, Eq.~\eqref{eq:structural-pattern-count} requires $L=\widetilde O(M)$ patterns per frame. Multiplying by the number of frames and samples per pattern yields
\begin{equation}\label{eq:structural-support-total}
    N_{\mathrm{supp}}=\widetilde O\!\left(\frac{3^kM^3}{A_{\mathrm{SPAM}}^2c_{\min}^{4}}\right).
\end{equation}
Each depth-$m$ sample uses Hamiltonian evolution time $mt$, while the reference uses none. Including this longer duration gives
\begin{equation}\label{eq:structural-support-time}
    T_{\mathrm{supp}}=\widetilde O\!\left(\frac{3^k\sqrt\kappa\,M^3}{A_{\mathrm{SPAM}}^2c_{\min}^{5}}\right).
\end{equation}
These unoptimized bounds establish the polynomial guarantee in Theorem~\ref{thm:sparse-Hamiltonian-support-learning}. Constructing the sensing matrices takes $O(kBLN_k)$ arithmetic and parity operations. The optimization in Eq.~\eqref{eq:structural-CS-estimator} is a second-order cone program, so it can be solved to the required numerical accuracy in polynomial time. For fixed $k$, $D=N_k+1=O(n^k)$, and the overall classical processing remains polynomial in $n$, $M$, $1/c_{\min}$, and the number of measurement records. All candidate coefficients in a frame use the same data, although classical reconstruction still involves the full set of candidates.

\section{Lindbladian reconstruction}\label{app:Lindbladian-reconstruction}

This appendix section gives full details on the Fourier transformation relationship of Lindbladian coefficients and their reconstruction.

\subsection{Pauli transfer functions and Fourier inversion}

Let $a,b\in\{0,1\}^{2n}$ label the Hermitian Pauli operators $P_a$, with addition $\oplus$ taken bitwise modulo two and $P_0=I$. Their products satisfy
\begin{equation}
    P_aP_b=\omega(a,b)P_{a\oplus b},
    \qquad \omega(a,b)\in\{\pm1,\pm i\},
\end{equation}
and their commutation signs are defined by $P_aP_b=\chi_{a,b}P_bP_a$, where $\chi_{a,b}\in\{\pm1\}$. Note that $\omega(a,b)=\omega(b,a)\chi_{a,b}=\omega(b,a)^{*}$ with $*$ denoting the complex conjugate, which is because $\chi_{a,b}=-1$ only when $\omega(a,b)$ is a imaginary number.

Consider a linear map of the form
\begin{equation}\label{eq:app-Lindbladian-definition}
    \mathcal L(\rho)=-i\sum_{a\neq0}c_a[P_a,\rho]
    +\sum_{i,j\neq0}\gamma_{i,j}
    \left(P_i\rho P_j-\frac12\{P_jP_i,\rho\}\right),
\end{equation}
where $c_0=0$ and the coefficients $c_a$ and $\gamma_{i,j}$ may be arbitrary complex numbers for the algebraic identities below. A physical Lindblad generator additionally has real $c_a$, Hermitian $\gamma$, and $\gamma\succeq0$. In the orthonormal Pauli basis $P_a/\sqrt{2^n}$, its transfer-matrix entries are $\mathcal L_{i,j}=2^{-n}\tr[P_i\mathcal L(P_j)]$. Define the Pauli transfer functions by
\begin{equation}\label{eq:app-Lindbladian-transfer-function}
    F_a(b):=\omega(a,b)^*\mathcal L_{a\oplus b,b}.
\end{equation}
The Pauli characters obey the orthogonality relation
\begin{equation}
    \frac{1}{4^n}\sum_{b\in\{0,1\}^{2n}}\chi_{a,b}\chi_{k,b}=\delta_{a,k}.
\end{equation}
Accordingly, define the Fourier coefficients and inverse transform of $F_a$ by
\begin{equation}
\begin{split}
    \widehat F_a(k)&=\frac{1}{4^n}\sum_{b\in\{0,1\}^{2n}}\chi_{k,b}F_a(b),\\
    F_a(b)&=\sum_{k\in\{0,1\}^{2n}}\chi_{k,b}\widehat F_a(k).
\end{split}
\end{equation}
All unspecified label sums below run over $\{0,1\}^{2n}$. Applying $\mathcal L$ to a normalized Pauli basis vector gives
\begin{equation}\label{eq:Lindbladian-PTM}
\begin{split}
    \mathcal{L}\dket{P_b}&=-i\sum_{a}c_a(1-\chi_{a,b})\omega(a,b)\dket{P_{a\oplus b}}\\
    &\quad+\sum_{i,j\neq 0}\gamma_{i,j}\left(\chi_{i,j}\chi_{b,j}-\frac{1}{2}\chi_{b,j}\chi_{b,i}-\frac{1}{2}\right)\omega(j,i)\omega(j\oplus i, b)\dket{P_{j\oplus i\oplus b}}
\end{split}
\end{equation}
The second line is obtained because
\begin{equation}
\begin{split}
    P_iP_bP_j-\frac{1}{2}\{P_jP_i,P_b\}&=P_iP_bP_j-\frac{1}{2}P_jP_iP_b-\frac{1}{2}P_bP_jP_i\\
    &=P_jP_iP_b\left(\chi_{i,j}\chi_{b,j}-\frac{1}{2}-\frac{1}{2}\chi_{b,j}\chi_{b,i}\right)\\
    &=\omega(j,i)\omega(j\oplus i, b)P_{j\oplus i\oplus b}\left(\chi_{i,j}\chi_{b,j}-\frac{1}{2}-\frac{1}{2}\chi_{b,j}\chi_{b,i}\right).
\end{split}
\end{equation}
Set $a=i\oplus j$ in the dissipative sum. Since both jump labels are nonzero, the remaining index satisfies $j\notin\{0,a\}$. Factoring out $\omega(a,b)$ and using the Pauli phase and commutation identities gives
\begin{equation}\label{eq:Lindbladian-PT-function}
\begin{split}
    F_{a}(b)&=-ic_a(1-\chi_{a,b})+\sum_{j\neq\{0,a\}}\omega(j,j\oplus a)\gamma_{j\oplus a,j}\left(\chi_{j\oplus a,j}\chi_{b,j}-\frac{1}{2}-\frac{1}{2}\chi_{b,j}\chi_{b,j\oplus a}\right)\\
    &=-ic_a(1-\chi_{a,b})+\sum_{j\neq\{0,a\}}\omega(j\oplus a, j)\chi_{a,j}\gamma_{j\oplus a,j}\left(\chi_{a,j}\chi_{b,j}-\frac{1}{2}-\frac{1}{2}\chi_{b,j}\chi_{b,j}\chi_{b,a}\right)\\
    &=-ic_a(1-\chi_{a,b})+\sum_{j\notin\{0,a\}}\omega(j\oplus a,j)\gamma_{j\oplus a,j}\Bigl(\chi_{b,j}-\frac{\chi_{a,j}}{2}(1+\chi_{a,b})\Bigr).
\end{split}
\end{equation}
Here in the second line we use the identity that $\chi_{a\oplus b,c}=\chi_{a,c}\chi_{b,c}$ and $\chi_{a,b}^2=1$ for arbitrary $a,b$.

For brevity, define $z_{a,j}:=\omega(j\oplus a,j)\gamma_{j\oplus a,j}$ for $j\notin\{0,a\}$. Then
\begin{equation}\label{eq:Lindbladian-function-character-expansion}
        F_a(b)
        =-ic_a(1-\chi_{a,b})
        +\sum_{j\notin\{0,a\}}z_{a,j}
        \left(\chi_{b,j}-\frac{\chi_{a,j}}{2}(1+\chi_{a,b})\right),
    \end{equation}
Character orthogonality turns each character into a single Fourier mode. Writing $\delta_{u,v}$ for the Kronecker delta, we obtain
\begin{equation}\label{eq:Lindbladian-Fourier-general-form}
\begin{split}
        \widehat F_a(k)&=4^{-n}\sum_{b}\left(-ic_a(1-\chi_{a,b})\chi_{k,b}
        +\sum_{j\notin\{0,a\}}z_{a,j}
        \left(\chi_{b,j}-\frac{\chi_{a,j}}{2}(1+\chi_{a,b})\right)\chi_{b,k}\right)\\
        &=4^{-n}\sum_{b}\left(-ic_a(\chi_{k,b}-\chi_{a\oplus k,b})
        +\sum_{j\notin\{0,a\}}z_{a,j}
        \left(\chi_{b,j\oplus k}-\frac{\chi_{a,j}}{2}(\chi_{k,b}+\chi_{a\oplus k,b})\right)\right)\\
        &=-ic_a(\delta_{k,0}-\delta_{k,a})+\sum_{j\notin\{0,a\}}z_{a,j}
        \left(\delta_{k,j}
        -\frac{\chi_{a,j}}{2}(\delta_{k,0}+\delta_{k,a})\right).
\end{split}
\end{equation}
For $a\neq0$, the modes $0$ and $a$ are distinct, so this expression reads
\begin{equation}\label{eq:Lindbladian-Fourier-coefficients}
    \begin{split}
        \widehat{F}_{a}(k)=\left\{\begin{aligned}
            -ic_a-\frac{1}{2}\sum_{j\notin\{0,a\}}\chi_{a,j}&\omega(j\oplus a,j)\gamma_{j\oplus a,j},\quad k=0\\
            ic_a-\frac{1}{2}\sum_{j\notin\{0,a\}}\chi_{a,j}&\omega(j\oplus a,j)\gamma_{j\oplus a,j},\quad k=a\\
            \omega(k\oplus a,k)&\gamma_{k\oplus a,k},\qquad\qquad\quad~ k\notin\{0,a\}
        \end{aligned}\right..
    \end{split}
    \end{equation}
Their difference is $-2ic_a$, which proves the Hamiltonian reconstruction formula. For $i,j\neq0$ with $i\neq j$, take $a=i\oplus j$ and $k=j$. Then $k\notin\{0,a\}$, and the third case gives the dissipative reconstruction formula. To include $i=j$, observe that $a=0$ gives $F_0(b)=\sum_{j\neq0}\gamma_{j,j}(\chi_{b,j}-1)$ and hence
\begin{equation}\label{eq:Lindbladian-diagonal-Fourier-coefficients}
    \begin{split}
        \widehat F_0(k)&=\left\{\begin{aligned}
            -\sum_{j\neq0}\gamma_{j,j},\quad k=0,\\
            \gamma_{k,k},\qquad k\neq 0.
        \end{aligned}\right.
    \end{split}
    \end{equation}
Since $\omega(j,j)=1$, this proves the same inverse formula for the diagonal entries and completes the proof of Lemma~\ref{lemma:Lindbladian-Fourier-reconstruction}.

\subsection{Gauge transformation and the canonical coefficients}

Let $T$ be a valid real diagonal gauge, with $T\dket{P_b}=\lambda_{\pt(P_b)}\dket{P_b}$, and set $\mathcal L'=T\mathcal LT^{-1}$. Its transfer functions are
\begin{equation}\label{eq:app-Lindbladian-transfer-gauge}
    F'_a(b)=\frac{\lambda_{\pt(P_{a\oplus b})}}{\lambda_{\pt(P_b)}}F_a(b).
\end{equation}
The Fourier reconstruction formulas give the following candidate coefficients for the transformed map:
\begin{equation}\label{eq:app-gauge-transformed-coefficients-through-F}
    \begin{split}
        c'_a&=\frac{i}{2}\left(\widehat F'_a(0)-\widehat F'_a(a)\right)
        =\frac{i}{2\cdot4^n}\sum_b(1-\chi_{a,b})
        \frac{\lambda_{\pt(P_{a\oplus b})}}{\lambda_{\pt(P_b)}}F_a(b),\\
        \gamma'_{i,j}&=\omega(i,j)^*\widehat F'_{i\oplus j}(j)
        =\frac{\omega(i,j)^*}{4^n}\sum_b\chi_{b,j}
        \frac{\lambda_{\pt(P_{i\oplus j\oplus b})}}{\lambda_{\pt(P_b)}}F_{i\oplus j}(b),
    \end{split}
\end{equation}
where $a\neq0$, $i,j\neq0$, and the sums run over $b\in\{0,1\}^{2n}$. We next verify directly that these expressions are real and Hermitian in the required senses. Substituting Eq.~\eqref{eq:Lindbladian-PT-function} into Eq.~\eqref{eq:app-gauge-transformed-coefficients-through-F} gives, for $a\neq0$,
\begin{equation}\label{eq:gauge-transformed-Hamiltonian-coefficients}
    \begin{split}
        c'_a
        &=\frac{1}{4^n}\sum_b\frac{\lambda_{\pt(P_{a\oplus b})}}{\lambda_{\pt(P_{b})}}(1-\chi_{a,b})\left(c_a+\frac{i}{2}\sum_{k\notin\{0,a\}}\omega(k\oplus a,k)\gamma_{k\oplus a,k}\chi_{b,k}\right).
    \end{split}
    \end{equation}
For $i,j\neq0$, set $a=i\oplus j$. The dissipative coefficients become
\begin{equation}\label{eq:gauge-transformed-dissipative-coefficients}
    \begin{split}
        \gamma'_{i,j}
        &=\frac{\omega(i,j)^*}{4^n}\sum_b
        \chi_{b,j}
        \frac{\lambda_{\pt(P_{a\oplus b})}}{\lambda_{\pt(P_{b})}}
        \Biggl[-ic_a(1-\chi_{a,b})\\
        &\qquad+\sum_{k\notin\{0,a\}}
        \omega(k\oplus a,k)\gamma_{k\oplus a,k}
        \left(\chi_{b,k}-\frac{\chi_{a,k}}{2}(1+\chi_{a,b})\right)
        \Biggr].
    \end{split}
    \end{equation}

The first term in Eq.~\eqref{eq:gauge-transformed-Hamiltonian-coefficients} is real because $c_a$ and the gauge factors are real. To examine the remaining term, pair the summand indexed by $k$ with the one indexed by $k\oplus a$. Using $\chi_{b,k\oplus a}=\chi_{b,k}\chi_{a,b}$ and the fact that $1-\chi_{a,b}$ vanishes unless $\chi_{a,b}=-1$, this part can be written as
\begin{equation}\label{eq:gauge-transformed-Hamiltonian-imaginary-part}
    \frac{i}{4^{n+1}}\sum_b
    \frac{\lambda_{\pt(P_{a\oplus b})}}{\lambda_{\pt(P_b)}}(1-\chi_{a,b})
    \sum_{k\notin\{0,a\}}\omega(k\oplus a,k)
    \left(\gamma_{k\oplus a,k}-\chi_{a,k}\gamma_{k,k\oplus a}\right)\chi_{b,k}.
\end{equation}
Note that when $\chi_{k\oplus a,k}=1$, $\omega(k\oplus a,k)\in\{\pm 1\}$ is real and $\gamma_{k\oplus a,k}-\chi_{k\oplus a,k}\gamma_{k,k\oplus a}=2i\operatorname{Im}(\gamma_{k\oplus a,k})$ is purely imaginary. When $\chi_{k\oplus a,k}=-1$, $\omega(k\oplus a,k)\in\{\pm i\}$ is imaginary and $\gamma_{k\oplus a,k}-\chi_{k\oplus a,k}\gamma_{k,k\oplus a}=2\operatorname{Re}(\gamma_{k\oplus a,k})$ is real. Their product always remain imaginary. Note that there is a $i/4^{n+1}$ factor in the beginning of the equation. Therefore in either case, multiplication by $i$ makes each contribution real. Hence $c'_a\in\mathbb R$.

To verify Hermiticity, conjugate Eq.~\eqref{eq:gauge-transformed-dissipative-coefficients}. The gauge ratio and characters are real, while $c_a\in\mathbb R$ and $\gamma_{k,k\oplus a}=\gamma_{k\oplus a,k}^*$. Using $\chi_{b,j}=\chi_{b,i}\chi_{b,a}$, move the factor $\chi_{b,a}$ into the bracket. Since $\chi_{a,b}(1+\chi_{a,b})=1+\chi_{a,b}$, and then relabel $k'=k\oplus a$, we obtain
\begin{equation}\label{eq:gauge-transformed-gamma-hermiticity}
\begin{split}
    (\gamma_{i,j}')^{*}&=\frac{\omega(i,j)}{4^n}\sum_b\chi_{b,j}\frac{\lambda_{\pt(a\oplus b)}}{\lambda_{\pt(b)}}\Biggl[ic_a(1-\chi_{a,b})+\sum_{k\notin\{0,a\}}\omega(k,k\oplus a)\gamma_{k,k\oplus a}\left(\chi_{b,k}-\frac{\chi_{a,k}}{2}(1+\chi_{a,b})\right)\Biggr]\\
    &=\frac{\omega(i,j)}{4^n}\sum_b\chi_{b,j}\frac{\lambda_{\pt(a\oplus b)}}{\lambda_{\pt(b)}}\Biggl[-ic_a(1-\chi_{a,b})\chi_{a,b}+\sum_{k\notin\{0,a\}}\omega(k,k\oplus a)\gamma_{k,k\oplus a}\left(\chi_{b,k\oplus a}\chi_{b,a}-\frac{\chi_{a,k\oplus a}}{2}(1+\chi_{a,b})\chi_{a,b}\right)\Biggr]\\
    &=\frac{\omega(j,i)^{*}}{4^n}\sum_b\chi_{b,i}\frac{\lambda_{\pt(a\oplus b)}}{\lambda_{\pt(b)}}\Biggl[-ic_a(1-\chi_{a,b})+\sum_{k'\notin\{0,a\}}\omega(k'\oplus a,k')\gamma_{k'\oplus a,k'}\left(\chi_{b,k'}-\frac{\chi_{a,k'}}{2}(1+\chi_{a,b})\right)\Biggr]\\
    &=\gamma'_{j,i}.
\end{split}
\end{equation}
Here in the third line we use $k'=k\oplus a$ in order to obtain the same form as Eq.~\eqref{eq:gauge-transformed-dissipative-coefficients}. And we use the property that $\omega(i,j)=\omega(j,i)^{*}$ and $\chi_{b,j}\chi_{b,i\oplus j}=\chi_{b,i}$. Therefore $\gamma'$ is always a Hermitian matrix. 

These explicit calculations have a general explanation. Let $N=4^n-1$ and let $\operatorname{Herm}(N)$ denote the real vector space of Hermitian $N\times N$ matrices. The real-linear map from $\mathbb R^N\oplus\operatorname{Herm}(N)$ to superoperators defined by Eq.~\eqref{eq:Lindbladian-definition} is injective by Lemma~\ref{lemma:Lindbladian-Fourier-reconstruction}. Its domain has real dimension $N+N^2=N(N+1)$. The space of Hermiticity-preserving, trace-annihilating maps has the same dimension: its Pauli transfer matrix is real and has $N+1$ rows, with the identity-output row fixed to zero. The map is therefore surjective as well, so every such superoperator has a unique representation with real Hamiltonian coefficients and Hermitian $\gamma$.

The gauge-transformed map $\mathcal L'$ is Hermiticity preserving because both $T$ and $\mathcal L$ have real Pauli transfer matrices. It annihilates trace because $T$ is diagonal and $\lambda_0=1$, so the identity-output row remains zero. The general argument therefore guarantees that $\mathcal L'$ has a unique canonical representation. By the Fourier reconstruction lemma, its coefficients are exactly the expressions in Eq.~\eqref{eq:app-gauge-transformed-coefficients-through-F}. Finally, the canonical Lindblad characterization implies that $\mathcal L'$ generates a CPTP semigroup exactly when $\gamma'\succeq0$. This proves Theorem~\ref{thm:Lindbladian-reconstruction}.

\subsection{First-order learnability of Lindbladian Fourier modes}\label{app:first-order-Lindbladian-Fourier-proof}

\noindent\textbf{Proposition~\ref{prop:first-order-individual-Lindbladian-Fourier} (restated).}
\textit{Write $H_0=\sum_a c_a^0P_a$ and let $\mathcal A_0=\{a:c_a^0\neq0\}$ denote its Pauli support. Under the assumptions of Proposition~\ref{prop:first-order-connected-Lindbladian},}
\begin{equation}
    \widehat F_a^{\mathrm{err}}(k)\text{ is individually first-order learnable}
    \quad\Longleftrightarrow\quad
    a\notin\mathcal A_0\quad\text{or}\quad
    \chi_{a,k}=1\quad\text{or}\quad
    \pt(P_k)=\pt(P_{a\oplus k}).
\end{equation}

\begin{proof}
We recall the setting and notation needed for the calculation. The generator is $\mathcal L=\mathcal L_0+\mathcal E$, where $\mathcal L_0(\rho)=-i[H_0,\rho]$, $H_0=\sum_{a\neq0}c_a^0P_a$ is known, and $\mathcal E=O(\varepsilon)$ is a unital error generator. The ideal pattern-transfer graph is assumed connected on all nonzero patterns. Its edges are the nonzero Pauli transfer entries of $\mathcal L_0$, with each Pauli operator replaced by its support pattern. Connectedness ignores edge directions. We write $\mathcal A_0=\{a:c_a^0\neq0\}$. We set $c_0^0=0$, since an identity term in $H_0$ does not affect the generator.

The $n$-qubit Pauli operators $P_a$ are indexed so that $P_aP_b=\omega(a,b)P_{a\oplus b}$, where $\omega(a,b)\in\{\pm1,\pm i\}$ and $\oplus$ denotes addition of the binary Pauli labels. The identity has index $0$, and $\pt(P_b)\in\{0,1\}^n$ records the qubits on which $P_b$ is nonidentity. The character $\chi_{a,b}\in\{\pm1\}$ is defined by $P_aP_b=\chi_{a,b}P_bP_a$ and satisfies $\chi_{a\oplus b,k}=\chi_{a,k}\chi_{b,k}$. We identify a pattern with its set of occupied qubits when writing intersections, differences, and cardinalities.

For $\mathcal K\in\{\mathcal L_0,\mathcal E,\mathcal L\}$, define its transfer functions by
\begin{equation}
    \mathcal K\dket{P_b}=\sum_a\omega(a,b)F_a^{\mathcal K}(b)\dket{P_{a\oplus b}}.
\end{equation}
We abbreviate $F_a^{\mathcal L_0}=F_a^0$ and $F_a^{\mathcal E}=F_a^{\mathrm{err}}$, so $F_a^{\mathcal L}=F_a^0+F_a^{\mathrm{err}}$ and $F_a^{\mathrm{err}}(b)=O(\varepsilon)$. Their Fourier modes are
\begin{equation}
    \widehat F_a^{\mathrm{err}}(k)=\frac{1}{4^n}\sum_b\chi_{k,b}F_a^{\mathrm{err}}(b).
\end{equation}
All sums over Pauli indices run over the $4^n$ labels unless restricted explicitly.

The admissible diagonal gauges act as $T\dket{P_b}=\lambda_{\pt(P_b)}\dket{P_b}$, with $\lambda_0=1$. Within the connected nonidentity component, their first-order form is $\lambda_u=\kappa(1+\varepsilon s_u)+O(\varepsilon^2)$, where $\kappa\neq0$ may have either sign and the real potentials $s_u$ can vary independently modulo a common shift. A mode is first-order learnable when every valid gauge changes it only by $O(\varepsilon^2)$. The claim to be proved is that this holds exactly when $a\notin\mathcal A_0$, $\chi_{a,k}=1$, or $\pt(P_k)=\pt(P_{a\oplus k})$.

The generator transforms as $\mathcal L'=T\mathcal L T^{-1}$, with preparation states and measurement effects transformed by $T$ and its inverse, respectively, so that experimental probabilities are unchanged. A valid gauge must preserve the physicality of these objects as well as the assumed model. The gauge-ratio expansion below applies to transfers between nonidentity Paulis. Transfers with identity input or output vanish by unitality or trace preservation, respectively; we set $s_0=0$ when including these zero terms in sums over all Pauli labels.

    We first derive the gauge variation of an arbitrary Fourier mode from the full transfer function in Eq.~\eqref{eq:first-order-total-Lindbladian-transfer-function}. Within the connected nonidentity component, a first-order gauge has the form $\lambda_u=\kappa(1+\varepsilon s_u)+O(\varepsilon^2)$. Hence, for a transfer from pattern $\pt(P_b)$ to pattern $\pt(P_{a\oplus b})$,
    \begin{equation}\label{eq:first-order-Lindbladian-gauge-ratio}
        \frac{\lambda_{\pt(P_{a\oplus b})}}{\lambda_{\pt(P_b)}}
        =1+\varepsilon\left(s_{\pt(P_{a\oplus b})}-s_{\pt(P_b)}\right)+O(\varepsilon^2).
    \end{equation}
    The ideal Hamiltonian contributes
    \begin{equation}
        F_a^0(b)=-ic_a^0(1-\chi_{a,b}),
    \end{equation}
    and the gauge transformation of the full transfer function is
    \begin{equation}
    \begin{split}
        \bigl(F_a^{\mathcal L}\bigr)'(b)
        &=\frac{\lambda_{\pt(P_{a\oplus b})}}{\lambda_{\pt(P_b)}}
        \left[F_a^0(b)+F_a^{\mathrm{err}}(b)\right]\\
        &=F_a^0(b)+F_a^{\mathrm{err}}(b)
        +\varepsilon\left(s_{\pt(P_{a\oplus b})}-s_{\pt(P_b)}\right)F_a^0(b)
        +O(\varepsilon^2),
    \end{split}
    \end{equation}
    where the product of the $O(\varepsilon)$ gauge correction with $F_a^{\mathrm{err}}(b)=O(\varepsilon)$ has been absorbed into $O(\varepsilon^2)$. Since the ideal part remains $F_a^0$, subtracting Eq.~\eqref{eq:first-order-total-Lindbladian-transfer-function} gives the physical error variation
    \begin{equation}\label{eq:first-order-error-F-gauge-change}
        \Delta F_a^{\mathrm{err}}(b)
        :=\bigl(F_a^{\mathrm{err}}\bigr)'(b)-F_a^{\mathrm{err}}(b)
        =\varepsilon\left(s_{\pt(P_{a\oplus b})}-s_{\pt(P_b)}\right)F_a^0(b)+O(\varepsilon^2).
    \end{equation}

    A first-order error coefficient is learnable precisely when its variation is $O(\varepsilon^2)$. If $a\notin\mathcal A_0$, then $F_a^0=0$ and Eq.~\eqref{eq:first-order-error-F-gauge-change} is $O(\varepsilon^2)$, so the entire Fourier block is first-order invariant.

    Now suppose $a\in\mathcal A_0$. If $\chi_{a,k}=1$, then $\chi_{k,a\oplus b}=\chi_{k,a}\chi_{k,b}=\chi_{k,b}$. It follows that
    \begin{equation}
    \begin{split}
        \Delta\widehat{F}^{\mathrm{err}}_{a}(k)&=\frac{1}{4^n}\sum_{b}\varepsilon\chi_{k,b}(s_{\pt(P_{a\oplus b})}-s_{\pt(P_{b})})F_{a}^{0}(b)+O(\varepsilon^2)\\
        &=\frac{1}{2}\frac{1}{4^n}\sum_{b}\varepsilon\Bigl(\chi_{k,b}(s_{\pt(P_{a\oplus b})}-s_{\pt(P_{b})})F_{a}^{0}(b)+\chi_{k,a\oplus b}(s_{\pt(P_{b})}-s_{\pt(P_{a\oplus b})})F_{a}^{0}(a\oplus b)\Bigr)+O(\varepsilon^2)\\
        &=\frac{1}{2}\frac{1}{4^n}\sum_{b}\varepsilon\chi_{k,b}F_{a}^{0}(b)\Bigl((s_{\pt(P_{a\oplus b})}-s_{\pt(P_{b})})+(s_{\pt(P_{b})}-s_{\pt(P_{a\oplus b})})\Bigr)+O(\varepsilon^2)\\
        &=O(\varepsilon^2).
    \end{split}
    \end{equation}
    In the third line we used $F_a^0(a\oplus b)=F_a^0(b)$, since both sides are zero when $P_a$ commutes with $P_b$ and equal $-2ic_a^0$ when they anticommute. Hence $\widehat F_a^{\mathrm{err}}(k)$ is first-order learnable whenever $\chi_{a,k}=1$.

    It remains to consider $\chi_{a,k}=-1$. To collect the gauge parameters pattern by pattern, for nonzero patterns $u,v\in\{0,1\}^n$ choose any Pauli $P_p$ with $\pt(P_p)=v$ and define
    \begin{equation}\label{eq:fixed-pattern-character-sum}
        \Phi_u(v)=\sum_{b:\,\pt(P_b)=u}\chi_{p,b}
        =3^{|u\setminus v|}(-1)^{|u\cap v|},
    \end{equation}
    which depends only on $u$ and $v$, not on the choice of $P_p$. Using $(1-\chi_{a,b})/2$ to select the Paulis that anticommute with $P_a$ gives
    \begin{equation}\label{eq:fixed-pattern-anticommuting-character-sum}
        \sum_{\substack{b:\,\pt(P_b)=u\\ \chi_{a,b}=-1}}\chi_{k,b}=\frac12\sum_{b:\,\pt(P_b)=u}\chi_{k,b}(1-\chi_{a,b})=\frac12\left[\Phi_u\bigl(\pt(P_k)\bigr)-\Phi_u\bigl(\pt(P_{a\oplus k})\bigr)\right].
    \end{equation}
    Starting from the Fourier definition and using Eq.~\eqref{eq:first-order-error-F-gauge-change}, the case $\chi_{a,k}=-1$ can now be evaluated directly:
    \begin{equation}\label{eq:first-order-Fourier-pattern-gauge-change}
    \begin{split}
        \Delta\widehat F_a^{\mathrm{err}}(k)
        &=\frac{\varepsilon}{4^n}\sum_b\chi_{k,b}
        \left(s_{\pt(P_{a\oplus b})}-s_{\pt(P_b)}\right)F_a^0(b)+O(\varepsilon^2)\\
        &=-\frac{2i\varepsilon c_a^0}{4^n}
        \sum_{b:\,\chi_{a,b}=-1}\chi_{k,b}
        \left(s_{\pt(P_{a\oplus b})}-s_{\pt(P_b)}\right)+O(\varepsilon^2)\\
        &=\frac{4i\varepsilon c_a^0}{4^n}
        \sum_{b:\,\chi_{a,b}=-1}\chi_{k,b}s_{\pt(P_b)}+O(\varepsilon^2)\\
        &=\frac{4i\varepsilon c_a^0}{4^n}\sum_{u\neq0}s_u
        \sum_{\substack{b:\,\pt(P_b)=u\\ \chi_{a,b}=-1}}\chi_{k,b}+O(\varepsilon^2)\\
        &=\frac{2i\varepsilon c_a^0}{4^n}\sum_{u\neq0}s_u
        \left[\Phi_u\bigl(\pt(P_k)\bigr)-\Phi_u\bigl(\pt(P_{a\oplus k})\bigr)\right]+O(\varepsilon^2).
    \end{split}
    \end{equation}
    The third line follows by changing variables $b\mapsto a\oplus b$ in the term containing $s_{\pt(P_{a\oplus b})}$ and using $\chi_{k,a\oplus b}=-\chi_{k,b}$. The last line follows from Eq.~\eqref{eq:fixed-pattern-anticommuting-character-sum}.

    The necessity argument concerns learnability over the full unital error model. To disprove learnability, it suffices to exhibit a gauge-equivalent pair within this model, rather than a nontrivial gauge at every boundary point. Such pairs can be constructed around an interior model with Kossakowski matrix $\gamma=\varepsilon\Gamma$, where $\Gamma\succ0$ satisfies unitality (for example, a positive multiple of the identity), and with full-rank preparation states and strictly positive POVM effects. For any fixed real potential vector $s$, choose $\lambda_u=\exp(\varepsilon t s_u)$ on nonzero patterns and $\lambda_0=1$, where $t$ is a sufficiently small constant independent of $\varepsilon$. The transformed dissipative matrix has the form $\gamma'=\varepsilon[\Gamma+tK(s)+O(\varepsilon)]$ for a finite Hermitian matrix $K(s)$, so positivity is preserved for sufficiently small $|t|$ and then sufficiently small $\varepsilon$. Unitality and trace preservation are preserved exactly by this diagonal gauge, and the transformed error remains $O(\varepsilon)$. The SPAM operators remain physical by continuity, with normalization and POVM completeness preserved because $T$ fixes the identity and preserves trace. Thus each potential direction used in the necessity argument is realized by valid gauges near an interior model. No assertion of the same freedom at every boundary model is needed.

    The right-hand side vanishes if the two endpoint patterns agree. Conversely, if the patterns differ, choose $u$ to be the singleton pattern of a qubit contained in exactly one endpoint. The bracket is then $\pm4$, and the corresponding freely variable $s_u$ changes the Fourier mode. This proves Eq.~\eqref{eq:first-order-individual-Fourier-characterization}.
\end{proof}

\subsection{Learnable combinations of exceptional dissipative coefficients}\label{app:collective-Lindbladian-learnability}

Under the assumptions of Corollary~\ref{cor:first-order-individual-Lindbladian-coefficients}, every coherent coefficient and every imaginary part of the Kossakowski matrix is individually first-order learnable. The only possible ambiguities occur among real coefficients $\operatorname{Re}\gamma_{i,j}$ for which
\begin{equation}
    i\oplus j\in\mathcal A_0,\qquad
    \chi_{i,j}=-1,\qquad
    \pt(P_i)\neq\pt(P_j).
\end{equation}
Here $H_0=\sum_{a\neq0}c_a^0P_a$ is the known ideal Hamiltonian, $\mathcal A_0=\{a:c_a^0\neq0\}$, and $P_iP_j=\omega(i,j)P_{i\oplus j}$ with $P_iP_j=\chi_{i,j}P_jP_i$. The pattern $\pt(P_i)$ records the nonidentity support of $P_i$. We assume a general $O(\varepsilon)$ unital Lindbladian error and an ideal pattern-transfer graph connected on all nonzero patterns, as in the preceding subsection.

Choose one ordering of each exceptional unordered pair and collect $r_{i,j}=\operatorname{Re}\gamma_{i,j}$ into a real vector $r$. Only one ordering is needed because Hermiticity gives $\operatorname{Re}\gamma_{j,i}=\operatorname{Re}\gamma_{i,j}$. The aim is to find all real vectors $y$ for which $y^{\mathsf T}r$ is first-order learnable, even when its individual terms are not.

For a nonzero pattern $u$ and a pattern $v$, recall the character sum
\begin{equation}
    \Phi_u(v)=\sum_{b:\,\pt(P_b)=u}\chi_{p,b}
    =3^{|u\setminus v|}(-1)^{|u\cap v|},
    \qquad \pt(P_p)=v.
\end{equation}
It depends only on the two patterns. At a qubit in $u\setminus v$, all three nonidentity Paulis contribute $+1$; at a qubit in $u\cap v$, one commutes with $P_p$ and two anticommute, giving a total of $-1$. Multiplying these single-qubit sums gives the formula above.

Write the gauge factors as $\lambda_u=\kappa(1+\varepsilon s_u)+O(\varepsilon^2)$ on nonzero patterns. The Fourier variation derived in Appendix~\ref{app:first-order-Lindbladian-Fourier-proof}, together with $\gamma_{i,j}=\omega(i,j)^*\widehat F_{i\oplus j}^{\mathrm{err}}(j)$, gives
\begin{equation}\label{eq:first-order-exceptional-gamma-matrix}
    M_{(i,j),u}
    =\frac{2ic_{i\oplus j}^0\omega(i,j)^*}{4^n}
    \left[\Phi_u\bigl(\pt(P_j)\bigr)-\Phi_u\bigl(\pt(P_i)\bigr)\right],
    \qquad
    \Delta r=\varepsilon Ms+O(\varepsilon^2).
\end{equation}
The rows label the selected exceptional pairs and the columns label nonzero patterns. The matrix is real because the exceptional Paulis anticommute, so $\omega(i,j)=\pm i$. It follows that
\begin{equation}
    \Delta(y^{\mathsf T}r)=\varepsilon\langle M^{\mathsf T}y,s\rangle+O(\varepsilon^2).
\end{equation}
Consequently, $y^{\mathsf T}r$ is first-order learnable exactly when $M^{\mathsf T}y=0$. Necessity uses the admissible variations near an interior model described in the preceding subsection. The nullspace $\ker M^{\mathsf T}$ therefore gives all learnable combinations in this sector, while $\operatorname{im}M$ gives the leading gauge variations of the coefficients themselves.

The rank also has a direct interpretation. A common shift of the potentials leaves every gauge ratio unchanged, so it lies in $\ker M$. Conversely, if $Ms=0$, the exceptional real coefficients do not change at first order, and all other physical coefficients are already first-order invariant. The entire generator therefore has zero first-order gauge variation. On every ideal edge $u\to v$, that variation is $\varepsilon(s_v-s_u)g_e^0$, with $g_e^0\neq0$, so $s_v=s_u$. Connectedness forces $s$ to be constant on all $2^n-1$ nonzero patterns. Hence
\begin{equation}
    \ker M=\operatorname{span}\{\mathbf 1\},
    \qquad \operatorname{rank}M=2^n-2.
\end{equation}

In practice, one lists the exceptional pairs, constructs $M$ from the known ideal coefficients and Pauli patterns, and computes a basis of $\ker M^{\mathsf T}$. Each basis vector supplies a learnable linear combination of the exceptional real coefficients. Known linear relations imposed by unitality should also be taken into account when selecting independent quantities: a nullspace vector representing an identically vanishing combination supplies no additional unknown parameter to estimate.

\end{document}